\documentclass[headinclude=true, headheight=1cm, headsepline, abstract=on, titlepage=false, twoside=true]{scrbook}
\usepackage[T1]{fontenc}
\usepackage[utf8]{inputenc}
\usepackage[english]{babel}
\usepackage{amsmath}
\usepackage{amssymb}
\usepackage{amsthm}
\usepackage{physics}
\usepackage{siunitx}
\usepackage{mathrsfs}
\usepackage{enumerate}
\usepackage{appendix} 
\usepackage{scrlayer-scrpage}
\usepackage{csquotes}
\usepackage[super]{nth}
\usepackage{todonotes}
\usepackage[%
  paper=a4paper,
  left=25mm,
  right=25mm,
  top=25mm,
  bottom=25mm,
  bindingoffset=15mm,
  includeheadfoot,
  foot=30pt,
]{geometry}
\newenvironment{abstract}{}{}
\usepackage{abstract}
\usepackage{graphicx}
\usepackage{tikz}
\usetikzlibrary{shapes.geometric}
\usetikzlibrary{decorations.pathmorphing}
\usepackage{caption}
\usepackage{subcaption}
\usepackage[backend=biber,style=ieee, sorting=none, subentry]{biblatex}
\defbibentryset{Gryb}{Gryb2019,Gryb2019b,Gryb2019c}
\defbibentryset{Dittrich}{Dittrich2006,Dittrich2007}
\defbibentryset{Rovelli}{Rovelli1990,Rovelli1991}
\defbibentryset{Ricci}{DeWitt1957,Moss1988}
\defbibentryset{rsquare}{Narnhofer1974,Kunstatter2009}
\defbibentryset{Hoehn}{Hoehn2019,Hoehn2020}
\defbibentryset{Gotay}{Gotay1984,Gotay1996}
\defbibentryset{Singularity}{Ashtekar2008,Husain2004,Kiefer2010}
\defbibentryset{PIQ}{Batalin1981,Fradkin1997,Henneaux1985}
\defbibentryset{PIQcritics}{Govaerts,Govaerts1991}
\defbibentryset{PIQcosmo}{Dorronsoro2018,Lehners2022}
\defbibentryset{Ashtekar}{Ashtekar2006,Ashtekar2006b,Ashtekar2006c}
\defbibentryset{GHY}{York1971,Gibbons1977,York1986}
\defbibentryset{Ligo}{Abbott2021,Abbott2021b}
\defbibentryset{EHT}{EHT2019,EHT2022}
\defbibentryset{PlanckCMB}{Planck2020,Planck2020b}
\defbibentryset{BC}{Barrett1998,Barrett2000}
\defbibentryset{rps}{Hajicek2000,Hajicek2000b,Malkiewicz2016,Malkiewicz2017}
\defbibentryset{Gielen}{Gielen2016,Gielen2016a}
\defbibentryset{GW}{Baker2017,Sakstein2017}
\defbibentryset{Halliwell}{Halliwell1988,Halliwell1991}
\usepackage{hyperref}
\usepackage[nameinlink]{cleveref}

\crefname{equation}{}{}

\reversemarginpar

\clearpairofpagestyles
\ihead{\headmark}
\ohead{\pagemark}

\addtokomafont{disposition}{\rmfamily}
\addtokomafont{part}{\scshape}
\addtokomafont{chapter}{\scshape}

\newcommand{\R}{\mathbb{R}}
\newcommand{\N}{\mathbb{N}}
\newcommand{\C}{\mathcal{C}}
\newcommand{\Z}{\mathbb{Z}}
\newcommand{\Obs}{\mathcal{O}}
\newcommand{\Ham}{\mathcal{H}}

\DeclareMathOperator{\sgn}{sgn}
\DeclareMathOperator{\arsinh}{arsinh}
\DeclareMathOperator{\arcosh}{arcosh}
\DeclareMathOperator{\artanh}{artanh}
\DeclareMathOperator{\arcoth}{arcoth}
\newcommand{\poisket}[2]{\left\lbrace #1, #2 \right\rbrace}
\newcommand{\J}[2]{J_{#1}\left(\frac{\sqrt{#2}}{\hbar}v\right)}
\newcommand{\K}[2]{K_{#1}\left(\frac{\sqrt{#2}}{\hbar}v\right)}
\newcommand{\I}[2]{I_{#1}\left(\frac{\sqrt{#2}}{\hbar}v\right)}
\newcommand{\dl}{\frac{\dd \lambda}{2\pi \hbar}}
\newcommand{\dk}{\frac{\dd k}{2\pi}}
\newcommand{\dkap}{\frac{\dd \kappa}{2\pi}}
\newcommand{\hkin}{\mathscr{H}_{\text{kin}}}
\newcommand{\hphys}{\mathscr{H}_{\text{phys}}}
\newcommand{\D}{\mathcal{D}}

\DeclareFieldFormat{titlecase}{#1} 
\DeclareFieldFormat{sentencecase}{#1}
\renewbibmacro*{bbx:savehash}{}

\DeclareMathOperator{\dom}{Dom}
\DeclareMathOperator{\ran}{ran}
\theoremstyle{definition}
\newtheorem{definition}{Definition}[chapter]

\theoremstyle{remark}
\newtheorem*{remark}{Remark}

\newtheorem{theorem}{Theorem}

\title
{ \bfseries  \scshape
	{\Huge The Problem of Time in Quantum Cosmology}\\[10mm]
	{ \small Thesis submitted for the award of the degree of doctor of philosophy}\\[10mm]
	{\small \emph{University of Nottingham, School of Mathematics}}\\[20mm]
\includegraphics[scale=0.5]{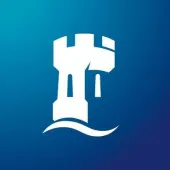}
\vspace{10mm}
}

\author{
	Student: Lucía Menéndez-Pidal de Cristina \\
	Supervisors: Prof. Jorma Louko, Dr. Steffen Gielen
	\\[10mm]	
} 
\addtokomafont{chapter}{\Huge}
\begin{document}
\maketitle
\newpage
\thispagestyle{empty}
\begin{abstract}
This thesis contains an analysis of the problem of time in quantum cosmology and its application to a cosmological minisuperspace model. In the first part, we introduce the problem of time and the theoretical foundations of minisuperspace models. In the second part, we focus on a specific minisuperspace universe, analyse it classically, and quantise it using the canonical quantisation method. The chosen model is a flat FLRW universe with a free massless scalar field and a perfect fluid. We explain how different types of perfect fluid can be accommodated in our model. We extract the Wheeler--DeWitt equation, and calculate its solutions. There are three dynamical variables that may be used as clock parameters, namely a coordinate $t$ conjugated to the perfect fluid mass, the massless scalar field $\varphi$, and $v$, a positive power of the scale factor. We define three quantum theories, each one based on assuming one of the previous dynamical quantities as the clock. This quantisation method is then compared with the Dirac quantisation. We find that, in each quantisation procedure, covariance is broken, leading to inequivalent quantum theories. In the third part, the properties of each theory are analysed. Unitarity of each theory is implemented by adding a boundary condition on the allowed states. The solutions to the boundary conditions are calculated and their properties are listed. Requiring unitarity is what breaks general covariance in the quantum theory. In the fourth part, we study the numerical properties of the wave functions in the three theories, paying special attention to singularity resolution and other divergences from the classical theory. The $t$-clock theory is able to resolve the singularity, the $\varphi$-clock theory presents some non trivial dynamics that can be associated with a resolution of spatial infinity, and the $v$-clock theory does not show significant deviations from the classical theory. In the last part, we expand our analysis in order to include another quantisation method: path integral quantisation, and finally, we conclude.

\end{abstract}

\tableofcontents
\listoffigures

\chapter*{Acknowledgements}
\thispagestyle{empty}
This thesis has not been easy. During this period I have lived through major political events, like the Brexit and the Covid-19 pandemic, and major personal events (if you know, you know). Nonetheless, despite all odds, the day has come, and I am very happy to present this work to the world.

I would like to thank my supervisors, Dr.~Steffen Gielen and Prof.~Jorma Louko, for guiding me throughout the process. Without their patience, good-willingness, and understanding this thesis would not have been possible. I cannot thank them enough for their guidance and encouragement. I have been extremely lucky to be supervised not only by rigorous scientists, but also by extraordinary people.

I am also very grateful to all the PhD students of the Mathematical Physics department, in particular, Simen Bruinsma, Eugenia Colafranceschi,  Tom Laird, Hans Nguyen, Marco Perin, Axel Polaczek and Giulia Ventagli. Thank you for all the coffee breaks, hang-outs, reading groups, climbing sessions, and in summary all the memories we made together. You made the office a safe place and the PhD journey less heavy. I am very happy to be able to call you my friends. Special thanks to Giulia for joining me in my crazy ideas, for example, joining a rugby club with no previous experience. 

This brings me to my next acknowledgments: all the unexpected friends I made along the way. I am especially grateful for my housemates Luisa Aldrete, Jahzeel Aguilera, Gabriela Durán, Blanca Franco, Gabriela Sandoval,  and Lucía Vázquez. Thank you for having the patience to live with me, and for helping me out in all kinds of situations, but more importantly, thank you for all the good moments we spent together! The dinners, cooking sessions, and trips made my stay in Nottingham so much more enjoyable. I want to give special thanks to Jahz and Gaby: I am very grateful for spending the first wave of Covid-19 alongside you. We managed to remain (moderately) sane when the outside world seemed to fall apart. I knew that whatever happened, I could always count on you, which is way more than what you would expect from an average housemate. Finally, I cannot forget my other friends like Annette Zhao, Eduardo Pernault, Teshan Rezel and Jessica Vidales. Thanks for staying close to me, especially when I was hitting rock bottom. 

This thesis would not have been possible without the support from my friends from Spain and Chile (even if some of you are not in Spain anymore!): Patricia Canton, Mateo Galdeano, Ana Gil, María Outeda, Marta Sánchez, Sebastián Chávez, \emph{et al}. It was not easy to leave my home country, but your closeness have shown me that true friendship is one of the constants of life: it remains unchanged even after a change of spacetime coordinates.

In addition to all my friends, I cannot acknowledge enough the support of my mum, dad and brother. Thank you for believing in me even after I lost faith in myself, thank you for your patience, your active listening, and your kind words. I know you did everything to make the distance feel less far and to make the hardships easier to carry. You made me the scientist and doctor I am today.

 I also want to thank Javier Carrón. Thank you for allowing me to grow both personally and academically by your side. I do not have words to express my feelings of gratitude towards you.
 
Last but not least, I am grateful to me. I never gave up, I stood up against everything, and here I am living the life I want to live.

\thispagestyle{empty}

\newpage
\thispagestyle{empty}

\begin{flushright}
\vspace*{5cm}
\emph{To my brother,}\\
\emph{Gabriel.}
\end{flushright}

\part{Foundations of quantum gravity and quantum cosmology}
\chapter{Notation and outline}
This thesis is a comprehensive analysis of the problem of time in a concrete minisuperspace example. Here, we present the principal assumptions and meaning of the mathematical notation used throughout the thesis. The reader can come back here to remind themselves of the general notation. The presentation order is roughly the same order in which the quantities are introduced in the thesis.

We use the sign convention $(-,+,+,+)$ and we assume the speed of light $c$ is 1. The abbreviations used in the thesis are:
\begin{itemize}
\item SR: special relativity.
\item GR: general relativity.
\item QM: quantum mechanics.
\item LQG, LQC: loop quantum gravity, loop quantum cosmology.
\end{itemize}

Next we present a mostly exhaustive list of the main symbols used. We tried our best to not repeat a same symbol for different quantities, but there are only so many letters in the Greek and Latin alphabet.
\begin{itemize}
\item $g_{ab}, \ g^{ab}$, $\sqrt{-g}$: general relativity metric, inverse metric and determinant. We use the units convention for which the metric is dimensionless. 
\item $\mathcal{S}$: classical action.
\item $R$: Ricci scalar.
\item $K$: Extrinsic curvature.
\item $\kappa=8\pi G$ where $G$ is the gravitational constant.
\item $\phi$, $\varphi$: free massless scalar field.
\item $\mathcal{G}_{ijkl}$: superspace metric.
\item $\rho$, $m$, $p$, $w$: perfect isentropic fluid energy density, pressure, mass and parameter in the state equation.
\item $J^a$: perfect isentropic fluid flux.
\item $\alpha^a$, $\beta_a$, $\vartheta$: isentropic perfect fluid action Lagrange multipliers.  
\item $N$, $N^i$: lapse function and shift vector.
\item $\mathcal{M}=\Sigma\times \R$: usual representation of the spacetime manifold.
\item $h_{ij}$, $V_0=\int_\Sigma \dd x^3 \sqrt{h}$: space metric and volume.
\item $\Lambda$, $\lambda$: cosmological constant or perfect fluid mass.
\item $\mathcal{H}$, $\mathcal{C}$: Hamiltonian and Hamiltonian constraint.
\item $\pi_x$: conjugated momentum of the variable $x$.
\item $v\propto a^{\frac{3(1-w)}{1-w}}$ will usually be referred as ``volume'' for short in an abuse of language.
\item $\tau$, $t$: coordinate time in the metric and conjugated variable to $\lambda$. 
\item $\lbrace \cdot, \cdot \rbrace$: Poisson braket. 
\item $\mathcal{O}$: Dirac observable.
\item $\square$: Laplace--Beltrami operator.
\item $\Psi$, $\Phi$: quantum wave functions.
\item $k$: quantum conjugated momentum to $\varphi$.
\item $J_{y}(x)$, $K_{y}(x)$, $I_{y}(x)$: ordinary and modified Bessel functions. 
\item $u=\log(\frac{v}{v_0})$ where $v_0$ is a constant used for dimensional reasons.
\item $\braket{\cdot}{\cdot}$: quantum inner product. The mathematical expression of the inner product varies from theory to theory. 
\item $\alpha(k,\lambda)$, $\beta(k,\lambda)$, $\gamma(k,\lambda)$, $\epsilon(k,\lambda)$ are parameters of the wave functions $\Psi$ and $\Phi$. When considering semiclassical states, the underscore $sc$ will be added to these parameters
\item $\theta(k)$, $\vartheta(\lambda)$ and $\kappa_0(\lambda)$ are self-adjoint extension parameters of the different theories.
\item $\mathscr{H}_{kin}$, $\mathscr{H}_{phys}$: kinematical and physical Hilbert space. 
\item $\hat{\mathcal{H}}$, $\hat{\mathcal{G}}$ and $\hat{\mathcal{F}}$ are the operators that specify time evolution through a Schrödinger or a Klein--Gordon equation in each of the theories. 
\item $\delta(x_0-x)$: Dirac delta distribution. $\delta(x_0-x)=\infty$ when $x=x_0$.
\item $\delta_{ab}$: Kronecker delta.
\item $\expval{x}_{\Psi}=\bra{\Psi}x\ket{\Psi}$ corresponds to the expectation value of the quantity $x$ with respect to the state $\Psi$.
\item $\sigma_x$, $\sigma^2_x$: standard deviation and variance of the variable $x$. 
\item $\psi(x)$: digamma function.
\item $\gamma$: Euler-Mascheroni constant.
\item $\Xi_v$, $\Xi_{v^2}$, $\epsilon$: cutoffs. 
\item $\D X$: Functional differential of the functional $X$.
\item $\delta X$: functional variation of the quantity $X$.
\item $\varrho$: gauge fixing of the lapse derivative in the path integral quantisation.
\item $\Pi$: Lagrange multiplier of the path integral quantisation.
\item $\varepsilon(\tau)$: infinitesimal coordinate transformation parameter. For the gauge fixing of the path integral quantisation it will be replaced by the anticommuting ghost fields $\omega c(\tau)$.
\item $\bar{c}$, $c$, $p$, $\bar{p}$: other anticommuting ghosts fields.

\end{itemize}

In general, a mathematical quantity with hats represents the quantisation of the equivalent classical quantity. The meaning of quantisation will be specified in a case by case basis. 

This work is divided in five parts. In the first part we present an analysis of the current landscape of quantum gravity and quantum cosmology and the necessary prerequisites for the rest of thesis. In the second part we introduce the model we want to analyse, first classically and then we quantise it using the methods described in the introduction. In the third part, we analyse the dynamics of the model using three different quantum clocks. In part IV, we show numerical results that provide more evidence to the analysis realised in the previous sections, and finally, in the last part, we present an alternative way of interpreting our model and the general conclusions of this work. 

This thesis is the result of the work produced in the scientific papers \cite{Gielen2020}, \cite{Gielen2021} and an essay \cite{Gielen2022}. In particular, the results presented in \cref{t-clock-sec}, \cref{v-clock-sec}, \cref{t-clock-dynamics} and \cref{v-clock-dynamics} are based on \cite{Gielen2020}, whereas \cref{dirac-sec}, \cref{phi-clock-sec}, and \cref{phi-clock-dynamics} show the results obtained in \cite{Gielen2021}. Other parts of the thesis, notably \cref{class-mod} may bear resemblance to both papers. It is impossible to present the results of \cite{Gielen2020} and \cite{Gielen2021} independently, as they are very intertwined together. We instead try to convene their message in a clear and logical order, expanding whenever possible the analysis already done. \Cref{pathin} is a more speculative chapter based on still ongoing work. All the original work is presented as such. 
\chapter{Introduction}
\section{Introduction to quantum gravity}

\subsection{The frontiers of physics}

In 1900 Lord Kelvin delivered his famous lecture about the two ``\nth{19} century clouds in Physics'', subsequently written as a paper \cite{Kelvin1901}. There, he points out the two principal unresolved issues at the time. We will call them the light cloud and heat cloud for short. 

The light cloud refers to the failure of the Michelson Morley experiment. Maxwell had recognised light as an electromagnetic wave and it was thus believed that it needed a propagation medium, called aether. The Earth would be embedded in the aether fluid, and due to its motion around the 
Sun, the Earth and the aether would have a relative motion with respect to each other, leading to differences in the light propagation speed. The experiment failed to observe this speed change, complicating substantially the properties such a medium should have.

The heat cloud deals with heat properties, more concretely it refers to the failure of the equipartition theorem to describe certain situations  where quantum effects were later found to be non negligible. For example, at low temperatures, the heat capacity of a solid was lower than expected, and black body radiation was incorrectly modelled, leading to the so called ``ultraviolet catastrophe''. 

The resolution of the first cloud came with the development of the theory of special relativity (SR) by Albert Einstein \cite{Einstein1905}, whereas the second cloud would lead to the birth of the theory of quantum mechanics (QM) whose father is often considered to be Max Planck \cite{Planck1900}. Coming back to the light cloud, SR states that the speed of light is always the same in all reference frames. However, if the speed of light is constant, then time associated with the different reference frames has to be relative. The theory was expanded to the general relativity theory (GR)  \cite{Einstein1915,Einstein1916}. The geometry of space and time is entangled and represented by a four dimensional Lorentzian manifold $\mathcal{M}$, called spacetime, with metric $g_{\mu\nu}$. Einstein's equations can be derived from the variation of the Einstein--Hilbert action \cite{Hilbert1915} 
\begin{equation}
\mathcal{S}_{EH}=\frac{1}{2\kappa}\int_\mathcal{M}\dd^4 x R \sqrt{-g}\, ,
\end{equation}
where $\kappa=8\pi G$ where $G$ is the Newtonian gravitational constant, $R$ is the Ricci scalar and $\sqrt{-g}$ is the determinant of the metric. For manifolds that have a boundary, it is necessary to add the Gibbons--Hawking--York boundary term \cite{GHY}
\begin{equation}
-\frac{1}{\kappa}\int_{\partial \mathcal{M}}\dd^3 x \sqrt{h}K\, ,
\end{equation}
where $K$ is the extrinsic curvature trace and $\sqrt{h}$ the determinant of the metric at the boundary.

The theory of GR explains the motion of slow and fast bodies under strong gravitational fields and has been thoroughly tested. The theory has a very vast range of applications, ranging from solar system dynamics to the interaction between neighbouring galaxies. Some examples of GR successes at different scales are the prediction of Mercury's perihelion precession \cite{Park2017}, pulsar dynamics \cite{Stairs2003}, and gravitational lensing of galaxies. Even given the success of GR, there have been many attempts to modify the theory only relying on classical physics, leading to a very wide family of theories often referred as modified gravity theories \cite{Clifton2011}.

The theory of GR predicts the existence of extreme objects, like black holes, introduced first by Schwarzschild as early as 1916 \cite{Schwarzschild1916}. Since their first theoretical postulation, black holes have been a major subject of research. The first black holes to be discovered were stellar mass black holes, like Cygnus-X, the first one to be identified in 1971 \cite{Webster1972}. Perhaps the ultimate confirmation test of GR is the direct observation of gravitational waves from black hole merging coming from the LIGO--Virgo--KAGRA collaborations \cite{Ligo}. The data provided by these collaborations has set strong constraints for theories of modified gravity coming from the ringdown of the gravitational waves \cite{Carullo2021}, or the difference between gravitational wave detection and gamma rays detection (in the case of a neutron star merger) \cite{GW}, for example.

However, there are other categories of black holes, like supermassive black holes that inhabit the centre of galaxies. The first quasars (quasi stellar objects), galaxies with an active supermassive black hole in their centre, were identified in the sixties \cite{Greenstein1964}, and strong indirect evidence of the presence of a supermassive black hole in the centre of our galaxy, Sagittarius A*, was found in 1998 \cite{Ghez1998}. The first pictures of two supermassive black holes, M87* (located at the centre of the galaxy M87) and Sagittarius A* were taken by the Event Horizon Telescope \cite{EHT}, setting a landmark for the study of black holes and GR. These observations set more constraints for theories of modified gravity \cite{Afrin2021}.

It seems that so far GR has been a success story, and that the theory is capable to make accurate predictions even in the most extreme environments. However, there are a couple of mysteries surrounding black holes. Mathematically, these objects have a mass $M$ concentrated in one point in spacetime; the metric $g_{\mu\nu}$ becomes singular. This singularity is often signalled by divergent tensorial quantities. Singularities, and in general divergent quantities, are a hint that a theory breaks down in certain regimes and should therefore be completed by another theory. In the black hole case, the singularity comes with an event horizon. While a rigorous definition of an event horizon would require pages of mathematical formalism, the pedestrian definition works fine for the purposes of this introduction. In short, the event horizon is a region surrounding the singularity such that all matter and light that has the misfortune of ``falling in'' is lost forever for an outside observer. The event horizon effectively divides spacetime in two regions: a singularity-free region, where we live, and a region that contains the singularity, and in which predictability is lost, among many other unpleasant consequences. The event horizon protects us from the singularities of black holes, as they are effectively unreachable for us (if we reach them we cannot come back). 

The singularity of black holes may be viewed an indicator that GR is not a complete theory. Copying Lord Kelvin's terminology, singularities in GR may be one of the \nth{21} century clouds in physics. Cherry on top, there are other types of even more challenging singularities. These singularities arise from the study of cosmological spacetimes. GR has been used to describe not only the spacetime around a massive object, but the universe as a whole. It is assumed that to a certain scale, the universe is homogenous and isotropic. These two assumptions together are often referred as the cosmological principle.  The content of the universe is supposed to be same everywhere (homogeneity), and there should be no preferred direction (isotropy). A solution to Einstein equation which is both isotropic and homogeneous is the Friedmann--Lemaître--Robertson--Walker (FLRW) family of metrics. The line element $\dd s^2$ of such metrics is usually represented as:
\begin{equation}
\dd s^2=-N(\tau)^2\dd \tau^2 +a(\tau)^2h_{ij}\dd x^i \dd x^j\, ,
\end{equation}
where $h_{ij}$ is an euclidean three metric whose coordinates not depend on $\tau$ and $x^i$ ($i=1,2,3$) are the spatial coordinates. $\tau$ is the time coordinate, the parameter $a$ is called scale factor and $N$ the lapse function. We will come back to the lapse function later. The metric $h_{ij}$ has constant curvature. These universes have been largely studied and provide a very good approximation for our own universe. Some generalisations to the FLRW metrics are the anisotropic and homogeneous Kantowski--Sachs models \cite{Kantowski1966} of metric
\begin{equation}
\dd s^2=-N(\tau)^2\dd \tau^2 +a(\tau)^2\dd r^2+b(\tau)^2(\dd \theta^2 + \sin^2\theta\dd \varphi^2)\, .
\end{equation}
The presence of two scale factors $a$ and $b$ breaks isotropy. There are other more complicated models like the Bianchi models. They are named after Bianchi's classification of 3 dimensional Lie algebras. Bianchi universes are homogenous cosmologies with a 3 dimensional group of isometries. Their line element can be written as:
\begin{equation}
\dd s^2=-N(\tau)^2\dd \tau^2 + h_{ij}(\tau) u^i\otimes u^j\, ,
\end{equation}
where $u^i$ are the invariant one-forms associated with the given isometry group. The simplest Bianchi universe is Bianchi I, whose isometry group is $\R^3$. Its line element is
\begin{equation}
\dd s^2=-N(\tau)^2\dd \tau^2 + a(\tau)^2\dd x^2+b(\tau)^2\dd y^2 +c(\tau)^2\dd z^2\, ,
\end{equation}
where now we have 3 scale factors $a$, $b$ and $c$. The most complicated (and interesting) Bianchi model is Bianchi IX, that is associated with $SO(3)$ symmetry. The one forms  $u^i$ can be written as:
\begin{align}
u^1&=-\sin\psi \dd \theta+\sin\theta\cos\psi \dd \varphi\, , \nonumber\\
u^2&=\cos\psi \dd \theta +\sin\theta\sin\psi\dd \varphi\, , \\
u^3&=\cos\theta\dd \varphi+\dd \psi \nonumber
\end{align}
where, $\varphi$, $\theta$ and $\psi$ are Euler angles on the sphere.

Most cosmological models share a common feature: a big bang and/or big crunch singularity. In most cases, this singularity is signalled by a vanishing scale factor, meaning that all matter in the universe is condensed in a single point in space. The terminology big bang\footnote{The word big bang was first coiled as a mocking term, because people believed that the universe must have been static and therefore eternal. Expanding universes and the presence of a ``beginning of time'' were (and are still) very unsettling.} vs big crunch describe whether the singularity lies in the past vs in the future of an observer. In fact, it has been demonstrated that under very generic conditions cosmological models present an initial big bang singularity \cite{Hawking1975}. This singularity is naked, i.e~it has no event horizon to protect us from itself. The discovery of the cosmic microwave background (CMB) points out that indeed, the universe seems to have emerged from a very dense state and then expanded to what we observe today. The latest data suggests a flat universe with several perfect fluids: matter (which most of it is non-relativistic dark matter), radiation, and a cosmological constant $\Lambda$, $\Lambda$CDM, that originated from a very condensed and warm state \cite{PlanckCMB}. 

We may think that the presence of singularities in such extreme environments is no coincidence. It is believed that quantum effects are supposed to play an important rôle, and therefore one should try to include them in the theory of quantum gravity. The \nth{19} century light cloud has been replaced with the unavoidable presence of singularities in GR, one of the frontiers of physics today. 

However, to understand better why a quantum theory of gravity might be the answer to the classical singularities, we need to introduce the solution of Lord Kelvin's heat cloud, i.e.~the theory of quantum mechanics. Max Planck was the first to propose that the electromagnetic energy was quantised  in ``quantas'' each having an energy $E$ of
\begin{equation}
E=h \nu\,,
\end{equation} 
where $\nu$ is the frequency of each and $h$ is the now called Planck constant \cite{Planck1900}. The existence of a ``light particle'' (the photon) was theorised by Einstein and earned him a Nobel prize for explaining the photoelectric effect \cite{Einstein1905b}. Quantisation of the energy of a system was a solution for most failures of the equipartition theorem, in particular the ultraviolet catastrophe in the black body heat spectrum. 

The theory of QM was developed by many talented people. Erwin Schrödinger introduced the ``wave formalism'' of QM, and he is considered the father of wave mechanics. His famous equation, the Schrödinger equation, may be written as
\begin{equation}
\hat{\mathcal{H}}\psi(t)=i \hbar \pdv{}{t}\psi(t)\, ,
\label{schreq}
\end{equation}
where $\psi(t)$ is a wave function that describes the system, $\hat{\mathcal{H}}$ is a differential operator, the Hamiltonian of the system, and $\hbar$ is the reduced Planck constant $\hbar=\frac{h}{2\pi}$. Werner Heisenberg developed an approach based on matrix operators, that turned out to be equivalent to the wave function formalism. The nature of the theory is probabilistic, the wave function $\psi$ representing the probability measure of the system. Paul Dirac found out the similarities between classical and quantum theories comparing Poisson brackets and Dirac brackets \cite{Dirac1925} and formalised the notion of Hilbert space. The first complete rigorous description of QM is attributed to John von Neumann \cite{vonNeumann}.

Needless to say, QM was not only a revolution in physics but also a philosophical revolution. GR challenges the notion of time and space, but QM defies determinism and the existence of our own reality. We recap briefly the postulates of QM. We assume the reader is familiar with the braket notation. We will use this notation for the postulates and throughout the thesis.
\begin{enumerate}
\item The state of an isolated quantum system is represented by a unitary state vector $\ket{\psi}$, belonging to a Hilbert space $\mathscr{H}$, the state space. If the system were to be composite, the resulting Hilbert space would be the tensor product of the individual state spaces.
\item Observables are self-adjoint operators acting on the Hilbert space $\mathscr{H}$. Measurements of an observable $\mathcal{O}$ on a state $\psi$ always result in an eigenvalue of $\mathcal{O}$, let's say $o$. After a measurement, the state $\ket{\psi}$ collapses into the subspace of eigenvectors of eigenvalue $o$. The collapse of the wave function is highly debated, but unfortunately, we do not have time to focus on this issue of QM.
\item Evolution of a state vector is governed by the Schrödinger equation   \cref{schreq} where $\hat{\mathcal{H}}$ is a self-adjoint operator. In other words, the evolution of a state can be described by a unitary transformation:
\begin{equation}
\ket{\psi(t)}=\hat{\mathcal{U}}\ket{\psi(t_0)}\, ,
\end{equation}
where $\ket{\psi(t_0)}$ is the initial state at time $t_0$, and $\hat{\mathcal{U}}=e^{-i t\frac{\hat{\mathcal{H}}}{\hbar}}$ is the time evolution operator.
\end{enumerate}
These postulates are of major importance, and we will come back to them. Despite its philosophically challenging nature, the theory of QM was also an amazing success and its applications lead to great breakthroughs in atomic physics.

However, to fully understand subatomic physics, or environments in which particles are not only subject to quantum mechanics, but also have relativistic speeds, we need another theory. The theory of quantum field theory (QFT) fills this gap. We can describe QFT as a theory combining classical field theory with quantum mechanics and special relativity. The major success of QFT is the theory of the standard model of particle physics, that explains the electromagnetic, weak and strong interactions, see \cite{Butterworth2016} (and references therein). The standard model, like GR, has been thoroughly tested and is a very successful theory. However, a few discrepancies signal that this theory may also be incomplete. For example, the tension over the magnetic momentum of the muon has been getting bigger recently \cite{Abi2021}, and evidence against lepton universality has been discovered \cite{Aaij2022}.

In addition to that, the biggest flaw in the standard model may be the incapacity to include gravity. The standard model explains the electro-magnetic, the strong and the weak forces with the exchange of bosons. When trying to apply the same technique and include a new boson carrier of the gravitational force, the graviton, the theory becomes perturbatively non-renormalisable. The presence of divergent integrals in QFT is a common feature (see any QFT book, for example \cite{Peskin}). The different methods for tackling these infinite quantities, usually referred as renormalisation and regularisation, are very interesting and would lead to enough material for another PhD thesis. We will point only the implications for our work. The standard model is considered to be a renormalisable theory: with the addition of a finite number of extra terms (called counter terms), the infinite quantities can be made finite. The theory of the standard model+graviton is perturbatively non-renormalisable: we would need an infinite amount of counter terms to cancel the infinite integrals \cite{Klauder1975,Shomer2007}. The non-renormalisability of a quantum field theory including gravity is a sign that QFT is also an incomplete theory, that fails to describe environment in which both quantum interaction and gravitational interactions are strong. The heat cloud observed by Lord Kelvin has transformed into the incompleteness of the standard model and QFT. 

In conclusion, both GR and QFT signal their incompleteness by the presence of infinities in the form of singularities and non-renormalisability. There have been many attempts to get rid of these infinite quantities, coming from the particle physics side and the more geometrical side of the spectrum; however physicists seem to agree on the following: both theories are incomplete and therefore need to be expanded. We hope that this non exhaustive list of the feats and failures of GR, QM and QFT has convinced the reader of this necessity. Of particular interest for us, are the attempts of finding a theory of quantum gravity, i.e.~a theory that is neither GR nor QFT and is able to describe environments in which both gravitational and quantum effects are important. We will explore the different proposals for such a theory of quantum gravity in the next section.

\subsection{What is really quantum gravity?}

The concept of quantum gravity is very slippery. There have been many attempts to reconcile GR and QM in very diverse environments and coming from different approaches. For example, people have tried to incorporate an invariant length scale into SR. This length scale would be important for stages of large energy. This approach is called doubly special relativity (DSR) \cite{Magueijo2002,Amelino-Camelia2010}. DSR would be a limiting case of a theory of quantum gravity, when gravity effects are less important in comparison to velocities and energy.

 There have also been efforts to expand the theory of general relativity using ``new physics'', most notably to introduce QFT, or effective field theory, in the analysis of black holes. One of the first attempts to do this was done by Hawking, who derived the black hole temperature \cite{Hawking1974}; black holes could now vanish due to quantum effects. This lead to black hole information paradox: what happens with the information that has been ``swallowed'' into a black hole? \cite{Hawking1976}. A series of non-singular bodies, called black hole mimickers, have been proposed, mainly coming from effective approaches to quantum gravity, see for example \cite{Mazza2021} (and references therein). None of these techniques are conclusive yet, but there is great effort in trying to understand black hole singularities.

Yet another approach linked to effective field theories is the idea of applying asymptotic safety to gravity \cite{Eichhorn2019}. Asymptotic safety is a theoretical paradigm that can be applied to any quantum field theory by extending it in the high energies regime. A quantum field theory including gravity could be made renormalisable via asymptotic safety by the presence of a nice enough ultraviolet fixed point. In this framework one works with the coupling constants of the theory and analyse their variations at high energies. For some examples to see how this can be applied to gravity see \cite{Shomer2007} (also for some criticism) and \cite{Niedermaier2006}, for example. 

All the previously mentioned techniques are very useful tools that have their range of application, but they cannot be called a theory of quantum gravity, with perhaps the exception of asymptotic safety. Indeed, asymptotic safety is often considered a quantum theory of its own right, as it is a renormalisable quantum field theory that can potentially include gravity. Nonetheless, we think that a theory of quantum gravity should come with the existence of drastically different physics and interactions to get rid of the black holes and big bang singularities. There are people that have attempted to find a new theoretical framework that would allow for a theory of quantum gravity. There are two main starting points that one can use to try to find a theory of quantum gravity. One is to start form QFT and particle physics, we will call this the particle physics perspective. The other one is to start with GR and modify it, we will call it the geometrical perspective. The two principal candidates coming from these perspectives are string theory \cite{Mukhi2011} from particle physics perspective, and loop quantum gravity (LQG) \cite{Ashtekar2021}, from the geometrical perspective.

String theory assumes that particles can be obtained by different vibrational modes on a one dimensional object, called string. We will not discuss this approach to quantum gravity in much detail, but we wanted to mention it to have a more complete picture. In popular science, string theory and LQG often appear to be opposed theories fighting for the same spot, but they have slightly different goals. String theory is an attempt of expanding the standard model to include gravitational interactions, and hence can be considered a ``theory of everything'', in the sense that all forces would be explained by the same formalism. LQG, and any other theory coming from the geometrical perspective, does not look at the other forces or interactions, rather focuses on how classical spacetime can be made a quantum object. The starting point is so different, that it is almost impossible to give a meaningful comparison between the two approaches.

Besides the particle physicist approach to quantum gravity, the first geometrical approach, namely the Wheeler--DeWitt quantisation \cite{DeWitt1967,DeWitt1967b}, was developed in the sixties. This is the quantisation scheme we will use in a cosmological toy model, and therefore we present here the principal characteristics of this approach. We follow \cite{Wiltshire1995} for the derivation of the equations. The first thing we have to do is to define GR in the so-called canonical formulation: we split spacetime in one time direction $\tau$ and 3 space directions $x^k$, that are related with an euclidean metric $h_{ij}$, the metric induced on the spatial hypersurface $\Sigma$ defined by $x^k$. This is possible in general if $\mathcal{M}$ is globally hyperbolic. The line element in this decomposition is
\begin{equation}
\dd s^2=g_{\mu\nu}\dd x^\mu \dd x^\nu=(-N^2+N^iN_i)\dd \tau^2+2N_i \dd \tau \dd x^i+h_{ij}\dd x^i\dd x^j\, ,
\end{equation}
where $N(\tau,x^k)$ and $N^i(\tau,x^k)$ are the lapse function and shift vector. In general, one can think of $N$ as the difference between the coordinate time $\tau$ and the proper time on curves normal to $\Sigma$. The shift vector measures how the normal changes from one hypersurface to another. The coordinates are said to be commoving if $N^i=0$. Note that we use $h_{ij}$ to lower and raise space indices: $N^iN_i=h_{ij}N^iN^j$ and $N_i=h_{ij}N^j$. 

The Einstein-Hilbert action with a cosmological constant is
\begin{equation}
\mathcal{S}=\frac{1}{2\kappa}\left(\int_{\mathcal{M}}\dd^4 x\sqrt{-g}(R-2\Lambda)-2\int_{\partial\mathcal{M}}\dd^3 x \sqrt{h}K\right)\, .
\end{equation}
In the canonical decomposition, the gravitational action (leaving the boundary term out) can be written with a Lagrangian density
\begin{equation}
\mathcal{S}=\int \dd \tau \mathcal{L}=\frac{1}{2\kappa}\int \dd \tau \dd^3x N\sqrt{h}\left\lbrace K_{ij}K^{ij}-K^2+{}^3R-2\Lambda\right\rbrace\, ,
\label{gravaction}
\end{equation}
where $K_{ij}$ is the extrinsic curvature, $K=K^i_i$, and ${}^3R$ is the Ricci scalar of the 3 dimensional induced metric. Thus, one can define a Hamiltonian form for the action
\begin{equation}
\mathcal{S}=\int \dd \tau \dd^3 x \left\lbrace \pi^0\dot{N}+\pi^i\dot{N}_i-N\mathcal{C}-N_i\mathcal{C}^i \right\rbrace\, ,
\end{equation}
where the canonical momenta are defined as
\begin{equation}
\pi^0=\fdv{\mathcal{L}}{\dot{N}}, \hspace*{4mm} \pi^i=\fdv{\mathcal{L}}{\dot{N}_i}\, .
\end{equation}
The constraints $\mathcal{C}$ and $\mathcal{C}^i$ are
\begin{align}
\mathcal{C}&=2\kappa \mathcal{G}_{ijkl}\pi^{ij}\pi^{kl}-\frac{\sqrt{h}}{2\kappa}({}^3R-2\Lambda)\, ,\nonumber\\
\mathcal{C}^i&=-2\pi^{ij}_{;j}\, ,
\label{classconstraints}
\end{align}
where the subindex $;j$ denotes the covariant derivative with respect to the 3-metric, the momentum $\pi^{ij}$ is the conjugated momentum to the 3-metric
\begin{equation}
\pi^{ij}=\fdv{\mathcal{L}}{\dot{h_{ij}}}=-\frac{\sqrt{h}}{2\kappa}(K^{ij}-h^{ij}K)\, ,
\end{equation}
and the tensor $\mathcal{G}_{ijkl}$ is known as the Wheeler--DeWitt metric and has expression
\begin{equation}
\mathcal{G}_{ijkl}=\frac{1}{2}h^{-1/2}(h_{ik}h_{jl}+h_{il}h_{jk}-h_{ij}h_{kl})\, .
\end{equation}

Variation with respect to $N$ and $N^i$ yields to the constraints
\begin{equation}
\mathcal{C}=0,  \hspace{4mm}\mathcal{C}^i=0\, ,
\label{constraints}
\end{equation}
These constraints, respectively called the Hamiltonian and the diffeomorphism constraints, are a consequence of the diffeomorphism invariance of GR, that can be viewed as a constrained system \cite{GaugeSystems}.

Let us now analyse the configuration space of GR. We are interested in the space of three metrics on the spatial hypersurface, but modulo diffeomorphism invariance, in order not to count several times the same configuration. This space is infinite dimensional and can be loosely written as:
\begin{equation}
\lbrace h_{ij}(x)|x\in \Sigma \rbrace/\text{Diff}_0(\Sigma)\, ,
\label{superspace}
\end{equation}
where $\text{Diff}_0(\Sigma)$ are the diffeomorphisms connected to the identity. The infinite dimensions come from the fact that the space accounts for the metrics over all points $x\in \Sigma$. This space is denoted as \emph{superspace}. The Wheeler--DeWitt metric is a metric on the superspace. If we label $A,B\in\lbrace
h_{11},h_{22},h_{33},h_{12},h_{13},h_{23}\rbrace$, i.e., as running over the independent components of $h_{ij}$ and write the Wheeler--DeWitt metric as
\begin{equation}
\mathcal{G}_{AB}=\mathcal{G}_{(ij)(kl)}\, .
\end{equation}
It can be seen that this metric has signature $(-,+,+,+,+,+)$ regardless of the signature of the spacetime metric $g_{\mu\nu}$.

To quantise this system, there are two main approaches, the \emph{canonical quantisation} and the \emph{path integral quantisation}. Canonical quantisation consists in changing \cref{constraints} to operators applied on a Hilbert space, in everyday language ``to put hats''. This implies replacing the momenta $\pi^0$, $\pi^i$ and $\pi^{ij}$ by derivatives to find
\begin{equation}
\hat{\mathcal{C}}\Psi=0,  \hspace{4mm}\hat{\mathcal{C}}^i\Psi=0\, ,
\end{equation}
where $\Psi$ is the wave function of the universe. The quantum constraint $\hat{\mathcal{C}}^i$ is easier to deal with and encodes the covariance of $\Psi$ under a coordinate change of $h_{ij}$ \cite{Higgs1958}. The quantisation of the Hamiltonian constraint leads to the famous \emph{Wheeler--DeWitt equation}. This equation was first written as:
\begin{equation}
\hat{\mathcal{C}}\Psi=\left(-2\hbar^2\kappa\mathcal{G}_{ijkl}\pdv[2]{}{h_{ij}}{h_{kl}} -\frac{\sqrt{h}}{2\kappa}({}^3R-2\Lambda)\right)\Psi=0\, .
\label{Wheeler}
\end{equation}
This comes from the replacement $\pi^{ij}\rightarrow i\hbar \pdv{}{h_{ij}}$ in the first equality of \cref{classconstraints}. Note that this is not a single equation, but rather one equation for every point $x$ in $\Sigma$. The operator $\hat{\mathcal{C}}$ is an operator acting on the superspace. This quantisation leads to the ordering problem. Indeed, while classically everything commutes, it is not clear which differential operator we should use as a replacement for $\pi^{ij}$. The resulting quantum theory depends on this choice. Our answer to this question is to use the following ordering: 
\begin{equation}
\mathcal{G}_{ijkl}\pi^{ij}\pi^{kl}\longrightarrow -\frac{\hbar^2}{\sqrt{-\mathcal{G}}}\pdv{}{h_{ij}}\left(\mathcal{G}_{ijkl}\sqrt{-\mathcal{G}}\pdv{}{h_{kl}}\right)\, .
\end{equation}
Here $\sqrt{-\mathcal{G}}$ is the determinant of the metric $\mathcal{G}_{ijkl}$. The right-hand side corresponds to the Laplace--Beltrami operator of the superspace metric. This choice makes the Wheeler--DeWitt equation covariant under a coordinate change of the superspace metric \cite{Hawking1985}. We call this ordering the Hawking and Page ordering. The study of the solutions to the Wheeler--DeWitt equation allows us to extract information about the possible wave functions of a given universe. In this work, we analyse a cosmological model using this approach to quantum gravity. Our model has a matter component, but matter can be straightforwardly added to the action \cref{gravaction} and quantised in a similar fashion, expanding $\mathcal{G}_{ijkl}$ to include also the matter variables. See \cite{Wiltshire1995} for an example of such construction with a massless scalar field. Although this quantisation scheme works with symmetry-reduced models, The Wheeler--DeWitt quantisation is not well-defined in the general case and its formulation remains formal today. This is why alternatives formulations have been proposed.

An obvious one is the \emph{path integral quantisation}. This quantisation method started to be explored in the seventies \cite{Gibbons1977}. The idea behind it consists in expanding Feynman QFT path integral representation to gravity. The path integral represents the probability amplitude to go from a universe with initial configuration $in$ to the final configuration $f$ via a functional integral of the form:
\begin{equation}
\braket{g_{\mu\nu,f},\phi_f}{g_{\mu\nu,in},\phi_{in}}=\int \mathcal{D} g_{\mu\nu} \D\phi_i \dots \exp[i \mathcal{S}(g_{\mu\nu},\phi_i, \dot{\phi_i},\dots)]\, .
\label{pi}
\end{equation}
Here $\phi$ represents the matter components of the universe in a very generic way and $\D$ is called functional differential. Here we integrate over all the possible paths of $g_{\mu\nu}$ and $\phi_i$. We will devote an entire chapter (\cref{pathin}) to the analysis of this quantisation method. The path integral \cref{pi} can potentially be a very powerful calculation tool (as the Feynman diagrams and Feynman rules testify), but it is a slippery object that is not entirely well defined mathematically. In this 3+1 canonical decomposition of gravity, the path integral quantisation may be written as an amplitude between initial and final hypersurface configurations $\Sigma_{in}$ and $\Sigma_f$
\begin{equation}
\braket{h_{ij,f},\phi_f,\Sigma_{f}}{h_{ij,in},\phi_{in},\Sigma_{in}}=\int \mathcal{D} g_{\mu\nu} \D\phi_i \dots \exp[i \mathcal{S}(g_{\mu\nu},\phi_i, \dot{\phi_i},\dots)]\, .
\end{equation}
The measure $\mathcal{D}g_{\mu\nu}\mathcal{D}\phi_i$ is ill-defined, and there are other issues (like the oscillations of the action $\mathcal{S}$) that difficult the obtention of a workable theory of quantum gravity from this quantisation scheme. However, this method can once again be applied to simplified models.

Last but not least, we ought to mention LQG. This geometrical approach is closely linked to the Wheeler--DeWitt approach we have briefly outlined. Indeed, the starting point is the Einstein--Hilbert action and the constraint equations \cref{classconstraints}. As explained in \cite{Ashtekar2021}, work we follow for this introduction, the form of the constraints are non polynomial in the metric $h_{ij}$ and its associated momentum $\pi^{ij}$. This makes them very complicated to handle mathematically, especially the Hamiltonian constraint. In order to simplify the constraints, GR is treated as a gauge theory by introducing new variables in terms of an $SU(2)$ connection $A^I_i$ and an $su(2)$ potential $E^i_I$ \cite{Ashtekar1985}. Here the uppercase letters $I,J$ etc refer to the internal $SU(2)$ indices, whereas the lowercase letters refer to the original spatial hypersurface indices. In these variables, the Hamiltonian density of GR can be written as function of the connection and the potential
\begin{equation}
\mathcal{H}(A,E)=\int_{\Sigma}\dd x^3(NS+N^iV_i+N^IG_I )\, ,
\end{equation}
where 
\begin{equation}
G_I=D_iE^i_I, \hspace{4mm} V_i=E^j_IF^I_{ij}, \hspace{4mm} S=\frac{1}{2}\epsilon^{IJ}_{\phantom{IJ}K}E^i_IE^j_JF^K_{ij}\,  ,
\end{equation}
again where $\epsilon^{IJ}_{\phantom{IJ}K}$ are the structure constants  related with the $SU(2)$ metric and $F^I_{ij}=2\partial_{[i}A^I_{j]}+\epsilon^I_{\phantom{I}JK}A^J_iA^K_j$ is called the field strength. The notation $D$ refers to the total (or covariant) derivative. $N$ and $N^i$ are the shift function and the lapse vector and $N^I$ is the generator of gauge rotations. To obtain the equations of motion one now need to impose three constraints:
\begin{equation}
G_{I}=0,\hspace{4mm} V_i=0, \hspace{4mm} S=0\, ,
\label{LQGconstraints}
\end{equation}
These constraints are often referred as the Gauss, vector and scalar constraints (which explains the notation). The first one is new, and comes from the addition of the new variables, but the other two are the equivalent Hamiltonian and diffeomorphism constraints. This theory is equivalent to GR, but now the constraints are written in a much simpler compact form. Now one can follow the canonical or path integral quantisation on this theory, resulting in the quantum theory of LQG. However, instead of directly quantising $A^I_i$ and $E^i_I$, one smears them  respectively around curves (or loops) $\ell$ parametrised by $t^I$ and surfaces $S$ with test fields $f^I$, resulting in the holonomies and fluxes
\begin{equation}
h_\ell(A)=\mathcal{P}\exp[\int_\ell \dd \ell^i A^I_it^I], \hspace{4mm} E_{f,S}=\int_S\dd ^2 S_i f^IE^i_I\, .
\end{equation}
One then proceeds to the introduction of abstract mathematical operators $\hat{h}_\ell(A)$ and $\hat{E}_{f,S}$ as a base for the canonical quantisation. One can view the resulting theory as a graph of links $\ell$ and vertices with intertwinners $i_n$. The details on this construction are rather technical, so we leave them out, but we can say that LQG gives a very unique vision of quantum spacetime. Several models like the Barrett--Crane model \cite{BC} and the EPRL (Engel--Livine--Pereira--Rovelli) model \cite{Engle2008} have been proposed from a path integral representation of LQG. They are often referred as spinfoam models

There is no experimental proof that none of the main theories of quantum gravity presented here, namely string theory, Wheeler--DeWitt quantisation, and LQG are the definitive theory of quantum gravity. The problem of recovering GR as a ``classical limit'' of these theories is still ongoing (the notion of classical limit in itself seems to be a conceptually hard problem). Nonetheless, all theories have promising features and applications. In the next section we discuss one of the many technical problems of the Wheeler--DeWitt (and LQG) quantisation and its application to cosmology.
\section{The problem of time in quantum gravity}
Let us come back for a moment to the Wheeler--DeWitt equation \cref{Wheeler}. This is in principle a nice second order partial differential equation. But on a closer look, a very non-trivial question arises: \emph{How do we account for evolution in this equation?} If we come even further back, and we want to compare it with the Schrödinger equation \cref{schreq} there is a striking difference: whereas the left-hand side is conceptually the same, the right-hand side is $i\hbar \pdv{}{t}\psi$ in the Schrödinger equation and 0 in the Wheeler--DeWitt equation. In short, quantum mechanics relies on the existence of an \emph{external} continuous parameter $t$ to account for time evolution, whereas such external variable does not exist in the Wheeler--DeWitt quantisation of GR. The reason behind this is that in GR time is a part of the spacetime manifold, and cannot be an external parameter. An equation like the Wheeler--DeWitt equation \cref{Wheeler} implies that $\Psi$ is frozen, it cannot evolve. This issue is known as \emph{the problem of time} and it is a central question in quantum gravity \cite{Isham1992,Kuchar2011,Anderson2012}.

In our case, with our Wheeler--DeWitt equation, there are several strategies one can attempt to give meaning to the evolution of the universe. They fall in three main categories \cite{Isham1992}:
\begin{enumerate}
\item \emph{Tempus ante quantum}: We choose an \emph{internal} degree of freedom to serve as clock \emph{before} quantisation. This usually involves solving the constraints classically and quantise the remaining ``true'' degrees of freedom. One example of this procedure is the reduced phase space quantisation \cite{rps}. This approach involves breaking the general covariance of GR at the classical level. Then, a question arise: which internal variable should we use? This is known as the multiple choice problem. The resulting theories stemming from this choice are generally inequivalent. Another early example of this procedure applied to a cosmological model can be found in \cite{Blyth1975}.

\item \emph{Tempus post quantum}: We first obtain the Wheeler--DeWitt equation (or an equivalent wave equation depending on the quantisation scheme) and choose one of the internal degrees of freedom to serve as a clock. We then build a Hilbert space with respect to the remaining variables. This approach is the one we will follow for our thesis. Although the idea is similar to the reduced phase space quantisation, as it involves choosing an internal parameter to be the clock, the execution is very different. The multiple choice problem also applies here and this is what we will try to analyse. An effective approach to the problem of time can be found in \cite{Bojowald2016}. In this setting, there is a conjecture established by Gotay and Demaret in \cite{Gotay}. They postulate that the principal features of the quantum theories, such as singularity resolution are inherited by the classical properties of the chosen clock. If the clock is classically slow, i.e.~reaches the singularity in a finite time, the resulting quantum theory will need a reflecting boundary condition to ensure unitarity and this will lead to singularity resolution. On the contrary, if the clock is fast, this boundary will not be needed. We will verify and expand their conjecture.

\item \emph{Tempus nihil est}: We first build a too large \emph{kinematical Hilbert space} using all the internal variables of the Wheeler--DeWitt equation, and this equation is then used to find the true \emph{physical Hilbert space}, where the wave function evolves. This approach is known as Dirac quantisation. As observables are frozen in time, dynamics must be expressed in a relational way. This implies taking observables corresponding to the value of quantity A when quantity B takes a given value \cite{Dittrich,Rovelli,Tambornino2011}. We will also briefly analyse Dirac quantisation of our cosmological model. Dirac quantisation, despite a priori looking better than the two other approaches as no choice of clock needs to be made, is not as straightforward as it seems. The way of obtaining a physical Hilbert space is complicated and requires an algebraic procedure, like group averaging. Group averaging has to be implemented in a case by case basis, see \cite{Tate1992} for examples of this. Of particular interest is the work of Höhn and collaborators \cite{Hoehn}, that presents a scheme for which one can change relational clocks in the Dirac quantisation scheme. 
\end{enumerate} 
Note that LQG also has a Hamiltonian (also named scalar) operator coming from \cref{LQGconstraints}, hence also presents the same issues, at least in the canonical quantisation scheme.

The problem of time rises many questions: should we choose an internal variable to serve as clock? Then which one? Should we choose it before or after quantising? What are the consequences of this choice? Are there alternative answers? It has been argued that the path integral quantisation scheme avoids the problem of time \cite{Kuchar2011}, and we will briefly mention this approach as well.

However, studying these questions within a given theory of quantum gravity can be very complicated mathematically. Our approach to the problem of time consists in analysing the problem in a much simpler setting, namely studying a cosmological toy model. On one side, We believe that the results found in these very simple toy models can be very useful to find insights applicable for the full theory of quantum gravity. On the other side, the study of these models can provide direct answers to the presence (or absence) of singularities, which might be the main reason to consider a theory of quantum gravity. In fact, the study of quantum cosmology is a very active sub-area of the work of quantum gravity, along with black hole analysis. We present the basis of quantum cosmology and our model in the following. 
\section{Introduction to quantum cosmology and our model}
Quantum cosmology is the result of the application of the different quantum theories to cosmological models. In the Wheeler--DeWitt quantisation the construction is straightforward: instead of working with the infinite dimensional superspace defined in \cref{superspace} we truncate most of the degrees of freedom to obtain a particular \emph{minisuperspace} (or \emph{midisuperspace} if there are more degrees of freedom than usual). An usual truncation is to assume a homogeneous metric, for example FLRW, (a calculation we will show explicitly in the next section). There is some debate on whether this approach is valid or it is a too brutal approximation \cite{Kuchar1989,Sinha1991,Mazzitelli1992}. Indeed, by reducing the infinite degrees of freedom to a finite number with the imposition of homogeneity, one may lose crutial aspects of the quantum theory. Such reduction may not be justified, especially taking into account that inhomegeneities could diverge at the singularity. In addition to that, comparison of minisuperspace models embedded in slightly bigger midisuperspace model suggests that minisuperspace behaviour cannot always be related to one of the midisuperspace. In any case, these minisuperspace models should be considered useful toy models, able to provide a testing ground for quantum gravity. 

Minisuperspaces are based on the Wheeler--DeWitt approach to quantum gravity, but there are other approaches to quantum gravity coming from other theories, see \cite{Bojowald2015} for a review. Notably, applying the prescripts of LQG to cosmological models leads to the well known theory of loop quantum cosmology (LQC) \cite{Ashtekar2011}. LQC leads to very interesting results regarding singularity resolution and other dynamics of the universe. Another approach to quantum cosmology related to LQG is group field theory (GFT). In general, GFT are field theories over a group manifold but there are in one-to-one correspondence with spinfoam models \cite{Reisenberger2001} and therefore this formalism can be applied to gravity and cosmology \cite{Oriti2006,Gielen2016b}.

In this work we study the problem of time and the multiple choice problem in a minisuperspace model. The chosen model is a flat FLRW universe with two matter components, namely a free massless scalar field and a perfect fluid. Different types of perfect fluid will be discussed throughout the thesis. Why did we choose this model? Other similar models have been analysed in the literature \cite{Blyth1975} but they usually contain only one matter degree of freedom. This degree of freedom is often used as clock. As our goal is to compare different clocks, having different matter components comes in handy. We also think it is important that the model tries to be as realistic as possible. Of course assuming isotropy and homogeneity simplifies the calculations by a lot, but having a field gives the opportunity for more complex dynamics. We will quantise the model using three clocks, a clock $t$ coming from the perfect fluid, the scalar field  clock $\varphi$, and a geometrical clock $v$ where $v$ is a positive power of the scale factor. 

This model has been previously studied from different perspectives with the different clocks. In their work \cite{Gryb}, Gryb and Thebault worked with a universe that contained a cosmological constant and used the $t$ clock. On the other hand Gielen and Turok \cite{Gielen} studied a universe filled with radiation and used a clock equivalent to our $v$ clock. A very similar model with a fixed cosmological constant has been analysed in the Wheeler--DeWitt and LQC quantisation using the $\varphi$ clock \cite{Pawlowski2011}. The same model with a fixed positive cosmological constant has been analysed semiclassically \cite{Bojowald2010b}. In this thesis we will reproduce and expand their results, and give a meaningful comparison between the three clocks.

\part{Minisuperspace model presentation}
\chapter{The classical model}
\label{class-mod}
\section{One model, many perfect fluids choices}
\label{energyint}
\subsection{Standard GR}

We wish to study the problem of time through a toy model. As introduced previously, the chosen universe is a homogenous and isotropic flat FLRW universe. The matter contents of this universe are a  massless and free scalar field $\phi$ and a perfect fluid characterised by its equation of state $ p=w \rho$ where $p$ is the pressure and $\rho$ the energy density. Some particularly interesting cases are radiation $(w=\frac{1}{3})$, dust $(w=0)$ and dark energy $(w=-1)$. We leave the specific choice of the perfect fluid for later.

We start by writing the action for our model. The dynamics of GR coupled to a free massless scalar and a perfect fluid are defined by an action
\begin{equation}
 \mathcal{S} = \int \dd^4 x \left\{\sqrt{-g}\left[\frac{R}{2\kappa}-\frac{1}{2}g^{ab}\partial_a\phi\partial_b\phi -\rho\left(\frac{|J|}{\sqrt{-g}}\right)\right]+J^a(\partial_a\vartheta+\beta_A\partial_a\alpha^A)\right\}\, . 
 \label{action}
\end{equation}
The perfect fluid action used here is Eq.~(6.10) of \cite{Brown1993}, which describes an isentropic fluid. An isentropic perfect fluid is a fluid with constant entropy per particle. The dynamical variables are the spacetime metric $g_{ab}$, scalar field $\phi$, densitised particle number flux $J^a$ and Lagrange multipliers $\vartheta$, $\beta_A$ and $\alpha^A$. For isentropic fluids, $\rho$, the energy density of the fluid, is a function of only $|J|=\sqrt{-g_{ab}J^aJ^b}$ and $\sqrt{-g}$, but does not depend on the entropy per particle. We have also defined $\kappa=8\pi G$ where $G$ is Newton's constant. The Lagrange multipliers $\alpha^A$ represent a coordinate system for the flow lines of the fluid. Given a spacetime hypersurface and a coordinate system $\alpha^A$, with $A=1,2,3$, a flow line can be labelled by the coordinate value of its intersection point with the hypersurface.

In order to avoid divergent integrals we define spacetime as a manifold with topology $\mathbb{R}\times \Sigma$ where $\Sigma$ is a bounded three-dimensional manifold. All the matter fields and the geometry are homogeneous and locally isotropic on each copy of $\Sigma$. The metric is
\begin{equation}
\dd s^2=-N(\tau)^2\dd \tau^2 + a(\tau)^2 h_{ij} \dd x^i \dd x^j \ .
\label{metric}
\end{equation}
Here $h_{ij}$ is a flat metric, $a(\tau)$ is the scale factor and $N(\tau)$ is the lapse function. The lapse function parametrises the freedom of choosing different time coordinates. This coordinate system represents the canonical decomposition of our model. Due to the FLRW symmetry, the shift vector $N^i=0$ everywhere. The choice $N=1$ means we are working in cosmological time, but if, for example, $N(\tau)=a(\tau)$ we are in conformal time. Due to homogeneity and isotropy, the field $\phi$ and the particle number density of the perfect fluid are only a function of $\tau$. Moreover the flux is of the form $J^{a}=a^3n\delta^a_\tau$ where $n$ is the particle number density. This means that the flux is proportional to the commoving number of particles. All Lagrange multipliers are also only functions of $\tau$. 

The reduced minisuperspace action, after an integration by parts, is then
\begin{equation}
\mathcal{S} = V_0\int_{\mathbb{R}} \dd \tau\left[-\frac{3\dot{a}^2a}{N\kappa}+\frac{a^3}{2N}\dot\phi^2 -Na^3\rho(n)+a^3 n(\dot\vartheta+\beta_A\dot\alpha^A)\right]\, .
\label{minisuperaction}
\end{equation}
$V_0=\int_{\Sigma} \dd^3 x\sqrt{h}$ is the coordinate volume of $\Sigma$. In order for $V_0$ to be finite, it is important that $\Sigma$ is bounded. For simplicity, we will suppose that $\Sigma$ has no boundary (in particular $\Sigma$ could be a three torus). However, the manifold $\mathcal{M}=\Sigma \times \R$ has a boundary. It is therefore necessary to include the Gibbons--Hawking--York term in \cref{minisuperaction}. In this simple case it is sufficient to add the following term to the action:
\begin{equation}
\mathcal{S}_{GHY}=-\frac{3V_0}{\kappa}\left[ \frac{\dot{a}(\tau){a^2(\tau)}}{N(\tau)}\right]_{\tau=-\infty}^{\tau=\infty}\,.
\end{equation}
Then why are there no boundary terms in the action \cref{minisuperaction}? To obtain the final expression \cref{minisuperaction} we have integrated the term coming from the Ricci scalar $R$, $-\frac{3\dot{a}^2a}{N\kappa}$, by parts. The Gibbons--Hawking--York term and the boundary term coming from the integration by parts cancel, leading to a well-defined variational principle.

The last term in \cref{minisuperaction} involving $\beta_A$ and $\alpha^A$ may now be dropped: variation with respect to $\vartheta$ imposes particle number conservation $\frac{\dd}{\dd \tau}(a^3 n) = 0$ and there is no further constraints from these other Lagrange multipliers. The other constraints, requiring the fluid flow to be directed along flow lines labelled by the $\alpha^A$, are trivial in a highly symmetric FLRW universe.

Now, it is time to specify a bit more our perfect fluid. For a perfect fluid with equation of state $p=w\rho$ where $w\neq -1$ we have $\rho=\rho_0n^{1+w}$ where $\rho_0$ is a fixed constant. We can then replace $n$ by $m$ such that $na^3=\left(\frac{m}{\rho_0}\right)^{\frac{1}{1+w}}$ and write the action as a function of $m$:
\begin{equation}
\mathcal{S}=V_0\int_{\mathbb{R}} \dd \tau \left[ -\frac{3 \dot{a}^2 a}{ \kappa N}+\frac{a^3}{2N}\dot{\phi}^2-N\frac{m}{a^{3w}}+m\dot{\chi}\right]\, .
\label{perfluidaction}
\end{equation}
Here we have changed Lagrange multiplier from $\vartheta$ to $\chi$ so that $\dot{\chi}=\frac{m^{\frac{-w}{1+w}}\dot{\vartheta}}{\rho_0^{\frac{1}{1+w}}}$ in an attempt to simplify the notation. Recall that $\dv{(a^3n)}{\tau}=0$ implies $\dv{m}{\tau}=0$, making $m$ a constant energy density (or mass) of the perfect fluid. A free field with no potential is a perfect fluid with $w=1$, which is in principle supported by this formalism. However, we already have a free field in our theory. Having multiple identical quantum clocks and being able to change between them is another very interesting problem that we leave for future work. In addition to that, We will see later that this case is fundamentally different. Finally, note that the quantities $m$ and $\chi$ appear as conjugated variables in the action \cref{perfluidaction}.

What happens if the perfect fluid is dark energy? In that case we cannot do the transformation that leads to \cref{perfluidaction}, as it is ill-defined. However using unimodular gravity we can describe a similar action for this type of perfect fluid.

\subsection{Unimodular Gravity}

The theory of unimodular gravity gained popularity with the different problems related to the cosmological constant. Today we know that in order to obtain the right dynamics of the universe the cosmological constant has to have a very small positive value. However, when we try to explain this value as vacuum energy coming from QFT there is a massive disagreement between the predicted value and the observed one. Unimodular gravity has been proposed as a solution of this problem \cite{Smolin2009, Ellis2011}. Moreover, unimodular gravity has also been studied  in the context of the problem of time \cite{Unruh1989,Kuchar1991b}.

It was first noted by Einstein that a \emph{unimodular} choice of coordinates, i.e., $\sqrt{-g}=1$ was a good way to partially fix a coordinate system \cite{Einstein1916}. Soon after, the first version of  unimodular gravity was introduced by fixing the determinant of the metric $\sqrt{-g}=1$ before deriving the Einstein's equation \cite{Einstein1919}. This theory is invariant under a smaller group than the full diffeomorphism group of GR, as $\sqrt{-g}$ has to remain unchanged under a coordinate transformation. A consequence of this reduction is that only the trace-free part of the Einstein equation is imposed as equation of motion. In the Hamiltonian formulation, a cosmological constant $\Lambda$ appears as a constant of integration corresponding to the value of the Hamiltonian on a given solution. We then recover the Einstein--Hilbert action of GR with a cosmological constant ``for free''.

However, in order to compare with \cref{perfluidaction}, we are interested in parametrised unimodular gravity. A parametrised version of unimodular gravity is obtained by introducing additional fields to restore full diffeomorphism invariance. This leads to the following action:
\begin{equation}
\mathcal{S}_{PUM}=\int \dd x^4 \left\lbrace\frac{\sqrt{-g}}{2\kappa}[R-2\Lambda]+\Lambda\partial_\mu T^\mu\right\rbrace\,  ,
\label{unimaction}
\end{equation}
where $\Lambda$ and $T^\mu$ are dynamical fields \cite{Henneaux1989}. One can obtain this action by promoting a special set of coordinates satisfying $\sqrt{-g}=\alpha$ to dynamical fields $X^A(x)$, where $x$ are now arbitrary coordinate labels \cite{Kuchar1991b}, thus restoring the full diffeomorphism invariance of the theory. The fields $T^\mu$ depend on the $X^A$'s. Variations with respect to $\Lambda$ in \cref{unimaction} fix $\sqrt{-g}=\kappa\partial_\mu T^\mu$ and variations with respect to $T^\mu$ imply $\Lambda$ is a constant. 

In our specific case we work with an FLRW universe with a massless scalar field, so the action is formed of the unimodular part \cref{unimaction} and the field part. Given the FLRW symmetries of the theory, the term $\Lambda\partial_\mu T^\mu$ reduces to $\Lambda \dot{T}$ giving the final form of the action
\begin{equation}
\mathcal{S}=\int_\R \dd \tau \left[-\frac{3V_0a}{\kappa N}\dot{a}^2+\frac{V_0a^3}{2N}\dot{\phi}^2-\frac{V_0a^3 N\Lambda}{\kappa}+\Lambda \dot{T} \right]\,.
\label{ccaction}
\end{equation}
Note that we have followed the same procedure that we used to obtain \cref{minisuperaction}: we add the Gibbons--Hawking--York term and perform an integration by parts. 

We see that the actions \cref{perfluidaction} and \cref{ccaction} are very similar. They both contain a $\dot{a}^2$ dependent term, a $\dot{\phi}^2$ dependent term and a linear term either on $\dot{\chi}$ or $\dot{T}$. $\Lambda$ and $m$ play the same rôle. In fact, if we write the action as a function of $\tilde{\Lambda}=\frac{\Lambda}{\kappa}$ and $\tilde{T}=\frac{\kappa}{V_0}T$ we find the exact same expression as \cref{perfluidaction}:
\begin{equation}
\mathcal{S}=V_0\int_{\mathbb{R}} \dd \tau \left[ -\frac{3 \dot{a}^2 a}{ \kappa N}+\frac{a^3}{2N}\dot{\phi}^2-N\frac{\tilde{\Lambda}}{a^{3w}}+\Tilde{\Lambda}\dot{\tilde{T}}\right]\, ,
\end{equation}
where $w=-1$. We have then derived an action expression valid also for dark energy. Recall that $\Lambda$ and $T$ (and $\tilde{\Lambda}$ and $\tilde{T}$) are a pair of conjugated momenta just as $m$ and $\chi$ were in \cref{perfluidaction}. We thus will consider that \cref{perfluidaction} is valid also in the dark energy case.

\section{The Hamiltonian formulation}
\label{hamform}
In order to quantise our model, we want to bring the action \cref{perfluidaction} to the Hamiltonian form. Later on, when quantising, we will identify the classical momenta with derivatives. The canonical momenta are
\begin{equation}
\pi_a=-V_0\frac{6 a}{\kappa N}\dot{a}, \hspace*{4mm} \pi_\phi=V_0\frac{a^3}{N}\dot{\phi}\,.
\end{equation}
Recall that $m$ is the conjugated momentum to $\chi$, satisfying the Poisson bracket relation $\lbrace \chi, m \rbrace=1$. The Hamiltonian of the model is
\begin{equation}
\mathcal{H}=N\left[-\frac{1}{12}\frac{\kappa \pi_a^2}{V_0 a}+\frac{1}{2}\frac{\pi_\phi^2}{V_0 a^3}+V_0\frac{m}{a^{3w}}\right]\, ,
\end{equation}
and the action can be written as 
\begin{equation}
\mathcal{S}=\int \dd \tau \left[\pi_a\dot{a}+\pi_\phi \dot{\phi}+m\dot{\chi}-\mathcal{H}\right]\, .
\label{hamaction}
\end{equation}
The lapse function $N$ is multiplying the whole expression of $\mathcal{H}$, making it a totally constraint Hamiltonian. The lapse function is a Lagrange multiplier of the theory, so from \cref{hamaction} we can deduce that taking the variation of the action will lead the usual Hamilton's equations and to the constraint:
\begin{equation}
-\frac{1}{12}\frac{\kappa \pi_a^2}{V_0 a}+\frac{1}{2}\frac{\pi_\phi^2}{V_0 a^3}+V_0\frac{m}{a^{3w}}=0\,.
\label{constraint}
\end{equation} 
This constraint is the Hamiltonian constraint of the system. Recall that, when expressing GR in a Hamiltonian way, we usually also find the diffeomorphism constraint \cite{Arnowitt1959}, but it is trivial in FLRW symmetry.

In order to simplify notation, we set $\kappa=1$, and we perform the following change of variables in \cref{perfluidaction}:
\begin{align}
v&=4\sqrt{\frac{V_0}{3}}\frac{a^{\frac{3(1-w)}{2}}}{1-w},  &\pi_v&=\sqrt{\frac{1}{12V_0}}\pi_a a^{\frac{3w-1}{2}}, &\\
\varphi&=\sqrt{\frac{3}{8}}(1-w)\phi, & \pi_\varphi&=\sqrt{\frac{8}{3}}\frac{\pi_\phi}{1-w}\,. &
\label{changevar}
\end{align}
The Hamiltonian takes the form
\begin{equation}
\mathcal{H}=\tilde{N}\left[-\pi_v^2+\frac{\pi_\varphi^2}{v^2}+\lambda\right]\, ,
\label{hamsimple}
\end{equation}
where we have also defined the lapse as $\tilde{N}=Na^{-3w}=N\left(\frac{16V_0}{3v^2(1-w)^2}\right)^{\frac{w}{1-w}}$ and $\lambda=V_0m$.  The Hamiltonian constraint looks very simple after this change:
\begin{equation}
\C=-\pi_v^2+\frac{\pi_\varphi^2}{v^2}+\lambda=0
\label{hamconst}
\end{equation}

In this form $w$ no longer appears explicitly. Note that this change is only valid if $w\neq 1$. Our description does not cover the case in which the perfect fluid is a free field, but as said previously, that is an entirely different problem as our model already has a free field $\phi$.  In the dark energy interpretation and working in the convention where the metric is dimensionless and the line element has units of length, the Hamiltonian has dimensions of length and $v$ has dimension of volume, whereas $\tilde{N}$ is dimensionless.

The expression \cref{hamsimple} has many interesting features. On the one hand, it has two quadratic momenta, but also a linear term, $\lambda$. This linear term is a distinguishing feature of our model in comparison to earlier models \cite{Blyth1975}. On the other hand, the two quadratic momenta can be written using the Rindler wedge metric
\begin{equation}
g^{AB}=\begin{pmatrix}
-1 & 0 \\
0 & \frac{1}{v^2}
\end{pmatrix}\, ,
\end{equation}
so that
\begin{equation}
\mathcal{H}=\tilde{N}\left[ g^{AB}\pi_A\pi_B + \lambda\right]\, .
\end{equation}
If we consider the coordinate $\varphi$ to be timelike and $v$ to be spacelike, the metric $g_{AB}$ can be used to describe a particle moving with constant acceleration. This spacetime, which is a section of Minkowski space, is referred as Rindler wedge. In this analogy $\lambda > 0$ would
correspond to a ``tachyon'',  or a particle with negative mass squared, whereas $\lambda<0$ would be a massive particle. On the contrary, if $\varphi$ is spacelike and $v$ is timelike, this metric is used to describe a different section of Minkowski spacetime, often referred as Milne wedge. The Milne wedge is used to represent the Milne universe, a hypothetical flat universe with 0 energy density \cite{Levy2012}.

For simplicity, from now on we choose to interpret the perfect fluid as dark energy. In this interpretation the coordinates and momenta are:
\begin{equation}
v=2\sqrt{\frac{V_0}{3}}a^3, \hspace{3mm} \pi_v=\sqrt{\frac{1}{12V_0}}\frac{\pi_a}{a^2}\, .
\end{equation}
Hence the $v$ coordinate is proportional to a ``scale volume'' of the universe, justifying its notation. Its conjugated momentum $\pi_v$ is proportional to the Hubble constant and $\lambda$ would play the rôle of cosmological constant. There is nothing privileged about this energy interpretation, but it is simpler to stick to a specific one in order to minimise confusion. Almost all the following results are independent of the energy interpretation and the ones that are not will be pointed out. As long as $w<1$, $v$ will always be a positive power of the scale factor.

In this theory, the lapse function can be chosen arbitrarily, but a particularly attractive choice is  $\tilde{N}=1$. In this case the Hamiltonian is of the form $\mathcal{H}=-\mathcal{H}_0+\lambda$ with $\mathcal{H}_0= -g^{AB}\pi_A\pi_B$. The Hamiltonian constraint becomes $\mathcal{H}_0=\lambda$, thus $\lambda$ plays the rôle of the energy of the system defined by $v$ and $\varphi$ and their conjugate momenta. Let $t=\frac{\chi}{V_0}$ be the conjugated momentum to $\lambda$; in this gauge, $\dv{t}{\tau}=1$ which implies that we can use $t$ to express the evolution of $\varphi$ and $v$. Note that whereas $\tau$ is just a label for a coordinate in the metric, $t$ is a dynamical variable of the system. For dark energy $w=-1$ and this corresponds to $N=a^{-3}$, but all perfect fluid choices similarly have a preferred time coordinate in which the dynamics take the simplest form. The $N=a^{-3}$ gauge is often called unimodular gauge. In this gauge we have that the determinant of the metric is one. The $\tilde{N}=1$ gauge has been extensively used for the $w=0$ (dust) case \cite{Brown1994, Husain2011}.
\section{Classical solutions}
\label{class-sol-sec}
Hamilton's equations can be solved straightforwardly. In this model we are interested in expressing evolution using different clocks. In principle, all monotonic variables are good choices of clock, which makes $t$, $\varphi$ and $v$ all viable candidates, with the subtlety that $v$ experiences a turnaround for $\lambda<0$ and hence is not a good clock everywhere. The loss of monotonicity can easily be avoided, by considering separately the evolution in the $v$ growing and decreasing branches. The canonical momentum $\pi_v$ would also be a good clock everywhere, as it is monotonic. A similar clock has been studied in a simpler model \cite{Blyth1975}. In this section, we present the equations of motion of the dynamical variables with respect to the classical clocks $t$, $\varphi$ and $v$. 

$\lambda$ and $\pi_\varphi$ are constants of motion, which means that they are constant along a given solution. We saw above that in the case where one thinks of a perfect fluid with $w=-1$, $\lambda$ is essentially the cosmological constant and thus could take either sign. For other types of perfect fluid one might assume that particle number density and energy density must be positive and only consider $\lambda>0$. The classical solutions we present are always well-defined for any interpretation of the perfect fluid matter. Recall that the main difference between a cosmological constant coming from unimodular gravity, rather than from GR, is that here solutions with different values for $\lambda$ are allowed, and $\lambda$ has its own conjugated coordinate. This makes $\lambda$ a dynamical variable, rather than the usual constant of nature.

\begin{figure}
\centering
\begin{subfigure}{1\textwidth}
\centering
\includegraphics[scale=0.75]{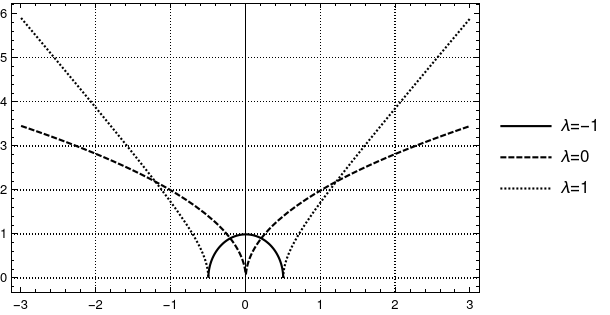}
\caption{$v$ as a function of $t$ for several values of $\lambda$.}
\end{subfigure}
\begin{subfigure}{1\textwidth}
\centering
\includegraphics[scale=0.75]{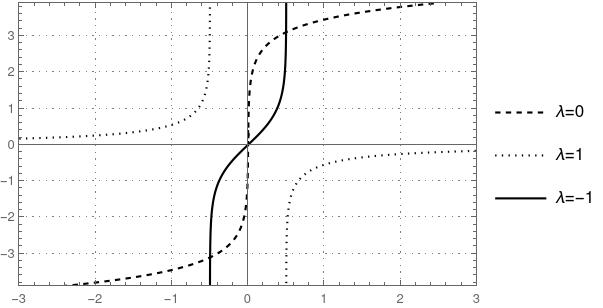}
\caption{$\varphi$ as a function of $t$ for several values of $\lambda$.}
\end{subfigure}
\caption[Classical solutions as functions of $t$.]{In this figure we observe the evolution of the principal dynamical variables using $t$ as clock. The chosen values of the integration parameters is $t_0=\varphi_0=0$. Note that $t_0=0$ is a singular value in the $\lambda=0$ case, but it can be approximated continuously. The conjugated momentum to $\varphi$, $\pi_\varphi$ has been chosen to have value 1. In these plots the big crunch and big bang singularities lie respectively at $t=\pm \frac{1}{2}$ for $\lambda\neq 0$ and at $t=0$ for vanishing $\lambda$. We see how the scalar field diverges at the singular points. We also see that for $\lambda<0$, $v$ is not a good clock everywhere.}
\label{clasplot-vt}
\end{figure}

We first present the solutions with respect to $t$. Remember that in the gauge we are working $\dv{t}{\tau}=1$, hence $t$ can be used as evolution parameter. For $\lambda\neq 0$ and $\pi_\varphi\neq 0$ we have
\begin{equation}
v(t)=\sqrt{-\frac{\pi_\varphi^2}{\lambda}+4\lambda(t-t_0)^2}, \hspace*{3mm} \varphi(t)=\frac{1}{2}\log\abs{\frac{\pi_\varphi-2\lambda(t-t_0)}{\pi_\varphi+2\lambda(t-t_0)}}+\varphi_0\, .
\label{eqsm-t-posneg}
\end{equation}
$t_0$ and $\varphi_0$ are integration constants. We see that $v=0$   when $\abs{t-t_0}=\frac{\pi_\varphi}{2\abs{\lambda}}$. For $\lambda=0$ the solutions are slightly different:
\begin{equation}
v(t)=2\sqrt{\abs{\pi_\varphi}\abs{t-t_0}},\hspace*{3mm}\varphi(t)=\frac{1}{2}\sgn(\pi_\varphi(t-t_0))\log\abs{\frac{t}{t_0}-1}+\varphi_0\,.
\label{eqsm-t-null}
\end{equation}
The scalar field diverges logarithmically when $v=0$ for all values of $\lambda$. The plots of $v(t)$ and $\varphi(t)$ are presented in figure \cref{clasplot-vt}. 

It is easy to see from  \cref{clasplot-vt} that $v$ reaches 0 at a finite value of $t$ and therefore there is a range of the parameter $t$ for which $v$ stops being well defined. However, this is not enough to prove that this universe presents a big bang and big crunch singularity. A quick way to acknowledge the big bang/big crunch singularity is to focus on commoving observers, i.e., $\dd x=\dd y=\dd z =0$. Since curvature invariants diverge as $v\rightarrow 0$, if $v=0$ can be reached in finite proper time $\tilde{\tau}$, it would be clear that there is a singularity at that point. We are working in the gauge for which $\tilde{N}=\frac{1}{2}\sqrt{\frac{3}{V_0}}vN=1$, hence $N=2\sqrt{\frac{V_0}{3}}v^{-1}$. Let us consider $\lambda>0$, the expanding branch of the universe ($t>0$), and  $t_{sing}=\frac{\abs{\pi_\varphi}}{2\lambda}$.  Using the fact that in this case $\dd\tilde{\tau}=N\dd t$, the proper time for commoving observer between $\tilde{\tau}_1>\tilde{\tau}_{sing}$ and $\tilde{\tau}_{sing}$ is
\begin{eqnarray}
\int_{\tilde{\tau}_{sing}}^{\tilde{\tau}_1}\dd \tilde{\tau}&=&2\sqrt{\frac{V_0}{3}}\int_{t_{sing}}^{t_1} \frac{\dd t}{v(t)}\nonumber\\&=&\sqrt{\frac{V_0}{3\lambda}}\log(\frac{2\lambda t_1}{\abs{\pi_\varphi}}+\sqrt{\frac{4\lambda^2t_1^2}{\pi_\varphi^2}-1})<\infty\,.
\end{eqnarray}
Note that we have set $t_0=0$ for simplicity. This calculation can be adapted to the $\lambda<0$ case. For the $\lambda=0$ case it is even easier:
\begin{equation}
\int_{\tilde{\tau}_{sing}}^{\tilde{\tau}_1}\dd \tilde{\tau}=2\sqrt{\frac{V_0}{3}}\int^{t_1}_{t_{sing}}\frac{\dd t}{v(t)}=2\sqrt{\frac{V_0 t_1}{3\abs{\pi_\varphi}}}<\infty\,.
\end{equation}
In conclusion, a commoving observer reaches $v=0$ in finite proper time, which implies the existence of a big bang singularity (in the past of the commoving observers) and a big crunch singularity (in the future of commoving observers).

As classically GR is covariant, we can express the dynamic quantities as functions of $\varphi$, which is a globally well-defined clock as long as $\pi_\varphi\neq 0$. We will assume this throughout the thesis. Once again we have to distinguish between the different signs of $\lambda$. For $\lambda>0$:
\begin{equation}
v(\varphi)=\frac{\abs{\pi_\varphi}}{\sqrt{\lambda}\abs{\sinh(\varphi-\varphi_0)}}, \hspace*{3mm} t(\varphi)=-\frac{\pi_\varphi}{2\lambda}\coth(\varphi-\varphi_0) + t_0\, ,
\end{equation} 
and for $\lambda<0$:
\begin{equation}
v(\varphi)=\frac{\abs{\pi_\varphi}}{\sqrt{-\lambda}\abs{\cosh(\varphi-\varphi_0)}}, \hspace*{3mm} t(\varphi)=-\frac{\pi_\varphi}{2\lambda}\tanh(\varphi-\varphi_0) + t_0\, .
\end{equation} 

\begin{figure}
\centering
\begin{subfigure}{1\textwidth}
\centering
\includegraphics[scale=0.75]{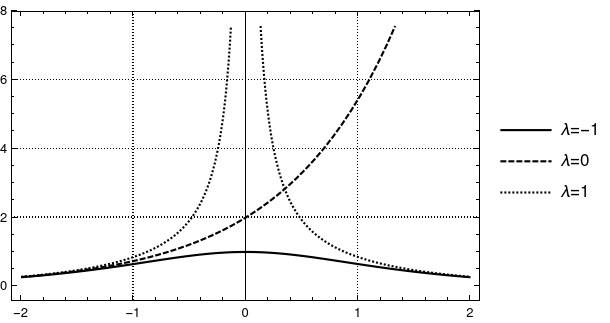}
\caption{$v$ as a function of $\varphi$ for several values of $\lambda$.}
\end{subfigure}
\begin{subfigure}{1\textwidth}
\centering
\includegraphics[scale=0.75]{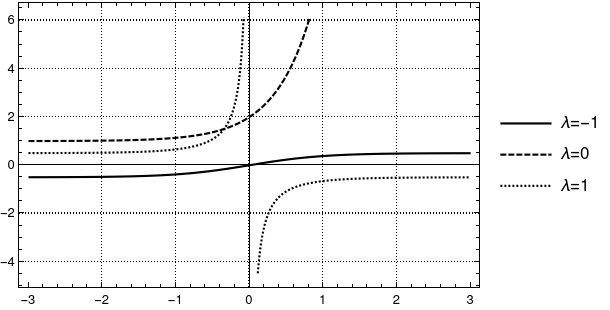}
\caption{$t$ as a function of $\varphi$ for several values of $\lambda$}
\end{subfigure}
\caption[Classical solutions as functions of $\varphi$.]{We have represented the evolution of the principal dynamical variables using $\varphi$ as a clock. The constants chosen are $t_0=\varphi_0=0$. It is interesting to see that the singularity ($v=0$) is pushed to $\infty$. In the $\lambda=0$ case, we have chosen to represent only the expanding curve, the collapsing one is symmetric with respect to the vertical axis. In this case $t_0=0$ is a singular value, so we have chosen $t_0=1$. In the limit $t_0=0$ the curves still have the same qualitative forms.}
\label{clasplot-vphi}
\end{figure}

The case $\lambda=0$ is rather different. In this case $\varphi$ takes all values from $-\infty$ to $\infty$ in a single branch (either the expanding or the contracting one). Hence we have to restrict ourselves to parametrise half of the full solution \cref{eqsm-t-null}. The reason behind this is that instead of reaching a constant value $\varphi_0$ at large volume, the field grows logarithmically when $\lambda=0$. The equation of motion for $v$ is
\begin{equation}
\dv{v}{\tau}=\lbrace v,\mathcal{H}\rbrace=-2\tilde{N}\pi_v\, .
\end{equation}
Given the fact that $\tilde{N}>0$, the sign of $\pi_v$ is what determines whether the universe is expanding or contracting. In particular, if $\lambda\geq 0$, the sign of $\pi_v$ does not change during the evolution. The two possible solutions for $\lambda=0$ are 
\begin{equation}
v(\varphi)=2\sqrt{\abs{t_0 \pi_\varphi}}e^{-\sgn(\pi_v\pi_\varphi)(\varphi-\varphi_0)}, \hspace*{3mm} t(\varphi)=t_0-\abs{t_0}\sgn(\pi_v)e^{-2\sgn(\pi_v\pi_\varphi)(\varphi-\varphi_0)}\, .
\end{equation}
These classical curves are represented in \cref{clasplot-vphi}. 

We also want to study the clock $v$. As we can see from \cref{eqsm-t-posneg}, $v$ is not a good clock everywhere. In fact, for $\lambda \geq 0$, $v$ is a good clock in each branch separately. If $\lambda<0$, the universe has a turnaround, hence $v$ is not a valid clock around that point. The expressions are different depending on the sign of $\lambda$. For $\lambda \neq 0$ we have
\begin{equation}
t(v)=t_0-\sgn(\pi_v)\frac{1}{2}\sqrt{\frac{v^2}{\lambda}+\frac{\pi_\varphi^2}{\lambda^2}}, \hspace{2mm} \varphi(v)=\varphi_0+ \log\abs{\frac{\pi_\varphi}{\sqrt{\abs{\lambda}}v}+\sqrt{\frac{\pi_\varphi^2}{\abs{\lambda} v^2}+\sgn{\lambda}}}\, .
\label{classol-v}
\end{equation}
Depending on the sign of $\pi_\varphi$, $\varphi(v)$ is either and increasing or decreasing. Note that for $\lambda<0$, the terms inside the square roots are only well-defined if $v<\frac{\abs{\pi_\varphi^2}}{\sqrt{\abs{\lambda}}}$ which is the maximum volume reached in the evolution. When $v$ reaches this value \cref{classol-v} is no longer well-defined and in any interval around that point $v$ fails to be a good clock (good in the sense of monotonic). Note that in the turnaround $\frac{\dd v}{\dd \tau}=0$ and hence $\pi_v=0$.  Once again the case $\lambda=0$ is slightly different:
\begin{equation}
\varphi(v)=\varphi_0-\sgn(\pi_\varphi\pi_v)\log(\frac{v}{2\sqrt{\abs{t_0\pi_\varphi}}}), \hspace{3mm} t(v)=t_0-\sgn(\pi_v)\frac{v^2}{4\abs{\pi_\varphi}}\,.
\label{classol-v2}
\end{equation}

\begin{figure}
\centering
\begin{subfigure}{1\textwidth}
\centering
\includegraphics[scale=0.75]{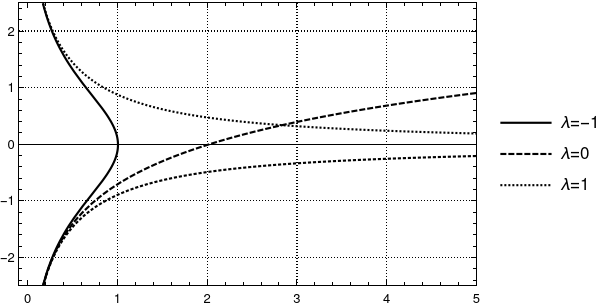}
\caption{$\varphi$ as a function of $v$ for different values of $\lambda$.}
\end{subfigure}
\begin{subfigure}{1\textwidth}
\centering
\includegraphics[scale=0.75]{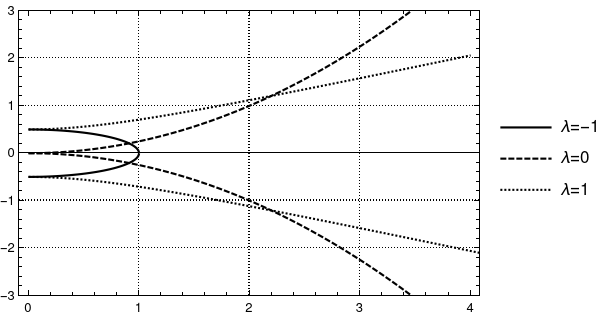}
\caption{$t$ as a function of $v$ for different values of $\lambda$.}
\end{subfigure}
\caption[Classical solutions as functions of $v$.]{Here are represented all the dynamical variables as a function of $v$. It is clear from these graphs that $v$ is not a good clock everywhere. We have chosen $\pi_\varphi=1$ and $t_0=\varphi_0=0$ except for the curve $\varphi(v)$ when $\lambda=0$, where we chose $t_0=1$. In this case, $t_0=0$ is not well-defined. For clarity, as $\varphi(v)$ takes values on the whole real line for any value of $\sgn{(\pi_v)}$ in the case $\lambda=0$, we have chosen to represent only one of the two resulting curves to avoid crossings.}
\label{classplot-phiv}
\end{figure}

The curves \cref{classol-v} and \cref{classol-v2} are plotted in figure \cref{classplot-phiv}. The $v$ clock is rather special: it is defined to take values from $0$ to $\infty$ and the sign of $\pi_v$ is what tell us whether we lie in the contracting or expanding universe.

\section{Dirac observables}
\label{diracobs}
Our system has a local gauge symmetry under time reparametrisations (a remanent of the full diffeomorphism symmetry of GR). Given that all observables must be gauge invariant, it seems that only constants of motion and quantities of the form ``the value of $x$ when my time coordinate has value $y$'', are the only possible local observers \cite{Unruh1989}. However, there is a way  to incorporate time evolution: building relational observables. These quantities express how some phase space variables vary as functions of other phase space variables. As our model is homogeneous, we only need one relational coordinate, which acts as a clock. This gives a notion of relational evolution \cite{Dittrich,Rovelli,Tambornino2011}.  In this section we define the Dirac observables that will be important when quantising our universe. As we will be working with three different clocks, we define Dirac observable with respect to all of them.

The Hamiltonian constraint \cref{hamconst} defines a constrained surface $\mathscr{C}$ where the solutions of the equations of motion evolve. Dirac observables $\mathcal{O}$ are functions of the phase space variables that are invariant under the flow generated by the constraint $\C$:
\begin{equation}
\poisket{\C}{\mathcal{O}}\approx 0\, ,
\end{equation}
where $\poisket{\cdot}{\cdot}$ is the Poisson bracket of the system and the notation $\approx$ is a standard notation that means that the equality only has to hold on the constrained surface $\mathscr{C}$ rather than on the whole phase space (in which case we would use the equal sign). This apparently subtle detail is what will allow the system to have non-trivial Dirac observables. Concretely,
\begin{equation}
\poisket{\C}{\mathcal{O}}\approx 0 \iff  -\frac{2\pi_\varphi^2}{v^3}\pdv{\mathcal{O}}{\pi_v}-\pdv{\mathcal{O}}{t}-\frac{2\pi_\varphi}{v^2}\pdv{\mathcal{O}}{\varphi}+2\pi_v\pdv{\mathcal{O}}{v} \approx 0\,.
\label{obsconst}
\end{equation}
Our Hamiltonian is of the form $\tilde{N}\C$ so $\poisket{\C}{\Ham}\approx 0$. From there we immediately deduce that any Dirac observer $\Obs$ is a constant of motion. \cref{obsconst} does not depend on $\lambda$ or $\pi_\varphi$, hence any function of these conserved quantities is a Dirac observable. However these observables are constant along a given trajectory, making them unable to account for the evolution of the system.

We are interested in the study of three clocks, $v$, $t$ and $\varphi$. For this reason we are interested in the relational observables $t(v=v_1)$, $\varphi(v=v_1)$, $v(t=t_1)$, $\varphi(t=t_1)$, $v(\varphi=\varphi_1)$, and $t(\varphi=\varphi_1)$. These observables are all constructed following the same fashion: we choose our relational  clock $x$ and we look at the values of the rest of the quantities when the clock takes the fixed value $x_1$. Because $x_1$ is fixed, all these observables are constants of motion, however by making $x_1$ vary, we can construct a complete family of observables that parametrises the evolution of the system. 

Let us begin by fixing a starting point $P$ in the constrained surface $\mathscr{C}$. $P$ has coordinates $(t_j,v_j,\varphi_j, \lambda, \pi_{v_j}, \pi_\varphi)$; this initial data is enough to determine uniquely a trajectory in the constraint surface. If we decide to use $t$ as clock, we can ask ourselves what is the value of the volume or the field when $t$ shows a value $t_1$. The volume is a very important observable because the classical singularity lies at $v=0$. The expression for this observable is:
\begin{equation}
v(t=t_1)=\left\lbrace \begin{array}{cc}
\sqrt{-\frac{\pi_\varphi^2}{\lambda}+4\lambda\left(t_1-t_{{j}}-\frac{v_{{j}}\pi_{v_{{j}}}}{2\lambda}\right)^2}, & \lambda<0 \ \textrm{or} \ \lambda >0\\[3mm]
 2\sqrt{|\pi_\varphi|\left|t_1-t_{{j}}+\frac{v_{{j}}^2}{4\pi_\varphi}\right|},& \lambda=0.
\end{array} \right.
\label{voft}
\end{equation} 
The first expression is only valid when the argument of the square root is positive; as we saw earlier, the classical solutions reach singularities at some finite value of $t$ beyond which they would not be defined. Note that, logically the observables are defined such that when $t_1=t_j$ $v(t=t_j)=v_j$.  The expression for $\varphi(t=t_1)$ is
\begin{equation}
\varphi(t=t_1)=\left\lbrace
\begin{array}{cc}
\arcoth\left(-\frac{2\lambda}{\pi_\varphi}\left(t_1-t_j-\frac{v_j\pi_{v_j}}{2\lambda}\right)\right)-\artanh(\frac{\pi_\varphi}{v_j\pi_{v_j}})+\varphi_i, & \lambda>0\\[3mm]
\artanh\left(-\frac{2\lambda}{\pi_\varphi}\left(t_1-t_j-\frac{v_j\pi_{v_j}}{2\lambda}\right)\right)-\artanh(\frac{v_j\pi_{v_j}}{\pi_\varphi})+\varphi_i,  & \lambda<0 \\[3mm]
-\frac{1}{2}\sgn(\pi_\varphi\pi_{v_j})\log(\frac{4\pi_{v_j}(t_1-t_j)}{v_j}+1)+\varphi_j, &\lambda=0
\end{array} \right.
\end{equation}
Note $\varphi(t=t_1)$ is defined only when the absolute value of what is inside the hyperbolic tangent/cotangent is smaller/bigger than 1, which is an equivalent condition than what we saw for $v$. Outside this range, this observable is simply not defined. A similar condition holds for the case $\lambda=0$ as the logarithm has to be positive.

However, we are also interested in other clocks, such as $v$. As we saw earlier, this clock is not valid throughout the entire evolution of the system. Therefore, to match the numerical analysis of the following section, we choose to focus on a Dirac observable valid in the expanding branch of the universe
\begin{equation}
t(v=v_1)=\left\lbrace \begin{array}{cc}
t_{{ j}}-\frac{v_{{j}}\abs{\pi_{v_{{ j}}}}}{2\lambda}+\frac{1}{2}\sqrt{\frac{v_1^2}{\lambda}+\frac{\pi_\varphi^2}{\lambda^2}}, & \lambda<0 \ \textrm{or} \ \lambda >0\\[3mm]
\frac{v_1^2-v_{{ j}}^2}{4|\pi_\varphi|}+t_{{ j}},& \lambda=0 .
\end{array} \right. 
\label{tofv}
\end{equation} 
and
\begin{equation}
\varphi(v=v_1)=\left\lbrace\begin{array}{cc}
\sgn(\pi_{v_j})\arsinh\left(\frac{\pi_\varphi}{v_j\sqrt{\lambda}}\right)-\sgn(\pi_{v_j})\arsinh\left(\frac{\pi_\varphi}{v_1\sqrt{\lambda}}\right)+\varphi_i, & \lambda>0 \\[3mm]
\sgn(\pi_{v_j}\pi_\varphi)\arcosh\left(\frac{\abs{\pi_\varphi}}{v_j\sqrt{-\lambda}}\right)-\sgn(\pi_{v_j}\pi_\varphi)\arcosh\left(\frac{\pi_\varphi}{v_1\sqrt{-\lambda}}\right)+\varphi_i, & \lambda<0\\[3mm]
-\sgn(\pi_\varphi\pi_{v_j})\log(\frac{v_1}{v_j})+\varphi_i, & \lambda=0
\end{array} \right.
\end{equation}
 Note that for $\lambda<0$ the expressions are only valid if $v_1\leq \frac{\abs{\pi_\varphi}}{\sqrt{-\lambda}}$ which is the upper bound of the volume. The sign of $\pi_{v_j}$ is what determines which curve we are at (see \cref{classplot-phiv}).

Finally, we present the relevant Dirac observables for the $\varphi$ clock:
\begin{equation}
v(\varphi=\varphi_1)=\left\lbrace \begin{array}{ll}
\displaystyle{\frac{\abs{\pi_\varphi}}{\sqrt{\lambda}\abs{\sinh\left(\varphi_1-\varphi_{{j}}+\sgn{\pi_{v_{{j}}}}{ \arsinh}\left(\frac{\pi_\varphi}{v_{{j}} \sqrt{\lambda}}\right)\right)}}}\,, & \lambda>0 \\
\displaystyle{\frac{\abs{\pi_\varphi}}{\sqrt{-\lambda}\,\cosh\left(\varphi_1-\varphi_{{j}}+\sgn{\pi_\varphi\pi_{v_{{j}}}}{\arcosh}\left(\frac{\abs{\pi_\varphi}}{v_{{j}} \sqrt{-\lambda}}\right)\right)}}\,, & \lambda<0 \\
\displaystyle{v_{{j}} e^{-\sgn{\pi_{\varphi}\pi_{v_{{j}}}}(\varphi_1-\varphi_{{j}})}}\,, & \lambda =0
\end{array} \right.
\label{vofphi}
\end{equation}
and 
\begin{equation}
t(\varphi=\varphi_1)=\left\lbrace \begin{array}{ll}
\displaystyle{-\frac{\pi_\varphi}{2\lambda}\coth\left(\varphi_1-\varphi_{{j}}+{\artanh}\left( \frac{\pi_\varphi}{\pi_{v_{{j}}} v_{{j}}} \right) \right)}+t_{{j}}+\frac{ \pi_{v_{{j}}}v_{{j}}}{2\lambda}\,, & \lambda>0 \\
\displaystyle{-\frac{\pi_\varphi}{2\lambda}\tanh\left(\varphi_1-\varphi_{{j}}+{\artanh}\left( \frac{\pi_{v_{{j}}}v_{{j}}}{\pi_\varphi} \right) \right)+t_{{j}}+\frac{\pi_{v_{{j}}}v_{{j}}}{2\lambda}}\,, & \lambda<0 \\
 \displaystyle{\left(t_{{j}}+\frac{v_{{j}}}{4\pi_{v_{{j}}}}\right)\left(1-\frac{v_{{j}}}{4t_{{j}}\pi_{v_{{j}}}+v_{{j}}}e^{-2\,\sgn{(\pi_\varphi\pi_{v_{{j}}})}(\varphi_1-\varphi_{{j}})}\right)}\,, & \lambda=0\,.
\end{array}\right.
\label{tofphi}
\end{equation}

All these Dirac observables have something in common that distinguish them from the simple constants $\pi_\varphi$ and $\lambda$. For $\pi_\varphi$ and $\lambda$ we have $\poisket{\C }{\pi_\varphi}=\poisket{\C}{\lambda}=0$, i.e., the Poisson bracket is 0 everywhere, not just in the constraint surface. Conversely, for $t(v=v_1)$, we find $\poisket{\C}{t(v=v_1)}=f_{t(v=v_1)}\C$, where $f_{t(v=v_1)}$ is a non-trivial function of the phase space variables. A similar non-trivial function may be found when doing the Poisson bracket of the Dirac observable defined above. Of course, in the constraint surface $\mathscr{C}$ all Poisson bracket vanish, but the Poisson brackets of \cref{voft}--\cref{tofphi} are not trivial on the whole phase space.

Let us do a quick summary of this chapter. First of all, we have deduced the action of our model from GR (in the case where the perfect fluid is not dark energy), and then from unimodular gravity (in the case where the perfect fluid is dark energy). Then, we have chosen coordinates such that the Hamiltonian constraint has two quadratic momenta parametrised with the Rindler metric and a linear term related to the perfect fluid energy density. After that, we were able to calculate the equations of motion of all the dynamical variables using different clocks and plot them. This allowed us to have a very good picture on how the model behaves classically. Finally, we have analysed the classical phase space of this model and found out the relevant Dirac observables. This concludes the classical analysis of the model, as we have all the tools we need to start analysing the quantisation of this universe. 
\chapter{Quantisation(s) of the model}
\label{quant}
\section{The Wheeler--DeWitt equation}
In this section we present the derivation of the Wheeler--DeWitt equation of the model.  Recall this quantisation scheme is one of the first methods used for cosmological models, being developed in the sixties \cite{DeWitt1967}, and still widely used today. In our case, the Wheeler--DeWitt equation is solvable analytically; in this chapter we introduce its solutions and analyse the resulting quantum theories focusing especially on unitarity.

In order to obtain the Wheeler--DeWitt equation of the model we need to recall the Hamiltonian constraint \cref{hamconst}
\begin{equation}
\C = -\pi_v^2+\frac{\pi^2_\varphi}{v^2}+\lambda=g^{AB}\pi_A\pi_B+\lambda=0\,  ,
\end{equation}
where $g^{AB}$ is the inverse metric on the Rindler wedge parametrised by $v$ and $\varphi$. To quantise our universe, we replace the term $g^{AB}\pi_A\pi_B$ with $-\hbar^2\square$ where $\square$ is the Laplace--Beltrami operator on the Rindler wedge:
\begin{equation}
\square=\frac{1}{\sqrt{-g}}\pdv{}{q^A}\left(g^{AB}\sqrt{-g}\pdv{}{q^B}\right)=-\frac{1}{v}\pdv{}{v}\left(v\pdv{}{v}\right)+\frac{1}{v^2}\pdv{}{\varphi^2}\, .
\end{equation}
Hence, the Wheeler--DeWitt equation is
\begin{equation}
\left( \hbar^2 \pdv{}{v^2} +\frac{\hbar^2}{v}\pdv{}{v}-\frac{\hbar^2}{v^2}\pdv[2]{}{\varphi}-i\hbar\pdv{}{t}\right)\Psi(v, \varphi, t)=0\, .
\label{wdw}
\end{equation}
Note that the term $\lambda$ has been replaced by  $-i\hbar\pdv{}{t}$. This equation is a version of \cref{Wheeler}, but applied to our model. Because we are assuming a form of the metric and of the matter components, we are left with only three degrees of freedom for our minisuperspace, one coming from the metric ($v$) and two coming from the matter components of the universe ($\varphi$ and $\lambda$). The other (infinite) degrees of freedom of the superspace have been frozen. This Wheeler--DeWitt equation is the same found in \cite{Gryb}.

Recall from the introduction that we are using the Hawking and Page ordering \cite{Hawking1985}. This makes the Wheeler--DeWitt equation covariant under a change of coordinates in the Rindler wedge parametrised by the pair $(v,\varphi)$\footnote{In general, this choice is unique up to the addition of a term $\hbar^2\xi R$ where $R$ is the Ricci scalar and $\xi$ a free parameter \cite{Ricci,Halliwell1988}. Nevertheless, in our case the minisuperspace metric is flat, hence $R=0$.}. However, in order to study the ambiguities in the Wheeler--DeWitt equation one may add free parameters that correspond to different orderings (see e.g., \cite{Steigl2005, Veira2020}), and study their impact on the resulting theory. In this thesis, we will only focus on the specific Wheeler--DeWitt equation \cref{wdw}, but how the resulting theories may depend on the ordering of the operators is another very interesting problem and open issue in quantum cosmology.

The Wheeler--DeWitt equation can be solved using a separation of variables ansatz $\Psi(v,\varphi,t)=\nu(\varphi)\psi(v)e^{i\lambda\frac{t}{\hbar}}$ which leads to two different equations:
\begin{equation}
\frac{\nu''(\varphi)}{\nu(\varphi)}=A\, ,
\label{exp}
\end{equation}
and 
\begin{equation}
v^2\psi''(v)+v\psi'(v)+\left(\frac{\lambda}{\hbar^2}v^2-A\right)=0\, .
\label{bessel}
\end{equation}
Note that all functions here are one dimensional. The prime notation corresponds to taking the derivative with respect to that parameter.

The first equation is straight forward to solve and its solutions are real exponentials $e^{\kappa\varphi}$ if $A>0$, and imaginary exponentials $e^{ik \varphi}$ if $A<0$. Equation \cref{bessel} is known as the Bessel equation and its solutions are Bessel functions $\J{\pm \abs{\kappa}}{\lambda}$ and $\J{\pm i\abs{k}}{\lambda}$, again depending on the sign of $A$. In conclusion, the general solution to the Wheeler--DeWitt equation \cref{wdw} is:
\begin{align}
\Psi(v,\varphi,t)&=&\int_{-\infty}^{\infty} \dl \int_{-\infty}^{\infty} \dk e^{ik\varphi}e^{i\lambda\frac{t}{h}}\left[ \alpha(k,\lambda)\J{i\abs{k}}{\lambda}+\beta(k,\lambda)\J{-i\abs{k}}{\lambda}\right] \nonumber \\
&+&\int_{-\infty}^{\infty} \dl \int_{-\infty}^{\infty} \dkap e^{\kappa\varphi}e^{i\lambda\frac{t}{h}}\left[ \gamma(\kappa,\lambda)\J{\abs{\kappa}}{\lambda}+\epsilon(\kappa,\lambda)\J{-\abs{\kappa}}{\lambda}\right]\, .
\label{general-sol}
\end{align}
$\alpha$, $\beta$, $\gamma$ and $\epsilon$ are free complex functions. For $\lambda<0$ we follow the convention
\begin{equation}
J_x\left(\frac{\sqrt{\lambda}}{\hbar}v\right)=J_x\left(\frac{i\sqrt{-\lambda}}{\hbar}v\right)=e^{\frac{ix\pi}{2}}I_x\left(\frac{\sqrt{-\lambda}}{\hbar}v\right)\, ,
\label{Iconvention}
\end{equation}
where $x$ is any possible order and $I_x$ is the modified Bessel function of the first kind. Functions like \cref{general-sol} will often be referred as wave functions of the universe.

An interesting feature of this model is that the universe is in a superposition of cosmological constants $\lambda$. This comes from the fact that we are working with unimodular gravity rather than simply GR. In other energy interpretations, like dust, this means that the universe is in a superposition of mass parameter $m$, where $m$ can have negative values (hence implying tachyons).

The Wheeler--DeWitt equation \cref{wdw} and its general solution \cref{general-sol} are not enough for a reasonable interpretation of the theory. In particular, we would like our universe to evolve in ``time''. As the Wheeler--DeWitt equation is simply a differential equation we first choose one of the internal dynamical variables to play the r\^ole of clock, just as we did classically, following the \emph{tempus post quantum} prescription. This will allow us to find a physical interpretation of the wave function of the universe. Nonetheless, this interpretation depends on the chosen clock variable. In the next section we present briefly the differences of the clock choices before diving deeper in the calculations.

\section{The different interpretations of the Wheeler--DeWitt equation and unitarity}
\label{unit}
Let us first introduce what we mean by choosing an internal variable as clock. We first quantise the universe by finding the Wheeler--DeWitt equation, and then we choose one of the variables as our clock. Once this is done, we build a Hilbert space on the remaining variables by choosing an appropriate inner product (the meaning of appropriate will be explained shortly). Finally, we restrict ourselves to the subspace of \cref{general-sol} whose norm is conserved under the chosen inner product. This step is key to have a well-defined notion of probability distribution. The clock choice gives a conditional time interpretation: knowing the value of the clock, what is the value of an observable quantity? In this way, despite being an internal variable of the system, the clock is treated like an external quantum mechanical time parameter, i.e., it has no uncertainty. As general relativity is a covariant theory, no clock has any advantage with respect to the others besides being globally monotonic rather than locally.

We first analyse $t$ as clock. By defining the Hamiltonian
\begin{equation}
\hat{\Ham} =\hbar^2\left(-\pdv[2]{}{v}-\frac{1}{v}\pdv{}{v}+\frac{1}{v^2}\pdv[2]{}{\varphi}\right)\, ,
\label{hams}
\end{equation}
the Wheeler--DeWitt equation looks like a Schr\"odinger equation using $t$ as clock:
\begin{equation}
i\hbar\pdv{}{t}\Psi(t, \varphi, t)=-\hat{\Ham} \Psi(t,\varphi,t)\, .
\label{schr}
\end{equation}
Replacing $-i\pdv{}{t}$ by $\lambda$ here, we see that the values of $\lambda$ are the eigenvalues of \cref{hams}. This structure suggests using a standard $L^2$ inner product 
\begin{equation}
\braket{\Psi}{\Phi}_t=\int_0^\infty \dd v \int_{-\infty}^\infty \dd \varphi \ v\bar{\Psi}(t,\varphi,v)\Phi(t,\varphi,v)\, .
\label{tinnerprod}
\end{equation}
The subindex $t$ will serve to distinguish this inner product to the ones from the other theories. As the Wheeler--DeWitt equation only contains first order derivatives in $t$, there are no time derivatives in the inner product. The Hilbert space considered here is then $L^2(\mathcal{R},\sqrt{-g} \dd v \dd \varphi )$ where $\mathcal{R}$ is the Rindler wedge in $v$ and $\varphi$ and $\sqrt{-g}=v$ the determinant of the Rindler wedge metric. This measure is covariant under changes of coordinates in the Rindler wedge. The interpretation of the Wheeler--DeWitt equation as a Schrödinger equation in unimodular gravity has been one of the reasons to propose that unimodular gravity as a solution to the problem of time \cite{Unruh1989}. Indeed, usually the action is quadratic in the momenta, which leads to second order derivates. Unimodular gravity is an easy way to add a linear term in the action. This term can later be interpreted as the energy of the Hamiltonian, like we are doing here. 

Now that we have constructed the basis of the quantum theory we make a crucial assumption: We would like this theory to be unitary. This means that inner products $\braket{\Psi}{\Phi}_t$ should be preserved over time, i.e., $\pdv{}{t}\braket{\Psi}{\Phi}_t=0$. Unitarity, as we recalled in the introduction, is one of the principles of QM, and it allows us to have a well-defined notion of probability interpretation of the quantum theory. Contrary to what we can think at first glance, our theory is not automatically unitary, rather we have to impose an extra condition:
\begin{equation}
\pdv{}{t}\braket{\Psi}{\Phi}_t=0 \iff \int_{-\infty}^\infty\dd \varphi \left[ v\left( \bar{\Psi}\pdv{}{v}\Phi-\Phi\pdv{}{v}\bar{\Psi}\right)\right]_{v=0}^{v=\infty}=0\, .
\label{bound1}
\end{equation}
Hence not all square integrable solutions of \cref{wdw} satisfy \cref{bound1}, one has to find a subspace of $L^2(\mathcal{R}, \sqrt{-g} \dd v \dd \varphi)$ where the boundary condition holds. This boundary condition is non-trivial only in the limit $v=0$ (more on this can be found in \cref{t-clock-sec}).

Why are we finding a boundary condition? The operator $\hat{\Ham}$ \cref{hams} is not self adjoint with respect to the inner product \cref{tinnerprod}. The technicalities around self-adjoint extensions of an operator are presented in \cref{selfadj}. Curious and mathematically inclined readers can consult them. The key thing to take out is that despite the fact that $\hat{\Ham}$ is not self-adjoint, it possesses a one parameter family of self-adjoint extensions. Hence we must restrict the solutions to the Wheeler--DeWitt equation to solutions to the boundary condition \cref{bound1}.

Under a change of coordinates $\Psi(v,\varphi, t)=v^{-\frac{1}{2}}e^{ik\varphi}e^{i\lambda\frac{t}{\hbar}}w(v)$ the Wheeler--DeWitt equation can be transformed in a Schr\"odinger equation with a radial $1/r^2$ potential:
\begin{equation}
-\hbar^2 \pdv[2]{}{v}w(v)-\hbar^2\frac{k^2+\frac{1}{4}}{v^2}w(v)=\lambda w(v)\, .
\label{schr-potential}
\end{equation}
The properties of this potential are well known \cite{rsquare}. Depending on the (dimensionless) strength of the potential one distinguishes different cases. We are in the strongly attractive case ($-(k^2+\frac{1}{4})<0$), which requires a boundary condition at $v=0$ to make $\hat{\Ham}$ self-adjoint. If we consider a particle moving under this potential, it would reach $v=0$ in a finite proper time. The normalisable solutions to \cref{wdw} that satisfy \cref{bound1} are very non-trivial to calculate, and they will be presented in \cref{t-clock-sec}.  

In this section we are interested in the different interpretations of the Wheeler--DeWitt equation \cref{wdw}, so let us choose another clock, like $\varphi$. If we multiply the Wheeler--DeWitt equation by a factor $v^2$, we obtain
\begin{equation}
\left(\hbar^2\pdv[2]{}{\varphi}-\hbar^2\left(\pdv{}{\log(v/v_0)}\right)^2 +i\hbar v^2 \pdv{}{t}\right)\Psi(v,\varphi,t)=0\, ,
\label{wdw2}
\end{equation}
where $v_0$ is a constant introduced to cancel the dimension of $v$. In principle, as equations \cref{wdw} and \cref{wdw2} have the same solutions, they are equivalent. \cref{wdw2} looks like a Klein Gordon equation with respect to $v$ and $\varphi$ with an added potential term. Contrary to Schr\"odinger equations, Klein--Gordon type equations require inner products with time (in this case $\varphi$) derivatives. Hence, the inner product we are working with is 
\begin{equation}
\braket{\Psi}{\Phi}_\varphi= i \int_{-\infty}^\infty \dd t \int_0^\infty \frac{\dd v}{v} \left(\bar{\Psi}(v,\varphi,t)\pdv{}{\varphi}\Phi(v,\varphi,t)-\Phi(v,\varphi,t)\pdv{}{\varphi}\bar{\Psi}(v,\varphi,t) \right)\, .
\label{phiinnerprod}
\end{equation}
Once again, the inner product is a function on the phase space variables $v$ and $t$, but not $\varphi$, which is the clock of the theory. However, the Hilbert space considered is not a simple $L^2$, rather the functions with finite norm with the chosen inner product. Note that this inner product is also invariant under a change of coordinates in $(t,v)$. The differential operator appearing in \cref{phiinnerprod} is $n^\mu\partial_\mu=\frac{1}{v}\partial_\varphi$ which is the normal to the $\varphi=$ const surfaces in the Rindler wedge metric. 

Once more, we require that our inner product is unitary, i.e., $\pdv{}{\varphi}\braket{\Psi}{\Phi}_\varphi=0$. This leads to another boundary condition
\begin{equation}
\pdv{}{\varphi}\braket{\Psi}{\Phi}_\varphi=0 \iff \int_{-\infty}^\infty \dd t\left[ v\left( \bar{\Psi}\pdv{}{t}\Phi-\Phi\pdv{}{t}\bar{\Psi}\right)\right]_{v=0}^{v=\infty}=0\, .
\label{bound2}
\end{equation}
This condition is very similar to \cref{bound1}, but has different consequences. This time, the boundary condition is non-trivial in the limit $v=\infty$. With the ansatz $\Psi(v,\varphi,t)=\psi(v,\varphi)e^{i\lambda\frac{t}{\hbar}}$ we can see that this boundary condition is equivalent to require self-adjointness of the operator
\begin{equation}
\hat{\mathcal{G}}=-\hbar^2\pdv[2]{}{u}-\lambda v_0^2e^{2u}\, ,
\label{ham2}
\end{equation}
with respect to a standard $(L^2,\dd u)$ inner product parametrised by the coordinate $\log(\frac{v}{v_0})=u$. For $\lambda>0$, this Hamiltonian contains an attractive potential such that classically, a particle can reach $u=\infty$ (and hence $v=\infty$) in a finite time. Thus, in the quantum theory we have to impose a boundary condition at $v=\infty$. The characterisation of this Hamiltonian and its self-adjoint extensions has already been studied in \cite{Fredenhagen2003, Kobayashi1996}. We will re-derive their results in detail in \cref{phi-clock-sec}. 

In the $\lambda<0$ case, the potential term in \cref{ham2} is repulsive, making the Hamiltonian already self-adjoint without the need for additional boundary conditions. This case has also been analysed \cite{DHoker1982}.

In a nutshell, we had to add an extra condition on the allowed wave functions in order to have a unitary theory for the clocks $t$ and $\varphi$. What happens with the remaining clock $v$?

The equation \cref{wdw2} can be seen as a Klein--Gordon equation in $v$ or rather $\log(v/v_0)$. Hence, to use $v$ as clock we define the following inner product
\begin{equation}
\braket{\Psi}{\Phi}_v= i \int_{-\infty}^\infty \dd t \int_{-\infty}^\infty \dd \varphi \left[ v\bar{\Psi}(v,\varphi,t)\pdv{}{v}\Phi(v,\varphi,t) -v\Phi(v,\varphi,t)\pdv{}{v}\bar{\Psi}(v,\varphi,t)\right]\, .
\label{vproduct}
\end{equation}
This time, the differential operator used is $n^\mu\partial_\mu=\partial_{\log(v/v_0)}$ which is normal to the $v=$ const surfaces. Conservation of the inner product is equivalent to self-adjointness of the operator
\begin{equation}
\hat{\mathcal{F}}=-\hbar^2\pdv[2]{}{\varphi}-i\hbar v^2\pdv{}{t}\, ,
\label{Fop}
\end{equation}
for an $(L^2, \dd t \dd \varphi)$ inner product. Note that contrary to the last two cases, $\hat{\mathcal{F}}$ is time (in this case $v$) dependent. However, this operator is already self-adjoint, so all solutions to the Wheeler--DeWitt equation have a time independent norm. 

Even before calculating the allowed states for each theory we appreciate how the choice of clock has many non-trivial effects. We can already see from here that covariance is broken from the choice of clock; the different boundary conditions mean that the quantum theories are inequivalent. It is worth mentioning that the different boundary conditions come from ensuring unitarity of the quantum theory. Unitarity is sometimes not applied as a fundamental concept when studying phenomenology of quantum universes: one writes the Wheeler--DeWitt equation, finds some solution and calculates relevant quantities. Unitarity may then be required to emerge at the semiclassical level. What we are finding here is that the study of the Hilbert space of the theory is important. In particular, it is important to mention that unitarity is not given, even constructing the Hilbert space in the most covariant and systematic way. If we believe that unitarity is a defining feature of a such cosmological models, we have to add it by hands. However, one could argue that this issue is an artefact of the framework we are working with. What happens when instead of choosing a clock we Dirac quantise our model?
\section{Dirac quantisation}
\label{dirac-sec}

The Dirac quantisation is often defined as the ``clock--neutral'' quantisation \cite{Hoehn}, as it supposed to treat all dynamical variables identically. Therefore, this approach is sometimes considered preferable with respect to the relational one. The Dirac quantisation programme works as follows: one builds a kinematical Hilbert space $\mathscr{H}_{\text{kin}}$, which corresponds to the quantisation of the system without imposing the constraints. As this Hilbert space is too big, the question is then how to implement the constraints, find the physically relevant wave functions, and provide the system with a notion of time evolution. This approach is historically important, as the canonical formulation of LQG is based on the Dirac quantisation \cite{Ashtekar2021}. The Dirac quantisation is also used for LQC models \cite{Ashtekar2011}.

The constraints are implemented as operators (how to build these operators is subject to the usual ordering ambiguities) and the physical states are those that are annihilated by the constraints (see \cite{Tate1992} for a comprehensive introduction). Constructing the physical Hilbert space and its corresponding inner product involves a careful analysis of the spectrum of the constraints. Depending on the system, finding explicitly the physical Hilbert space $\hphys$ is quite challenging mathematically and there is not an established answer to this problem \cite{Marolf1995,Marolf2000}.

Usually, in order to implement the quantisation process, the constraints have to be self-adjoint operators on $\hkin$. For an example of this requirement in a very similar model to ours see \cite{Pawlowski2011}. This model features a universe with scalar field and a fixed positive cosmological constant (as opposed to ours where the cosmological constant can vary between two different trajectories in phase space). In this model, the Hamiltonian was required to be self-adjoint on the kinematical Hilbert space leading to a boundary condition very similar to ours. (We will see in the next sections that these two models have also very similar dynamics).

Let us consider again the Wheeler--DeWitt equation \cref{wdw}:
\begin{equation}
\left( \hbar^2 \pdv[2]{}{v} +\frac{\hbar^2}{v}\pdv{}{v}-\frac{\hbar^2}{v^2}\pdv[2]{}{\varphi}-i\hbar\pdv{}{t}\right)\Psi(v, \varphi, t)=0\, .
\label{wdw-re}
\end{equation}
This is the quantum version of the Hamiltonian constraint and must be viewed as the operator $\hat{\mathcal{C}}_1$ acting on the space of solutions:
\begin{equation}
\hat{\mathcal{C}}_1\Psi(v,\varphi,t)=0, \hspace{4mm} \hat{\mathcal{C}}_1:=\hbar^2 \pdv{}{v^2} +\frac{\hbar^2}{v}\pdv{}{v}-\frac{\hbar^2}{v^2}\pdv[2]{}{\varphi}-i\hbar\pdv{}{t}
\end{equation}
The kinematical Hilbert space is $L^2(\mathcal{M}, v \ \dd t \dd \varphi \dd v)$ where $\mathcal{M}=\mathcal{R}\times \R$ with $\mathcal{R}$ the Rindler wedge in $\varphi$ and $v$ and the measure $v \dd t \dd \varphi \dd v$ is the one inherited from the Rindler wedge metric. With this measure, the kinematical inner product is invariant under a change of coordinates. The kinematical inner product is then straight forward
\begin{equation}
\braket{\Psi}{\Phi}_{\text{kin}_1}=\int_{-\infty}^\infty \dd t \ \dd \varphi \int_0^\infty \dd v\ v \bar{\Psi}(v, \varphi, t) \Phi(v,\varphi,t) \, .
\end{equation}
Square integrable functions under this inner product are functions of the form $\Psi=v^{-\frac{1}{2}}\Upsilon$ where $\Upsilon$ is square integrable in the standard $L^2$ inner product on $\R^2\times \R_+$. By doing this change of coordinates and Fourier transforming in $\varphi$ and $t$, we find that self-adjointness of $\hat{\mathcal{C}}_1$ is equivalent to require self-adjointness of the operator
\begin{equation}
\hat{\mathcal{D}}_1=\hbar^2\left(-\pdv[2]{}{v}-\frac{\frac{1}{4}+k^2}{v^2} \right)\, 
\end{equation}
with respect to the standard inner product on $L^2(\R_+, \dd v)$. This operator is exactly the same we had to impose self-adjointness to in \cref{unit} when using the $t$ clock i.e., \cref{schr-potential}. Hence, we are finding the same issue in the two approaches. Moreover, in this case Dirac quantisation and relational lead to the same theory: as the constraint is linear in $\lambda$, the momentum conjugated to $t$, the group averaging procedure would remove the integral $t$ of the measure leaving us with the inner product \cref{tinnerprod}\footnote{The group averaging procedure is a way to find a inner product for the physical Hilbert space from the inner product of the kinematical Hilbert space by (in a hand-wavy manner) applying a ``Dirac delta'' of the constraints $\int \dd \xi e^{i\hat{\mathcal{C}}\xi}$ to the kinematical inner product.}.

However, recall that in order to use $v$ and $\varphi$ as clocks, we had to multiply the Hamiltonian constraint by a phase space function, $v^2$. We obtained another Hamilton constraint
\begin{equation}
\left(\hbar^2\pdv[2]{}{\varphi}-\hbar^2\pdv[2]{}{u}+i\hbar v_0^2e^{2u} \pdv{}{t}\right)\Psi(v,\varphi,t)=0\, ,
\label{wdw2-re}
\end{equation}
where we have applied the change of variable $u=\log(v/v_0)$ so that all kinematical variables are valued over the entire real axis. Here, the operator considered is
\begin{equation}
\hat{\mathcal{C}}_2:=\hbar^2\pdv[2]{}{\varphi}-\hbar^2\pdv[2]{}{u} +i\hbar v_0^2e^{2u} \pdv{}{t}\, .
\end{equation} 
We impose $\hat{\mathcal{C}}_2\Psi=0$, the same constraint used in \cite{Pawlowski2011}, where they analysed the same model but with a fixed (non unimodular) cosmological constant with respect to the $\varphi$ clock. Naively, we would say that multiplying the Hamilton constraint by a phase function does not affect the theory as it does not affect its solutions. The second order derivatives in $\varphi$ and $u$ look like a flat Laplacian in Cartesian coordinates in a $(1+1)$ space, hence motivating the kinematical inner product
\begin{equation}
\braket{\Psi}{\Phi}_{\text{kin}_2}=\int_{-\infty}^\infty \dd t \dd \varphi \dd u \ \bar{\Psi}(u,\varphi,t)\Phi(u,\varphi,t)\, ,
\end{equation}
where we assumed that the metric on $u$ and $\varphi$ is the flat metric $\eta^{AB}$, again, by interpreting $\hat{\mathcal{C}}_2$ as the Laplace--Beltrami operator of a metric (in this case flat). This difference of metric in the two kinematical inner products comes from the rescaling induced by multiplying the Wheeler--DeWitt equation by $v^2$. 

Again by taking the Fourier transform in $\lambda$ and $\varphi$, we see that requiring that $\hat{\mathcal{C}}_2$ is self-adjoint is equivalent to asking for self-adjointness of the operator
\begin{equation}
\hat{\mathcal{D}}_2 =-\hbar^2 \pdv[2]{}{u}-\lambda v_0^2 e^{2u}
\end{equation} 
with respect to an $L^2(\R, \dd u)$ inner product for any given value of $\lambda$. Note that this operator is the same as $\hat{\mathcal{G}}$, defined in \cref{ham2}. When using $\varphi$ as clock we have to impose self-adjointness of this operator by hand.

We can conclude from this analysis that the different notions of unitarity with respect to different clocks do not arise from the relational quantisation, but from writing the Wheeler--DeWitt equation in the forms \cref{wdw-re} or \cref{wdw2-re}. In the classical theory these different Wheeler--DeWitt equations correspond to different choices of lapse function. Indeed, in \cref{hamform}, the Hamiltonian was written as 
\begin{equation}
\Ham=\tilde{N}\left[-\pi_v^2+\frac{\pi_\varphi^2}{v^2}+\lambda \right]\, ,
\end{equation}
so that \cref{wdw} corresponds to the choice $\tilde{N}=1$ and \cref{wdw2-re} to $\tilde{N}=v^2$. Classically this does not make any difference, it just corresponds to different choices of time coordinate in the metric. However, this classical symmetry of time reparametrisation is broken in the quantum theory in a subtle way: changing the lapse leads to a different kinematical inner product and then to different criteria to self-adjointness of the Hamiltonian constraints, and finally to different boundary conditions (that will lead to different dynamics as we will see in the following chapters).

Recently, a framework for clock changing in quantum gravity has been developed \cite{Hoehn}. Our findings are not in contradiction with theirs as in these papers the Hamiltonian constraint and the kinematical Hilbert space are taken as given and therefore not changed throughout the work. In our case, we have changed the constraint form $\hat{\mathcal{C}}$ to $\hat{\mathcal{N}}\hat{\mathcal{C}}$. Using the same inner product, it might be possible that these two constraints are self-adjoint at the same time if one chose to quantise $\mathcal{N}\mathcal{C}$ as $\hat{\sqrt{\mathcal{N}}}\hat{\mathcal{C}}\hat{\sqrt{\mathcal{N}}}$. Doing this is not the path we have chosen, and moreover is not guaranteed to work as $\sqrt{\mathcal{N}}=v_0e^u$ might not become a self-adjoint operator. Our main point is that $\hat{\mathcal{C}}\Psi=0$ has the same solutions that $\hat{\mathcal{N}}\hat{\mathcal{C}}\Psi=0$. The ambiguity in the choice of the Wheeler--DeWitt equation seems to be unavoidable if one maintains reparametrisation invariance in the classical theory. The $v$-clock theory seems to have no analogue in the Dirac quantisation scheme.

Before we proceed to the following chapters, it is important to recall the basics to our approach to the problem of time. In our case, we first quantised the theory using the Wheeler--DeWitt equation and then chose a clock (\emph{tempus post quantum}). This first step is already different from approaches to quantisation, like deparametrisation and reduced quantisation (\emph{tempus ante quantum}). In the reduced quantisation framework, one solves the constraint at the classical level by identifying the constraint surface, and then one quantise the resulting theory, which possesses only true degrees of freedom. This identification, or gauge choice, is not unique and hence the quantum theories are not equivalent \cite{rps}. We also considered the chosen clock to act as an external parameter, and hence not forming part of the Hilbert space of the quantum theory. In the Page--Wooters formalism, another tempus post quantum approach, one considers models that are quantised as a tensor product between the clock Hilbert space and the system Hilbert space. The clock is hence also quantised and clock measurements have the usual uncertainties that come with quantum mechanics. The way to measure dynamics in these cases is via conditional probabilities of the form `what is the probability that the operator $\hat{\Obs}$ has value $o$ if the clock operator $\hat{\mathcal{T}}$ has value $\tau$?' This approach has the benefit of considering clocks like quantum systems, but then the notion of time evolution becomes blurry. Moreover, it has been shown that in some simple settings the Page--Wooters formalism, reduced quantisation and Dirac quantisation, are equivalent \cite{Hoehn}. Lastly, another approach to the problem of time is through path integral quantisation. In this setting, the histories are sums over space geometries. We will discuss further about that in \cref{pathin}. 

In summary, in this chapter we have computed the Wheeler--DeWitt equation of our model and calculated its solutions. Following this, we have built a quantum theory around all the clock choices we are interested in, and interpreted unitarity as self-adjointness of a specific operator. Finally, we have shown how the issue of unitarity is a direct consequence of time reparametrisation invariance rather than a consequence of the quantisation used. In the next chapters, we will focus more on the dynamics of the different quantum theories, but we want to highlight the crucial rôle of unitarity. It is really the key assumption that leads to all the different theories we have introduced. One might be tempted to assume that unitarity is not a fundamental property of quantum gravity, in order to obtain an equivalent theory for all clock choices, but this assumption is very hard to motivate as it is one of the most fundamental principles of quantum mechanics. Hence, we believe in the need of a more fundamental approach to these minisuperspace models before we making big claims such as singularity resolution. In the following part, we will dive deeper in the allowed wave functions in each relational quantum theory and calculate some important expectation values in order to analyse the early and late times behaviour of these theories. 
\part{Theoretical analysis of the different theories}
\chapter{The $t$-clock theory}
\label{t-clock-sec}

Let us start to analyse the $t$-clock theory. We recall the Wheeler--DeWitt equation \cref{wdw}
\begin{equation}
\left( \hbar^2 \pdv{}{v^2} +\frac{\hbar^2}{v}\pdv{}{v}-\frac{\hbar^2}{v^2}\pdv[2]{}{\varphi}-i\hbar\pdv{}{t}\right)\Psi(v, \varphi, t)=0\, ,
\label{wdw-t}
\end{equation}
and its solutions \cref{general-sol}
\begin{align}
\Psi(v,\varphi,t)&=&\int_{-\infty}^{\infty} \dl \int_{-\infty}^{\infty} \dk e^{ik\varphi}e^{i\lambda\frac{t}{h}}\left[ \alpha(k,\lambda)\J{i\abs{k}}{\lambda}+\beta(k,\lambda)\J{-i\abs{k}}{\lambda}\right] \nonumber \\
&+&\int_{-\infty}^{\infty} \dl \int_{-\infty}^{\infty} \dkap e^{\kappa\varphi}e^{i\lambda\frac{t}{h}}\left[ \gamma(\kappa,\lambda)\J{\abs{\kappa}}{\lambda}+\epsilon(\kappa,\lambda)\J{-\abs{\kappa}}{\lambda}\right]\, .
\label{general-sol-1}
\end{align}
In \cref{unit} defined the inner product \cref{tinnerprod}
\begin{equation}
\braket{\Psi}{\Phi}_t=\int_0^\infty \dd v \int_{-\infty}^\infty \dd \varphi \ v\bar{\Psi}(t,\varphi,v)\Phi(t,\varphi,v)\, ,
\label{inner-prod-t1}
\end{equation}
which involves integration over the variables $v$ and $\varphi$ but not the clock $t$. Not all functions of the form \cref{general-sol-1} have a time (in this case $t$) independent norm. In order to obtain a unitary theory we need to restrict ourselves to solutions that fulfil the boundary condition \cref{bound1}
\begin{equation}
\int_{-\infty}^\infty\dd \varphi \left[ v\left( \bar{\Psi}\pdv{}{v}\Phi-\Phi\pdv{}{v}\bar{\Psi}\right)\right]_{v=0}^{v=\infty}=0\, .
\label{boundt}
\end{equation}
This condition affects the specific expressions of the coefficients $\alpha(k,\lambda)$, $\beta(k,\lambda)$, $\gamma(k,\lambda)$ and $\epsilon(k,\lambda)$. 

But firstly, we have to see which solutions \cref{general-sol-1} are normalisable under the inner product \cref{inner-prod-t1}. We can consider the $\kappa$ and the $k$ sectors separately, so for simplicity we will consider first wave functions such that $\alpha(k,\lambda)=\beta(k,\lambda)=0$. When applying \cref{inner-prod-t1} to such states we find that we have to solve integrals of the type
\begin{equation}
\int_{-\infty}^\infty \frac{\dd \kappa_1}{2\pi}\int_{-\infty}^\infty \frac{\dd \kappa_2}{2\pi}\int_{-\infty}^\infty \dd \varphi \ e^{(\kappa_1+\kappa_2)\varphi} \left(\text{combinations of real Bessel functions} \right)\, .
\end{equation}
The $\varphi$ integral is always divergent unless $\kappa_1=\kappa_2$, but this set has measure zero in the $(\kappa_1,\kappa_2)$ space. In conclusion, to have normalisable states we must enforce $\gamma(\kappa,\lambda)=\epsilon(\kappa,\lambda)=0$. 

In order to calculate the solutions to the boundary condition \cref{boundt}, the imaginary Bessel functions can be divided into two sectors, $\lambda<0$ and $\lambda>0$. For each sector we find the one parameter family of solutions to the boundary condition. The calculations of the following sections are rather complicated and some results involve distribution theory. In order to not overflow the section we have moved the most technical calculations to \cref{bessel-int}. We will cite it when necessary.

\section{ The $\lambda>0$ sector}

We now proceed to give explicit the explicit solutions to \cref{boundt}. Let us consider wave functions of the form
\begin{equation}
\Psi_+(v,\varphi,t)=\int_0^\infty  \dl \int_{-\infty}^\infty \dk e^{ik\varphi}e^{i\lambda\frac{t}{\hbar}}\left[\alpha(k,\lambda)\J{i\abs{k}}{\lambda}+\beta(k,\lambda)\J{-i\abs{k}}{\lambda} \right]\, .
\label{phiplus}
\end{equation}
When evaluating the boundary condition \cref{boundt} where $\Psi$ and $\Phi$ take the form \cref{phiplus}, the integral over $\varphi$ simplifies to $2\pi\delta(k_1-k_2)$ giving 
\begin{align}
&\int_{-\infty}^\infty \dk \int_0^\infty \frac{\dd \lambda_1}{2\pi\hbar} \frac{\dd \lambda_2}{2\pi\hbar}  e^{i(\lambda_2-\lambda_1)\frac{t}{\hbar}}\left[v \left(\bar{\alpha}_1 \J{-i\abs{k}}{\lambda_1}+\bar{\beta}_1\J{i\abs{k}}{\lambda_1}\right)\right.\times\nonumber\\
&\left(\alpha_2 \partial_v\J{i\abs{k}}{\lambda_2}+\beta_2\partial_v\J{-i\abs{k}}{\lambda_2}\right)-v\left(\bar{\alpha}_1 \partial_v\J{-i\abs{k}}{\lambda_1}+\bar{\beta}_1\partial_v\J{i\abs{k}}{\lambda_1}\right)\times \nonumber \\
 &\left.\left(\alpha_2 \J{i\abs{k}}{\lambda_2}+\beta_2\J{-i\abs{k}}{\lambda_2}\right)\right]_{v=0}^{v=\infty}=0\, ,
 \label{cond}
\end{align}
where $\alpha_j=\alpha(k,\lambda_j)$ and $\beta_j=\beta(k,\lambda_j)$, $j=1,2$. The inside part of the integral can be rearranged into 4 terms:
\begin{align}
&v\bar{\alpha}_1\alpha_2\left(\J{-i\abs{k}}{\lambda_1}\partial_v\J{i\abs{k}}{\lambda_2}-\partial_v\J{-i\abs{k}}{\lambda_1}\J{i\abs{k}}{\lambda_2}\right)\label{cond1}\\
&+v\bar{\beta}_1\beta_2\left(\J{i\abs{k}}{\lambda_1}\partial_v\J{-i\abs{k}}{\lambda_2}-\partial_v\J{i\abs{k}}{\lambda_1}\J{-i\abs{k}}{\lambda_2}\right)\label{cond2} \\
&+v\bar{\alpha}_1\beta_2\left(\J{-i\abs{k}}{\lambda_1}\partial_v\J{-i\abs{k}}{\lambda_2}-\partial_v\J{-i\abs{k}}{\lambda_1}\J{-i\abs{k}}{\lambda_2}\right)\label{cond3}\\
&+v\bar{\beta}_1\alpha_2\left(\J{i\abs{k}}{\lambda_1}\partial_v\J{i\abs{k}}{\lambda_2}-\partial_v\J{i\abs{k}}{\lambda_1}\J{i\abs{k}}{\lambda_2}\right)\label{cond4}
\end{align}

To solve the boundary condition we must analyse these terms in the limits $v=\infty$ and $v=0$. We study both limits separately. In the limit $v=0$ the Bessel functions have the form:
\begin{equation}
\J{\pm i \abs{k}}{\lambda_j}\underset{v\rightarrow 0}{\longrightarrow} \frac{e^{\pm i\abs{k}\log(\frac{\sqrt{\lambda_j}}{2\hbar}v)}}{\Gamma(\pm i\abs{k} +1)}\, .
\label{asympt-small}
\end{equation}
By direct substitution into \cref{cond} we find that \cref{cond3} and \cref{cond4} vanish and \cref{cond1} and \cref{cond2} lead to
\begin{equation}
\bar{\alpha}_1(\lambda_1,k)\alpha_2(\lambda_2,k)e^{-i\abs{k}\log\sqrt{\frac{\lambda_1}{\lambda_2}}}-\bar{\beta}_1(\lambda_1,k)\beta_2(\lambda_2,k)e^{i\abs{k}\log\sqrt{\frac{\lambda_1}{\lambda_2}}}=0
\label{condi}
\end{equation}
for all values of $k$, $\lambda_1$ and $\lambda_2$. By setting $\lambda_1=\lambda_2$ we can see that $\abs{\alpha(k,\lambda)}=\abs{\beta(k,\lambda)}$. The general solution to \cref{condi} is
\begin{equation}
\beta(k,\lambda)=\alpha(k,\lambda)e^{i\theta(k)}e^{i\abs{k}\log\left(\frac{\lambda}{\lambda_0}\right)}\, ,
\end{equation}
where $\theta(k)$ is an arbitrary real function of $k$ and $\lambda_0$ is a reference scale. This free function represents the degree of freedom of the one dimensional self-adjoint extension of \cref{hams}. This concludes the analysis of the $v=0$ limit.

When $v$ is large we use the asymptotic expression of the $J$-Bessel functions
\begin{equation}
\J{\pm i\abs{k}}{\lambda_j}\underset{{v\rightarrow\infty}}{\longrightarrow}\sqrt{\frac{2\hbar}{\pi\sqrt{\lambda_j}v}}\cos(\frac{\sqrt{\lambda_j}v}{\hbar}\mp \frac{i\abs{k}\pi}{2}-\frac{\pi}{4})\, .
\label{asympt-large}
\end{equation}
When taking the derivative with respect to $v$ and substituting in \cref{cond1}--\cref{cond4} we can discard all terms that depend on $\frac{1}{v}$ as they go to zero. Hence, the terms \cref{cond1} and \cref{cond2} vanish immediately. However, \cref{cond3} and \cref{cond4} do not vanish so easily; they become respectively
\begin{align}
&\frac{\bar{\alpha_1}\beta_2}{\pi(\lambda_1\lambda_2)^{\frac{1}{4}}}\left((\sqrt{\lambda_1}+\sqrt{\lambda_2})\sin((\sqrt{\lambda_1}-\sqrt{\lambda_2})\frac{v}{\hbar})\right.-  \\
&\left.(\sqrt{\lambda_1}-\sqrt{\lambda_2})\left(\cosh(\abs{k}\pi)\cos((\sqrt{\lambda_1}+\sqrt{\lambda_2})\frac{v}{\hbar})-i\sinh(\abs{k}\pi)\sin((\sqrt{\lambda_1}+\sqrt{\lambda_2})\frac{v}{\hbar})\right)\nonumber\right)
\end{align}
for \cref{cond3} and
\begin{align}
&\frac{\bar{\beta_1}\alpha_2}{\pi(\lambda_1\lambda_2)^{\frac{1}{4}}}\left(-(\sqrt{\lambda_1}+\sqrt{\lambda_2})\sin((\sqrt{\lambda_1}-\sqrt{\lambda_2})\frac{v}{\hbar})\right.+  \\
&\left.(\sqrt{\lambda_1}-\sqrt{\lambda_2})\left(\cosh(\abs{k}\pi)\cos((\sqrt{\lambda_1}+\sqrt{\lambda_2})\frac{v}{\hbar})+i\sinh(\abs{k}\pi)\sin((\sqrt{\lambda_1}+\sqrt{\lambda_2})\frac{v}{\hbar})\right)\nonumber\right)
\end{align}
for \cref{cond4}. The second line of each term is formed of trigonometric functions of argument $(\sqrt{\lambda_1}+\sqrt{\lambda_2})\frac{v}{\hbar}$; according to \cref{delta-triglim}, when taking the limit $v=\infty$ these will result in Dirac delta distributions $\delta(\sqrt{\lambda_1}+\sqrt{\lambda_2})$ (or zero) that vanish as $\lambda_1$ and $\lambda_2$ are always positive. Moreover, also using \cref{delta-triglim}, the first line of terms results in $(\lambda_1-\lambda_2)\delta(\sqrt{\lambda_1}-\sqrt{\lambda_2})$  that is always zero due to the prefactor. In conclusion, the limit $v=\infty$ does not add anything to the allowed states.

We can now say that the allowed wave functions (of positive $\lambda$) are
\begin{equation}
\Psi_+(v,\varphi,t)=\int_0^\infty\dl \int_{-\infty}^{\infty} \dk \ e^{ik\varphi}e^{i\lambda\frac{t}{\hbar}}\alpha(k,\lambda)\Re\left[e^{i\theta(k)-i\abs{k}\log\sqrt{\frac{\lambda}{\lambda_0}}}\J{i\abs{k}}{\lambda}\right]\, .
\label{phipos}
\end{equation}
Here we have chosen to write these states as a combination of real Bessel functions. Recalling the small argument approximation of the Bessel functions \cref{asympt-small}, we can see that for small values of $v$ the Bessel functions look like plane waves either coming to the singularity (for Bessel functions of order $-i\abs{k}$), or going out of the singularity (for Bessel functions of order $i\abs{k}$). Wave functions of the universe like \cref{phipos} are always an equal weight combination of incoming and outgoing waves from the singularity. This superposition will play a key rôle when studying the dynamics of our universe. 

States of the form \cref{phipos} have a time independent norm, but they are not normalised, so we proceed here to find the normalisation of these states, a quite long and technical calculation. Readers not interested in the derivation of these states can directly jump to formula \cref{normstatest}. For simplicity, we label
\begin{equation}
\psi_{k,\lambda}(v,\varphi)=e^{ik\varphi}\Re\left[e^{i\theta(k)-i\abs{k}\log\sqrt{\frac{\lambda}{\lambda_0}}}\J{i\abs{k}}{\lambda}\right]\, .
\end{equation}
These states are eigenfunctions of the (now self-adjoint) Hamiltonian \cref{hams} of eigenvalue $-\lambda$. Thus, as $\lambda>0$ lies in the continuum spectrum of that operator, we should look to modify $\alpha(k,\lambda)$ such that $\braket{\psi_{k_1,\lambda_1}}{\psi_{k_2,\lambda_2}}=(2\pi)^2\hbar \delta(k_1-k_2)\delta(\lambda_1-\lambda_2)$. Because $\psi_{k,\lambda}$ are eigenstates of the operator \cref{hams}, we know they are automatically orthogonal. In addition to that, the $\varphi$ part is already normalised as $\int \dd \varphi \ e^{i(k_1-k_2)\varphi}=2\pi \delta(k_1-k_2)$. Hence, it is enough to evaluate the following integral:
\begin{align}
\int_0^\infty \dd v & \ v\bar{\psi}_{k,\lambda_1}\psi_{k,\lambda_2} =\int_0^\infty \dd v \ v \bigg\lbrace \bar{\alpha}_1\alpha_2 \times  \nonumber \\
&\left.\Re\left[e^{i\theta(k)-i\abs{k}\log\sqrt{\frac{\lambda_1}{\lambda_0}}}\J{i\abs{k}}{\lambda_1}\right] \Re\left[e^{i\theta(k)-i\abs{k}\log\sqrt{\frac{\lambda_2}{\lambda_0}}}\J{i\abs{k}}{\lambda_2}\right]\right\rbrace \nonumber\\
&=\frac{1}{4}\bar{\alpha}_1\alpha_2\int_0^\infty \dd v \ v \left\lbrace e^{-2i\theta(k)+i\abs{k}\log\sqrt{\frac{\lambda_1\lambda_2}{\lambda_0^2}}}\J{-i\abs{k}}{\lambda_1}\J{-i\abs{k}}{\lambda_2} \right. \nonumber \\
& \left. +e^{i\abs{k}\log{\sqrt{\frac{\lambda_1}{\lambda_2}}}}\J{-i\abs{k}}{\lambda_1}\J{i\abs{k}}{\lambda_2} + \text{complex conjugate}\right\rbrace\, ,
\label{norm1}
\end{align}
We already know that this integral yields a result of the form $f(\lambda_1)\delta(\lambda_1-\lambda_2)$. The $v$ integral converges in a distributional sense, as shown in \cref{P-sec}. The $v=0$ contribution to the integral is
\begin{align}
\frac{1}{4}\bar{\alpha}_1\alpha_2\left\lbrace 2i\frac{\sinh(\pi\abs{k})}{\pi(\lambda_1^2-\lambda_2^2)}-2i\frac{\sinh(\pi\abs{k})}{\pi(\lambda_1^2-\lambda_2^2)} \right\rbrace=0\, ,
\end{align}
as shown in \cref{zero-contr}. On the other side, the upper limit $v=\infty$ results in a combination of Dirac deltas $\delta(\sqrt{\lambda_1}\pm\sqrt{\lambda_2})$ and some prefactors (see \cref{P-int} for the specific formula). Using that $\lambda_1$ and $\lambda_2$ are always positive (and hence $\delta(\sqrt{\lambda_1}+\sqrt{\lambda_2})$ is always 0), and that the final result is real we obtain
\begin{align}
&\int_0^\infty\dd v \ v \bar{\psi}_{k,\lambda_1}\psi_{k,\lambda_2}= \nonumber \\
&=\frac{\hbar^2\abs{\alpha_1}^2 }{\sqrt{\lambda_1}}\left(\cos(\abs{k}\log\frac{\lambda_1}{\lambda_0}-2\theta(k))+\cosh(\abs{k}\pi) \right)\delta(\sqrt{\lambda_1}-\sqrt{\lambda_2}) \nonumber \\
&=\hbar^2\abs{\alpha_1}^2\left(\cos(\abs{k}\log\frac{\lambda_1}{\lambda_0}-2\theta(k))+\cosh(\abs{k}\pi) \right)\delta(\lambda_1-\lambda_2)\, .
\end{align}
In the second step we set $\lambda_1=\lambda_2$. In conclusion, given $\int\dk \dl \abs{\alpha(k,\lambda)}^2=1$ an orthonormal basis for the eigenstates of positive energy is given by
\begin{equation}
\psi_{k,\lambda}(v,\varphi)=e^{ik\varphi}\frac{\sqrt{2\pi}e^{ik\varphi}\Re\left[e^{i\theta(k)-i\abs{k}\log\sqrt{\frac{\lambda}{\lambda_0}}}\J{i\abs{k}}{\lambda}\right]}{\sqrt{\hbar\cos\left(-2\theta(k)+\abs{k}\log\frac{\lambda}{\lambda_0}\right)+\hbar\cosh(\abs{k}\pi)}}\, .
\label{eigen-l}
\end{equation}

These results are in accordance with \cite{Gryb} (note that their different definition of $\Lambda$ accounts for the different factors of 2 and $\hbar$). Using again the large argument approximation of the Bessel functions \cref{asympt-large} we find that at large $v$
\begin{align}
\psi_{k,\lambda}(v,\varphi)\propto \frac{F(k)}{v}&\left[ e^{-\frac{k\pi}{2}}\cos(\frac{\sqrt{\lambda}}{\hbar}v-\frac{\pi}{4}-\theta(k)+\abs{k}\log\sqrt{\frac{\lambda}{\lambda_0}})\right.\nonumber \\
&+ \left. e^{\frac{k\pi}{2}}\cos(\frac{\sqrt{\lambda}}{\hbar}v-\frac{\pi}{4}+\theta(k)-\abs{k}\log\sqrt{\frac{\lambda}{\lambda_0}})\right]\, ,
\end{align}
where $F(k)$ represents all remaining $k$ dependence. This wave function can be interpreted as a combination of plane waves with phase difference $\Theta(k,\lambda)=\pm \frac{\pi}{2}+2\theta(k)-k\log(\frac{\lambda}{\lambda_0})$. Hence, we can interpret these states as being scattered and experiencing a phase shift across $v=0$. For a more in detail discussion around this fact, consult \cite{Gryb}.

\section{The $\lambda<0$ sector}

Let us consider the negative part of the spectrum
\begin{equation}
\Psi_-(v,\varphi,t)=\int_{-\infty}^0  \dl \int_{-\infty}^\infty \dk e^{ik\varphi}e^{i\lambda\frac{t}{\hbar}}\left[\alpha(k,\lambda)\J{i\abs{k}}{\lambda}+\beta(k,\lambda)\J{-i\abs{k}}{\lambda} \right]\, .
\end{equation}
To not overwhelm the notation we stick to the expressions $\alpha(k,\lambda)$ and $\beta(k,\lambda)$ for the prefactors, but as we are in a different part of the spectrum of $\lambda$, the relations we find previously do not hold here. In order to simplify the notation it is better to work with the modified Bessel functions
\begin{equation}
K_{\alpha}(z)= \frac{\pi}{2}\frac{I_{-\alpha}(z)-I_{\alpha}(z)}{\sin(\alpha\pi)}\, ,
\label{Kdef}
\end{equation}
along with the $I$-Bessel functions defined in \cref{Iconvention}. The $K$ and $I$-Bessels can be used to form a complete set of solutions to the Wheeler--DeWitt equation \cref{wdw}, and hence we can redefine $\alpha$ and $\beta$ such that
\begin{equation}
\Psi_-(v,\varphi,t)=\int_{-\infty}^0  \dl \int_{-\infty}^\infty \dk e^{ik\varphi}e^{i\lambda\frac{t}{\hbar}}\left[\alpha(k,\lambda)\K{i\abs{k}}{-\lambda}+\beta(k,\lambda)\I{i\abs{k}}{-\lambda} \right]\, .
\label{phineg}
\end{equation}
At large arguments, the asymptotic behaviour of the $I$-Bessel is 
\begin{equation}
\I{\pm i\abs{k}}{-\lambda}\underset{v\rightarrow\infty}{\longrightarrow}\frac{\sqrt{\hbar}e^{\frac{\sqrt{-\lambda}}{\hbar}v}}{\sqrt{2\pi\sqrt{-\lambda}v}}\, .
\end{equation}
Due to the real exponential contribution, the $I$-Bessel functions diverge in the large $v$ limit. Hence, in order to obtain normalisable states one should consider wave functions where $\beta(k,\lambda)\equiv 0$. The large argument asymptotic form of the $K$-Bessel functions is 
\begin{equation}
\K{ i\abs{k}}{-\lambda} \underset{v\rightarrow\infty}{\longrightarrow} \sqrt{\frac{\hbar \pi}{2\sqrt{-\lambda}v}}e^{-\frac{\sqrt{-\lambda}v}{\hbar}}\, ,
\label{Kbig}
\end{equation}
which tends to zero, so wave functions that contain these Bessel functions are all normalisable. In addition to that, evaluating the boundary condition \cref{boundt} becomes trivial; we thus find no non-trivial contribution to the states from this limit. On the other hand, the asymptotic form of these Bessel functions for small arguments is
\begin{equation}
\K{ i\abs{k}}{-\lambda} \underset{v\rightarrow 0}{\longrightarrow} \frac{1}{2}\left(\Gamma(- i\abs{k})e^{ i \abs{k} \log(\frac{\sqrt{-\lambda}}{2\hbar}v)}+\Gamma( i\abs{k})e^{- i\abs{ k} \log(\frac{\sqrt{-\lambda}}{2\hbar}v)}\right)\, .
\label{Ksmall}
\end{equation}
Introducing this form in the $v=0$ limit of \cref{boundt}, we find that two negative values $\lambda_1$ and $\lambda_2$ have to satisfy
\begin{equation}
\log\left( \sqrt{\frac{\lambda_1}{\lambda_2}}\right)=\frac{\pi n}{\abs{k}}, \hspace{3mm} n\in \Z\, .
\end{equation}
In other words, for a given $k$, the only allowed $\lambda$ are of the form
\begin{equation}
\lambda^k_n=\lambda^k_{R}e^{\frac{-2\pi n}{\abs{k}}},\hspace{3mm}n\in \Z\, , 
\label{lambdaneg}
\end{equation}
for some $\lambda_R<0$, where $\lambda_R$ is a reference scale. $\lambda_R$ does not define a ground energy because the Hamiltonian \cref{hams} is unbounded. All these requirements restrict the form of \cref{phineg} to 
\begin{equation}
\Psi_-(v,\varphi,t)=\sum_{n=-\infty}^\infty   \int_{-\infty}^\infty \dk e^{ik\varphi}e^{i\lambda\frac{t}{\hbar}}\alpha(k,\lambda_n^k)\K{i\abs{k}}{-\lambda_n^k}\, .
\label{phineg1}
\end{equation}
The integral over $\lambda$ has been replaced by a sum. If we consider $\hat{\Ham}_k$, i.e.~the Fourier transform in $\varphi$ of the Hamiltonian \cref{hams}
\begin{equation}
\hat{\Ham}_k=\hbar^2\left(-\pdv[2]{}{v}-\frac{1}{v}\pdv{}{v}-\frac{1}{v^2}k^2 \right)\, ,
\label{hamsk}
\end{equation}
we see that for each $k$, the allowed values of $\lambda$ are discrete. Thus, $\lambda<0$ forms part of the discrete spectrum of $\hat{\Ham}_k$ and $\lambda>0$ forms part of its continuum spectrum .

So far, we have analysed what is the boundary condition restriction on states that are in a superposition of $\lambda$ of only one sign. However, it is necessary to check that the boundary condition \cref{boundt} is also fulfilled for $\Psi_+$ and $\Psi_-$ of the form \cref{phipos} and \cref{phineg1}, that is
\begin{equation}
\int_{-\infty}^\infty\dd \varphi \left[ v\left( \bar{\Psi}_+\pdv{}{v}\Psi_--\Psi_-\pdv{}{v}\bar{\Psi}_+\right)\right]_{v=0}^{v=\infty}=0\, .
\end{equation}
The limit $v=\infty$ is trivially zero (as can be seen remembering that the modified $K$-Bessel functions tend to 0). Nonetheless, the limit $v=0$ is not trivial and leads to 
\begin{equation}
e^{2i\theta(k)+i\abs{k}\log\left(\frac{v^2\lambda}{\hbar^2}\right)}+e^{i\abs{k}\log\frac{\lambda}{\lambda_0}+i\abs{k}\log\left(\frac{-v^2\lambda_n^k}{h^2}\right)}=0\, ,
\end{equation}
where $\lambda>0$ comes from $\Psi_+$ and $\lambda_n^k$ follows \cref{lambdaneg}. This condition has to hold for every $k$. The dependence in $v$ and $\lambda$ vanish after simplification leaving us with
\begin{equation}
e^{2i\theta(k)+i\abs{k}\log\left(-\frac{\lambda_0}{\lambda_n^k}\right)}=-1
\end{equation}
Finally, using $-1=e^{i(2n+1)\pi}$, we have
\begin{equation}
\lambda_n^k=-\lambda_0e^{-\frac{(2n+1)\pi}{\abs{k}}+\frac{2\theta(k)}{\abs{k}}}\, .
\end{equation}
Thus,
\begin{equation}
\lambda_R^k=-\lambda_0e^{-\frac{\pi}{k}+\frac{2\theta(k)}{k}}\, .
\end{equation}
In conclusion, the two reference scales are related. Note that changing $\lambda_0$ to $\lambda_0'$ and $\theta(k)$ to $\theta(k) +k\log\sqrt{\frac{\lambda_0}{\lambda_0'}}$ leaves $\lambda^k_R$ invariant.

Finally, we calculate the norm of the negative $\lambda$ eigenstates. Consider
\begin{equation}
\phi_{k,\lambda^k_n}(v,\varphi)=\alpha(k,\lambda_n^k)e^{ik\varphi}\K{i\abs{k}}{-\lambda^k_n}\, . 
\end{equation}
We would like to modify $\alpha(k,\lambda_n^k)$ so that $\braket{\phi_{k_1,\lambda^{k_1}_{n_1}}}{\phi_{k_2,\lambda^{k_2}_{n_2}}}=2\pi\delta(k_1-k_2)\delta_{n_1,n_2}$. Notice that this time the normalisation requires the use of a Dirac delta for the continuous $k$ part and a Kronecker delta for the discrete $\lambda$ part. As for the positive $\lambda$ sector, these states are already orthogonal, and the $\varphi$ part is already normalised, hence it is enough to analyse the integral
\begin{equation}
\int_0^\infty \dd v \ v \bar{\phi}_{k_1,\lambda^{k_1}_{n_1}}\phi_{k_2,\lambda^{k_2}_{n_2}}=\bar{\beta}_1\beta_2 \int^\infty_0 \dd v \ v \K{i\abs{k}}{-\lambda^k_{n_1}}\K{i\abs{k}}{-\lambda_{n_2}^k}\, .
\end{equation}
Recall that the modified Bessel functions are real, even for imaginary order and note that we are using the shorthand notation $\beta_i\equiv \beta(k,\lambda^k_{n_i})$. These integrals are known in the literature (see p.\ 658, formula 6.521(3) in \cite{Integrals}):
\begin{equation}
\int_0^\infty \dd x \ x K_{\nu}(a x)K_\nu(bx)=\frac{\pi (ab)^{-\nu}(a^{2\nu}-b^{2\nu})}{2\sin(\nu \pi)(a^2-b^2)}\, .
\label{formula}
\end{equation}
In our specific case, taking the limit $a=b$ we have that \cref{formula} vanishes for $n_1\neq n_2$, while for $n_1=n_2$
\begin{equation}
\int_0^\infty \dd v \ v \abs{\phi_{\lambda_n^k,k}}^2=\abs{\beta(k,\lambda_n^k)}^2\frac{\pi\hbar^2\abs{k}}{-2\lambda^k_n\sinh(\abs{k}\pi)}\, .
\end{equation}
Therefore, the correct normalisation for the negative energy eigenstates is
\begin{equation}
\phi_{\lambda^k_n,k}(v,\varphi)=\frac{1}{\hbar}\sqrt{\frac{-2\lambda^k_n\sinh(\abs{k}\pi)}{\pi \abs{k}}}e^{ik\varphi}\beta(k,\lambda^k_n)\K{i\abs{k}}{-\lambda^k_n}\, .
\end{equation}

We are now in condition to write the general wave function of the universe in terms of the orthonormal basis $\lbrace\phi_{k,\lambda^k_n}, \psi_{k,\lambda}\rbrace$:
\begin{align}
\Psi(v,\varphi,t)&=\int_{-\infty}^\infty \dk e^{ik\varphi}\left[\sum_{n=-\infty}^\infty e^{i\lambda^k_n \frac{t}{\hbar}} \frac{1}{\hbar}\sqrt{\frac{-2\lambda^k_n\sinh(\abs{k}\pi)}{\pi \abs{k}}}\alpha(k,\lambda^k_n)\K{i\abs{k}}{-\lambda^k_n} \right. \nonumber \\
& \left.+\int_0^\infty \dl e^{i\lambda\frac{t}{\hbar}}\frac{\sqrt{2\pi}e^{ik\varphi}\Re\left[e^{i\theta(k)-i\abs{k}\log\sqrt{\frac{\lambda}{\lambda_0}}}\J{i\abs{k}}{\lambda}\right]}{\sqrt{\hbar\cos\left(-2\theta(k)+\abs{k}\log\frac{\lambda}{\lambda_0}\right)+\hbar\cosh(\abs{k}\pi)}}\alpha(k,\lambda) \right]\, ,
\label{normstatest}
\end{align}
where $\alpha(k,\lambda)$ and $\alpha(k,\lambda^k_n)$ satisfy $\int \dk  \left[ \sum_{n=-\infty}^\infty\abs{\beta(k,\lambda_n^k)}^2+ \int_0^\infty \dl \abs{\alpha(k,\lambda)}^2\right]=1$. We remind here that the self-adjoint extension is parametrised by the free function $\theta(k)$, $\lambda_0>0$ is an arbitrary reference scale and $\lambda_n^k=-\lambda_0e^{-\frac{(2n+1)\pi}{\abs{k}}+\frac{2\theta(k)}{\abs{k}}}$ are the allowed negative $\lambda$ values. 

We see that requiring unitarity leads to non-trivial conditions over the allowed states. The most important consequence is that the combination of Bessel functions must be real, i.e., in a superposition such that the two modes $\pm i \abs{k}$ have the same weight (this is the case for both the $K$-Bessels and the $J$-Bessels). For small values of $v$, close to the classical singularity, one can say that the wave function of the universe is in a superposition of plane waves incoming to and outgoing from the singularity. The consequences of this superposition will be explored in \cref{numerics}, but it is already clear that this will have an effect on the classical singularity.

Looking closely at \cref{normstatest} we see that this expression is well-defined for all values of $t$; this is in opposition to the classical solutions \cref{eqsm-t-posneg} that vanish when $\abs{t-t_0}<\frac{\abs{\pi_\varphi}}{2\abs{\lambda}}$, and it is a direct consequence of the way we built our theory. Indeed, we started from a classical theory defined with a \emph{slow clock} at the singularity, i.e., a clock that does not tick fast enough to push the singularity to $\pm \infty$, and then we built a unitary quantum theory around that clock\footnote{In fact, what makes $t$ a slow clock, is that for every chosen finite $t_c$, there is a solution of the equation of motion for which $t_c$ lies in the classically forbidden interval $\abs{t-t_0}<\frac{\abs{\pi_\varphi}}{2\abs{\lambda}}$. This classically forbidden interval always lies within the initial domain of $t$, $\R$. For other clocks like $v$ the initial domain is $(0,\infty)$.}. The unitarity demand introduces the reflecting boundary condition \cref{boundt}, which changes the behaviour of the quantum solutions around the classical theory, thus obtaining a theory well defined everywhere in $t$ space. Recall that, in their works \cite{Gotay}, Gotay and Demaret theorise that all (unitary) quantum universes build from a slow clock (at the singularity) lead to a well-defined and non-singular quantum theory. On the contrary, given a \emph{fast clock} at the singularity, i.e., a clock that pushes the singularity to its domain boundaries, no extra boundary condition is required. The $t$-clock theory is consistent with this conjecture.

A last important thing to discuss around \cref{normstatest} is the rôle of the parameter $\theta(k)$. This function corresponds to the one parameter freedom in choosing the self-adjoint extension of the operator $\hat{\Ham}$ \cref{hams} (or equivalently \cref{hamsk}). One could (and should!) ask which features of the theory are sensitive to this parameter. We have not analysed this in our model, and when dealing with the numerical analysis of our model we will make the choice of $\theta$ that leads to the simplest calculations. However, the dependence of the self-adjoint extension parameter has been analysed in \cite{Gryb} and it was found that despite making a difference quantitatively, it did not make any difference qualitatively. Our results, including the normalisation of the wave functions of the universe are compatible with their work.
\chapter{The $v$-clock theory}
\label{v-clock-sec}
\section{Normalisation and positivity of the inner product}

The choice of $t$ as clock is motivated by the fact that $t$ is monotonic classically and that the Wheeler--DeWitt equation is a Schrödinger equation in $t$, but other choices are possible. In particular, one can use $v$ as clock, the only catch being that in the $\lambda<0$ case it is not valid throughout the entire evolution. Indeed, the volume experiences a turnaround. We can also use $v$ as clock before or after reaching the maximum value, even if we would have to use another clock around this maximum. This choice was studied in a similar model \cite{Gielen}. In this section we build a unitary quantum theory using this clock.

As we have seen previously, to use this clock it is better to multiply the Wheeler--DeWitt equation by $v^2$ to obtain \cref{wdw2}:
\begin{equation}
\left(\hbar^2\pdv[2]{}{\varphi}-\hbar^2\left(\pdv{}{\log(v/v_0)}\right)^2 +i\hbar v^2 \pdv{}{t}\right)\Psi(v,\varphi,t)=0\, .
\end{equation}
Multiplying the original equation by a phase space function does not change the solutions of the Wheeler--DeWitt equation. However, due to the second order derivatives in $\log(v/v_0)$ it makes possible to interpret it as a Klein--Gordon equation with a potential in the variable $\log(v/v_0)$, where $v_0$ is a constant needed to match the units of $v$. In an abuse of language we will refer to $v$ as the clock whereas the true clock is $\log(v/v_0)$. Remember that in this set up, the chosen inner product \cref{vproduct} is
\begin{equation}
\braket{\Psi}{\Phi}_v= i \int_{-\infty}^\infty \dd t \int_{-\infty}^\infty \dd \varphi \left[ v\bar{\Psi}(v,\varphi,t)\pdv{}{v}\Phi(v,\varphi,t) -v\Phi(v,\varphi,t)\pdv{}{v}\bar{\Psi}(v,\varphi,t)\right]\, .
\label{vinnerprod}
\end{equation}

The associated Hilbert space is composed of the normalisable functions under this inner product, but it is not of the  standard $L^2$ square integrable form. Regardless, a common feature of Klein--Gordon theories is the non positive definitiveness of the inner product. This case is not  an exception: the general norm of a state of the form
\begin{equation}
\Psi(v,\varphi,t)=\int_{-\infty}^\infty  \dl \int_{-\infty}^\infty \dk e^{ik\varphi}e^{i\lambda\frac{t}{\hbar}}\left[\alpha(k,\lambda)\J{i\abs{k}}{\lambda}+\beta(k,\lambda)\J{-i\abs{k}}{\lambda} \right]\, ,
\label{wf1}
\end{equation}
is
\begin{equation}
\norm{\Psi}^2_v= \frac{2}{\pi}\int_{-\infty}^\infty \dk \int_{-\infty}^\infty \dl \left[ -\abs{\alpha(k,\lambda)}^2+\abs{\beta(k,\lambda)}^2 \right]\sinh(\abs{k}\pi)\, .
\label{normv1}
\end{equation}
Even if this norm is not positive in general, it is straightforward to redefine the inner product to have a positive norm. From \cref{normv1} we see that we can decompose the wave functions \cref{wf1} into negative and positive frequencies: $\Psi=\Psi_-+\Psi_+$ where
\begin{align}
\Psi_-&=\int_{-\infty}^\infty \dk \int_{-\infty}^\infty \dl e^{i\lambda\frac{t}{\hbar}t}e^{ik\varphi}\alpha(k,\lambda)\J{i\abs{k}}{\lambda}\nonumber \\
\Psi_+&=\int_{-\infty}^\infty \dk \int_{-\infty}^\infty \dl e^{i\lambda\frac{t}{\hbar}}e^{ik\varphi}\beta(k,\lambda)\J{-i\abs{k}}{\lambda}\, .	
\end{align}
These states are such that $\norm{\Psi_-}_v\leq 0$, $\norm{\Psi_-}_v\geq 0$ and $\braket{\Psi_-}{\Psi_+}_v=0$, making positive and negative frequency states decoupled. Note that here frequency and the subindices $\pm$ do not refer to an eigenvalue of $v\pdv{}{v}$ but to the sign of the inner product $\braket{\cdot}{\cdot}_v$. Due to this decoupling, it is possible to build a consistent ``single universe'' quantum theory from the positive frequency sector only and no need of ``third quantisation'' \cite{McGuigan1988}, in which $\Psi$ would be promoted to a quantum field.

In our case we will consider both the $\Psi_+$ and $\Psi_-$ sectors and construct a positive definite inner product from $\braket{\cdot}{\cdot}_v$. It is enough to redefine the inner product of the negative frequency modes by adding a global minus sign in front of the $\Psi_-$ contribution:
\begin{equation}
\norm{\Psi}_{v'}=\norm{\Psi_+}_{v}-\norm{\Psi_-}_{v}\, ,
\end{equation}
or more explicitly, 
\begin{equation}
\norm{\Psi}^2_{v'}= \frac{2}{\pi}\int_{-\infty}^\infty \dk \int_{-\infty}^\infty \dl \left[ \abs{\alpha(k,\lambda)}^2+\abs{\beta(k,\lambda)}^2 \right]\sinh(\abs{k}\pi)\, .
\label{normv}
\end{equation}
This inner product treats positive and negative $\lambda$ modes in exactly the same way (contrary to the $t$ inner product) despite the $\lambda<0$ modes falling off exponentially at large $v$ rather than oscillating like the $\lambda>0$ ones.

But what happens to wave functions of the form
\begin{equation}
\Psi(v,\varphi,t)=\int_0^\infty  \dl \int_{-\infty}^\infty \dkap e^{\kappa\varphi}e^{i\lambda\frac{t}{\hbar}}\left[\gamma(\kappa,\lambda)\J{\abs{\kappa}}{\lambda}+\epsilon(\kappa,\lambda)\J{-\abs{\kappa}}{\lambda} \right]\, ?
\end{equation}
Identically to the $t$-clock case, the integral $\int \dd \varphi$ is divergent making these states non renormalisable. We can hence focus only on functions of the form \cref{wf1}.

As we can see from \cref{normv}, this inner product is time ($v$) independent, meaning that no extra condition is required in this theory (recall that \cref{Fop} is unitary). The Hilbert space hence contains all regular functions of the form \cref{wf1}.

\section{Semiclassical interpretation}

We have seen that the solutions to the Wheeler--DeWitt equation have a time independent norm already without the need of boundary conditions. Following the Gotay and Demaret conjecture \cite{Gotay}, we can explain is this by the fact that $v$ is a fast clock everywhere: the singularity is at the boundary of the domain of $v$ ($v=0$) and hence no extra condition is needed. Moreover, there is no Hamiltonian that can be interpreted as the generator of time evolution. This leads to an interesting semiclassical interpretation. Indeed, let us recall the operator $\hat{\mathcal{F}}$:
\begin{equation}
\hat{\mathcal{F}}=\left( -\hbar^2 \pdv[2]{}{\varphi} -i\hbar v_0e^{2u}\pdv{}{t}\right)\, ,
\end{equation}
where $u=\log{\frac{v}{v_0}}$. The wave functions of this theory have to satisfy the second order differential equation
\begin{equation}
-\hbar^2\pdv[2]{}{u}\Psi(u,\varphi,t)=\hat{\mathcal{F}}\Psi(u,\varphi,t)
\label{KGu}
\end{equation}

If $\hat{\mathcal{F}}$ was independent of $u$ and had only positive eigenvalues, this equation could be replaced by two Schrödinger equations
\begin{equation}
i\hbar \pdv{}{u}\Psi_{\pm}(u,\varphi,t)=\pm\sqrt{\hat{\mathcal{F}}}\Psi_{\pm}(u,\varphi,t)\, .
\end{equation}
In this case, self-adjointness and positivity would be enough to ensure that $\sqrt{\hat{\mathcal{F}}}$ is a well-defined operator. Moreover, in this case $\sqrt{\hat{\mathcal{F}}}$ could be interpreted as the Hamiltonian of the theory. 

However, $\hat{\mathcal{F}}$ is not time independent; if we want to rewrite equation \cref{KGu} as a Schrödinger equation, we have to introduce the operator $\hat{\Ham}_u$
\begin{equation}
i\hbar\pdv{}{u}\Psi(u,\varphi,t)=\hat{\Ham}_u\Psi(u,\varphi,t)\, ,
\label{sch-eq-u}
\end{equation}
where $\hat{\Ham}_u$ would be the generator of time evolution and it is a solution of the equation
\begin{equation}
\hat{\Ham}_u^2+i\hbar\pdv{\hat{\Ham}_u}{u}=\hat{\mathcal{F}}\Psi(u,\varphi,t)\, .
\end{equation}
There is no self-adjoint operator $\hat{\Ham}_u$ that is a solution of this equation and this can be easily verified by taking expectation values (assuming states that are well-behaved in both sides of the equation):
\begin{equation}
\expval{\hat{\Ham_u^2}}+i\hbar\pdv{}{u}\expval{\hat{\Ham_u}}=\expval{\hat{\mathcal{F}}}\, .
\label{expvalcond}
\end{equation}
The right-hand side always takes real values whereas the left-hand side does not unless $\pdv{\hat{\Ham}}{u}=0$. That would imply that $\hat{\mathcal{F}}$ must also be time independent, thus ultimately leading to a contradiction. Notice also that, even classically, the constraint
\begin{equation}
\mathcal{C}=-\pi_v^2+\frac{\pi_\varphi^2}{v^2}+\lambda\, ,
\end{equation}
does not admit a splitting $\mathcal{C}=\pi_v^2+\Ham_v^2$ such that $\Ham_v$ is a Dirac observable. 

In a nutshell, there is no link between our theory and a Schrödinger theory with a self-adjoint Hamiltonian. Still, if one tries to interpret the $v$ theory as an effective Schrödinger theory in the semiclassical regime we require the use of a complex Schrödinger time. Similar analysis have been done already, we summarise here the method used in \cite{Bojowald2011}. More examples of models that acquire a complex semiclassical time can be found in \cite{Bojowald2010}.

The main idea is to relax the requirement that \cref{sch-eq-u} is a Schrödinger equation in the same time variable $u$ than \cref{KGu}, instead we introduce a variable $\sigma$ such that $\pdv{\sigma}=\pdv{u}$. Assuming $\delta$ is small, we can thus write $\sigma$ as $u+\delta$ and rewrite the condition over expectation values \cref{expvalcond} as
\begin{equation}
\expval{\hat{\Ham}_u}^2+\expval{\hat{\Ham}_u\pdv{\hat{\Ham_u}}{u}+\pdv{\hat{\Ham}_u}{u}\hat{\Ham}_u}\delta+i\hbar\pdv{\expval{\hat{\Ham}_u}}{u}=\expval{\hat{\mathcal{F}}}\, ,
\label{expvalcond2}
\end{equation} 
where we have expanded $\expval{\Ham_u}^2$ at first order in $\delta$. Now, if we assume that the expectation values are over semiclassical states for which covariances are small, we can do the following approximation:
\begin{equation}
\expval{\hat{\Ham}_u\pdv{\hat{\Ham}_u}{u}}=\expval{\pdv{\hat{\Ham}_u}{u}\hat{\Ham}_u}=\expval{\hat{\Ham}_u}\pdv{\expval{\hat{\Ham}_u}}{u}\, .
\end{equation}
If we set $\delta=-\frac{i \hbar}{2\expval{\hat{\Ham_u}}}$ (which is purely imaginary), then at first order of perturbation $\hat{\Ham}_u$ can be considered the square root of $\hat{\mathcal{F}}$ and hence the theory can be seen as a Schrödinger theory. Notice that $\sigma=u-\hbar \frac{i}{2\expval{\hat{\Ham}_u}}$ can be considered a semiclassical expansion due to the factor of $\hbar$. Thus, the only way write our theory as a first-order theory semiclassically is by introducing a complex time variable.

\begin{figure}
\centering
\begin{subfigure}{0.49\textwidth}
\centering
\includegraphics[width=\textwidth]{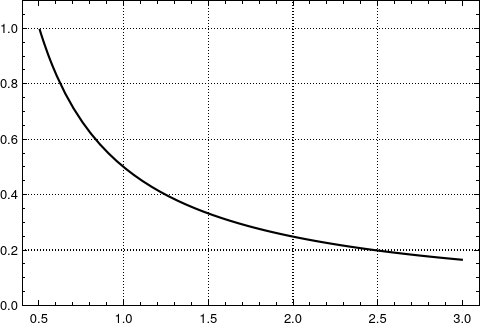}
\caption{$\abs{\delta(t)}$ for a solution in which $\hbar=\pi_\varphi=1$, $\lambda=1/2$, and $t_0=0$ for values of $t$ between $1/2$ and 3. The singularity is at $t=1/2$ and $\delta(1/2)=1$ which is a finite value.}
\end{subfigure}
\begin{subfigure}{0.49\textwidth}
\includegraphics[width=\textwidth]{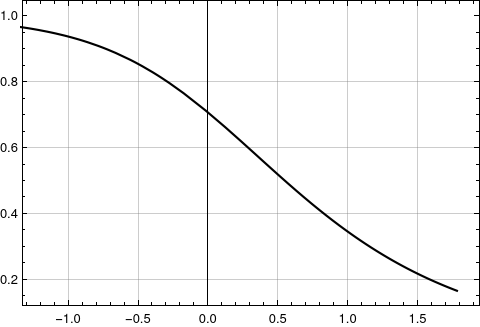}
\caption{Parametric plot of the real part of $\sigma(t)$ and the imaginary part of $\sigma(t)$ for a solution in which $\hbar=\pi_\varphi=v_0=1$, $\lambda=1/2$, and $t_0=0$ for values of $t$ between $1/2$ and 3.}
\end{subfigure}
\caption{Analysis of the imaginary contribution $\delta$.}
\label{delta-fig}
\end{figure}

An interesting exercise is to calculate the imaginary part of the effective time $\sigma$ on a classical solution. To do so, we replace $\expval{\hat{\Ham}_u}$ by its classical limit $\pi_u=v\pi_v$. Recall from \cref{class-sol-sec} the classical trajectories (for non vanishing $\lambda$):
\begin{equation}
v(t)=\sqrt{-\frac{\pi_\varphi^2}{\lambda}+4\lambda(t-t_0)^2}, \hspace{4mm} \pi_v(t)=\frac{2\lambda(t_0-t)}{\sqrt{-\frac{\pi_\varphi^2}{\lambda}+4\lambda(t-t_0)^2}}\, ,
\end{equation}
so that $v\pi_v=2\lambda(t_0-t)$ and 
\begin{equation}
\delta =-i\frac{\hbar}{4\lambda(t_0-t)}\, .
\end{equation}
For an expanding solution with $\lambda>0$ we see that $\delta$ reaches a maximum value of $\frac{\hbar}{\abs{\pi_\varphi}}$ at the classical singularity and tends to 0 as $t$ goes to infinity, although, for macroscopic $\pi_\varphi$, i.e., $\abs{\pi_\varphi}\gg \hbar$ the imaginary part remains very small. The plots of $\delta$ alone as a function of $t$ and of $\delta$ compared with $u$ for the same value of $t$ are presented in \cref{delta-fig}. In this comparison we had to set the parameters $\lambda$, $\hbar$, $\pi_\varphi$, and $v_0$ to arbitrary values. In fact the numerical values of the plots are irrelevant, the most important feature is the general shape which is overall conserved for different choices of the parameters.

From these plots, we can visualise how the imaginary part of $\sigma$ is relevant near the classical singularity but tends to 0 as the volume $v$ grows to infinity, confirming the general expectation that far away from the singularity the quantum theory is well described by a real time Schrödinger theory, whereas close to the singularity, this approximation not longer holds. Interestingly, the quantum behaviour of the perfect bounce model \cite{Gielen} was also captured by semiclassical complex trajectories in conformal time, leading to the avoidance of the classical singularity in the complex plane, similarly to quantum tunnelling.

Coming back to the Gotay and Demaret conjecture, we see here that as  $v$ is fast everywhere, no boundary condition is needed in this theory, and thus near the classical singularity we do not observe the superposition of plane waves present in the $t$-clock theory. This leads to the suspicion (later confirmed by the numerical analysis) that this theory does not resolve the singularity. It is now time to centre ourselves around the last possible theory, the $\varphi$-clock theory.
\chapter{The $\varphi$-clock theory}
\label{phi-clock-sec}
The last theory we study here is the $\varphi$-clock theory. As seen previously, we can rewrite the Wheeler--DeWitt equation by multiplying it by $v^2$ to find
\begin{equation}
\left(\hbar^2\pdv[2]{}{\varphi}-\hbar^2\left(\pdv{}{\log(v/v_0)}\right)^2 +i\hbar v^2 \pdv{}{t}\right)\Psi(v,\varphi,t)=0\, .
\end{equation}
By using the ansatz $\Psi(v,\varphi,t)=\psi(v,\varphi)e^{i\lambda\frac{t}{\hbar}}$ we obtain
\begin{equation}
\left(\hbar^2\pdv[2]{}{\varphi}-\hbar^2\left(\pdv{}{\log(v/v_0)}\right)^2 -v^2\lambda\right)\psi(v,\varphi)=0\, .
\label{wdwv2}
\end{equation}
In \cref{unit} we introduced the $\varphi$-inner product \cref{phiinnerprod} which we recall here:
\begin{equation}
\braket{\Psi}{\Phi}_\varphi=i\int_{-\infty}^\infty\dd t \int_0^\infty \frac{\dd v}{v}\left( \bar{\Psi}(v,\varphi,t)\pdv{}{\varphi}\Phi(v,\varphi,t)-\Phi(v,\varphi,t)\pdv{}{\varphi}\Psi(v,\varphi,t) \right)\, ,
\label{phiinnerproduct}
\end{equation}
Remember that $n^\mu\partial_\mu=\frac{1}{v}\partial_\varphi$, where here $n^\mu$ is the normal to the $\varphi=$ const surfaces in the Rindler wedge metric. 

In \cref{unit} we saw that our theory is not unitary unless the correct boundary conditions are satisfied. In fact, introducing the variable $u=\log(v/v_0)$ we verified that the condition $\partial_\varphi\braket{\Psi}{\Phi}_\varphi=0$ is equivalent to the self-adjointness problem for the operator \cref{ham2}
\begin{equation}
\hat{\mathcal{G}}=-\hbar^2\pdv[2]{}{u}-\lambda v_0^2e^{2u}\, ,
\label{Gop}
\end{equation}
in an $L^2(\R\times \R^+, \frac{\dd v}{v}\dd t)$, or equivalently $L^2(\R^2, \dd u\dd t)$, inner product, which is then also equivalent to imposing the boundary condition \cref{bound2}
\begin{equation}
\int \dd t \left[ v \bar{\Psi}\pdv{}{v}\Phi-v\Phi\pdv{}{v}\bar{\Psi} \right]_{v=0}^{v=\infty}=0\, .
\label{boundphi}
\end{equation}
The operator $\hat{\mathcal{G}}$ has a very different behaviour depending on the sign of $\lambda$. If we consider its classical theory, (i.e.~replacing $i\hbar \partial_u$ by $p_u$ and treating $p_u$ as a classical momentum of a particle), $\mathcal{G}$ is of the form $T+V$ where $V=-\lambda v_0e^{2u}$ is a potential term and $T$ is a kinetic term (note the absence of hats in $\mathcal{G}$). If $\lambda$ is positive, this potential is attractive enough to accelerate a particle to reach $u=\infty$ in a finite time, yet for $\lambda<0$ the potential is repulsive and this acceleration does not occur. The classical behaviour signals two facts: on the one side, the operator $\hat{\mathcal{G}}$ is not self-adjoint for positive $\lambda$ and we would need a reflective boundary condition around $v=\infty$. On the other side, it is self-adjoint for negative $\lambda$ so no boundary condition is needed.

There is a one parameter family of linear subspaces of the wave functions that satisfy the boundary condition \cref{boundphi}, as derived first (to our knowledge) in \cite{Kobayashi1996} and in very comprehensive way in \cite{Fredenhagen2003}. The work \cite{Fredenhagen2003} was done in the context of S-branes and $\lambda$ was a fixed quantity. In our case, as $\lambda$ is not a fixed parameter but a dynamical variable, the one parameter freedom of choosing subspace depends on $\lambda$. In this section, we will reproduce the results of \cite{Fredenhagen2003} but our normalisation will be different because we are working with a Klein--Gordon rather than a Schrödinger inner product. For the $\lambda<0$ case, the boundary solution is trivially satisfied, and the eigenstates of $\hat{\mathcal{G}}$ are also known, as shown in \cite{DHoker1982}. We will re-derive them using in our context.

Yet, before we start these calculations, there is a major point to take into account. As we saw for the $v$-clock theory, Klein--Gordon inner products such as $\braket{\cdot}{\cdot}_\varphi$ are not expected to be positive definite and  hence these inner products cannot be used for a consistent (Born) probability interpretation. However, the explicit construction of a positive inner product over the whole space of solutions of the Wheeler--DeWitt equation is neither needed (because we will only work on the subspaces solving the boundary condition \cref{boundphi}) nor straightforward. In fact, solutions that do not satisfy the boundary condition \cref{boundphi} have no physical interpretation in this setting: no meaningful probability distribution can be associated to a state whose norm is not conserved over time. By first imposing unitarity, the resulting wave functions naturally split into orthonormal positive, null, and negative norm subspaces, which makes the redefinition of the inner product possible for such subspaces. In conclusion, we will build a positive definite inner product after finding the family of self-adjoint extensions of the theory and only on the subspace of solutions of the boundary condition. 

In the following sections we derive the form of the wave functions that are both normalisable and satisfy the boundary condition \cref{boundphi}. Normalisability puts further restrictions on the states. Our results are compatible with the analysis of \cite{Pawlowski2011} where a quantum theory in $\varphi$ time was constructed for a cosmological model with massless scalar field and fixed $\lambda>0$. In this work, the authors define a Dirac quantisation of the kinematical Hilbert space rather than fixing a clock variable before quantisation, but as we have seen in \cref{dirac-sec}, the same self-adjointness problems arise in this framework. Another subtlety is that they do not use the volume $v$ but a dual representation in terms of a variable $b$ corresponding to the Hubble parameter, hence even if their results are not directly comparable to ours, we reach similar conclusions.

\section{The $\lambda>0$ sector}

Remember that the general solution to the Wheeler--DeWitt equation \cref{wdwv2} is
\begin{align}
\Psi(v,\varphi,t)&=&\int_{-\infty}^{\infty} \dl \int_{-\infty}^{\infty} \dk e^{ik\varphi}e^{i\lambda\frac{t}{h}}\left[ \alpha(k,\lambda)\J{i\abs{k}}{\lambda}+\beta(k,\lambda)\J{-i\abs{k}}{\lambda}\right] \nonumber \\
&+&\int_{-\infty}^{\infty} \dl \int_{-\infty}^{\infty} \dkap e^{\kappa\varphi}e^{i\lambda\frac{t}{h}}\left[ \gamma(\kappa,\lambda)\J{\abs{\kappa}}{\lambda}+\epsilon(\kappa,\lambda)\J{-\abs{\kappa}}{\lambda}\right]\, .
\end{align}
We discuss real and imaginary order Bessel functions separately, as they have very different asymptotic behaviour, hence we start by considering a wave function with $\gamma(\kappa,\lambda)=\epsilon(\kappa,\lambda)=0$:
\begin{equation}
\Psi_1(v,\varphi,t)=\int_{-\infty}^{\infty} \dl \int_{-\infty}^{\infty} \dk e^{ik\varphi}e^{i\lambda\frac{t}{h}}\left[ \alpha(k,\lambda)\J{i\abs{k}}{\lambda}+\beta(k,\lambda)\J{-i\abs{k}}{\lambda}\right] \, .
\label{psi1}
\end{equation}
The inner product of these states is
\begin{align}
\braket{\Psi_1}{\Psi_1}&=-\int \frac{\dd k_1 \dd k_2}{(2\pi)^2}\dl \frac{\dd v}{v}e^{i(k_2-k_1)\varphi}(k_1+k_2)\times \\
&\left[ \bar{\alpha_1}\alpha_2\J{-i\abs{k_1}}{\lambda}\J{i\abs{k_2}}{\lambda}+\bar{\alpha}_1\beta_2\J{-i\abs{k_1}}{\lambda}\J{-i\abs{k_2}}{\lambda} \right.\nonumber \\
&\left. + \bar{\beta}_1\alpha_2 \J{i\abs{k_1}}{\lambda}\J{i\abs{k_2}}{\lambda}+\bar{\beta}_1\beta_2\J{i\abs{k_1}}{\lambda}\J{-i\abs{k_2}}{\lambda} \right]\, , \nonumber
\end{align}
where we use again the abbreviation $\alpha_i$ for $\alpha(k_i,\lambda)$ and $\beta_i$ for $\beta(k_i,\lambda)$. This expression is clearly $\varphi$ dependent. In order to simplify it further, we make use of the explicit expressions for the integral of the two products of Bessel functions of \cref{bessel-int}. Here we use equation \cref{2-Bessel-int-1} to find
\begin{align}
\braket{\Psi_1}{\Psi_1}=&- \int \frac{\dd k_1 \dd k_2}{(2\pi)^2}\dl  e^{i(k_2-k_1)\varphi}(k_1+k_2)\times \\
&\left\lbrace \text{PV}\frac{2i}{\pi(k_1^2-k_2^2)}\left(\sinh((\abs{k_1}+ \abs{k_2})\frac{\pi}{2})\right.\left[\bar{\alpha}_1\alpha_2-\bar{\beta}_1\beta_2 \right] \right. \nonumber \\
&+ \left.\sinh\left((\abs{k_1}-\abs{k_2})\frac{\pi}{2}) \right)\left[\bar{\alpha}_1\beta_2-\bar{\beta}_1\alpha_2 \right]\right) \nonumber \\
&+2\frac{\sinh\left((\abs{k_1}+\abs{k_2})\frac{\pi}{2} \right)}{\abs{k_1}+\abs{k_2}}\delta(\abs{k_1}-\abs{k_2})\left[ \bar{\alpha}_1\alpha_2+\bar{\beta}_1\beta_2\right] \nonumber\\
&\left.+2\frac{\sinh\left((\abs{k_1}-\abs{k_2})\frac{\pi}{2} \right)}{\abs{k_1}-\abs{k_2}}\delta(\abs{k_1}+\abs{k_2})\left[ \bar{\alpha}_1\beta_2+\bar{\beta}_1\alpha_2\right]\right\rbrace\, , \nonumber
\end{align}
where PV denotes the Cauchy principal value, a definition of the integral in terms of a symmetric limit around the singular point $k_1=k_2$. We need to analyse these terms one by one. The two last terms come from the $v=0$ limit. In particular, the last term, depending on $\delta(\abs{k_1}+\abs{k_2})$, will not contribute to the final result. The Dirac delta $\delta(\abs{k_1}-\abs{k_2})$ can be simplified using:
\begin{equation}
(k_1+k_2)\delta(\abs{k_1}-\abs{k_2})=(k_1+k_2)[\delta(k_1+k_2)+\delta(k_1-k_2)]=2k_1\delta(k_1-k_2)\, .
\label{absdelta}
\end{equation}
When integrating over $k_1$ or $k_2$, the Dirac delta ensures $k_1=k_2$ leading $e^{i(k_2-k_1)\varphi}=1$ and thus the contribution of this term is independent of $\varphi$. 

The first two terms come from the limit $v=\infty$ and they cannot be further simplified; thus they must vanish. Hence, the boundary condition \cref{boundphi} is non-trivial only in the $v=\infty
$ limit. Due to the presence of the principal value, the integral over $k$ of these factors only depends on the antisymmetric part with respect to the point $k_1=k_2$. This leads to the condition
\begin{equation}
Y(k_1,k_2,\lambda,\varphi)-Y(2k_2-k_1,k_2,\lambda,\varphi)=0\, ,
\end{equation}
(which is odd with respect to the reflection around the singular point $k_1=k_2$), where 
\begin{align}
Y(k_1,k_2,\lambda,\varphi)=e^{i(k_2-k_1)\varphi}&\sinh((\abs{k_1}+ \abs{k_2})\frac{\pi}{2})\left[\bar{\alpha}_1\alpha_2-\bar{\beta}_1\beta_2 \right]\nonumber\\
+&\sinh\left((\abs{k_1}-\abs{k_2})\frac{\pi}{2} \right)\left[\bar{\alpha}_1\beta_2-\bar{\beta}_1\alpha_2 \right]\, .
\label{Y-cond}
\end{align}
To solve this condition we expand it around the singular point $k_1=k_2$ up to linear order to find:
\begin{align}
&\alpha(k,\lambda)\bar{\beta}(k,\lambda)-\bar{\alpha}(k,\lambda)\beta(k,\lambda)-\cosh(k\pi)\left(\abs{\alpha(k,\lambda)}^2-\abs{\beta(k,\lambda)}^2\right) \label{unitarity-cond}\\
&+\frac{2}{\pi}\sinh(k\pi)\left(\beta(k,\lambda)\pdv{\bar{\beta}(k,\lambda)}{k}-\alpha(k,\lambda)\pdv{\bar{\alpha}(k,\lambda)}{k}+i\varphi\left(\abs{\alpha(k,\lambda)}^2 -\abs{\beta(k,\lambda)}^2\right) \right)=0 \nonumber
\end{align}
The last term depends on $\varphi$, so it has to vanish independently, hence
\begin{equation}
\abs{\alpha(k,\lambda)}^2=\abs{\beta(k,\lambda)}^2 \implies e^{i\chi(k,\lambda)}\beta(k,\lambda)=\alpha(k,\lambda)\, ,
\label{alpha-beta-cond}
\end{equation}
where $\chi(k,\lambda)$ is an unspecified function taking values between $[-\pi,\pi)$. Using this information in \cref{unitarity-cond} we find the differential equation
\begin{equation}
\pi\sin(\chi(k,\lambda))+\sinh(k\pi)\pdv{}{k}\chi(k,\lambda)=0\, .
\label{diffeqchi}
\end{equation} 
The general solution to this equation is
\begin{equation}
\chi(k,\lambda)=-2\arctan\left[\vartheta(\lambda)\coth\left(\frac{\abs{k}\pi}{2}\right) \right]\, ,
\label{chi}
\end{equation}
where $\vartheta(\lambda)$ is a free function. By looking at \cref{diffeqchi} we see that we have the apparent freedom to multiply $\chi(k,\lambda)$ by a overall minus sign which can be different for positive and negative $k$ (note that $\chi(k,\lambda)$ is ill-defined for $k=0$). However, this sign is fixed by imposing that the solution to \cref{diffeqchi} must also be a solution to \cref{unitarity-cond}. Given that $\chi(k,\lambda)$ as defined in \cref{chi} solves \cref{unitarity-cond} we have found the most general solution to the boundary condition.

Let us analyse the free function $\vartheta(\lambda)$ deeper. Two obvious solutions to \cref{Y-cond} are $\beta(k,\lambda)=\pm \alpha(k,\lambda)$ for some (or all) values of $\lambda$. The ``$+$'' solution corresponds to $\chi(k,\lambda)=\vartheta(\lambda)=0$, at least for these values of $\lambda$. However, the ``$-$'' solution is $\chi(k,\lambda)=-\pi$ which formally corresponds to $\vartheta(\lambda)=\infty$, which must hence be included as a possibility. The function $\vartheta(\lambda)$, defined to take values over $\R\cup\lbrace\infty\rbrace$, is the degree of freedom of the one dimensional self-adjoint extension of the operator \cref{Gop}, analogously to the free parameter $\theta(k)$ present in the $t$-clock theory. A similar free parameter was found in \cite{Fredenhagen2003} and \cite{Kobayashi1996}.

Now, the inner product of any state where $\alpha(k,\lambda)$ and $\beta(k,\lambda)$ follow \cref{alpha-beta-cond} with $\chi(k,\lambda)$ defined as in \cref{chi} gives
\begin{equation}
\braket{\Psi_1}{\Psi_1}_\varphi=-\int_{-\infty}^\infty \dk \int_0^\infty \dl \frac{2\sinh(k\pi)}{\pi}\abs{\alpha(k,\lambda)}^2\, .
\label{innerprod1}
\end{equation}
In consequence, this inner product is not positive definite, the $k>0$ modes give a negative contribution to the norm. We follow here the same approach from the $v$-clock theory, we redefine the inner product \cref{phiinnerproduct} so that it becomes positive definite. This allows all $k$ modes to be considered as physical states. This redefinition can be done with ease as the $k>0$ and $k<0$ modes are decoupled after the imposition of the boundary condition. We thus modify the inner product on these states
\begin{equation}
\braket{\Psi_1}{\Psi_1}_{\varphi'}=\braket{\Psi_1}{\Psi_1}_{\varphi,k<0}-\braket{\Psi_1}{\Psi_1}_{\varphi,k>0}\, ,
\label{newphiinnerprod}
\end{equation}
where the notation $k<0$ and $k>0$ refer to the value of integration of $k$ in \cref{innerprod1}. Explicitly this gives
\begin{equation}
\braket{\Psi_1}{\Psi_1}_{\varphi'}=\int_{-\infty}^\infty \dk \int_0^\infty \dl \frac{2\sinh(\abs{k}\pi)}{\pi}\abs{\alpha(k,\lambda)}^2\, ,
\end{equation}
which is manifestly positive definite. Note that even after this redefinition norm of states is still conserved, i.e., the theory is still unitary.

In conclusion, a normalised solution to the Wheeler--DeWitt equation satisfying the boundary condition \cref{boundphi} built using only imaginary order Bessel function can be expressed as
\begin{equation}
\Psi_1(v,\varphi,t)=\int_{-\infty}^\infty \dk \int_0^\infty \dl e^{ik\varphi}e^{i\lambda\frac{t}{\hbar}}\alpha(k,\lambda)\sqrt{\frac{2\pi}{\sinh(\abs{k}\pi)}}\Re\left[ e^{i\frac{\chi(k,\lambda)}{2}}\J{i\abs{k
}}{\lambda}\right]\, ,
\label{psi11}
\end{equation}
where $\int_{-\infty}^\infty \dk \int_0^\infty \abs{\alpha(k,\lambda)}^2=1$. Note the similarity between this solution and the expression of $\Psi_+$, \cref{phipos}, found in the $t$-clock theory. In both cases the combination of Bessel functions must be real and in both cases there is a free parameter that acts like a phase. However, in the $t$-theory the free parameter $\theta(k)$ is a function of $k$, whereas in the $\varphi$-clock theory $\vartheta(\lambda)$ is a function of $\lambda$.

However, the expression for $\chi(k,\lambda)$ presented in \cref{chi} can be inconvenient to work with later in the numeric section and the comparison to \cite{Fredenhagen2003} is not clear, so we chose to rewriting the arctangent in terms of logarithms such that:
\begin{equation}
e^{i\frac{\chi(k,\lambda)}{2}}=\sqrt{\frac{\sinh(\frac{\abs{k}\pi}{2})-i\vartheta(\lambda)\cosh(\frac{\abs{k}\pi}{2})}{\sinh(\frac{\abs{k}\pi}{2})+i\vartheta(\lambda)\cosh(\frac{\abs{k}\pi}{2})}}\, .
\end{equation}
One can further simplify this expression by introducing a new function $\kappa_0(\lambda)$ such that
\begin{equation}
\vartheta(\lambda)=\tan\left(\kappa_0(\lambda)\frac{\pi}{2} \right)\, ,
\label{vartheta}
\end{equation}
where $\kappa_0(\lambda)$ takes values in the interval $[0,2)$. In this notation $\kappa_0(\lambda)=1$ corresponds to the possible choice $\vartheta(\lambda)=\infty$ discussed earlier. Using this new function we can rewrite  \cref{psi11}
\begin{align}
\Psi_1(v,\varphi,t)=\int_{-\infty}^\infty \dk \int_0^\infty \dl e^{i k \varphi}e^{i\lambda\frac{t}{\hbar}}\alpha(k,\lambda)\sqrt{\frac{2\pi}{\sinh(\abs{k}\pi)}}\times \nonumber\\
\Re\left[ \sqrt{\frac{\sinh(\abs{k}-i\kappa_0(\lambda)\frac{\pi}{2})}{\sinh(\abs{k}+i\kappa_0(\lambda)\frac{\pi}{2})}}\J{i\abs{k}}{\lambda} \right]\, .
\label{psi12}
\end{align}
This is the form used in \cite{Fredenhagen2003} but with a different normalisation (as they are using another inner product).

Now that we have found normalisable states with imaginary order Bessel functions, let us turn our attention to real order Bessel functions by considering states with $\alpha(k,\lambda)=\beta(k,\lambda)=0$
\begin{equation}
\Psi_2(v,\varphi,t)=\int^\infty_0 \dl \int_{-\infty}^\infty \dkap e^{\kappa\varphi}e^{i\lambda\frac{t}{\hbar}}\left[\gamma(k,\lambda)\J{\abs{\kappa}}{\lambda}+\epsilon(\kappa,\lambda)\J{-\abs{\kappa}}{\lambda} \right]\, .
\end{equation}
The norm squared of such a state is
\begin{align}
\braket{\Psi_2}{\Psi_2}&=i\int \frac{\dd \kappa_1 \dd \kappa_2}{(2\pi)^2}\frac{\dd \lambda}{2\pi \hbar}\frac{\dd v}{v}e^{(\kappa_1+\kappa_2)\varphi}(\kappa_2-\kappa_1) \times   \\
& \left[ \bar{\gamma}_1\gamma_2\J{\abs{\kappa_1}}{\lambda}\J{\abs{\kappa_2}}{\lambda}+\bar{\gamma}_1\epsilon_2\J{\abs{\kappa_1}}{\lambda}\J{-\abs{\kappa_2}}{\lambda} \right. \nonumber \\
& \left. +\bar{\epsilon}_1\gamma_2\J{-\abs{\kappa_1}}{\lambda}\J{\abs{\kappa_2}}{\lambda}+\bar{\epsilon}_1\epsilon_2\J{-\abs{\kappa_1}}{\lambda} \J{-\abs{\kappa_2}}{\lambda}\right] \nonumber \, ,
\end{align}
where as usual all functions are evaluated at the same $\lambda$ due to the appearance of the factor $\int \dd t \ e^{i(\lambda_2-\lambda_1) \frac{t}{\hbar}}=2\pi\hbar\delta(\lambda_2-\lambda_1)$, and we use the abbreviation $\gamma_i=\gamma(\kappa_i,\lambda)$ and $\epsilon_i=\epsilon(\kappa_i,\lambda)$. We have studied the integral over $v$ in \cref{bessel-int} and found that it does not converge (neither in the usual sense nor in a distributional sense), unless the order of the two Bessel functions are strictly positive. In this case, the expression is given by \cref{2-Bessel-int-2}. Consequently, we must make $\epsilon(\kappa,\lambda)=0$ in order to obtain a normalisable state. We then find
\begin{equation}
\braket{\Psi_2}{\Psi_2}=-\frac{2i}{\pi} \int \frac{\dd \kappa_1 \dd \kappa_2}{(2\pi)^2}\frac{\dd \lambda}{2\pi \hbar}e^{(\kappa_1+\kappa_2)\varphi}\frac{\sin((\abs{\kappa_1}-\abs{\kappa_2})\frac{\pi}{2})}{\kappa_1+\kappa_2}\bar{\gamma}(\kappa_1, \lambda)\gamma(\kappa_2,\lambda)\, .
\end{equation}
The only way to ensure that this inner product is time independent is by setting
\begin{equation}
\sin((\abs{\kappa_1}-\abs{\kappa_2})\frac{\pi}{2})=0 \implies \abs{\kappa_1}-\abs{\kappa_2}=2n, \ n \in \Z\, .
\end{equation}
This condition can be solved separately for each value of $\lambda$; but given a fixed $\lambda$ only a discrete set of values for $\kappa$ is allowed, namely those satisfying
\begin{equation}
\abs{\kappa}=\kappa_0'(\lambda)+2n \ \text{for some} \ n\in \N\cup \lbrace 0\rbrace\, ,
\end{equation}
where $\kappa_0'(\lambda)$ is an arbitrary function of $\lambda$ which we can choose to take values in $[0,2)$. Thus the wave functions that have a preserved norm under the inner product are
\begin{align}
\Psi_2(v,\varphi,t)=\int_0^\infty \frac{\dd \lambda}{2\pi\hbar}&e^{i\lambda\frac{t}{\hbar}}\left[ \sum_{n=0}^\infty \left(\gamma^+_n(\lambda)e^{(\kappa_0'(\lambda)+2n)\varphi}+\gamma^-_n(\lambda)e^{-(\kappa_0'(\lambda)+2n)\varphi} \right)\times \right. \nonumber \\
&\left. \J{\kappa_0'(\lambda)+2n}{\lambda} \right]\, .
\label{psi2}
\end{align}
These functions have norm zero, which is indeed $\varphi$ independent. These states have no classical analogue; one would interpret them as configuration for which $\pi_\varphi^2<0$ in the classical constraint \cref{hamconst}, similar to the tunnelling solutions under a potential barrier in quantum mechanics. In quantum cosmology similar states that may ``decay'' in relational time have been briefly discussed as ``quantum puff'' universes by Misner \cite{Misner1972}, however their interpretation is not clear at all. Such states are also mentioned in Blyth's PhD thesis \cite{Blyth1974}, once again with difficulties to find a good interpretation. Luckily, in our case they have norm zero, so they have no influence in the probabilistic interpretation of our theory. We will not consider universes composed only of such states because the fact that they have norm zero makes it impossible to calculate expectation values of observables.

As we did for the $t$-clock theory, we must now calculate the inner product of two states $\Psi_1$ and $\Psi_2$ given by \cref{psi11} and \cref{psi2}
\begin{align}
\braket{\Psi_1}{\Psi_2}=&i\int \dk \dl \frac{\dd v}{v}\sum_{n=0}^\infty \left( e^{(\kappa_0'(\lambda)+2n-ik)\varphi}\bar{\alpha}(k,\lambda)\gamma^+_n(\lambda)(\kappa_0'(\lambda)+2n+ik) \right. \nonumber \\
+&\left. e^{(-\kappa_0'(\lambda)-2n-ik)\varphi}\bar{\alpha}(k,\lambda)\gamma^-_n(\lambda)(-\kappa_0'(\lambda)-2n+ik) \right)\times \nonumber \\
&\sqrt{\frac{2\pi}{\sinh(\abs{k}\pi)}}\Re\left[e^{i\frac{\chi(k,\lambda)}{2}}\J{\kappa_0'(\lambda)+2n}{\lambda}\J{i\abs{k}}{\lambda} \right]\, .
\end{align}
It is better for now to use the form with $\chi(k,\lambda)$ for $\Psi_1$ to simplify the calculations. The $v$ integral has been calculated in the \cref{bessel-int}, and the exact formula is given in \cref{2-Bessel-int-3}
\begin{align}
\braket{\Psi_1}{\Psi_2}&=\frac{2i}{\pi}\int \dk \dl \sum_{n=0}^\infty \left( e^{(\kappa'_0(\lambda)+2n-ik)\varphi}\frac{\bar{\alpha}(k,\lambda)\gamma_n^+(\lambda)}{\kappa_0'(\lambda)+2n -ik}\right. \nonumber\\
&-\left. e^{(\kappa_0'(\lambda)-2n-ik)\varphi}\frac{\bar{\alpha}(k,\lambda)\gamma^-_n(\lambda)}{\kappa'_0(\lambda)+2n+ik}\right)\nonumber\\
&\times \sqrt{\frac{2\pi}{\sinh(\abs{k}\pi)}}\Re\left[e^{i\frac{\chi(k,\lambda)}{2}}\sin((\kappa_0'(\lambda)+2n-i\abs{k})\frac{\pi}{2}) \right]\, .
\end{align}
To ensure this is $\varphi$-independent we demand $\Re\left[e^{i\frac{\chi(k,\lambda)}{2}}\sin((\kappa_0'(\lambda)+2n-i\abs{k})\frac{\pi}{2}) \right]=0$. This is equivalent to
\begin{equation}
\tan(\kappa'_0(\lambda)\frac{\pi}{2})=-\tan(\frac{\chi(k,\lambda)}{2})\tanh(\frac{\abs{k}\pi}{2})\, .
\end{equation}
Using \cref{vartheta} and \cref{chi} we find that 
\begin{equation}
\tan(\kappa'_0(\lambda)\frac{\pi}{2})=\tan(\kappa_0(\lambda)\frac{\pi}{2}) .
\end{equation}
Hence, if we choose $\kappa'_0(\lambda)=\kappa_0(\lambda)+2n$ where $n$ is a integer for all $\lambda$, the condition $\pdv{}{\varphi}\braket{\Psi_1}{\Psi_2}=0$ is satisfied. We impose $n=0$ as we defined both $\kappa_0$ and $\kappa_0'$ to take values between 0 and 2. Thus, given a state composed of both real and imaginary Bessel functions, there is no cross contribution to its norm. Note that other choices like  are possible as the tangent function is $\pi$-periodic, but we choose the simplest one.

We are now in condition to write the most generic normalised wave function that satisfies the boundary condition \cref{boundphi} of the $\lambda>0$ sector:
\begin{align}
\Psi(v,\varphi,t)&=\int_{-\infty}^\infty \dk \int_{0}^\infty \dl \ e^{ik\varphi}e^{i\lambda\frac{t}{\hbar}}\alpha(k,\lambda)\sqrt{\frac{2\pi}{\sinh(\abs{k}\pi)}}\times \nonumber\\
&\Re\left[\sqrt{\frac{\sinh((\abs{k}-i\kappa_0(\lambda))\frac{\pi}{2})}{\sinh((\abs{k}+i\kappa_0(\lambda))\frac{\pi}{2})}}\J{i\abs{k}}{\lambda} \right] \nonumber\\
&+\int_0^\infty \dl \ e^{i\lambda\frac{t}{\hbar}}\left[\sum_{n=0}^\infty \left(\gamma^+_n(\lambda)e^{(\kappa_0(\lambda)+2n)\varphi}+\gamma^-_n(\lambda)e^{-(\kappa_0(\lambda)+2n)\varphi} \right)\times \right. \nonumber \\
&\left. \J{\kappa_0(\lambda)+2n}{\lambda} \right]\, ,
\end{align}
where $\int_{-\infty}^\infty \dk \int_0^\infty \abs{\alpha(k,\lambda)}^2=1$. Note that there is no normalisation condition for $\gamma^\pm_n(\lambda)$ because they do not contribute to the total norm of the state.

\section{The $\lambda<0$ sector}
Now that we have analysed in great detail the states with $\lambda>0$, we can turn our attention to the $\lambda<0$ ones. Recall that classically the behaviour of the operator $\hat{\mathcal{G}}=-\hbar^2\pdv[2]{}{u}-\lambda v_0^2e^{2u}$, \cref{Gop} is very different depending on the sign of $\lambda$. In particular, for $\lambda<0$ we will see that no boundary condition is needed. 

Like we did previously, we start by analysing purely imaginary order Bessel functions working with the modified Bessel functions $\I{i\abs{k}}{\lambda}$ and $\K{i\abs{k}}{\lambda}$
\begin{equation}
\Psi_3(v,\varphi,t)=\int_{-\infty}^0 \dl \int_{-\infty}^\infty \ e^{ik\varphi}e^{i\lambda\frac{t}{\hbar}}\left[ \alpha(k,\lambda)\K{i\abs{k}}{-\lambda}+\beta(k,\lambda)\I{i\abs{k}}{-\lambda} \right]\, .
\end{equation}
Note that $\alpha(\lambda,k)$ and $\beta(\lambda,k)$ are not the parameters we found in the $\lambda>0$ case but generic functions. The asymptotic behaviour of the $I$-Bessel functions is
\begin{equation}
\I{\pm i\abs{k}}{-\lambda}\underset{v\rightarrow\infty}{\longrightarrow}\frac{\sqrt{\hbar}e^{\frac{\sqrt{-\lambda}}{\hbar}v}}{\sqrt{2\pi\sqrt{-\lambda}v}}\, ,
\label{Iasympt}
\end{equation}
Given the inner product \cref{phiinnerproduct}, the norm of a state containing $I$-Bessel functions diverges. This feature is common both in the $t$-clock theory and in the $\varphi$-clock theory. Hence, we restrict ourselves to states
\begin{equation}
\Psi_3(v,\varphi,t)=\int^0_{-\infty}\dl \int_{\infty}^\infty \dk \alpha(k,\lambda)\K{i\abs{k}}{-\lambda}\, .
\label{psi3}
\end{equation}
The inner product of these states is
\begin{equation}
\braket{\Psi_3}{\Psi_3}_\varphi=-\int \dl \frac{\dd k_1\dd k_2}{(2\pi)^2}\frac{\dd v}{v}(k_1+k_2)e^{i(k_2-k_1)\varphi}\bar{\alpha}_1\alpha_2 \K{i\abs{k_1}}{-\lambda}\K{i\abs{k_2}}{-\lambda}\, .
\end{equation}
Here we are using the notation $\alpha_i=\alpha(k_i,\lambda)$, and we simplified one $\lambda$ integral using $\int \dd t e^{i(\lambda_1-\lambda_2)\frac{t}{\hbar}}=2\pi \hbar\delta(\lambda_1-\lambda_2)$. The integral over $v$ converges in the distributional sense and its expression can be found in \cref{kintegral}, resulting in
\begin{equation}
\braket{\Psi_3}{\Psi_3}_\varphi=-\int \dl \frac{\dd k_1 \dd k_2}{(2\pi)^2}
e^{i(k_2-k_1)\varphi}\frac{\pi^2(k_1+k_2)\bar{\alpha}_1\alpha_2}{2\abs{k_1}\sinh(\abs{k_1}\pi)}\left[\delta(\abs{k_1}-\abs{k_2})+\delta(\abs{k_1}+\abs{k_2}) \right]\, .
\end{equation}
The term $\delta(\abs{k_1}+\abs{k_2})$ does not contribute to the integral and the other delta can be simplified using \cref{absdelta} to find
\begin{equation}
\braket{\Psi_3}{\Psi_3}_\varphi=-\frac{\pi^2}{2}\int \dl \dk \frac{\abs{\alpha(k,\lambda)}^2}{\sinh(k\pi)}\, .
\label{psi3norm1}
\end{equation}

There are a few things to comment about this result. First of all, during the derivation of \cref{kintegral} we saw there was no contribution to the result from the $v=\infty$ limit, which is another way of seeing that the boundary condition \cref{boundphi} is trivial in this case. The triviality of the boundary condition can also be verified directly by substituting there a state like \cref{psi3} and using the small and large argument asymptotic forms of the $K$-Bessel functions \cref{Kbig} and \cref{Ksmall} in each limit. In a nutshell, the operator $\hat{\mathcal{G}}$ \cref{Gop} is self-adjoint in the $\lambda<0$ case. Secondly, we see that in \cref{psi3norm1} positive $k$ modes have a negative contribution; the inner product is again not positive definite. Hence, given that the $k$ modes are decoupled in this inner product we can define a new inner product for these modes by
\begin{equation}
\braket{\Psi_3}{\Psi_3}_{\varphi'}=\braket{\Psi_3}{\Psi_3}_{\varphi,k<0}-\braket{\Psi_3}{\Psi_3}_{\varphi,k>0}\, ,
\end{equation}
where the subindices $k\lessgtr 0$ refer to the interval of integration in $k$. This is exactly the same modification we did for \cref{newphiinnerprod}. With this new inner product the squared norm of states $\Psi_3$ is
\begin{equation}
\braket{\Psi_3}{\Psi_3}_{\varphi'}=\frac{\pi^2}{2}\int^0_{-\infty}\dl \int_{-\infty}^\infty \dk \frac{\abs{\alpha(k,\lambda)}^2}{\sinh(\abs{k}\pi)}\geq 0\, .
\end{equation}

Finally, we need to consider real order Bessel functions for $\lambda<0$. Given that the asymptotic behaviour of the $I$-Bessel function \cref{Iasympt} does not depend on the order of the Bessel function, real order $I$-Bessel functions are not normalisable. We thus define
\begin{equation}
\Psi_4(v,\varphi,t)=\int_{-\infty}^0 \dl \int_{-\infty}^\infty \dkap e^{\kappa\varphi}e^{i\lambda\frac{t}{\hbar}}\alpha(k,\lambda)\K{\abs{\kappa}}{-\lambda}\, ,
\end{equation}
as a candidate for normalised function. The calculation of $\braket{\Psi_4}{\Psi_4}$ involves the integral \cref{2-KBessels-int-2} which diverges for all values of $\kappa$, hence there are no normalisable real order Bessel function states for $\lambda<0$.

In conclusion, we are now in a position to give the most general normalised solution to the Wheeler--DeWitt equation which solves the boundary condition \cref{boundphi}:
\begin{align}
\Psi(v,\varphi,t)&=\int_{-\infty}^\infty \dk \int_0^\infty \dl e^{i k\varphi}e^{i\lambda\frac{t}{\hbar}}\alpha(k,\lambda)\sqrt{\frac{2\pi}{\sinh(\abs{k}\pi)}}\times \label{phigeneralstate}  \\
& \Re\left[\sqrt{\frac{\sinh(\frac{\pi}{2}(\abs{k}-i\kappa_0(\lambda)))}{\sinh(\frac{\pi}{2}(\abs{k}+i\kappa_0(\lambda)))}}\J{i\abs{k}}{\lambda} \right]  \nonumber\\
&+\int_0^\infty \dl e^{i\lambda\frac{t}{\hbar}}\left[\sum_{n=0}^\infty \left(\gamma^+_n(\lambda)e^{(\kappa_0(\lambda)+2n)\varphi}+\gamma_n^-(\lambda) e^{-(\kappa_0(\lambda)+2n)\varphi}\right)\times \right.  \nonumber\\ 
&  \left. \J{\kappa_0(\lambda)+2n}{\lambda} \right] \nonumber \\
& + \int_{-\infty}^\infty \dk \int_{-\infty}^0 \dl e^{ik\varphi}e^{i\lambda\frac{t}{\hbar}}\sqrt{\frac{2\sinh(\abs{k}\pi)}{\pi^2}}\eta(k,\lambda)\K{i\abs{k}}{-\lambda}\, , \nonumber
\end{align}
where $\int_0^\infty \dl \int_{-\infty}^\infty \dk \abs{\alpha(k,\lambda)}^2+\int_{-\infty}^0 \dl \int_{-\infty}^\infty \dk \abs{\eta(k,\lambda)}^2 =1$. As we have discussed previously, the function $\kappa_0(\lambda)\in [0,2)$ represents the $\hat{\mathcal{G}}$ \cref{Gop} self-adjoint extension choice, in a similar fashion than the parameter $\theta(k)$ in the $t$-theory. Thus, the most obvious question is what is the rôle of $\kappa_0(\lambda)$ in the dynamics of the theory? We have not studied this explicitly, but we are confident that $\kappa_0(\lambda)$ does not change substantially the behaviour of relevant observables. In the closely related model \cite{Pawlowski2011} it is shown that the impact of choosing different parameters for a fixed value of $\lambda$ is negligible. 

Regarding the physical interpretation of the theory, we found that the modes with $\lambda>0$, that correspond to classical solutions which can reach infinity in a finite time, must satisfy a boundary condition. This boundary condition may be interpreted as a reflection around $v=\infty$. Identically to what happened to the $t$-theory, states of the form \cref{phigeneralstate} are continuous everywhere. When analysing the classical theory (see \cref{clasplot-vphi}), we saw that classical solutions $v(\varphi)$ and $t(\varphi)$ are not continuous everywhere, rather they present a discontinuity when $\varphi=\varphi_0$ (where $v$ reaches $\infty$). The quantum boundary condition ensures continuity past the threshold $\varphi=\varphi_0$ as the $\lambda>0$ modes are reflected from $v=\infty$. The $\varphi$ clock is indeed slow at $v=\infty$, but contrary to the $t$ clock, it is fast at the singularity. In the numerical analysis we will see that this reflecting boundary condition implies that the solutions experience a turn around at finite maximum volume rather than reaching infinity, thus expanding the Gotay and Demaret conjecture to include also ``infinity resolution''.

In conclusion, we have three theories for which the allowed wave functions look very different. From \cref{t-clock-sec} to \cref{phi-clock-sec} we have illustrated the breaking of covariance coming from the clock choice. As explained in \cref{dirac-sec}, the $t$-clock and $\varphi$-clock theories can also be obtained by the Dirac quantisation of the same model multiplying the Hamiltonian by a non-trivial phase space function. These results trigger the question: is there a preferred quantum theory? From the clock choice perspective there is no preference, all the chosen clocks are valid and equivalent choices, and all of them have been explored in the literature. However, one can argue that the theory that gives the desired dynamics should be considered over the rest. But, what is the ``desired dynamics''? In the context of quantum cosmology (and black hole spaces), the ultimate goal for a quantum universe is to resolve the singularity (the definition of ``resolving the singularity'' comes with its own caveats that we will analyse shortly). It is then interesting to study the dynamics of the three theories we found. In the following chapter we will calculate numerically expectation values of different observables and build testable criteria for singularity resolution for all theories.
\part{Numerical analysis of the different theories}
\chapter{Dynamics of the three theories}
\label{numerics}
One of the main reasons to consider quantum gravity, and in particular, quantum cosmology, is to resolve singularities. The most striking ones are perhaps the big bang/big crunch singularities, as they are not hidden by an event horizon. So far, we have done a complete analysis of the three different theories obtained considering different dynamical variables as quantum clocks, and we have seen that they are not equivalent. In consequence, we expect different behaviours towards the big bang/big crunch singularity. Nonetheless, defining conditions for singularity resolution is not trivial, and we must first specify them. In fact, several criteria are used in the literature, see \cite{Singularity}. A strong criterion is to demand that the energy density (that classically becomes infinite at the big bang/big crunch) or other divergent physical quantities have a universal upper bound satisfied for all states; see \cite{Ashtekar2008} for an example of energy density upper bound. 

We use a simpler and less constraining criterion for our theory, similar to what was proposed by Gryb and Thébault \cite{Gryb}, i.e, we demand that expectation values of classically singular quantities, like the volume or the scale factor, are always non singular for a given semiclassical state. However, the meaning of ``non-singular behaviour'' and ``semiclassical state'' has to be specified in a case by case basis. We will give more details about singularity resolution criteria in the following sections.

In the rest of the chapter we will construct semiclassical states for each of the three theories and give explicit values for expectation values of $v$ and $t$ with respect to these semiclassical states, compare them to the relevant classical Dirac observables calculated in \cref{diracobs}, and analyse whether these semiclassical universes are non-singular. After we have understood the dynamics of each theory, we will analyse their causal structure to highlight their differences. 

\section{Dynamics of the $t$-clock theory}
\label{t-clock-dynamics}

We start by analysing the dynamics of the Schrödinger theory. Let us recall the general solution of the Wheeler--DeWitt equation and the boundary condition:
\begin{align}
\Psi(v,\varphi,t)&=\int_{-\infty}^\infty \dk e^{ik\varphi}\left[\sum_{n=-\infty}^\infty e^{i\lambda^k_n \frac{t}{\hbar}} \frac{1}{\hbar}\sqrt{\frac{-2\lambda^k_n\sinh(\abs{k}\pi)}{\pi \abs{k}}}\alpha(k,\lambda^k_n)\K{i\abs{k}}{-\lambda^k_n} \right. \nonumber \\
& \left.+\int_0^\infty \dl e^{i\lambda\frac{t}{\hbar}}\frac{\sqrt{2\pi}\Re\left[e^{i\theta(k)-ik\log\sqrt{\frac{\lambda}{\lambda_0}}}\J{i\abs{k}}{\lambda}\right]}{\sqrt{\hbar\cos\left(-2\theta(k)+\abs{k}\log\frac{\lambda}{\lambda_0}\right)+\hbar\cosh(\abs{k}\pi)}}\alpha(k,\lambda) \right]\, ,
\end{align}
where $\theta(k)$ is a free parameter. The first line corresponds to the $\lambda<0$ modes and the second line to $\lambda>0$. 

Building a semiclassical state amounts to choosing a form for $\alpha(k,\lambda)$, $\alpha(k,\lambda^k_n)$, and $\theta(k)$. Our choices are the following:
\begin{align}
\theta(k)&=0, \\
\alpha(k,\lambda^k_n)&=0 , \\
\alpha(k,\lambda)&=\alpha_{sc}(k,\lambda)=C\frac{2\sqrt{\hbar \pi}}{\sqrt{\sigma_k\sigma_\lambda}}e^{-\frac{(k-k_c)^2}{2\sigma^2_k}-\frac{(\lambda-\lambda_c)^2}{2\sigma^2_\lambda}}\, . \label{alphasc}
\end{align}
The choice $\theta(k)=0$ is the simplest choice of self-adjoint extension. We work under the assumption that the qualitative results are independent of the self-adjoint extension choice, as studied in \cite{Gryb}, in particular \cite{Gryb2019b}. However, any given function of $k$ (even non-continuous functions) is allowed. It is neither possible, nor needed, to study the effect of all possible self-adjoint extensions to analyse singularity resolution in this theory. 

Our choice $\alpha(k,\lambda_n^k)=0$ comes from the analysis of our universe. Let us recall the physical interpretation of the parameter $\lambda$. In \cref{energyint} we saw that it can be related to a perfect fluid energy density, and hence its values will depend on the energy density interpretation. In particular, $\lambda<0$ implies a negative energy density, which would be considered as exotic matter in the case where the perfect fluid is dust or radiation. In this work we have focused on the ``dark energy interpretation'' of $\lambda$, i.e. we interpret $\lambda$ as a cosmological constant. In this case negative values are possible, but according to the data we have, it is very clear that our universe has a small positive cosmological constant. Therefore, we will only consider positive $\lambda$ modes.

Our last choice is standard in the study of minisuperspaces. We consider a normalised wave functions centred around the classical values $\lambda_c$ and $k_c$. Again, to simplify some calculations, we will only consider $k>0$ modes and therefore the constant $C$ plays the rôle of a normalisation constant; it ensures $\int_0^\infty \dl \int_0^\infty \dk \abs{\alpha_{sc}(k,\lambda)}^2=1$. The parameters $\sigma_k$ and $\sigma_\lambda$ are the standard deviations of the Gaussians. A state with smaller expectation values can be considered as more ``classical''. For most states we consider $k_c$ and $\lambda_c$ are large enough with respect to $\sigma_k$ and $\sigma_\lambda$, that $C$ can be approximated to 1.

In a nutshell, our semiclassical state is 
\begin{equation}
\Psi_{sc,t}(v,\varphi,t)=\int_0^\infty \dk \int_0^\infty \dl e^{ik\varphi}e^{i\lambda\frac{t}{\hbar}}\alpha_{sc}(k,\lambda)\frac{\sqrt{2\pi}\Re\left[ e^{ik\log\sqrt{\frac{\lambda}{\lambda_0}}}\J{ik}{\lambda}\right]}{\sqrt{\hbar \cos(k\log\frac{\lambda}{\lambda_0})+\hbar \cosh(k\pi)}}\, .
\end{equation}
Here the subindex $t$ is a label to distinguish the semiclassical states from the rest of the theories. Note again the equal weight combination of Bessel functions of $\pm i k$ index. Recall that $\lambda_0$ is a constant needed for dimensional reasons that will be set to 1 to obtain the numerical results of this section.

Now that we have an expression for the semiclassical states, we can calculate the expectation values of different observables. In this theory the most interesting observable probably is the volume
\begin{equation}
\bra{\Psi_{sc,t}(v,\varphi,t)}v\ket{\Psi_{sc,t}(v,\varphi,t)}=\expval{v(t)}_{\Psi_{sc,t}}\, .
\end{equation}
The label $t$ refer to the fact that this calculation is done using the $t$-clock inner product. The expression for this observable is
\begin{align}
\expval{v(t)}_{\Psi_{sc,t}}&=\int_0^\infty \dd v \ v^2\int \frac{\dd \lambda_1 \dd \lambda_2}{(2\pi\hbar)^2}\dk e^{-i(\lambda_1-\lambda_2)\frac{t}{\hbar}}\alpha_{sc}(k,\lambda_1)\alpha_{sc}(k,\lambda_2)\times \nonumber\\
& \frac{2\pi \Re\left[e^{-i k\log \sqrt{\frac{\lambda_1}{\lambda_0}}}\J{ik}{\lambda_1}\right]\Re\left[e^{-i k\log \sqrt{\frac{\lambda_2}{\lambda_0}}}\J{ik}{\lambda_2}\right]}{\sqrt{\hbar \cos(k\log\frac{\lambda_1}{\lambda_0})+\hbar\cosh(k\pi)}\sqrt{\hbar \cos(k\log\frac{\lambda_2}{\lambda_0})+\hbar\cosh(k\pi)}}\, .
\label{vtsemiclass}
\end{align}
These integrals cannot be simplified by the techniques we used in our appendix. As $\alpha_{sc}(k,\lambda)$ is chosen to be real $\expval{v(t)}_{\Psi_{sc,t}}$ is symmetric with respect of time reversal $t\longrightarrow -t$, which simplifies the numerical evaluation. Indeed, one can evaluate $\expval{v(t)}_{\Psi_{sc,t}}$ for positive values of $t$, and  then deduce the $\expval{v(t)}_{\Psi_{sc,t}}$ for the corresponding negative values of $t$. Given that numerical evaluation is very slow, this allows us to save a lot of time. 

Nevertheless, the evaluation of \cref{vtsemiclass} is a challenge numerically. To simplify it we will assume that the Gaussian over $k$ is sharply peaked, i.e.~$\frac{\sigma_k}{k_c}\ll 1$. Consequently, $\alpha_{sc}(k,\lambda_1)\alpha_{sc}(k,\lambda_2)=K(k)L(\lambda_1)L(\lambda_2)$ where $K(k)$ can be approximated to a Dirac $\delta$ distribution, $K(k)\approx \delta(k-k_c)$ up to a certain degree, hence removing the $k$ integration from the calculation. The $v$ integral is still a challenge, so we have verified that $\braket{\Psi_{sc}}{\Psi_{sc}}=1$ to very high precision.

Let us recall the characteristics of the $t$-clock theory. In \cref{class-sol-sec} we have seen that the classical trajectories terminate at a finite $t_{sing}$ where $v(t_{sing})=0$. The classical evolution cannot be continued beyond $t=t_{sing}$, however, our quantum theory is by construction unitary so the state is well-defined along the entire $t$ axis. Hence, our criterium for singularity resolution is that if for a specific $\Psi_{sc,t}$, there exists a constant $C_{\Psi_{sc,t}}$ such that $\expval{v(t)}_{\Psi_{sc,t}}\geq C_{\Psi_{sc,t}}> 0$, this state resolves the singularity. A stronger criterium for singularity resolution would be to require that the constant $C_{\Psi_{sc,t}}$ is independent of the state, but we will not do this in this work. Some other quantum behaviours that we might observe are $\expval{v(t_p)}_{\Psi_{sc,t}}=\infty$ for a finite $t_p$ or $\expval{v(t)}_{\Psi_{sc,t}}\longrightarrow 0$ when $t\longrightarrow \pm \infty$, however these behaviours can still be seen as singularities in the quantum theory.

\begin{figure}
\centering
\begin{subfigure}{1\textwidth}
\centering
\includegraphics[width=0.8\textwidth]{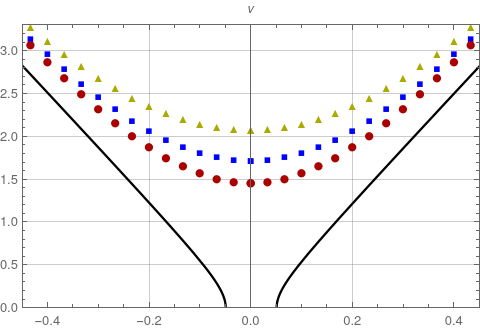}
\caption{The dotted lines represent $\expval{v(t)}_{\Psi_{sc,t}}$ for values $k_c=1$, $\lambda_c=10$, $\hbar=1$, and $\sigma_\lambda=3$ (red circles), $\sigma_\lambda=2.5$ (blue squares) and $\sigma_\lambda=2$ (yellow triangles). The black thick line corresponds to the classical trajectory with the same values of $k_c$ and $\lambda_c$. We see that the quantum trajectory is well-defined for all $t$'s and reaches a positive minimum value $V_{min}$.}
\end{subfigure}
\begin{subfigure}{1\textwidth}
\centering
\includegraphics[width=0.8\textwidth]{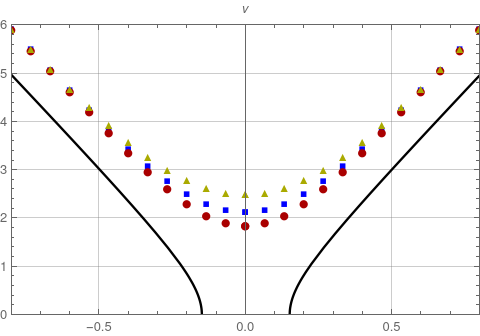}
\caption{The dotted lines represent $\expval{v(t)}_{\Psi_{sc,t}}$ for values $k_c=3$, $\lambda_c=10$, $\hbar=1$, and $\sigma_\lambda=3$ (red circles), $\sigma_\lambda=2.5$ (blue squares) and $\sigma_\lambda=2$ (yellow triangles). The black thick line corresponds to the classical trajectory with the same values of $k_c$ and $\lambda_c$. We see that the quantum trajectory is well-defined for all $t$'s and reaches a positive minimum value $V_{min}$.}
\end{subfigure}
\caption{Comparison between $\expval{v(t)}_{\Psi_{sc,t}}$ and $v_c(t)$ for different values of the parameters.}
\label{vtplots}
\end{figure}

We compare $\expval{v(t)}_{\Psi_{sc,t}}$ \cref{vtsemiclass} to the classical curve
\begin{equation}
v_c(t)=\sqrt{4\lambda_ct^2-\frac{\hbar^2k_c^2}{\lambda_c}}\, ,
\label{vclasst}
\end{equation}
where $k_c$ and $\lambda_c$ take the same values as in $\alpha_{sc}(k,\lambda)$ \cref{alphasc}. \Cref{vtplots} shows the comparison between $\expval{v(t)}_{\Psi_{sc,t}}$ and $v_c(t)$ for different values of the parameters. We see clearly that $\expval{v(t)}_{\Psi_{sc,vt}}$ reaches a minimum value strictly above zero and slowly tends to the classical curve as $t$ increases. The quantum expectation value is always above the classical one and it is well-defined before the big bang and after the big crunch singularity. This behaviour is exactly what we would expect from a nonsingular universe.

\Cref{vtplots} also shows solutions for various values of $\sigma_\lambda$ with the rest of the parameters unchanged. To our surprise, the states with greater standard deviations $\sigma_\lambda$ seem to agree better with the classical theory at late times and seem to have a smaller minimum value. In other words, states with a larger quantum spread seem to have a more abrupt transition between the two classical branches. In fact, the minimum value $\expval{v(t=0)}_{\Psi_{sc,t}}=V_{min}$ is a function of the parameters $\lambda_c$, $k_c$ and $\sigma_\lambda$, and it appears that it is decreasing in $\sigma_\lambda$. For values of the parameters $k_c=1$, $\lambda_c=10$ and $\hbar=1$ we have
\begin{equation}
V_{min,\sigma_\lambda=2}=2.08\pm 0.04 ,\hspace{4mm} V_{min,\sigma_\lambda=2.5}=1.72\pm 0.04, \hspace{4mm} V_{min,\sigma_\lambda=3}=1.47\pm 0.02,
\label{vmin}
\end{equation}
where the errors are the estimation of the integration error of Mathematica. This quantity does not have a classical analogue, hence the fact that $V_{min}$ decreases with $\sigma_\lambda$ is not problematic \emph{per se} for the quantum theory. In \cite{Gryb}, the authors found a very similar behaviour in a limit in which the contribution of the cosmological constant dominates over the scalar field. They found that $V_{min}\propto 1/\sigma_\lambda$. Our results do not assume this limit and deviate from this relation but not very strongly.

One interpretation to this behaviour is the following: despite that $\lambda$ and $t$ are conjugated variables, in some sense $\lambda$ is also conjugated to $v$ (recall the small argument asymptotic behaviour of the Bessel functions $\J{ik}{\lambda}\propto e^{i k \log\frac{\sqrt{\lambda}v}{2\hbar}}$), so that a greater spread in $\lambda$ would imply a smaller spread in $t$ and $v$, bringing the quantum expectation value closer to the classical solution.

The errors in $V_{min}$ tell us that the integrals of this section are non-trivial to implement numerically. The $v$ integral in \cref{vtsemiclass} poses the most problems: The lower end contribution is always zero as the prefactor of the Bessel functions cancels the infinite oscillations, but the upper end contribution is harder to deal with. The Bessel functions behave like trigonometric functions for large values of $v$ and these oscillating integrals are a challenge, for this reason the estimated error of the numerical results is non negligible. There are a couple of secondary factors that one should also take into account when doing these numeric calculations. First of all, increasing $k_c$ mean that the Bessel function oscillate more rapidly, interfering further in the calculations. This is why we decided to study only modes with $k_c\leq 3$. Furthermore, the term $e^{-i(\lambda_1-\lambda_2)\frac{t}{\hbar}}$ adds more oscillations, making the results at large $t$ more prone to errors. Finally, decreasing $\sigma_\lambda$, makes $\alpha_{sc}(k,\lambda)$ more peaked, and functions with larger gradients tend to be more difficult to integrate (this explains the errors in \cref{vmin}). It is almost impossible to establish an exact parameter range in which the numerical integration gives a reliable enough result, but we consider that within the ranges we are plotting in \cref{vtplots} numerical errors are manageable.

We are aware \cref{vtplots} may not convey the fact that $\expval{v(t)}_{\Psi_{sc,t}}$ tends to the classical trajectory, at least when $k=3$. We had to find a compromise where the behaviour around $v=0$ is clearly visible, the difference between the different quantum curves noticeable, and the plot range wide enough to see the dynamics at larger $t$. In order to check that the classical and quantum curves agree at larger times ($t$) we have checked the relative difference between the two in a safe range of the parameters:
\begin{equation}
\Delta_{rel}(t)=\left(1-\frac{\expval{v(t)}_{\Psi_{sc,t}}}{v_c(t)} \right)\, .
\end{equation}
As the quantum curve seems to be always above the classical one, $\Delta_{rel}$ is negative. Looking at \cref{Deltarelt} we see that in all cases, $\Delta_{rel}$ approaches 0 when $t$ increases and smaller values of $\sigma_\lambda$ give larger numerical errors. Larger values of $k_c$ seem to slow down convergence, but we do not consider this to be of worry. 

\begin{table}
\begin{center}
\begin{tabular}{c|c|c|c|c|c}
$\lambda_c$ & $k_c$ & $\sigma_\lambda $ & $\Delta_{rel}(0.2)\ (\%)$ & $\Delta_{rel}(0.5)\ (\%)$ & $\Delta_{rel}(1)\ (\%)$  \\
\hline
\hline
10 & 1& 3&$-53.15\pm 3.73$ &$-10.51\pm 2.44$& $-4.76\pm 2.03$\\
10 &1 & 2.5 &$-68.00\pm 2.49$&$-11.80\pm 2.58$&$-4.921\pm 0.901$ \\
10 &1 &2 &$-92.59\pm 5.37$&$-15.13\pm 3.39$&$-5.15\pm 1.73$ \\
10& 3&3 &$-173.80\pm 1.98$ &$-32.025 \pm 0.943$ & $-14.870 \pm 0.990$ \\
10 & 3 & 2.5 & $-96.37\pm 7.53$ &$-33.07\pm 1.14$ & $-14.97\pm 1.05$ \\
10 & 3 & 2 & $-233.6\pm 12.4$ & $-36.34\pm 3.70$ & $-15.03\pm 1.82$ 
\end{tabular}
\end{center}
\caption{Relative difference between the classical solution and quantum expectation values of $\lambda_c$, $k_c$ and $\sigma_\lambda$. The errors are calculated using Mathematica integration error estimate and standard error calculation.}
\label{Deltarelt}
\end{table}

Finally, it is interesting to study the variance of the states throughout the bounce. If a state maintains a small variance throughout the evolution, it stays semiclassical. This semiclassicality occurs in models of LQC \cite{Singularity}. The variance of our model is:
\begin{equation}
\sigma^2{\expval{v(t)}_{\Psi_{sc,v}}}=\expval{v^2(t)}_{\Psi_{sc,v}}-\expval{v(t)}^2_{\Psi_{sc,v}}\, 
\end{equation}
where,
\begin{align}
\expval{v^2(t)}_{\Psi_{sc,t}}&=\int_0^\infty \dd v \ v^3\int \frac{\dd \lambda_1 \dd \lambda_2}{(2\pi\hbar)^2}\dk e^{-i(\lambda_1-\lambda_2)\frac{t}{\hbar}}\alpha_{sc}(k,\lambda_1)\alpha_{sc}(k,\lambda_2)\times \nonumber\\
& \frac{2\pi \Re\left[e^{-i k\log \sqrt{\frac{\lambda_1}{\lambda_0}}}\J{ik}{\lambda_1}\right]\Re\left[e^{-i k\log \sqrt{\frac{\lambda_2}{\lambda_0}}}\J{ik}{\lambda_2}\right]}{\sqrt{\hbar \cos(k\log\frac{\lambda_1}{\lambda_0})+\hbar\cosh(k\pi)}\sqrt{\hbar \cos(k\log\frac{\lambda_2}{\lambda_0})+\hbar\cosh(k\pi)}}\, .
\end{align}
This time, the numerical integration of this quantity poses even more problems than $\expval{v(t)}_{\Psi_{sc,t}}$ due to the divergent integral $\int \dd v \ v^3$. We have tried several parameter combinations, and we consider that the integration errors are too high, hence we cannot present them in this section.

Overall, we consider that the results we obtained are accurate enough to prove that this theory resolves the singularity and the universe experiences a bounce. For a deeper numerical analysis (in particular regarding the dependence in the self-adjoint extension parameter) we refer to \cite{Gryb}. We confirm the fact that a slow clock ( as defined in \cite{Gotay}) solves the singularity generally.

\section{Dynamics of the $v$-clock theory}
\label{v-clock-dynamics}

Among all the studied theories, the $v$-theory has always been the outsider, and in this section this will be no different. In \cref{v-clock-sec} we have seen that this theory is unitary without boundary condition. The expression for a normalised state is
\begin{align}
\Psi(v,\varphi,t)&=\int_{-\infty}^\infty \dl \int_{-\infty}^\infty \dk e^{ik\varphi}e^{i\lambda\frac{t}{\hbar}}\sqrt{\frac{\pi}{2\sinh(\abs{k}\pi)}}\times \nonumber \\
&\left[\alpha(k,\lambda)\J{i\abs{k}}{\lambda}+\beta(k,\lambda)\J{-i\abs{k}}{\lambda} \right]\, .
\end{align}
The small argument asymptotic of the Bessel function is comparable to a plane wave: $\lim_{v\rightarrow 0} \J{\pm i\abs{k}}{\lambda}\propto e^{\pm i \abs{k} \frac{\sqrt{\lambda}}{2\hbar}}$, the $+$ sign are waves going into the singularity and the $-$ sign are waves going out of the singularity. We want to compare the quantum expectation values with a classical universe that expands from the big bang singularity (see \cref{clasplot-vt}), we consider solutions that are composed of only outgoing waves, i.e. $\alpha(k,\lambda)=0$. To construct semiclassical states, we make very similar assumptions to the $t$ theory, namely:
\begin{itemize}
\item We limit ourselves to the range $\lambda>0$, in order to work with universes with a perfect fluid positive energy density. 
\item We set $\beta(k,\lambda):=\alpha_{sc}(k,\lambda)$ as defined in \cref{alphasc}, i.e., a double normalised Gaussian centred in the classical values $k_c$ and $\lambda_c$
\end{itemize}
The expression of a semiclassical state is then
\begin{equation}
\Psi_{sc,v}(v,\varphi,t)=\int_0^\infty \dl \int_{-\infty}^\infty \dk e^{i k\varphi}e^{i\lambda\frac{t}{\hbar}}\sqrt{\frac{\pi}{2\sinh(\abs{k}\pi)}}\alpha_{sc}(k,\lambda)\J{-i\abs{k}}{\lambda}\, .
\end{equation}
The sub-indices $sc,v$ refer to the fact that this state is specifically the semiclassical state of the $v$-clock theory.

\begin{figure}
\centering
\includegraphics[width=0.75\textwidth]{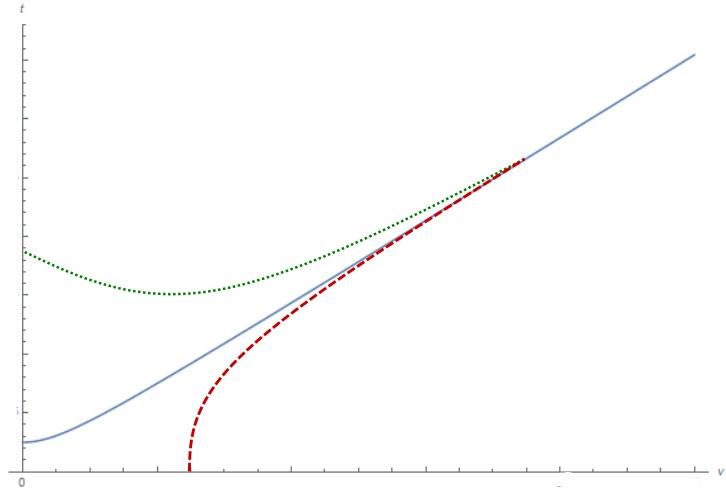}
\caption[Singularity resolution scenarios in the $v$-clock theory]{Possible singularity resolution in the $v$-clock theory. The blue  solid line represents a classical trajectory $t(v)$ for $\lambda>0$. The green dotted line represents a hypothetical trajectory of $\expval{t(v)}_{\Psi_{sc,v}}$ where $\partial_{v}\expval{t(v)}_{{\Psi_{sc,v}}}=0$ and then $\expval{t(v)}_{\Psi_{sc,v}}$ experiences a turnaround. The red dashed line corresponds to a $\expval{v(t)}_{\Psi_{sc,v}}$ trajectory that becomes ill-defined. Both hypothetical trajectories would be considered singularity resolution even if the green dotted curve reaches $v=0$.}
\label{tv-nonsing}
\end{figure}

In this case, studying singularity resolution is not as straightforward as for the other theories, because the volume of the universe $v$ is now the clock variable. However, we can study the expression of other important observables, like $\bra{\Psi_{sc,v}}t\ket{\Psi_{sc,v}}=\expval{t(v)}_{\Psi_{sc,v}}$. Classically, $v(t)$ and $t(v)$ are invertible quantities and classical trajectories, $t(v)$  cannot be defined past $t=t_{sing}$ as $v$ cannot be negative. Consequently, we consider that $\expval{t(v)}_{\Psi_{sc,v}}$ avoids the singularity if we observe one of the following behaviours: either $\expval{t(v)}_{\Psi_{sc,v}}$ becomes ill-defined as $\abs{\partial_v\expval{t(v)}_{\Psi_{sc,v}}}\rightarrow \infty$ for some value of $v$, or $\expval{t(v)}_{\Psi_{sc,v}}$ starts moving backwards, i.e., $\partial_v\expval{t(v)}_{\Psi_{sc,v}}=0$ somewhere. These hypothetical behaviours are presented in \cref{tv-nonsing}.

It is possible to simplify the expression for $\expval{t(v)}_{\Psi_{sc,v}}$. In fact,
\begin{align}
\expval{t(v)}_{\Psi_{sc,v}}&=\bra{\Psi_{sc,v}}t\ket{\Psi_{sc,v}} \nonumber\\
&=i\frac{\pi v}{2}\int \frac{\dd \lambda_1 \dd \lambda_2}{(2\pi)^2}\dk \dd t \ t e^{-i(\lambda_1-\lambda_2)\frac{t}{\hbar}}\frac{\bar{\alpha}_{sc}(k,\lambda_1)\alpha_{sc}(k,\lambda_2)}{\sinh(\abs{k}\pi)}\times \nonumber \\
&\left[\J{i\abs{k}}{\lambda_1}\partial_v\J{-i\abs{k}}{\lambda_2}-\J{-i\abs{k}}{\lambda_2}\partial_v\J{i\abs{k}}{\lambda_1} \right]\, .
\end{align}
Here we simplified the integral $\int \dd\varphi e^{-i(k_1-k_2)\varphi}=2\pi\delta(k_1-k_2)$. We cannot use the same trick for the $t$ integral, however this integral can be solved using:
\begin{align}
\int \frac{\dd \lambda_1\dd \lambda_2}{(2\pi\hbar)^2}& \dd t \ t e^{-i(\lambda_1-\lambda_2)\frac{t}{\hbar}}F(\lambda_1,\lambda_2) \nonumber \\
&=-\frac{i\hbar}{2}\int \dl \left. \left(\pdv{}{\lambda_1}F(\lambda_1,\lambda_2)-\pdv{}{\lambda_2}F(\lambda_1,\lambda_2) \right)\right|_{\lambda_1=\lambda_2=\lambda}\, .
\label{deltader}
\end{align}
This leads to
\begin{align}
\expval{t(v)}_{\Psi_{sc,v}}&=\int \dl \dk \abs{\alpha_{sc}(k,\lambda)}^2f(v,k,\lambda)\nonumber\\
 &+\frac{i\hbar}{2}\int \dl \dk \left[\bar{\alpha}_{sc}(k,\lambda)\partial_\lambda \alpha_{sc}(k,\lambda)-\partial_\lambda\bar{\alpha}_{sc}(k,\lambda)\alpha_{sc}(k,\lambda) \right]\, , 
\label{tvsemiclass}
\end{align}
where
\begin{align}
f(v,k,\lambda)=&\frac{\pi}{4\hbar\sinh(\abs{k}\pi)}\left[v^2\abs{\J{-1+i\abs{k}}{\lambda}}^2+\left(\frac{2\hbar^2k^2}{\lambda}+v^2 \right)\abs{\J{i\abs{k}}{\lambda}}^2  \right. \nonumber \\
-&\frac{i\hbar \abs{k}v}{\sqrt{\lambda}}\left[\J{i\abs{k}}{\lambda}\J{-1-i\abs{k}}{\lambda}-\text{c.c.} \right]\, ,
\label{f}
\end{align}
where the notation c.c.~means complex conjugate. Two things stand out from this expression. The first one is that \cref{tvsemiclass} does not depend on any approximation or expression for $\alpha_{sc}$ yet, hence this expression is not only valid on the semiclassical states we want to study, but also for general states. The second one is that the expression \cref{tvsemiclass} is manifestly real, as it should be.

Let us now consider $\alpha_{sc}(k,\lambda)$ as in \cref{alphasc}. Because $\alpha_{sc}(k,\lambda)$ is real, the second line of $\expval{t(v)}_{\Psi_{sc,v}}$ is trivially zero. In the limit $\sigma_k\rightarrow 0$ and $\sigma_\lambda\rightarrow 0$ we can approximate the integral:
\begin{equation}
\expval{t(v)}_{\Psi_{sc,v}}= \int \dl \dk \abs{\alpha_{sc}(k,\lambda)}^2f(v,k,\lambda)\approx f(v,k_c,\lambda_c)\, .
\label{tvapprox}
\end{equation}
This limit can be considered as a classical limit, as we have $\frac{\sigma_k}{k_c}\ll 1$ and $\frac{\sigma_\lambda}{\lambda_c}\ll 1$, and very small standard deviation implies small dispersion of the states. We would like to compare \cref{tvapprox} with the classical solution
\begin{equation}
t_c(v)=\frac{1}{2}\sqrt{\frac{\hbar^2k_c^2}{\lambda_c^2}+\frac{v^2}{\lambda_c}}\, ,
\label{tclass}
\end{equation}
where the subindex $c$ refers to the fact that this is a classical solution. $\lambda_c$ and $k_c$ should have the same values in the two expressions for a meaningful comparison. Expanding \cref{f} and \cref{tclass} around $v=0$ (the classical singularity) we find
\begin{align}
t_c(v)&= \frac{\hbar \abs{k_c}}{\lambda_c}+\frac{v^2}{4\hbar \abs{k_c}}-\frac{\lambda_c v^4}{16\hbar^3\abs{k_c}^3}+\frac{\lambda_c^2v^6}{32\hbar^5\abs{k_c}^5}+O(v^8) \\
f(v,k_c,\lambda_c)&=\frac{\hbar \abs{k_c}}{2\lambda_c}+\frac{v^2}{4\hbar \abs{k_c}}-\frac{\lambda_cv^4}{16\hbar^3(\abs{k_c}+\abs{k_c}^3)}+\frac{\lambda_c^2v^6}{32\hbar^5(4\abs{k_c}+5\abs{k_c}^3+\abs{k_c}^5)}+O(v^8)
\label{taylort}
\end{align}
There is a very close agreement between these two expressions, the first difference comes at order $v^4$. These expressions reveal an interesting interpretation of the parameter $k_c$: in the large $k_c$ limit, the two solutions agree, hence $k_c$ can be considered as another measure of semi-classicality. It is therefore insightful to study solutions with different values of $k_c$\footnote{As $k$ grows, the Bessel functions become increasingly harder to numerically integrate, hence we will not study functions with $k>10$.}. 

Already at this stage, we can confirm that this  quantum theory and the classical theory are in very close agreement. This backs up the theoretical analysis done in \cref{v-clock-sec}. There are no signs of singularity resolution as described around \cref{tv-nonsing}. However, we would like to plot $t_c(v)$ and $\expval{t(v)}_{\Psi_{sc,v}}$. In order to do that effectively, we use one last approximation. Assuming the limit is well-defined, we will only consider Gaussians \cref{alphasc} that are very peaked in $k$, i.e~$\frac{\sigma_k}{k_c}\ll 1$. Thus, $\abs{\alpha_{sc}(k,\lambda)}^2=K(k)L(\lambda)$ where $K(k)$ can be approximated to a delta distribution $K(k)\approx 2\pi\delta(k-k_c)$ allowing us to drop $k$ integral in \cref{tvsemiclass}. We are aware that this is a tremendous simplification, but without this approximation the numerical implementation of \cref{tvsemiclass} is almost impossible due to the non-triviality of the integral of the Bessel functions. The parameter $\sigma_\lambda$ remains and different values of it represent different states: if $\frac{\sigma_\lambda}{\lambda_c}$ is small, the wave function has less dispersion in $\lambda$ and hence can be considered as a ``more classical'' state.

\begin{figure}
\centering
\begin{subfigure}{1\textwidth}
\centering
\includegraphics[width=0.8\textwidth]{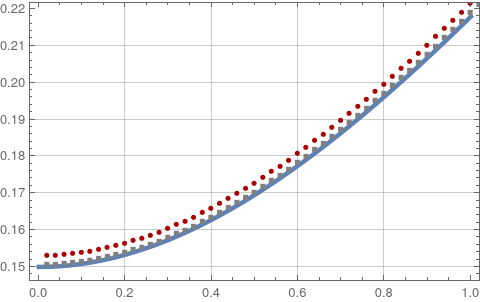}
\caption{The dotted lines represent $\expval{t(v)}_{\Psi_{sc,v}}$ for values of $k_c=3$, $\lambda_c=10$, $\hbar=1$ and $\sigma_\lambda=2$, (red circles) and $\sigma_\lambda=1$ (grey squares). The blue thick line corresponds to the classical trajectory $t_c(v)$ for the same values of $k_c$ and $\lambda_c$. We can see that the classical and quantum curves are very close to each other.}
\end{subfigure}
\begin{subfigure}{1\textwidth}
\centering
\includegraphics[width=0.8\textwidth]{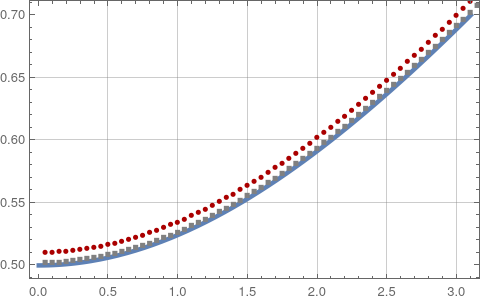}
\caption{The dotted lines represent $\expval{t(v)}_{\Psi_{sc,v}}$ for values of $k_c=10$, $\lambda_c=10$, $\hbar=1$, and $\sigma_\lambda=2$, (red circles) and $\sigma_\lambda=1$ (grey squares). The blue thick line corresponds to the classical trajectory $t_c(v)$ for the same values of $k_c$ and $\lambda_c$. Once again, the classical and quantum curves are very close to each other.}
\end{subfigure}
\caption{Comparison between $\expval{t(v)}_{\Psi_{sc,v}}$ and $t_c(v)$ for different values of the parameters.}
\label{fig-tv}
\end{figure}

In \cref{fig-tv} we present $\expval{t(v)}_{\Psi_{sc,v}}$ with respect to different values of $k_c$ and $\sigma_\lambda$. We see that the quantum expectation values follow the classical trajectory to the singularity, and there are no apparent signs of singularity resolution as the ones considered in \cref{tv-nonsing}. This behaviour is expected as $v$ is a fast clock everywhere. A few things stand out from the figure: the classical and quantum curves are still in very close agreement for higher values of $v$, and the quantum expectation value is always greater than the classical curve, in accordance to \cref{taylort}. Indeed, the term of order $v^3$ (which is the first term that differs from the classical and quantum curve) is bigger for $\expval{t(v)}_{\Psi_{sc,v}}$. In addition to that, the curves with smaller $\sigma_\lambda$ are in closer agreement with the classical theory, contrary to what happened in the $t$-clock theory. 

It is interesting to study the variance of such expectation values, to verify whether the state $\Psi_{sc,v}$ remains semiclassical throughout the evolution. In this case
\begin{equation}
\sigma^2{\expval{t(v)}_{\Psi_{sc,v}}}=\expval{t^2(v)}_{\Psi_{sc,v}}-\expval{t(v)}^2_{\Psi_{sc,v}}\, 
\end{equation}
should remain small throughout the evolution. The calculation of $\expval{t^2(v)}_{\Psi_{sc,v}}$ is very similar to the $\expval{t(v)}_{\Psi_{sc,v}}$ one, namely
\begin{align}
\expval{t^2(v)}_{\Psi_{sc,v}}&=\bra{\Psi_{sc,v}}t^2\ket{\Psi_{sc,v}} \nonumber\\
&=i\frac{\pi v}{2}\int \frac{\dd \lambda_1 \dd \lambda_2}{(2\pi)^2}\dk \dd t \ t^2 e^{-i(\lambda_1-\lambda_2)\frac{t}{\hbar}}\frac{\bar{\alpha}_{sc}(k,\lambda_1)\alpha_{sc}(k,\lambda_2)}{\sinh(\abs{k}\pi)}\times \nonumber \\
&\left[\J{i\abs{k}}{\lambda_1}\partial_v\J{-i\abs{k}}{\lambda_2}-\J{-i\abs{k}}{\lambda_2}\partial_v\J{i\abs{k}}{\lambda_1} \right]\, ,
\end{align}
where we have again simplified the integral $\int \dd\varphi e^{-i(k_1-k_2)\varphi}=2\pi\delta(k_1-k_2)$. This time, we use the formula
\begin{align}
\int \frac{\dd \lambda_1\dd \lambda_2}{(2\pi\hbar)^2}&\dd t \ t^2 e^{-i(\lambda_1-\lambda_2)\frac{t}{\hbar}}F(\lambda_1,\lambda_2)\nonumber \\
&=-\frac{\hbar^2}{2}\int \dl \left.\left(\pdv[2]{}{\lambda_1}F(\lambda_1,\lambda_2)+\pdv[2]{}{\lambda_2}F(\lambda_1,\lambda_2) \right)\right|_{\lambda_1=\lambda_2=\lambda}\, ,
\end{align}
which comes from the second derivative of a Dirac $\delta$ distribution. Thus,
\begin{align}
\expval{t^2(v)}_{\Psi_{sc,v}}&=\frac{1}{4}\int \dl \dk \abs{\alpha(k,\lambda)}^2\left[ \frac{\hbar^2k^2}{\lambda^2}+\frac{v^2}{\lambda}\right] \nonumber \\
&-\frac{i\pi v \hbar^2}{2}\int \dl \dk \frac{\alpha_{sc}(k,\lambda)\partial_\lambda\bar{\alpha}_{sc}(k,\lambda)}{\sinh(\abs{k}\pi)}\partial_{\lambda_1}h(\lambda_1,\lambda_2)|_{\lambda_1=\lambda_2=\lambda} \nonumber\\
&-\frac{i\pi v \hbar^2}{2}\int \dl \dk \frac{\bar{\alpha}_{sc}(k,\lambda)\partial_\lambda\alpha_{sc}(k,\lambda)}{\sinh(\abs{k}\pi)}\partial_{\lambda_2}h(\lambda_1,\lambda_2)|_{\lambda_1=\lambda_2=\lambda} \nonumber\\
&-\frac{\hbar^2}{2}\int \dl \dk \left[\bar{\alpha}_{sc}(k,\lambda)\partial^2_\lambda \alpha_{sc}(k,\lambda)+\partial^2_\lambda\bar{\alpha}_{sc}(k,\lambda)\alpha_{sc}(k,\lambda) \right]\, ,
\end{align}
where 
\begin{equation}
h(\lambda_1,\lambda_2)=\J{i\abs{k}}{\lambda_1}\partial_v\J{-i\abs{k}}{\lambda_2}-\J{-i\abs{k}}{\lambda_2}\partial_v\J{i\abs{k}}{\lambda_1}.
\end{equation}
In our case $\alpha_{sc}(k,\lambda)$ is real, so the two intermediate terms cancel due to the symmetry of $h(\lambda_1,\lambda_2)$. Moreover, it is straightforward to see that the first term can be directly related to the classical solution $t_c(v)$ \cref{tclass}, thus in our case
\begin{align}
\expval{t^2(v)}_{\Psi_{sc,v}}&=\int \dl \dk \abs{\alpha_{sc}(k,\lambda)}^2t_c^2(v) \nonumber \\
&-\hbar^2\int \dl \dk {\alpha}_{sc}(k,\lambda)\partial^2_\lambda \alpha_{sc}(k,\lambda)\, . 
\end{align}
Note that contrary to what happened in $\expval{t(v)}_{\Psi_{sc,v}}$, the second term does not vanish.
In the limit $\frac{\sigma_k}{k}\ll 1$, the integral over $\lambda$ of the second term can be done yielding to
\begin{equation}
\expval{t^2(v)}_{\Psi_{sc,v}}\approx\int \dl \abs{\alpha_{sc}(k_c,\lambda)}^2 t_c^2(v)+\frac{\hbar^2}{2\sigma_\lambda^2}
\label{tsquare}
\end{equation}
This is the expression we will use for the plots. We can already see from this expression that the term $\expval{t^2(v)}_{\Psi_{sc,v}}$ remains close to and slightly above the classical curve throughout the evolution. Indeed, as the first contribution is the classical expression modulated by a Gaussian and the second term is a small positive contribution.

\begin{figure}
\centering
\includegraphics[width=0.75\textwidth]{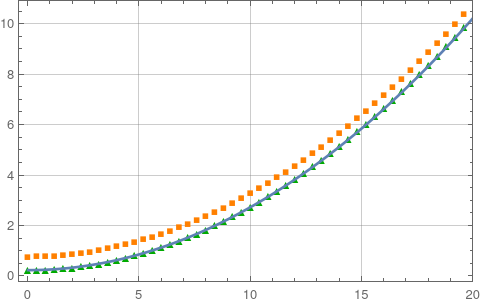}
\caption[Comparison between $\expval{t^2(v)}_{\Psi_{sc,v}}$, $\expval{t(v)}^2_{\Psi_{sc,v}}$ and $t_c(v)^2$.]{Comparison between $\expval{t^2(v)}_{\Psi_{sc,v}}$, $\expval{t(v)}^2_{\Psi_{sc,v}}$ and the classical curve $t_c(v)^2$ for $\lambda_c=k_c=10$, $\hbar=1$ and $\sigma_\lambda=1$. The orange squares represent the curve for $\expval{t^2(v)}_{\Psi_{sc,v}}$, whereas the green triangles $\expval{t(v)}^2_{\Psi_{sc,v}}$. The classical curve $t_c(v)^2$ is represented by a thick blue line. We see that $\expval{t^2(v)}_{\Psi_{sc,v}}$ and $\expval{t(v)}^2_{\Psi_{sc,v}}$ are close throughout the evolution of the state.}
\label{tsquarev}
\end{figure}

We represented $\expval{t^2(v)}_{\Psi_{sc,v}}$ and $\expval{t(v)}^2_{\Psi_{sc,v}}$ in \cref{tsquarev}.

Despite their apparent good shape, and the simplicity of these results in comparison to the previous section, we ought to take the numerical calculations and plots of this section with a grain of salt. The expressions \cref{tvsemiclass} and \cref{tsquare} present a number of challenges regarding their numerical integration:
\begin{itemize}
\item The numerical results for $\expval{t(v)}_{\Psi_{sc,v}}$ are only valid for small enough $v$. This is due to the fact that the Bessel functions in \cref{f} oscillate and the bigger $v$ gets, the more oscillations are contained in the $\lambda$ integration range (this becomes clear looking at the large argument asymptotic form \cref{asympt-large}). Mathematica is a very powerful computational tool but it does not deal well with very oscillatory integrals, as seen in the previous section already. In general, we will not trust the values for ``large $v$'s'' but the exact definition of what large $v$ really means involves not only an excellent knowledge of the behaviour of all the Bessel functions involved but also a very deep understanding of the mechanism of numerical integration Mathematica uses, and this goes well beyond the scope of this thesis. In our plots we have decided to stay in a range that remains as close as possible from $v=0$ but is large enough to show interesting behaviours, such as the classical and quantum curves not diverging from each other
\item  The numerical results for $\expval{t(v)}_{\Psi_{sc,v}}$ are only valid for small $k$. For larger $k$, the Bessel functions have larger values and more pronounced oscillations which difficult the numerical integration. We find that for $k>10$, Mathematica starts sending a lot of error messages (warning  us from the rapid oscillations) and delivering unreliable results, so we resolved to stick to values $k\leq 10$.
\item Both $\expval{t^2(v)}_{\Psi_{sc,v}}$ and $\expval{t(v)}_{\Psi_{sc,v}}$ are only valid within a certain range of $\lambda$. The expressions \cref{tvsemiclass} and \cref{tsquare} are divergent when $\lambda\rightarrow 0$. We resolve this by reducing the range of the integration over $\lambda$ to $(\lambda_c-3\sigma_\lambda,\lambda_c+3\sigma_\lambda)$. This range is wide enough to cover almost all the dispersion in $\lambda$ while avoiding the divergent part of the integral as long as $\lambda_c-3\sigma_\lambda \gg 0$. In general, $\lambda=10$ and $\sigma_\lambda\leq 3$ are `acceptable' values of the parameters. Theorically any set of parameter where $\lambda_c-3\sigma_\lambda>0$ should be valid, as the integration over such range is finite. However, in practice, when a function has a very steep gradient numerical integration becomes much harder to implement, so the set of acceptable values of the parameters has to be defined one more time case by case.
\end{itemize}
Despite these difficulties we still consider that the numerical results we found are meaningful and that they complement the theoretical predictions made in previous chapters. 

In conclusion, this theory does not solve the singularity resolution and the divergences of the quantum expectation values remain small. This is in agreement with what we studied in \cref{quant} and \cref{v-clock-sec}. These results confirm the conjecture that a classical fast clock cannot resolve the singularity when going to the quantum theory. 

\section{Dynamics of the $\varphi$-clock theory}
\label{phi-clock-dynamics}

The $\varphi$-theory is very interesting because the clock $\varphi$ behaves quite uniquely. It is slow (using the terminology of \cite{Gotay}) at $v=\infty$, and in consequence we expect to observe divergences from the classical theory around $\varphi=\varphi_0$, where $v(\varphi_0)=\infty$ (we refer to \cref{class-sol-sec} for a recap of the classical theory). A general solution of the Wheeler--DeWitt equation and of the boundary condition \cref{phigeneralstate} is
\begin{align}
\Psi(v,\varphi,t)&=\int_{-\infty}^\infty \dk \int_0^\infty \dl e^{i k\varphi}e^{i\lambda\frac{t}{\hbar}}\alpha(k,\lambda)\sqrt{\frac{2\pi}{\sinh(\abs{k}\pi)}}\times   \\
& \Re\left[\sqrt{\frac{\sinh(\frac{\pi}{2}(\abs{k}-i\kappa_0(\lambda)))}{\sinh(\frac{\pi}{2}(\abs{k}+i\kappa_0(\lambda)))}}\J{i\abs{k}}{\lambda} \right]  \nonumber\\
&+\int_0^\infty \dl e^{i\lambda\frac{t}{\hbar}}\left[\sum_{n=0}^\infty \left(\gamma^+_n(\lambda)e^{(\kappa_0(\lambda)+2n)\varphi}+\gamma_n^-(\lambda) e^{-(\kappa_0(\lambda)+2n)\varphi}\right)\times \right.  \nonumber\\ 
&  \left. \J{\kappa_0(\lambda)+2n}{\lambda} \right] \nonumber \\
& + \int_{-\infty}^\infty \dk \int_{-\infty}^0 \dl e^{ik\varphi}e^{i\lambda\frac{t}{\hbar}}\sqrt{\frac{2\sinh(\abs{k}\pi)}{\pi^2}}\eta(k,\lambda)\K{i\abs{k}}{-\lambda}\, , \nonumber
\end{align}
where $\kappa_0(\lambda)$ is the self-adjoint extension parameter, and $\alpha(k,\lambda)$ and $\eta(k,\lambda)$ are normalised: $\int_0^\infty \dl \int_{-\infty}^{\infty}\dk \abs{\alpha(k,\lambda)}^2+\int_{-\infty}^0\int_{-\infty}^\infty \dk \abs{\eta(k,\lambda)}^2=1$. We consider that semiclassical states are wave functions that fulfil the following:
\begin{itemize}
\item $\gamma_n^+(\lambda)=\gamma_n^-(\lambda)=0$. We have seen that $\J{\kappa_0(\lambda)+2n}{\lambda}$ states have norm zero and do not have an analogue in the classical theory. Consequently, they cannot be considered as semiclassical modes. 
\item $\eta(k,\lambda)=0$. As in the two previous sections we suppose the universe is in a superposition of positive cosmological constant modes (or positive perfect fluid energy density modes).
\item $\alpha(k,\lambda)=\alpha_{sc}(k,\lambda)$, i.e, we only consider Gaussian states as defined in \cref{alphasc}.
\end{itemize}
Finally, we also consider states for which the self-adjoint extension parameter $\kappa_0(\lambda)=0$. This is not a condition of semiclassicality, rather a choice to make the numerical integration simpler, and it is exactly the same procedure we followed in the $t$-clock theory. We know from \cite{Pawlowski2011}, where they studied the same model for a fixed cosmological constant, that the self-adjoint extension parameter does not change the presence or absence of big bang/big crunch singularity.  In conclusion, the states we work with in this section are:
\begin{align}
\Psi_{sc,\varphi}(v,\varphi,t)&=\int_{-\infty}^\infty \dk \int_0^\infty \dl e^{i k\varphi}e^{i\lambda\frac{t}{\hbar}}\alpha_{sc}(k,\lambda)\sqrt{\frac{2\pi}{\sinh(\abs{k}\pi)}}
\Re\left[\J{i\abs{k}}{\lambda} \right]\, . \nonumber\\
\end{align}
For simplicity, we only consider modes for which $k>0$, but maintain the notation with absolute value of $k$ in case someone wants to repeat or expand our calculations to other ranges of $k$. Once again, the subindex $sc,\varphi$ refers to the fact that this wave function is a semiclassical state in the $\varphi$-clock theory 

\subsection{Results for $\expval{v(\varphi)}_{\Psi_{sc,\varphi}}$}

There are two observables that can be calculated explicitly: $\bra{\Psi_{sc,\varphi}}v\ket{\Psi_{sc,\varphi}}=\expval{v(\varphi)}_{\Psi_{sc,\varphi}}$ and $\bra{\Psi_{sc,\varphi}}t\ket{\Psi_{sc,\varphi}}=\expval{t(\varphi)}_{\Psi_{sc,\varphi}}$. The criterion for singularity resolution is again $\expval{v(\varphi)}_{\Psi_{sc,\varphi}}>C_{\Psi_{sc,\varphi}}>0$ for a positive (and probably state dependent) constant $C_{\Psi_{sc,\varphi}}$. We interpret any failure to observe this behaviour as singular. Starting by $\expval{v(\varphi)}_{\Psi_{sc,\varphi}}$ we have
\begin{align}
\expval{v(\varphi)}_{\Psi_{sc,\varphi}}=&\int_0^\infty \frac{\dd k_1 \dd k_2}{(2\pi)^2} \int_0^\infty \dl \int_0^\infty \dd v \frac{2\pi e^{i(k_2-k_1)\varphi}(k_1+k_2)}{\sqrt{\sinh(\abs{k_1}\pi)\sinh(\abs{k_2}\pi)}} \nonumber \\
& \bar{\alpha}_{sc}(k_1,\lambda)\alpha_{sc}(k_2,\lambda)\Re\left[\J{i\abs{k_1}}{\lambda} \right]\Re\left[\J{i\abs{k_2}}{\lambda} \right]\, .
\label{expv}
\end{align}
This time we can calculate the $v$ integral analytically, which greatly speeds up the numerical analysis. However, the integral over $v$ is divergent. This is a priori not good news, as one cannot make sense of the quantity $\expval{v(\varphi)}_{\Psi_{sc,\varphi}}$, but we follow the same attitude particle physicists have regarding infinite quantities. When loop corrections make a scattering amplitude divergent, instead of giving up and ditching the full theory, what has been done is renormalise and/or regularise the results. In short, when a scattering amplitude is divergent in the theory, new contributions (usually called counterterms) are added to the theory so that the result of the scattering amplitude matches the (finite) quantity observed experimentally. This process is called renormalisation and it does not imply the existence of new physics. Renormalisation techniques have been a common practice since the seventies (see \cite{Wilson1975} for an early reference on the renormalisation group). Another option for dealing with these divergences would be to not compute directly $\expval{v(\varphi)}_{\Psi_{sc,\varphi}}$, but rather try to find a finite expectation value for a function of $v$. In \cite{Pawlowski2011}, the authors used $\tan(v)$ instead of $v$ to avoid similar issues. 

Our case is slightly different, because we are implying the existence of new  physics from the start (namely that spacetime is quantum) and moreover there is no experimental (in our case observational) quantity $\expval{v(\varphi)}_{obs}$ we can compare with. Nevertheless, we require that the $v$ integral in \cref{expv} is finite. Indeed, if we assume that this quantum  universe has a classical limit, namely the one studied in \cref{class-sol-sec}, we would like the classical quantity $\expval{v(\varphi)}_{\Psi_{sc,\varphi}}$ to agree with the classical curve, at least for a subset of the range of $\varphi$. Otherwise, it is impossible to recover the classical limit. Again, this assumption, as radical as it sounds is recurrent in particle physics and QFT. 

However, in order to make \cref{expv} finite, we need to regularise the integral. This can be done implementing a cutoff, which is in a way implies the existence of ``more additional physics'' that act at large $v$ and make to final result of the integral finite\footnote{We are using the word regularisation rather than renormalisation because in particle physics, the procedure of renormalisation does not assume the existence of forces and particles beyond the standard model. In this case we have to assume the existence of some additional phenomena or scale in which our theory is not valid anymore in order to implement a cutoff.}. There are different ways of regularising an integral. In this work we have focused on the two easiest ones (as to our knowledge). The relevant formulas are \cref{digamma_integral} and \cref{digamma_integral_2}. The first one diverges logarithmically, hence, in order to obtain a finite result, we introduce a finite cutoff $\Xi_v$. With this cutoff, the expectation value is 
\begin{align}
\expval{v(\varphi)}_{\Psi_{sc,\varphi}}&\approx -\int \frac{\dd k_1 \dd k_2}{(2\pi)^2} \dl \frac{\hbar}{\sqrt{\lambda}} \frac{e^{i(k_2-k_1)\varphi}(k_1+k_2)\cosh(\frac{k_1\pi}{2})\cosh(\frac{k_2\pi}{2})}{\sqrt{\sinh(\abs{k_1}\pi)\sinh(\abs{k_2}\pi)}} \times \nonumber \\
&\alpha_{sc}(k_1,\lambda)\alpha_{sc}(k_2,\lambda)\left\lbrace \log\left(\frac{4\hbar^2}{\lambda\Xi_v^2}\right)+\psi\left(\frac{1}{2}(1-i(\abs{k_1}-\abs{k_2})) \right) \nonumber \right. \nonumber \\
&\left. +\psi\left(\frac{1}{2}(1+i(\abs{k_1}-\abs{k_2}))\right)+\psi\left(\frac{1}{2}(1+i(\abs{k_1}+\abs{k_2}))\right) \right. \nonumber \\
&+\left. \psi\left(\frac{1}{2}(1-i(\abs{k_1}+\abs{k_2}))\right) +2\gamma \right\rbrace\, ,
\label{vphiexp}
\end{align}
where $\gamma$ refers to the Euler-Mascheroni constant and $\psi(x)$ to the digamma function. This cutoff can be interpreted as adding an extra term of the form $\int_{\frac{\sqrt{\lambda}\Xi_v}{h}}^\infty\dd v \frac{2}{\Xi_v}$. When, integrated, this contribution will cancel the logarithmic divergence. This addition of an extra term resembles the Pauli-Villars regularisation method that consists in adding a fictitious mass term to certain divergent propagators (see chapter 7 of \cite{Peskin}).   

The second method of regularisation, based on \cref{digamma_integral_2} gives:
\begin{align}
\expval{v(\varphi)}_{\Psi_{sc,\varphi}}&\approx -\int \frac{\dd k_1 \dd k_2}{(2\pi)^2} \dl \frac{\hbar}{\sqrt{\lambda}} \frac{e^{i(k_2-k_1)\varphi}(k_1+k_2)\cosh(\frac{k_1\pi}{2})\cosh(\frac{k_2\pi}{2})}{\sqrt{\sinh(\abs{k_1}\pi)\sinh(\abs{k_2}\pi)}} \times \nonumber \\
&\alpha_{sc}(k_1,\lambda)\alpha_{sc}(k_2,\lambda)\left\lbrace-\frac{2}{\epsilon} +\log\left(\frac{4\hbar^2}{\lambda}\right)+\psi\left(\frac{1}{2}(1-i(\abs{k_1}-\abs{k_2})) \right) \nonumber \right. \nonumber \\
&\left. +\psi\left(\frac{1}{2}(1+i(\abs{k_1}-\abs{k_2}))\right)+\psi\left(\frac{1}{2}(1+i(\abs{k_1}+\abs{k_2}))\right) \right. \nonumber \\
&+\left. \psi\left(\frac{1}{2}(1-i(\abs{k_1}+\abs{k_2}))\right) +2\gamma \right\rbrace\, ,
\label{vphiexp2}
\end{align}
where now $\epsilon\rightarrow 0$. The previous $\Xi_v$ logarithmic divergence is ``hidden'' in the regulator $\epsilon$, which comes from a dimensional regularisation approach. However, the two methods give the same non divergent part. As the divergent parts are comparable, indeed $\log[\Xi_v]=1/\epsilon$, we have decided to use \cref{vphiexp} for the numerical calculations. There is a quantitative difference in a logarithmic and a $1/\epsilon$ divergence, but the debate on which regularisation method is better goes beyond the scope of this thesis (as so many other interesting issues). We thus compare different values of the cutoff and take it to values as high as $10^{10}$. As we will see in more detail later in the section, the cutoff choice has a non-trivial influence in the results, especially around $\varphi=0$, but it is clear that qualitative features of the expectation values (which we are mostly interested here) are not too sensitive to it. The numerical evaluation of the $v$ integral (without using \cref{digamma_integral}) would require a cutoff similar to $\Xi_v$ leading to the same ambiguity. 

On a final note, the integrals involved in \cref{expv} are not absolutely convergent, but might be convergent in a weaker sense, where separating the three integrals and compute them independently does not make sense. This, of course does not solve the issue of implementing it numerically, but makes the divergences less problematic. 

The quantum expectation value \cref{vphiexp} is symmetric under $\varphi\rightarrow -\varphi$ and therefore we compare it to the classical solution
\begin{equation}
v_c(\varphi)=\frac{\hbar k_c}{\sqrt{\lambda_c}\abs{\sinh(\varphi)}}
\end{equation}

Even after simplifying the $v$ integration, \cref{vphiexp} is still very hard to compute numerically. The integrals over $\lambda$ are
\begin{equation}
\frac{2\hbar\sqrt{\pi}}{\sigma_\lambda}\int \dl \frac{\hbar}{\sqrt{\lambda}} e^{-\frac{(\lambda-\lambda_c)^2}{\sigma_\lambda}}\, , \hspace{4mm} \frac{2\hbar\sqrt{\pi}}{\sigma_\lambda}\int \dl \frac{\hbar}{\sqrt{\lambda}} e^{-\frac{(\lambda-\lambda_c)^2}{\sigma_\lambda^2}}\log\left(\frac{4\hbar^2}{\lambda\Xi_v^2} \right)\, .
\end{equation}
These integrals are finite (although they have a complicated expression) and it can be verified that, in the limit $\sigma_\lambda\rightarrow 0$, they reduce to the value of the integrand evaluated at $\lambda=\lambda_c$. All the calculations we show are in this limit. This approximation is analogous to $\sigma_k\rightarrow 0$ in the previous sections since in the inner product of this theory, the different $\lambda$ sectors are decoupled. 

\begin{figure}
\centering
\includegraphics[width=0.8\textwidth]{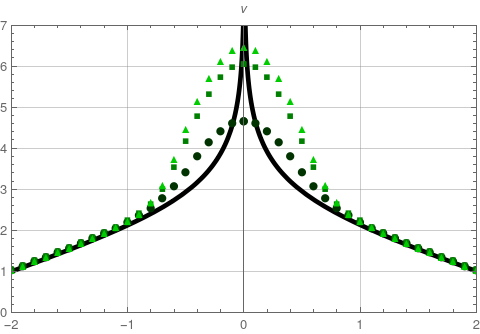}
\caption[Comparison between $\expval{v(\varphi)}_{\Psi_{sc,\varphi}}$ and $v_c(\varphi)$ for different values of the cutoff.]{Comparison between $\expval{v(\varphi)}_{\Psi_{sc,\varphi}}$ and $v_c(\varphi)$ for different values of the cutoff $\Xi_v$. The values of the cutoff are $\Xi_v=10^3$ (dark green circle), $\Xi_v=10^7$ (green squares), and $\Xi_v=10^{10}$ (light green triangles). The values for the rest of parameters are $k_c=10$, $\lambda_c=1$ and $\sigma_k=3$.}
\label{cutoff}
\end{figure}

\begin{figure}
\centering
\includegraphics[width=0.8\textwidth]{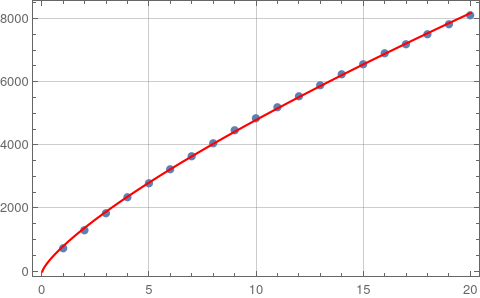}
\caption[Values of $V_{max}$ as a function of $\sigma_k$.]{Values of $V_{max}$ (vertical axis) as a function of $\sigma_k$ (horizontal axis). The rest of the parameters are $k_c=100$, $\lambda_c=1$. The curve has been fitted with a fit function of the form $V_{max}=a\sigma_k^b$ where $a=821.487$ and $b=0.767681$}
\label{sigmas}
\end{figure}

\begin{figure}
\centering
\begin{subfigure}{1\textwidth}
\centering
\includegraphics[width=0.8\textwidth]{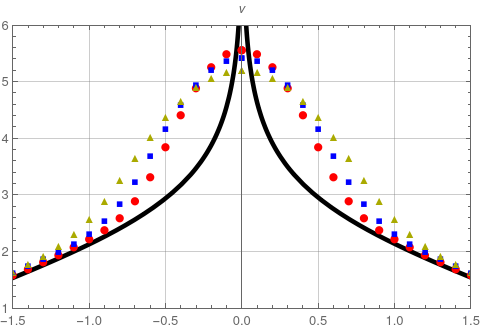}
\caption{The dotted lines represent $\expval{v(\varphi)}_{\Psi_{sc,\varphi}}$ for values $\lambda_c=1$, $k_c=10$, $\hbar=1$, $\Xi_v=10^{5}$ and $\sigma_k=3$ (red circles), $\sigma_k=2.5$ (blue squares) and $\sigma_k=2$ (yellow triangles). The black thick line corresponds to the classical trajectory with the same values of $k_c$ and $\lambda_c$. We see that the quantum trajectory is well defined for all $\varphi$'s and reaches a positive maximum value $V_{max}$.}
\end{subfigure}
\begin{subfigure}{1\textwidth}
\centering
\includegraphics[width=0.8\textwidth]{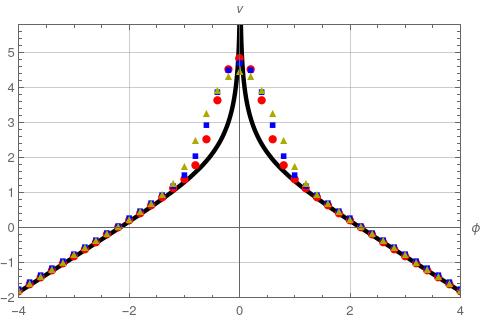}
\caption{The dotted lines represent $\expval{v(\varphi)}_{\Psi_{sc,\varphi}}$ for values $\lambda_c=5$, $k_c=10$, $\hbar=1$, $\Xi_v=10^{5}$ and $\sigma_k=3$ (red circles), $\sigma_k=2.5$ (blue squares) and $\sigma_k=2$ (yellow triangles). The black thick line corresponds to the classical trajectory with the same values of $k_c$ and $\lambda_c$. We see that the quantum trajectory is well defined for all $\varphi$'s and reaches a positive maximum value $V_{max}$.}
\end{subfigure}
\caption[Comparison between $\expval{v(\varphi)}_{\Psi_{sc,\varphi}}$ and $v_c(\varphi)$ for different values of $\sigma_k$ and $\lambda_c$.]{Comparison between $\expval{v(\varphi)}_{\Psi_{sc,\varphi}}$ and $v_c(\varphi)$ for different values of the parameters.}
\label{kplotsigmas}
\end{figure}

The numerical results are presented in \cref{cutoff,sigmas,kplotsigmas}. Figure \cref{cutoff} shows the classical curve and quantum expectation values with different values of the cutoff, figure \cref{sigmas} shows the maximum value of the volume $V_{max}=\expval{v(0)}_{\Psi_{sc,\varphi}}$ for different values of $\sigma_k$ and figure \cref{kplotsigmas} shows the classical curve and expectation values with different values of $\sigma_k$. There are several things to comment about these figures. First of all, we see that the values of the classical curve and the quantum expectation values are very close to each other for $\abs{\varphi}$ large enough (this specific value depends on the rest of the parameters). Moreover, in this limit, all quantum curves are very close to each other, meaning that the value of the cutoff is important only for small values of $\abs{\varphi}$, where the classical curves start diverging from the classical one and separating from each other. The quantum expectation value grows with $\varphi$ and reaches a maximum value $V_{max}=\expval{v(0)}_{\Psi_{sc,\varphi}}$ and this value grows weakly with $\Xi_v$. Then, as $\varphi>0$ the quantum expectation values decreases and gets closer to the classical curve. This behaviour is what we would expect from the analysis of \cref{phi-clock-sec} since we had to impose a reflective boundary condition at $v=\infty$, which corresponds to $\varphi=0$ in this example. Independently of the value of the cutoff, we see that the quantum corrections are small, but  start growing and lead to a faster expansion than in the classical theory, before slowing down and stopping completely at $\varphi=0$. This first phase of rapid expansion is in agreement with the results of \cite{Bojowald2011}, where a systematic expansion into higher order quantum fluctuations around the classical trajectory was studied in a similar model. In their work one sees explicitly how quantum fluctuations diverge as the volumes grows, which in our case trigger the recollapse later on. As interesting as these results are, we cannot say that they resolve the singularity: the quantum expectation value $\expval{v(\varphi)}_{\Psi_{sc,\varphi}}$ goes to zero at large $\varphi$. This is in accordance to the results obtained in \cite{Pawlowski2011} for a fixed positive cosmological constant.

There is another, more quantitative argument to show that the classical singularity is not resolved in this theory. Substituting the limit at large $\abs{\varphi}$ \cref{delta-1} into \cref{vphiexp} we conclude that $\lim_{ \varphi\rightarrow \pm \infty} \expval{v(\varphi)}_{\Psi_{sc,\varphi}}=0$. We consider this argument strong enough to replace the $\Delta_{rel}$ table made for the $t$-clock theory, in which we could not simplify the $v$ integration. 

In figures \cref{sigmas,kplotsigmas} we see the influence of the parameter $\sigma_k$ in the theory. The maximum value $V_{max}$ grows with $\sigma_k$, hence, bigger $\sigma_k$ trigger a more abrupt transition between the two classical branches and the quantum expectation value remains close to the classical solution for longer. We observed a very similar counter intuitive result in the $t$-clock theory regarding $V_{min}$. This behaviour can be explained using the same argument: $k$ and $\varphi$ (the clock) are conjugated variables, hence a greater spread in $k$ implies smaller uncertainty in $\varphi$ leading to the observed curves.

It is interesting to complement these results with the variance $\expval{v^2(\varphi)}_{\Psi_{sc,\varphi}}$ as we did in the $v$-clock theory. The integral we are interested in is
\begin{align}
\expval{v^2(\varphi)}_{\Psi_{sc,\varphi}}=&\int_0^\infty \frac{\dd k_1 \dd k_2}{(2\pi)^2} \int_0^\infty \dl \int_0^\infty \dd v \ v \frac{2\pi e^{i(k_2-k_1)\varphi}(k_1+k_2)}{\sqrt{\sinh(\abs{k_1}\pi)\sinh(\abs{k_2}\pi)}} \nonumber \\
& \bar{\alpha}_{sc}(k_1,\lambda)\alpha_{sc}(k_2,\lambda)\Re\left[\J{i\abs{k_1}}{\lambda} \right]\Re\left[\J{i\abs{k_2}}{\lambda} \right]\, .
\label{expv2}
\end{align}
As for $\expval{v(\varphi)}_{\Psi_{sc,\varphi}}$, the $v$ integral is divergent and needs to be regularised. If we implement a cutoff $\Xi_{v^2}$, using \cref{vphi2int}, we find
\begin{align}
&\expval{v^2(\varphi)}_{\Psi_{sc,\varphi}}\approx \int_0^\infty \frac{\dd k_1 \dd k_2}{(2\pi)^2} \int_0^\infty \dl \frac{\pi e^{i(k_2-k_1)\varphi}(k_1+k_2)}{\sqrt{\sinh(\abs{k_1}\pi)\sinh(\abs{k_2}\pi)}} \bar{\alpha}_{sc}(k_1,\lambda)\alpha_{sc}(k_2,\lambda)\times \nonumber\\
& \left(\frac{\hbar^2(k_2^2-k_1^2)}{4\lambda}\left( \coth(\frac{\pi}{2}(k_1+k_2))\sinh(\frac{\pi}{2}(k_1-k_2))+\coth(\frac{\pi}{2}(k_1-k_2))\sinh(\frac{\pi}{2}(k_1+k_2))\right)\right. \nonumber \\
&\left.+\hbar^2\frac{\left(2\frac{\sqrt{\lambda}}{\hbar} \Xi_{v^2}-\cos(2\frac{\sqrt{\lambda}}{\hbar}\Xi_{v^2})\right)}{\pi\lambda}\cosh(\frac{k_1\pi}{2})\cosh(\frac{k_2\pi}{2})\right)\, .
\label{variance1}
\end{align}
This time the divergence is linear, which is much worse than the logarithmic divergence in \cref{vphiexp}. We can regularise the integral with dimensional regularisation using \cref{vphi2int2}. This time, the divergent terms vanish living only the finite contribution
\begin{align}
&\expval{v^2(\varphi)}_{\Psi_{sc,\varphi}}\approx \int_0^\infty \frac{\dd k_1 \dd k_2}{(2\pi)^2} \int_0^\infty \dl \frac{\pi e^{i(k_2-k_1)\varphi}(k_1+k_2)}{\sqrt{\sinh(\abs{k_1}\pi)\sinh(\abs{k_2}\pi)}} \bar{\alpha}_{sc}(k_1,\lambda)\alpha_{sc}(k_2,\lambda)\times \nonumber\\
& \frac{\hbar^2(k_2^2-k_1^2)}{4\lambda}\left( \coth(\frac{\pi}{2}(k_1+k_2))\sinh(\frac{\pi}{2}(k_1-k_2))+\coth(\frac{\pi}{2}(k_1-k_2))\sinh(\frac{\pi}{2}(k_1+k_2))\right)\, .
\label{variance2}
\end{align}

It seems that the $v$ integral in $\expval{v^2(\varphi)}_{\Psi_{sc,\varphi}}$ diverges for one regularisation method and is finite with another one. This may appear contradictory, but different regularisation methods may yield to different results. After a preliminary numerical analysis it appears that \cref{variance2} would be valid at large $\varphi$, whereas for small $\varphi$, only \cref{variance1} gives sensible (positive) results. We thus use \cref{variance1} for the calculation. The expression is coherent with \cref{vphiexp}, where we used the same regularisation method. 

Last but not least, we might be tempted to ask ourselves whether it makes sense that the cutoffs $\Xi_{v}$ and $\Xi_{v^2}$ should be the same. We assume that, if $\Xi_v$ represents a scale for which the theory is not valid anymore, it should be the same for both results. 

\cref{variance1} gives a measure for the error of the results through the standard deviation
\begin{equation}
\sigma{\expval{v(\varphi)}_{\Psi_{sc,\varphi}}}=\sqrt{\expval{v^2(\varphi)}_{\Psi_{sc,\varphi}}-\expval{v(\varphi)}^2_{\Psi_{sc,\varphi}}}\, .
\label{error}
\end{equation}
This quantity is a measure of semiclassicality: if it is small throughout the evolution, it indicates that the state remains semiclassical, but if it grows larger when approaching $\varphi=0$ it means that the state cannot be considered semiclassical around $V_{max}$. If we had to take a measurement of the volume at any point in evolution, we would like to have a small deviation as this would mean a higher chance to measure a quantity close to $\expval{v(\varphi)}_{\Psi_{sc,\varphi}}$. 

\begin{figure}
\centering
\includegraphics[scale=1]{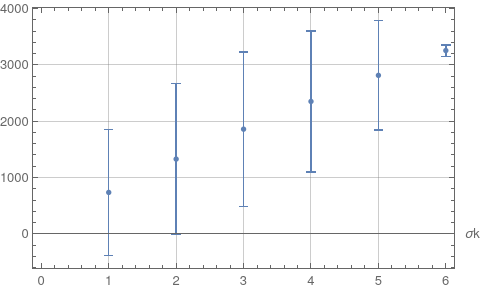}
\caption[Values for $V_{max}$ as a function of $\sigma_k$ with error bars.]{Values for $V_{max}$ (vertical axis) as a function of $\sigma_k$ (horizontal axis) with error bars calculated using \cref{error}. The rest of the parameters are $k_c=100$, $\lambda=1$ and $\Xi_v=\Xi_{v^2}=10^5$.}
\label{varianceplot}
\end{figure}

 It appears that the integrals involved in \cref{variance1} are even more difficult to handle than the ones in \cref{vphiexp}. They depend non-trivially on the deviation of the Gaussian $\sigma_k$, the cutoff $\Xi_{v^2}$, and are only valid in a limited range of $\varphi$, where $\varphi$ is small, otherwise the oscillations coming from the term $\exp(i(k_2-k_1)\varphi)$ render the expression extremely hard to compute numerically, even after all the approximations taken. Doing a complete analysis on the validity of these results would require again tremendous efforts, therefore we focus on trying to answer the most important question: Is $\sigma\expval{v(0)}_{\Psi_{sc,\varphi}}$ big? In the work \cite{Pawlowski2011}, where a similar model with a fixed positive cosmological constant was analysed from a Wheeler--DeWitt and the LQC perspective, it was shown that the dispersion of similar semiclassical states at late time (here large $\varphi$) is small, indicating that the state is indeed semiclassical. However, in our model, $\sigma{\expval{v(\varphi)}_{\Psi_{sc,\varphi}}}$ becomes very large when $\varphi=0$ as indicated in \cref{varianceplot}. 

\Cref{varianceplot} shows several things. It indicates, for fixed values of the rest of the parameters, a range in $\sigma_k$ in which the numerical integration gives a coherent result. Indeed, we see that the error is more or less the same for $\sigma_k=1,\dots, 5$ and then diminishes (before becoming complex and hence not reliable). If we limit ourselves to the interval $\sigma_k\in (1,5)$, we can also say that the error is quite large, and even compatible with a priori negative values for some cases. This is a sign, that when going through the quantum recollapse, the wave function does not remain semiclassical. This is different to what we observed in the $v$-clock theory, where $\expval{t^2(v)}_{\Psi_{sc,v}}$ remained small throughout all evolution (recall \cref{tsquarev}). However, in the $v$-theory there were no quantum effects, that could generate this dispersion. Looking at these results, we expect a similar dispersion to happen in the $t$-clock theory, concerning $\expval{v^2(t)}_{\Psi_{sc,t}}$. We could not verify this conjecture due to the intractability of the integrals involved. In \cite{Pawlowski2011}, in the LQC universe, the singularity was replaced by a quantum bounce, but the dispersion of the states remained small throughout the bounce. This is not what happens here in the reflection from infinity, this transition has a non-trivial effect in the dispersion of the state. 

The fluctuations of $V_{max}$ can be explained by the fact that the state is going through a reflection. In a non rigorous sense, part of state is expanding and part is contracting, hence generating what we observe in \cref{varianceplot}. In this case the fluctuations might be more sever because classically the volume grows to arbitrarily large values that might also contribute to the final result.

\subsection{Results for $\expval{t(\varphi)}_{\Psi_{sc,\varphi}}$}

To complete the analysis of the theory, we also study $\expval{t(\varphi)}_{\Psi_{sc,\varphi}}$. In this case the relevant formula is
\begin{align}
\expval{t(\varphi)}_{\Psi_{sc,\varphi}}&=\int_0^\infty \frac{\dd k_1 \dd k_2}{(2\pi)^2}\int_0^\infty\frac{\dd \lambda_1 \dd \lambda_2}{(2\pi\hbar)^2}\int_0^\infty \frac{\dd v}{v} \int_{-\infty}^\infty \dd t \ \frac{2\pi e^{i(k_2-k_1)\varphi}(k_1+k_2)}{\sqrt{\sinh(\abs{k_1}\pi)\sinh(\abs{k_2}\pi)}}\times \nonumber \\
& te^{i(\lambda_2-\lambda_1)\frac{t}{\hbar}}\bar{\alpha}_{sc}(k_1,\lambda_1)\alpha_{sc}(k_2,\lambda_2)\Re\left[\J{i\abs{k_1}}{\lambda_1} \right]\Re\left[ \J{i\abs{k_2}}{\lambda_2}\right]\, .
\end{align}
To simplify the $\lambda$ integral in the same way we did in the $v$-clock theory for the calculation of $\expval{t(v)}_{\Psi_{sc,v}}$, using \cref{deltader}. Hence,
\begin{align}
\expval{t(\varphi)}_{\Psi_{sc,\varphi}}&=-i\pi\hbar\int \frac{\dd k_1 \dd k_2}{(2\pi)^2}\dl \frac{\dd v}{v}\frac{e^{i(k_2-k_1)\varphi}(k_1+k_2)}{\sqrt{\sinh(\abs{k_1}\pi)\sinh(\abs{k_2}\pi)}} \times \nonumber \\
&\left\lbrace \left( \alpha_{sc}(k_2,\lambda)\partial_\lambda\bar{\alpha}_{sc}(k_1,\lambda)-\bar{\alpha}_{sc}(k_1,\lambda) \partial_\lambda\alpha_{sc}(k_2,\lambda)\right)\times \phantom{\Re\left[\J{i\abs{k_1}}{\lambda} \right]} \right. \nonumber \\ 
& \left.\Re\left[\J{i\abs{k_1}}{\lambda} \right]\Re\left[\J{i\abs{k_2}}{\lambda} \right]+ \bar{\alpha}_{sc}(k_1,\lambda)\alpha_{sc}(k_2,\lambda)H(k_1,k_2,\lambda,v) \right\rbrace
\end{align}
where
\begin{align}
H(k_1,k_2,\lambda,v)=&\frac{v}{4\hbar\sqrt{\lambda}}\left\lbrace\Re\left[ \J{i\abs{k_2}}{\lambda}\right]\Re\left[ \J{1+i\abs{k_1}}{\lambda}-\J{-1+i\abs{k_1}}{\lambda}\right]\right. \nonumber \\
-&\left.\Re\left[\J{i\abs{k_1}}{\lambda}\right]\Re\left[\J{-1+i\abs{k_2}}{\lambda}-\J{1+i\abs{k_2}}{\lambda} \right]\right\rbrace \, .
\end{align}
The terms containing the $\lambda$ derivative of $\alpha_{sc}(k,\lambda)$ vanish in our case, since $\alpha_{sc}(k,\lambda)$ is real and separable in $k$ and $\lambda$. Hence, only the term containing $H(k_1,k_2,\lambda,v)$ is important. This term is a combination of 16 integrals over $v$ that are slight variations from \cref{H-int-1} and \cref{H-int-2}. After performing every integral and regrouping the 16 terms one finds that all terms multiplying $\delta(\abs{k_1}\pm \abs{k_2})$ (that come from the contribution in the limit $v=0$) cancel, hence only the $v=\infty$ limit plays a rôle. Some terms diverge as $\log(\frac{4\hbar^2}{\lambda v^2})$, but these also cancel. Note that for these integrals we do not have a formula like \cref{generalform}, which allows for a different regularisation method. Finally, the terms containing digamma functions also simplify, leaving us with the extraordinarily simple expression
\begin{align}
\int \frac{\dd v}{v} H(k_1,k_2,\lambda,v)&=\frac{1}{4\lambda}\left(\coth((\abs{k_1}+\abs{k_2})\frac{\pi}{2})\sinh((\abs{k_1}-\abs{k_2})\frac{\pi}{2}) \right.\nonumber \\
&+\coth((\abs{k_1}-\abs{k_2})\frac{\pi}{2})\sinh((\abs{k_1}+\abs{k_2})\frac{\pi}{2})\, .
\end{align}

In conclusion the final expression of the expectation value $\expval{t(\varphi)}_{\Psi_{sc,\varphi}}$ is
\begin{align}
\expval{t(\varphi)}_{\Psi_{sc,\varphi}}=-\frac{i\pi\hbar}{4}\int \frac{\dd k_1 \dd k_2}{(2\pi)^2}\frac{\dd \lambda}{2\pi\hbar}\frac{e^{i(k_2-k_1)\varphi}(k_1+k_2)}{\lambda\sqrt{\sinh(\abs{k_1}\pi)\sinh(\abs{k_2}\pi)}}\alpha_{sc}(k_1,\lambda)\alpha(k_2,\lambda) \times \nonumber \\
\left(\coth(\frac{k_+\pi}{2})\sinh(\frac{k_-\pi}{2})+\coth(\frac{k_-\pi}{2})\sinh(\frac{k_+\pi}{2}) \right)\, , \, k_\pm :=\abs{k_1}\pm \abs{k_2}.
\label{exptphi}
\end{align}
This expression is antisymmetric with respect to the change $\varphi\rightarrow - \varphi$ which motivates comparing this quantum expectation value to the classical solution
\begin{equation}
t_c(\varphi)=-\frac{\hbar k_c}{2\lambda_c}\coth(\varphi)\, .
\end{equation}

\begin{figure}
\centering
\begin{subfigure}{1\textwidth}
\centering
\includegraphics[width=0.8\textwidth]{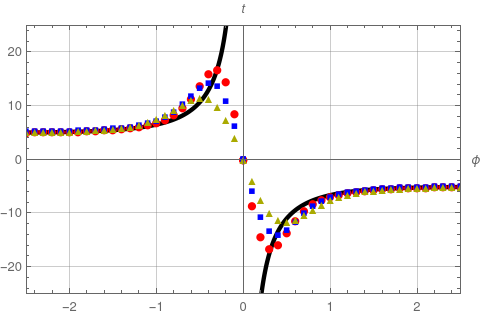}
\caption{The dotted lines represent $\expval{t(\varphi)}_{\Psi_{sc,\varphi}}$ for values $\lambda_c=1$, $k_c=10$, $\hbar=1$, $\Xi_v=10^{5}$ and $\sigma_k=3$ (red circles), $\sigma_k=2.5$ (blue squares) and $\sigma_k=2$ (yellow triangles). The black thick line corresponds to the classical trajectory with the same values of $k_c$ and $\lambda_c$. We see that the quantum trajectory is well defined for all $\varphi$'s and is not monotonic anymore.}
\end{subfigure}
\begin{subfigure}{1\textwidth}
\centering
\includegraphics[width=0.8\textwidth]{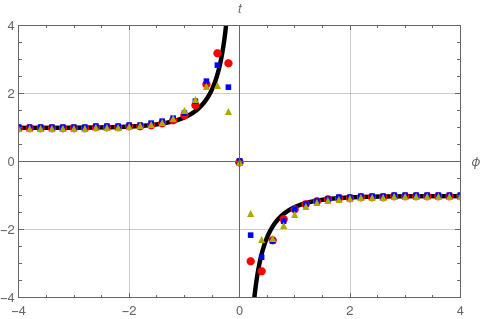}
\caption{The dotted lines represent $\expval{t(\varphi)}_{\Psi_{sc,\varphi}}$ for values $\lambda_c=5$, $k_c=10$, $\hbar=1$, $\Xi_v=10^{5}$ and $\sigma_k=3$ (red circles), $\sigma_k=2.5$ (blue squares) and $\sigma_k=2$ (yellow triangles). The black thick line corresponds to the classical trajectory with the same values of $k_c$ and $\lambda_c$. We see that the quantum trajectory is well defined for all $\varphi$'s and is not monotonic anymore.}
\end{subfigure}
\caption{Comparison between $\expval{t(\varphi)}_{\Psi_{sc,\varphi}}$ and $t_c(\varphi)$ for different values of the parameters.}
\label{expvaltphi}
\end{figure}

We evaluate \cref{exptphi} in figure \cref{expvaltphi} for the several values of the parameters. As in the case of $\expval{v(\varphi)}_{\Psi_{sc,\varphi}}$, at large $\abs{\varphi}$ the classical and quantum curves agree closely. For small $\abs{\varphi}$, the quantum expectation values reaches an extremum and then goes to zero, to transition smoothly between the classical expanding and contracting branch. We also observe the same behaviour with respect to changes in $\sigma_k$: for greater $\sigma_k$, the classical and quantum solutions agree more closely, and the transition between the expanding and contracting branch is more abrupt. These figures show that the expectation value $\expval{t(\varphi)}_{\Psi_{sc,\varphi}}$ is no longer monotonic with respect to $\varphi$ in any of the two sectors $t<0$ and $t>0$ and experiences a turnaround, unlike what happens classically. Note that contrary to $\expval{v(t)}_{\Psi_{sc},t}$, this observable does not need any cutoff and also signals the divergences between the classical and quantum theories.

Given the formula \cref{delta-1}, we can calculate analytically the limit $\expval{t(\varphi)}_{\Psi_{sc},\varphi}$ at large $\abs{\varphi}$, finding
\begin{equation}
\lim_{\varphi\rightarrow \pm \infty} \expval{t(\varphi)}_{\Psi_{sc},\varphi}=\mp \frac{\hbar}{2}\int \dk \dl \frac{k}{\lambda}\abs{\alpha_{sc}(k,\lambda)}^2\, ,
\end{equation}
Note that $\alpha_{sc}(k,\lambda)$ is a Gaussian in $k$ and $\lambda$ with means $k_c$ and $\lambda_c$. The integral over $k$ gives simply the mean $k_c$. The integral over $\lambda$ requires a regularisation at $\lambda=0$ (whose implementation is beyond the scope of this thesis), but after this regularisation, we can assume that the result of this integral is $\frac{1}{\lambda_c}$. Hence,
\begin{equation}
\lim_{\varphi\rightarrow \pm \infty} \expval{t(\varphi)}_{\Psi_{sc},\varphi}=\mp\frac{\hbar k_c}{2\lambda_c}=\lim_{\varphi\rightarrow\pm \infty} t_c(\varphi)\, .
\end{equation}
In conclusion, analytically the expression $\expval{t(\varphi)}_{\Psi_{sc},\varphi}$ tends to the classical solution. 

Since large values of $\abs{\varphi}$ correspond to the classical big bang/big crunch singularity, the behaviour of $\expval{t(\varphi)}_{\Psi_{sc,\varphi}}$ illustrates again that there is no singularity resolution in this theory. The quantum expectation values follows the classical curve all the way up to the singularity. 

\begin{figure}
\centering
\includegraphics[width=0.8\textwidth]{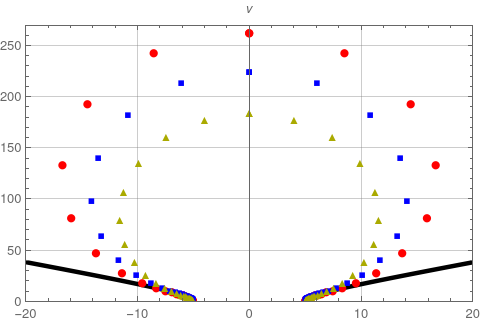}
\caption[Parametric plot of $\expval{v(\varphi)}_{\Psi_{sc,\varphi}}$ and $\expval{t(\varphi)}_{\Psi_{sc,\varphi}}$ in comparison to $v_c(t)$.]{Parametric plot of $\expval{v(\varphi)}_{\Psi_{sc,\varphi}}$ and $\expval{t(\varphi)}_{\Psi_{sc,\varphi}}$ in comparison to $v_c(t)$. The vertical axis accounts for the values of $v$ (and $\expval{v(\varphi)}_{\Psi_{sc,\varphi}}$) and the horizontal axis for values of $t$ (and $\expval{t(\varphi)}_{\Psi_{sc,\varphi}}$). The red circles correspond to trajectories with $\sigma_k=3$, the blue squares $\sigma_k=2.5$ and the yellow triangles $\sigma_k=2$. The rest of the parameters are $k_c=10$, $\lambda_c=1$ and $\Xi_v=10^5$.}
\label{parametric}
\end{figure}

Given that we have calculated $\expval{v(\varphi)}_{\Psi_{sc,\varphi}}$ and $\expval{t(\varphi)}_{\Psi_{sc,\varphi}}$ we can plot parametrically these two quantities for the same value of $\varphi$ and compare this plot with the classical trajectory $v_c(t)$ \cref{vclasst}. The result of this comparison can be found in figure \cref{parametric}. We see clearly how the universe emerges from the classical singularity and is very close to the classical $v_c(t)$, evolving forward in $t$. However, then $t$ experiences a turnaround and starts going backwards. The universe reaches a maximum volume when $t=0$ and finally approaches the contracting classical solution while $t$ starts going forward again. Form this picture it is clear that the big bang and big crunch singularities are not resolved.

In summary, this theory does not resolve the singularity, but contrary to the $v$-clock theory, it shows major divergences from the classical theory. These divergences are in accordance with the analysis performed in \cref{phi-clock-sec}. In particular, as the $\varphi$ clock is slow at $v=\infty$, the quantum expectation values show the ``infinity resolution'' that was theorised at the end of the chapter. This brings further evidence to the Gotay and Demaret conjecture. We believe the study of this minisuperspace model is particularly interesting because each of the possible quantum clocks shows a very particular behaviour regarding singularity resolution and divergences from the classical theory.

\section{Causal structure of the different theories}

In the previous sections we have seen how the expectation values $\expval{v(t)}_{\Psi_{sc,t}}$, $\expval{v(\varphi)}_{\Psi_{sc,\varphi}}$ and $\expval{t(v)}_{\Psi_{sc,v}}$ behave. We can use this information to study the causal structure of each of our theories with help of conformal diagrams. Conformal diagrams are a very popular tool in general relativity. To construct a conformal diagram explicitly, one needs to apply a conformal transformation to the metric $\dd s^2=-N(\tau)^2\dd^2 \tau + a(\tau)^2h_{ij}\dd x^i \dd x^j$ so that spacetime is mapped to a finite region. One can reduce a spherically symmetric spacetime to create a two-dimensional picture of the universe. Usually, the vertical axis is related to the timelike parameter of the metric and the horizontal axis to the remaining spacelike parameter. Every point in the picture represents a two-dimensional sphere. Hence, in conformal diagrams, lengths are modified, but angles are conserved. We can thus extrapolate the light cones of the conformal diagrams from the original theory. The concept of conformal infinity and conformal diagram was first introduced by Penrose \cite{Penrose1963}. These diagrams are often referred as Penrose diagrams, or Penrose--Carter diagrams. In our model, the expression of the conformal transformation depends on the energy interpretation of the model, i.e.~it depends on the equation of state of the perfect fluid $p=w\rho$, more concretely in the parameter $w$. 

We will only consider ``standard'' and non exotic perfect fluids in this work. In particular, we will not consider perfect fluids with $w<-1$, that would generate a ``big rip'' and perfect fluids with $w=\frac{1}{2}$ (a massless free scalar field). As expressed previously, our model is not well defined for two massless scalar fields. Of particular interest are the cases $w=-1$ (dark energy), $w=0$ (dust), and $w=\frac{1}{3}$ (light). Although in the previous chapters we have focused on the dark energy interpretation, we will expand our analysis to other perfect fluids to make it more exhaustive. 

In this section, we are not looking to define a conformal transformation valid on the whole spacetime, hence we consider only the asymptotic regimes of large and small volume $v$. At small $v$, close to the big bang/big crunch singularity, which is spacelike, the dynamics are dominated by the scalar field (assuming that $\pi_\varphi\neq 0$). Then, at large $v$, we can distinguish two cases. A universe with a perfect fluid that generates an accelerated expansion (like dark energy), is asymptotically de Sitter, whereas a universe with dust or light is asymptotically Minkowski. The reason why the asymptotic behaviour is perfect fluid dependent, is that different perfect fluids lead to a different lapse $\tilde{N}$ in \cref{hamsimple}. Examples of studies of conformal diagrams of asymptotically Minkowski spacetimes and asymptotically de Sitter spacetimes can be consulted in \cite{Frauendiener2000} and \cite{Spradlin2001}.

Now, the quantum theories are the result of symmetry reduction at the classical level, so the connection between a particular choice of time coordinate and the spacetime metric (as expressed by the lapse) is no longer obvious at the quantum level. Our way out is to assume that a particular time coordinate has the same interpretation in the classical and the quantum theory, so that the form of the lapse is unchanged.  
This is justified since in any case these conformal diagrams we draw can only represent expectation values in a quantum state, and clearly only make sense in a semiclassical regime. We believe the conformal diagrams illustrate very well where the corrections to the classical geometry are, and allow us to have a better understanding of the quantum universes.

\begin{figure}
\begin{center}
\begin{tikzpicture}
\node (I) at (-3.8,0) {};
\node (II) at (3.8,0) {};

\path 
(I)+ (45:4) coordinate (Itr) 
+(135:4) coordinate (Itl)
+ (225:4) coordinate  (Ibl);

\path
(-3.8,-0.20) + (225:4) node[below]  (Infty) {$v=\infty$};

\draw
 (Itl) -- (Ibl) ;
 
\draw[line width=2pt] (Ibl) --node[midway, above] {I} (Itr);

\draw[decorate,decoration=zigzag] (Itl) --node[midway,above] {$v=0$} (Itr);

\fill  (Ibl) circle (2pt);

\path 
(II) +(135:4) coordinate (IItl)
+ (225:4) coordinate  (IIbl)
+(-45:4) coordinate (IIbr);

\path
(3.8,0.20) + (135:4) node[above]  (Infty2) {$v=\infty$};

\draw
 (IItl) -- (IIbl);

\draw[line width=2pt]
   (IItl) --node[midway,below] {II} (IIbr);

\draw[decorate,decoration=zigzag] (IIbr) --node[midway, below] {$v=0$} (IIbl);

\fill  (IItl) circle (2pt);

\end{tikzpicture}
\caption[Conformal diagram for classical solutions, dust or radiation interpretation.]{Conformal diagram for classical solutions assuming the perfect fluid is dust or radiation. The zigzag line represents the singularity and the dot represents timelike infinity ($\iota^-$ or $\iota^+$). The thicker line represents $\mathcal{I}^{\pm}$, where $v=\infty$ too. I is a contracting universe and II is an expanding universe. There is no trajectory linking I and II. This universe is asymptotically Minkowski.}
\label{classth-cd2}
\end{center}
\end{figure}
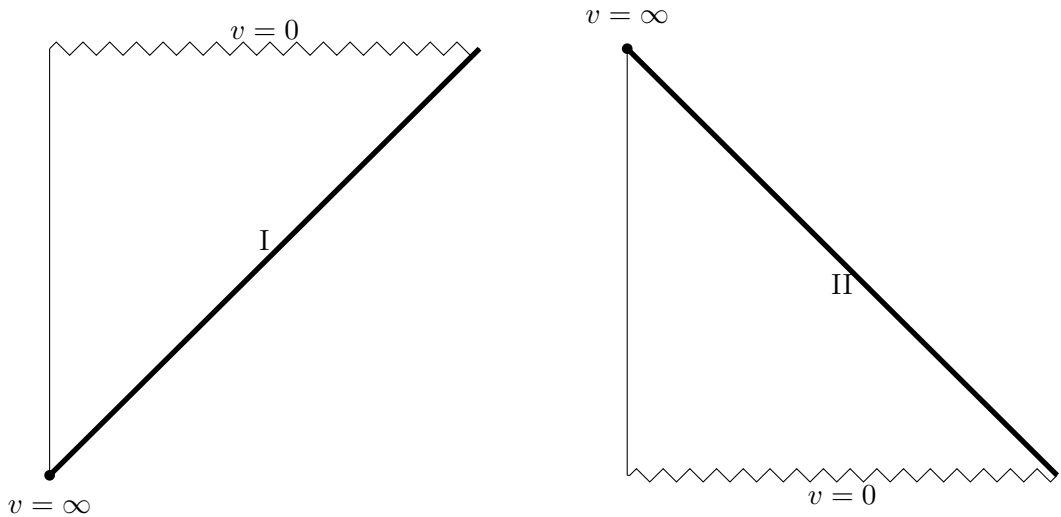

The conformal diagrams of the classical theory are represented in \cref{classth-cd} and \cref{classth-cd2}. Both in the asymptotically Minkowski and in the asymptotically de Sitter case there are two possible solutions, a contracting universe I, and an expanding universe II. Light cones are determined by taking lines at $\pm \ang{45}$, thus a timelike trajectory is a curve whose tangent vector at any point always stays inside the light cone at that point. In the contracting universe, the singularity is in the future light cone of every observer. In contrary, the reverse happens in the expanding universe, the singularity is in the past light cone of every observer. The classical universe is always composed of two disconnected parts.

We have seen in \cref{v-clock-dynamics} that the $v$-clock theory does not present big deviations from the classical theory. In fact, $\expval{t(v)}_{\Psi_{sc,v}}$ is very close to $t_c(v)$ all the time. Therefore, we assume that the conformal diagrams of this quantum theory are identical to  \cref{classth-cd} and \cref{classth-cd2} for semiclassical states $\Psi_{sc,v}$. The universe is composed of two disconnected regions and there is no trajectory that joins the two.

\begin{figure}
\begin{center}
\begin{tikzpicture}
\node (I) at (-3.8,0) {I};
\node (II) at (3.8,0) {II};

\path 
(I)+ (45:4) coordinate (Itr)
+(135:4) coordinate (Itl)
+ (225:4) coordinate  (Ibl)
+(-45:4) coordinate (Ibr);

\draw
 (Itl) -- (Ibl) (Ibr) -- (Itr);

\draw[decorate,decoration=zigzag] (Itr) --node[midway, above] {$v=0$} (Itl);

\draw[line width=2pt] (Ibl)  --node[midway, below] {$v=\infty$} (Ibr);

\path 
(II)+ (45:4) coordinate (IItr)
+(135:4) coordinate (IItl)
+ (225:4) coordinate  (IIbl)
+(-45:4) coordinate (IIbr);

\draw (IItr) -- (IIbr);

\draw
 (IIbl) -- (IItl);

\draw[decorate,decoration=zigzag] (IIbr) --node[midway, below] {$v=0$} (IIbl);

\draw[line width=2pt] (IItl) --node[midway, above] {$v=\infty$} (IItr);
\end{tikzpicture}
\caption[Conformal diagram for classical solutions, dark energy interpretation.]{Conformal diagram for classical solutions assuming that the perfect fluid is dark energy. The zigzag line represents the singularity and the thicker line represents spacelike infinity (and $\mathcal{I}^-$ or $\mathcal{I}^+$ depending on the diagram). I is a contracting universe and II is an expanding universe. There is no trajectory linking I and II. This universe is asymptotically de Sitter}
\label{classth-cd}
\end{center}
\end{figure}
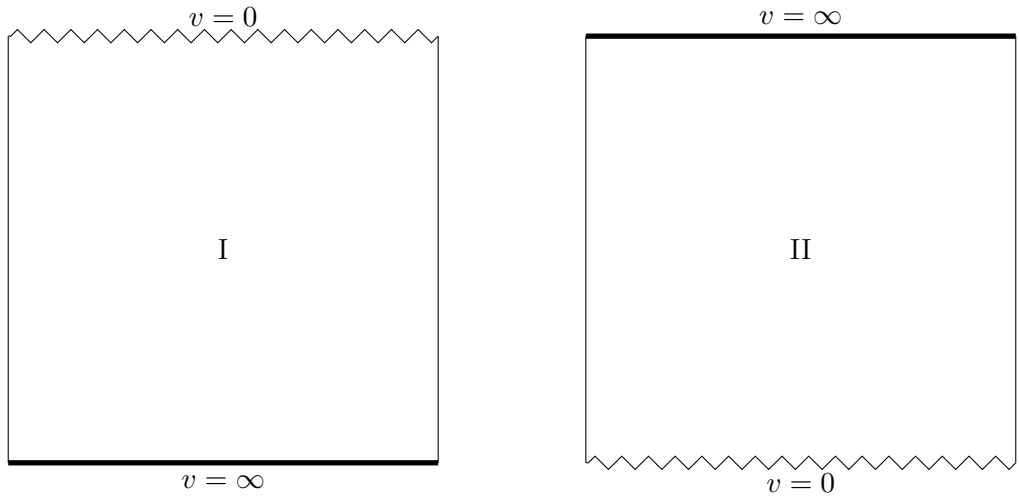

\begin{figure}
\begin{center}
\begin{tikzpicture}[scale=1]
\tikzset{trapezium stretches=true}
\node (I) at (-3.8,-2) {II};
\node (II) at (-3.8,2) {I};

\path 
(I)+ (45:4) coordinate (Itr)
+(135:4) coordinate (Itl)
+ (225:4) coordinate  (Ibl)
+(-45:4) coordinate (Ibr);

\path 
(II)+ (45:4) coordinate (IItr)
+(135:4) coordinate (IItl);

\draw (Ibr)--(Itr)--(IItr) (Ibl)--(Itl)--(IItl);
\draw[line width=2pt]  (Ibr)--node[midway, below] {$\expval{v}=\infty$} (Ibl);
\draw[line width=2pt] (IItr)--node[midway, above] {$\expval{v}=\infty$} (IItl);
\fill[gray, fill opacity=0.3] (-6.6,--0.5) rectangle (-1,-0.5);

\node (III) at (3.8,-2) {II};
\node (IV) at (3.8,2) {I};
\coordinate (IIItr) at (6.5,0);
\node (Infty) at (1,5.1) {$\expval{v}=\infty$};
\node (Infty2) at (1,-5.1) {$\expval{v}=\infty$};
\node[trapezium,trapezium left angle=90, trapezium right angle=83.5, minimum width=5.55cm, minimum height=0.5cm, fill, gray, fill opacity=0.3] at (3.44,0.25) {};
\node[trapezium,trapezium left angle=90, trapezium right angle=97.4, minimum width=5.55cm, minimum height=0.5cm, fill, gray, fill opacity=0.3] at (3.44,-0.25) {};

\path 
(III)
+(225:4) coordinate (IIIbl);

\path 
(IV)+ (135:4) coordinate (IVtl);

\fill  (IVtl) circle (2pt)

(IIIbl) circle (2pt);

\draw (IVtl)--(IIIbl)--(IIItr)--(IVtl);

\end{tikzpicture}
\caption[Conformal diagram of the $t$-clock theory.]{Conformal diagram of the $t$-clock theory. The left diagram corresponds to a universe in which the perfect fluid is dark energy, and the right diagram corresponds to a universe in which the perfect fluid is radiation or dust. Now the contracting region I lies to the future of the expanding region II. The singularity is still present, but the expectation value of the  volume remains finite. At large $v$ there is a region in which quantum fluctuations dominate and where the volume reaches its maximum expectation value $\expval{v}=V_{min_\Psi}$. It is impossible to talk about a ``classical trajectory''.}
\label{t-clock-cd}
\end{center}
\end{figure}
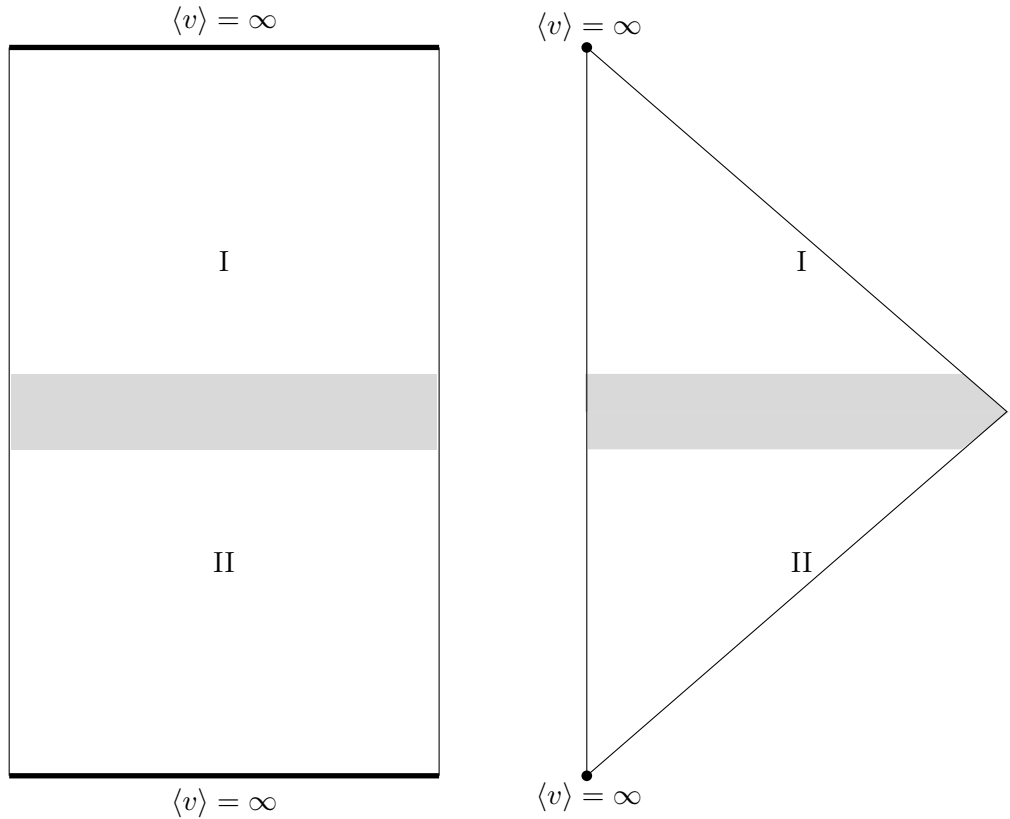

The conformal diagrams for semiclassical states in the $t$-clock theory are presented in \cref{t-clock-cd}; we replace $v_c(t)$ by $\expval{v(t)}_{\Psi_{sc,t}}$. The left diagram is the asymptotically de Sitter space and the right diagram is the asymptotically Minkowski space. As this expectation value does not go to zero, rather has a strictly minimum value before growing to infinity, we have glued universes I and II together removing the classical singularity. This singularity is replaced by a region in which quantum fluctuations are large, and where classical trajectories are not well defined. We have labelled this region in grey. However, the region near infinity is identical to the classical theory, both in the Minkowski and de Sitter case. Note that we do not know exactly where the limit between the highly quantum are and the semiclassical area lies, but we know that it is state dependent. The main difference between this theory and the classical universe is that now region I and II are connected, hence it is possible to cross between the two regions.

\begin{figure}
\begin{center}
\begin{tikzpicture}[scale=1]
\tikzset{trapezium stretches=true}
\node (I) at (-3.8,-2) {II};
\node (II) at (-3.8,2) {I};

\path 
(I)+ (45:4) coordinate (Itr)
+(135:4) coordinate (Itl)
+ (225:4) coordinate  (Ibl)
+(-45:4) coordinate (Ibr);

\path 
(II)+ (45:4) coordinate (IItr)
+(135:4) coordinate (IItl);

\draw (Ibr)--(Itr)--(IItr) (Ibl)--(Itl)--(IItl);
\draw[decorate,decoration=zigzag] (Ibr)--node[midway, below] {$\expval{v}=0$} (Ibl);
\draw[decorate,decoration=zigzag] (IItr)--node[midway, above] {$\expval{v}=0$} (IItl);
\fill[gray, fill opacity=0.3] (-6.6,--0.5) rectangle (-1,-0.5);

\node (III) at (3.8,0) {};
\node (i) at (5,1.3) {I};
\node (ii) at (2.6,-1.3) {II};

\path 
(III)+ (45:4) coordinate (IIItr)
+(135:4) coordinate (IIItl)
+ (225:4) coordinate  (IIIbl)
+(-45:4) coordinate (IIIbr);

\path (IIItl)+(0:0.5) coordinate (A);
\path (IIIbr)+(90:0.5) coordinate (B);
\path (IIItl)+(-90:0.5) coordinate (C);
\path (IIIbr)+(180:0.5) coordinate (D);

\draw (IIIbr)--(IIItr) (IIIbl)--(IIItl);

\draw[decorate,decoration=zigzag] (IIIbr)--node[midway, below] {$\expval{v}=0$} (IIIbl);

\draw[decorate,decoration=zigzag] (IIItr)--node[midway, above] {$\expval{v}=0$} (IIItl);

\fill[gray,opacity=0.3] (A)--(B)--(IIIbr)--(IIItl);
\fill[gray,opacity=0.3] (C)--(D)--(IIIbr)--(IIItl);

\end{tikzpicture}
\caption[Conformal diagram of the $\varphi$-clock theory.]{Conformal diagram of the $\varphi$-clock theory, with contracting region I and expanding region II. The left diagram corresponds to a universe in which the perfect fluid is dark energy, and the right diagram corresponds to a universe in which the perfect fluid is radiation or dust. Infinity replaced by the shaded area, where quantum fluctuations are large. In this state-dependent region the volume reaches a minimum expectation value $\expval{v}=V_{max_\Psi}$ and it is impossible to talk about a ``classical trajectory''.}
\label{phi-clock-cd}
\end{center}
\end{figure}
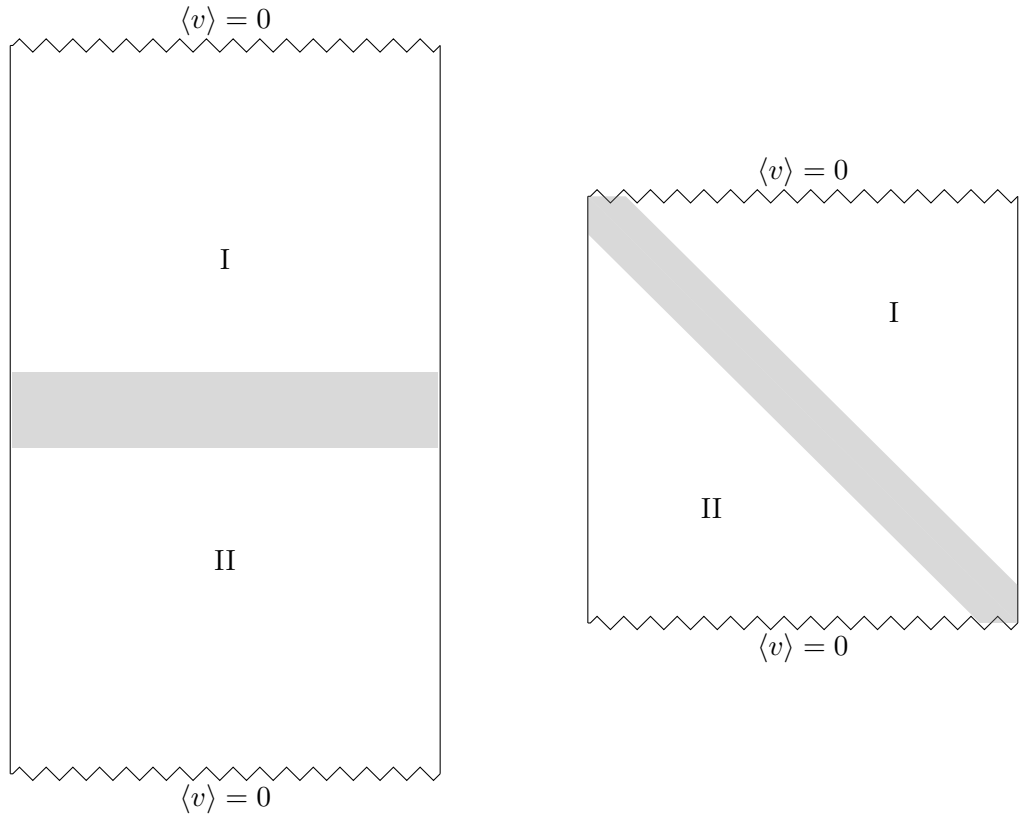

The conformal diagrams of the $\varphi$-clock theory can be seen in \cref{phi-clock-cd}. The right diagram corresponds to a universe in which the perfect fluid is dark energy, and the left diagram corresponds to a universe in which the perfect fluid is radiation or dust. We represent the expectation value $\expval{v(\varphi)}_{\Psi_{sc,\varphi}}$ instead of the classical $v_c(\varphi)$. Here the classical singularity remains, but infinity (which is either de Sitter or Minkowski) is ``resolved'' . Hence, regions I and II are glued the opposite way, infinity is replaced  but the classical singularity is unchanged. Once more, there is a region (represented in grey) in which quantum fluctuations dominate. If we assume that the perfect fluid is dark energy (left diagram of \cref{phi-clock-cd}), the theory has some similarities with Penrose's conformal cyclic cosmology \cite{Penrose}. Indeed, spacelike infinity is no longer seen as the future end point of a cosmological constant dominated universe, but it becomes a transition point into the new universe. However, in our case, the subsequent ``aeon'' is contracting, rather than expanding again, and the origin of the transition lies in quantum fluctuations. If we assume the perfect fluid is radiation of dust, the two parts are glued from the diagonal. (provided one has been rotated). The $\varphi$-theory is very puzzling because the quantum effects arise at late times, instead of arising close to the big bang/big crunch. It could be possible to find a fully cyclic cosmology if time evolution was controlled by a ``slow'' clock both at the singularities and at spacelike infinity. Unitarity would then enforce resolution of the singularity and the replacement of spacelike infinity by a quantum recollapse. 
A universe with such clock would be eternal.

After this numerical analysis, we can quantify how each clock affects the resulting quantum theory. The reflecting boundary conditions of the wave functions changes the causal structure of the universe and prevent or generate singularity resolution in semiclassical states. This motivates the following questions: Could other approaches to quantum cosmology give some insight with respect to the problem of time? What meaningful conclusions can we extract from this analysis? We will try to answer these questions in the next sections.
\part{Other approaches to the problem of time and conclusions}
\chapter{Path integral quantisation}
\label{pathin}

As there is not (yet) a fully satisfactory theory of quantum gravity, the problem of time appears in very different contexts and with many nuances. A very popular approach to quantisation of general relativity, besides the ones we have already seen, is the so called path integral quantisation procedure. The idea of this method is to extend the concept of path integral quantisation used in quantum mechanics and QFT to a whole spacetime. The object of interest is 
\begin{equation}
\int \mathcal{D} g_{\mu\nu} \D\phi_i \dots \exp[i \mathcal{S}(g_{\mu\nu},\phi_i, \dot{\phi_i},\dots)]\, ,
\label{pathinteq}
\end{equation}
Where $g_{\mu\nu}$ is the spacetime metric and $\phi_i$ represent the matter fields. The fields can be as complicated as one wishes, i.e., not only massless scalar fields. The action $\mathcal{S}$ may depend on the metric, the fields, their derivatives, and maybe higher derivatives if a version of modified gravity is considered. Recall that the notation $\D$ denotes a functional differential, where we are integrating over all paths. In general, people use Lagrangian actions in the path integral formulation, but Hamiltonian actions, in which one replaces the derivatives $\dot{\phi}_i$ with the canonical momenta $\pi_{\phi_i}$, are also possible. We will be working in a Hamiltonian formulation of the action. 

The path integral as written in \cref{pathinteq} represents roughly (the exact meaning of the path integral needs additional assumptions to be specified) the probability for a spacetime of initial configuration $g_{\mu\nu}=g_{\mu\nu,in}$ and $\phi_i=\phi_{i,in}$ to transition to a spacetime of configuration $g_{\mu\nu}=g_{\mu\nu,f}$ and $\phi_{i,f}$, where one has integrated over all possible momenta, or if one is working in the momentum representation, the probability to go from configuration $g=g_{in}$, $\pi_{\phi_i}=\pi_{\phi_i,in}$ to the configuration $g=g_{f
}$, $\pi_{\phi_i}=\pi_{\phi_i,f}$.  

The big difference between the path integral quantisation and the Wheeler--DeWitt quantisation that we used in the previous chapters is that the former is centred in the sum of histories, and the latter is centred in the wave function of the universe $\Psi$ \cite{Sorkin1997}. In the previous chapters, the wave function and the Hilbert space were clear, and we measured observable quantities, like the expectation value of the volume $v$. In this approach these characteristics appear to be hidden inside the integral. However, path integral quantisation offers a new perspective on the problem of time. We do not have the problem of the frozen formalism, at least initially \cite{Kuchar2011}. The most powerful feature of this approach of quantum gravity is perhaps that it can incorporate spacetime topology changes, simply by considering a sum of terms of the form \cref{pathinteq} where the spacetime manifold $\mathcal{M}$ has different topologies. In fact, the initial conditions can come from one spacetime topology and the final conditions from another. As the path integral is an object that allows us to go from some initial state to a final state of the universe, one can ask questions like what are the initial conditions of the universe? rather explicitly. The famous no-boundary proposal \cite{Hartle1983}, that postulates that the universe starts from a state that has no boundaries, was motivated in this setting. Path integral quantisation has been used in very different ways, included minisuperspace models and has been linked to the Wheeler--DeWitt equation \cite{Halliwell1988,Henneaux1985}. In this section we mainly follow the approach of \cite{Halliwell1988} in order to present the path integral quantisation of our model. In his work, Halliwell presents a universe with a quadratic Hamiltonian. Our Hamiltonian, however possesses an extra linear term. Mathematically a linear term poses no problem (contrary to what would happen with higher powers of the momenta).

We would like to devote the next few pages in the analysis of the path integral quantisation of our model, in order to relate this approach to the previous ones we have analysed and get more insights on diffeomorphism invariance and the problem of time. For completeness, rather than refer the reader to the classical section of the model, we will recap the important equations and concepts here. The metric of our universe was defined in \cref{metric}
\begin{equation}
\dd s^2=-N(\tau)^2\dd \tau^2 + a(\tau)^2 h_{ij}\dd x^i \dd x^j\, .
\end{equation}
The topology of our universe is $\R\times \Sigma$ where $\Sigma$ is a bounded three-dimensional manifold, for example a three torus. For concrete amplitudes, the topology of $\Sigma$ has to be specified, but we will not do this here.  The metric $h_{ij}$ is flat.  The matter components of the models are a (free) massless scalar field  $\phi$ and a perfect fluid with equation of state $p=w\rho$ where $p$ is the pressure and $\rho$ is the energy density. We have previously discussed that we can choose our perfect fluid to be a standard matter component like pressureless dust ($w=0$), radiation ($w=\frac{1}{3}$), but for the inclusion of dark energy ($w=-1$) we need to work with a parametrised version of unimodular gravity (recall that the ``parametrised'' part is very important to maintain full diffeomorphism invariance), see \cite{Henneaux1989} for more information about the different actions of unimodular gravity. The distinction between parametrised and non parametrised unimodular gravity plays a non-trivial rôle in the path integral quantisation formalism, for an example of path integral quantisation of a non parametrised version of unimodular gravity see \cite{Daughton1998}. As in the previous sections, for simplicity we identify the perfect fluid as being dark energy, but all the results we present are valid for all (standard) perfect fluids interpretation, unless specified otherwise.

After the change of variables \cref{changevar}
\begin{align}
v&=4\sqrt{\frac{V_0}{3}}\frac{a^{\frac{3(1-w)}{2}}}{1-w},  &\pi_v&=\sqrt{\frac{1}{12V_0}}\pi_a a^{\frac{3w-1}{2}}, &\\
\varphi&=\sqrt{\frac{3}{8}}(1-w)\phi, & \pi_\varphi&=\sqrt{\frac{8}{3}}\frac{\pi_\phi}{1-w}\,. &
\end{align}
the Hamiltonian of the model can be written as
\begin{equation}
\mathcal{H}_{class}=\tilde{N}\left[-\pi_v^2+\frac{\pi_\varphi^2}{v^2}+\lambda\right]\, ,
\label{hclass}
\end{equation}
where $\tilde{N}=Na^{-3w}=N\left(\frac{16V_0}{3v^2(1-w)^2}\right)^{\frac{w}{1-w}}$, $\lambda=V_0m$ where $m$ is the energy density of the fluid and $V_0=\int \dd^3 x \sqrt{h}<\infty$. The Hamiltonian constraint is
\begin{equation}
\C_{class}=-\pi_v^2+\frac{\pi_\varphi^2}{v^2}+\lambda=0\, .
\end{equation} 
The Hamiltonian action of this model is 
\begin{equation}
\mathcal{S}_{class}=\int_{\tau_{in}}^{\tau_f} \dd \tau \left\lbrace \pi_v \dot{v}+\pi_\varphi \dot{\varphi}+ \lambda\dot{t}-\tilde{N}\left[-\pi_v^2+\frac{\pi_\varphi^2}{v^2}+\lambda\right]\right\rbrace\, ,
\label{Sclass}
\end{equation}
where $\cdot=\frac{\dd }{\dd \tau}$. Note that in $\tau_{in}<\tau_f$. The coordinates $v$, $\varphi$ and $t$ are fixed at the boundaries by $v_{in}=v(\tau_{in})$, $\varphi_{in}=\varphi(\tau_{in})$ and $v_f=v(\tau_f)$, $\varphi_f=\varphi(\tau_f)$. The momenta $\pi_v$, $\pi_\varphi$ and $\lambda$ are free. If we write $q^\mu=(v,\varphi, t)$ and $\pi_\mu=(\pi_v,\pi_\varphi,\lambda)$ the equations of motion are:
\begin{equation}
\dot{q}^\mu=\tilde{N}\left\lbrace q^\mu, \mathcal{C}_{class} \right\rbrace, \hspace{4mm} \dot{\pi}_\mu=\tilde{N}\lbrace \pi_\mu, \mathcal{C}_{class} \rbrace
\end{equation}
with the Poisson bracket being defined by:
\begin{equation}
\lbrace F,G \rbrace=\pdv{F}{q^\mu}\pdv{G}{\pi_\mu}-\pdv{G}{q^\mu}\pdv{F}{\pi_\mu}
\end{equation}
The action \cref{Sclass} is diffeomorphism invariant. This can be seen by applying an infinitesimal coordinate transformation parametrised by $\varepsilon(\tau)$. This transforms the coordinates of the action by:
\begin{equation}
\delta q^\mu=\varepsilon(\tau)\lbrace q^\mu, \mathcal{C}_{class}\rbrace , \hspace{3mm} \delta \pi_\mu=\varepsilon(\tau)\lbrace \pi_\mu, \mathcal{C}_{class}\rbrace, \hspace{3mm} \delta \tilde{N}=\dot{\varepsilon}(\tau)\, .
\label{coordtrans}
\end{equation}
The variation of the action under the transformation \cref{coordtrans} is
\begin{equation}
\delta \mathcal{S}_{class}=\left[ \varepsilon(\tau)\left(\pi^\mu\pdv{}{\pi^\mu}\mathcal{C}_{class}-\mathcal{C}_{class}\right)\right]_{\tau_{in}}^{\tau_f}
\end{equation}
For the momenta that appear quadratically in $\mathcal{C}_{class}$, the variation of the action is 0 if and only if $\varepsilon(\tau)$ vanishes at the end points, i.e., $\varepsilon(\tau_{in})=\varepsilon(\tau_f)=0$. However, for the linear term $\lambda$, yields to a contribution of the form $\lambda$-$\lambda=0$, hence this term does not contribute for $\epsilon(\tau)$ boundary conditions. We will come back to the condition  $\varepsilon(\tau_{in})=\varepsilon(\tau_f)=0$ later.

The path integral we are interested in is
\begin{equation}
G(v_f,\varphi_f,t_f|v_{in},\varphi_{in},t_{in})=\int\D v \D\varphi \D t \ \D\pi_v\D\pi_\varphi \D\lambda\ \D \tilde{N} \exp[i\mathcal{S}_{class}]\, ,
\end{equation}
where $v_f=v(\tau_f)$, $\varphi_f=\varphi(\tau_f)$, and $t_f=t(\tau_f)$, is the final configuration and $v_{in}=v(\tau_{in})$, $\varphi_{in}=\varphi(\tau_{in})$, and $t_{in}=t(\tau_{in})$ is the initial configuration of the system. Diffeomorphism invariance of $\mathcal{S}_{class}$ poses a problem, as physically identical trajectories might be counted several times. It is then necessary to gauge fix the action. The gauge fixing procedure involves a complex process of adding some gauge fixing and additional ghosts fields in the action. A reader not interested in this procedure can jump directly to formula \cref{gfpathint}, we include it for completeness. Gauge fixing amounts to fixing the derivative of $\tilde{N}$:
\begin{equation}
\dot{\tilde{N}}=\varrho(q^\mu,\pi_\mu,\tilde{N})
\label{gaugeN}
\end{equation}
Hence, we add a gauge fixing term of the form
\begin{equation}
\mathcal{S}_{gf}=\int_{\tau_{in}}^{\tau_f}\dd \tau \ \Pi(\dot{\tilde{N}}-\varrho)\, ,
\end{equation}
to the classical action $\mathcal{S}_{class}$. $\Pi$ is another Lagrange multiplier that satisfies 
\begin{equation}
\Pi(\tau_{in})=0 \ \text{and} \ \Pi(\tau_{f})=0\, ,
\label{bcPi}
\end{equation} 
The equations of motion obtained variating the action with respect to $q^\mu$ and $\pi_\mu$ are now
\begin{equation}
\dot{q}^\mu=\tilde{N}\left\lbrace q^\mu, \mathcal{C}_{class} \right\rbrace+\Pi\lbrace q^\mu,\varrho \rbrace, \hspace{4mm} \dot{\pi}_\mu=\tilde{N}\lbrace \pi_\mu, \mathcal{C}_{class} \rbrace+ \Pi\lbrace \pi_\mu,\varrho \rbrace\, .
\label{eqmotion}
\end{equation}
We have integrated by parts and used the boundary conditions on $q^\mu$ to obtain the second equality. Variation with respect to $\tilde{N}$ leads to
\begin{equation}
\dot{\Pi}+\mathcal{C}_{class}=0\, ,
\label{eqmotionN}
\end{equation}
and variation with respect to $\Pi$ gives us the gauge condition.
Differentiating \cref{eqmotionN} and using \cref{eqmotion} (and the fact that the Hamiltonian is at most quadratic in the momenta) we find
\begin{equation}
\ddot{\Pi}+\dot{\mathcal{C}}_{class}=0 \implies \ddot{\Pi}+\lbrace \varrho, \mathcal{C}_{class} \rbrace\Pi=0\, .
\end{equation}
As $\Pi$ vanishes at $\tau_{in}$ and $\tau_f$ the unique solution to this equation is $\Pi\equiv 0$. If $\lbrace \varrho, \mathcal{C}_{class} \rbrace\neq 0$ there are non-trivial solutions to the equation, but we will consider $\varrho=0$ which makes $\lbrace \varrho, \mathcal{C}_{class} \rbrace= 0$. Thus the action $\mathcal{S}_{class}+\mathcal{S}_{gf}$ leads to the correct equations of motion and is gauge invariant.

However, the we need to make the path integral independent of the choice of gauge (or the choice of $\varrho$ in our case). The usual method for this is to add some additional anticommuting ghost fields and extend the phase space. The most general method (as to our knowledge) was developed by Batalin, Frankin and Vilkovisky (BFV) \cite{PIQ}. There are some criticisms to the BFV method \cite{PIQcritics}, mainly regarding the generality of this approach, as it may not be independent of the gauge fixing function $\varrho$ for all choices of $\varrho$. In our case, choosing $\varrho=0$ fixes most possible issues. The action resulting from the BFV method is invariant under Becchi-Rouet-Stora-Tyuting (BRST) symmetry.

 The BFV method involves replacing $\varepsilon(\tau)$ in the coordinate transformation \cref{coordtrans} by a parameter $\omega c(\tau)$ where $\omega$ is an anticommuting constant and $c(\tau)$ is an anticommuting field. Anticommuting variables are a very useful tool for QFT, we will assume that the reader is familiarised with them. If this is not the case, the basis of anticommuting variables are presented in \cref{anticom}. One wants to impose the relation $\dot{c}(\tau)=p(\tau)$ by adding a term $\bar{p}(\dot{c}-p)$. The variable $p$ is made dynamical by adding a term $\bar{c}\dot{p}$. The anticommuting fields $c$ and $\bar{c}$ are anticommuting variables, and $p$ and $\bar{p}$ are anticommuting momenta. These variables are called ghosts, as they will not appear in the final action. They are added to ensure that the final path integral action is independent of the choice of gauge. We expand the phase space from $(q^\mu,\pi_\mu, \tilde{N})$ to $(q^\mu,\pi_\mu, \tilde{N}, \Pi, c,\bar{c},p,\bar{p})$ and hence also expand the Poisson bracket to include the new anticommuting variables. The ghost action is:
\begin{equation}
\mathcal{S}_{gh}=\int_{\tau_{in}}^{\tau_f}\dd \tau \left\lbrace \bar{p}\dot{c}+\bar{c}\dot{p}-\bar{p}p+c\lbrace\varrho,\mathcal{C}_{class}\rbrace\bar{c}+p\pdv{\varrho}{\tilde{N}}\bar{c} \right\rbrace\, ,
\end{equation}
The total action is invariant: $\delta \mathcal{S}_{class}+\delta \mathcal{S}_{gf}+\delta\mathcal{S}_{gh}=0$ under the BRST transformation
\begin{align}
&\delta q^\mu=\omega c\lbrace q^\mu, \mathcal{C}_{class}\rbrace , \hspace{3mm} \delta \pi_\mu=\omega c\lbrace \pi_\mu, \mathcal{C}_{class}\rbrace, \hspace{3mm} \delta \tilde{N}=\omega p\, , \nonumber\\
&\delta\Pi=\delta c=\delta p=0\,, \nonumber \\
&\delta \bar{c}=-\omega\Pi, \hspace{4mm} \delta\bar{p}=-\omega\mathcal{C}_{class}\, ,
\label{BRST}
\end{align}
if we impose the additional boundary conditions:
\begin{equation}
c(\tau_{in})=c(\tau_f)=0,\ \text{and} \ \bar{c}(\tau_{in})=\bar{c}(\tau_f)=0\, ,
\label{bc}
\end{equation}
so the ghost fields vanish at the end points. Note that the first line of \cref{BRST} is just a normal coordinate transformation \cref{coordtrans} where $\varepsilon=\omega c$ and the other two lines are some extra conditions needed on the ghost fields that ensure the invariance of the total action. The first line of \cref{bc} comes from the original condition  $\varepsilon(\tau_{in})=\varepsilon(\tau_f)=0$. When calculating the path integral the ghost fields will disappear, so these extra conditions do not restrict our original theory. For more details on the proof of the action invariance under this BRST transformation see \cite{Halliwell}.

The path integral quantisation over the extended phase space is
\begin{align}
G(v_f,\varphi_f,t_f|v_{in},\varphi_{in},t_{in})=\int\D v \D\varphi \D t  \ \D\pi_v\D\pi_\varphi \D\lambda\ \D \tilde{N} \D \Pi \ & \D\bar{p}\D c\D p \D \bar{c}  \nonumber \\
& \exp[i(\mathcal{S}_{class}+\mathcal{S}_{gf}+\mathcal{S}_{gh})]\, .
\label{extendedPI}
\end{align}

In the case where the gauge fixing function $\varrho$ vanishes, the ghost integral 
\begin{equation}
\int \D\bar{p}\D c\D p \D \bar{c} \exp[i\mathcal{S}_{gh}]\, ,
\end{equation}
can be computed separately. It is important to stress that the notation $\D$ hides all the potentially problematic things of the functional integral, because it does not specify the ``weight'' of each trajectory. In more scientific terms, in order to do calculations we need to specify a measure. The choice of measure is usually a really non-trivial task, but as these fields have been added by hand, we have the freedom of choosing a measure that works for ur purposes. Specification of the integral is done by splitting the $\tau$ interval in $n+1$ intervals of length $\epsilon$ such that $\tau_f-\tau_{in}=\epsilon(n+1)$, where $\epsilon>0$ is a small quantity that will be taken to zero in the limit. This involves changing the integral over $\tau$ by a sum times $\epsilon$ and the continuous quantities $c$, $\bar{c}$, $p$ and $\bar{p}$ are discretised such that the derivative of the coordinates $c$ and $\bar{c}$ are replaced by a difference, and the momenta $p$ and $\bar{p}$ are replaced by a constant value. The simplest choice of measure, also referred as skeletonisation, is putting equal weight to all trajectories to find:
\begin{align}
\int \D\bar{p}\D c\D p \D \bar{c} \exp[i\mathcal{S}_{gh}]&=\int \dd p_{1/2} \dots \dd p_{n+1/2}\int\dd \bar{p}_{1/2} \dots \dd \bar{p}_{n+1/2} \nonumber \\
&\int \dd c_1 \dots \dd c_n \int \dd \bar{c}_1 \dots \dd \bar{c}_n \exp\left[i\sum_{k=0}^n\left\lbrace \bar{p}_{k+1/2}(c_{k+1}-c_k)\right.\right. \nonumber \\
&\left.+p_{k+1/2}(\bar{c}_{k+1}-\bar{c}_k)-\epsilon \bar{p}_{k+1/2}p_{k+1/2} \right\rbrace\Bigg]\, .
\label{berenzin}
\end{align}
There are a few things to note here. Firstly, the term $\int \dd \tau \ \bar{c}\dot{p}$ has been integrated by parts using the boundary condition \cref{bc} to find $-\int \dd  \tau \ \dot{\bar{c}}p=\int \dd \tau \ p\dot{\bar{c}}$. Secondly, there are $n+1$ integrals for each of the two momentum variables and $n$ integrals for each of the coordinate variables, and the boundary conditions \cref{bc} transform into $c_0=c_{n+1}=\bar{c}_0=\bar{c}_{n+1}=0$. Finally, the integrals appearing in the right-hand side are standard Berezin integrals, and they can be performed to lead to the final result:
\begin{equation}
\int \D\bar{p}\D c\D p \D \bar{c} \exp[i\mathcal{S}_{gh}]=\epsilon(n+1)=\tau_{f}-\tau_{in}
\label{cpint}
\end{equation}
For more explanations of the intermediate steps of this result, we refer again to \cite{Halliwell} and for the standard rules of Berezin integrals to \cref{anticom}. Now, all the ghost fields have vanished of the action.

We can also do the integral over $\D \Pi$ and $\D \tilde{N}$. This time, $\tilde{N}$ is integrated over whereas $\Pi$ is fixed at the end points by the boundary condition \cref{bcPi}. Using again the simplest possible measure we have:
\begin{align}
\int \D \tilde{N} \D \Pi \exp[i\int_{\tau_{in}}^{\tau_f} \dd \tau \ \Pi\dot{\tilde{N}}]&=\int \dd \tilde{N}_{1/2} \dots \dd \tilde{N}_{n+1/2}\frac{1}{(2\pi)^n}\int \dd \Pi_1 \dots \dd \Pi_n \\
&\times \exp[i \sum_{k=1}^n \Pi_k(\tilde{N}_{k+1/2}-\tilde{N}_{k-1/2})]\, .
\end{align}
Recall that once again there are $n+1$ $\tilde{N}$ integrals and $n$ $\Pi$ integrals. The $\Pi$ integrals can be done first and yield to standard Dirac delta functions leaving us with
\begin{equation}
\int \D \tilde{N} \D \Pi \exp[i\int_{\tau_{in}}^{\tau_f} \dd \tau \ \Pi\dot{\tilde{N}}]=\int \dd \tilde{N}_{1/2} \dots \dd \tilde{N}_{n+1/2}\prod_{k=1}^n \delta(\tilde{N}_{k+1/2}-\tilde{N}_{k-1/2})\, .
\end{equation}
Recall that the factors $1/(2\pi)$ are part of the measure so that they cancel the factors $2\pi$ coming from the Dirac delta functions. Such factors where not needed working with Berezin integrals. As we have $n$ delta functions and $n+1$ integrations, we are left with a single $\tilde{N}$ integral, which we can rename $\tilde{N}$ for simplicity leading to
\begin{equation}
\int \D \tilde{N} \D \Pi \exp[i\int_{\tau_{in}}^{\tau_f} \dd \tau \ \Pi\dot{\tilde{N}}]=\int \dd \tilde{N}
\label{piint}
\end{equation}

Combining \cref{cpint} and \cref{piint} we obtain the final formula
\begin{equation}
G(v_f,\varphi_f,t_f|v_{in},\varphi_{in},t_{in})=\int \dd \tilde{N} (\tau_f-\tau_{in})\int \D v \D \varphi \D t \D \pi_v \D \pi_\varphi \D \lambda \exp[i \mathcal{S}_{class}]
\label{gfpathint}
\end{equation}
Sometimes, a change of variables $T=\tilde{N}(\tau_f-\tau_{in})$ is performed, and the integral $\int \dd \tilde{N} (\tau_f-\tau_{in})$ is directly presented as $\int \dd T$ \cite{Henneaux1985}. Note that the integral \cref{gfpathint}, although specified and calculated for our specific model, is a general formula valid for all Hamiltonians as long as they are at most quadratic in the momenta  \cite{Halliwell}. 

The formula \cref{gfpathint} is the starting point to specify the path integral quantisation of our model. The Hamiltonian \cref{hclass} has the same form of the parametrised non relativistic point particle, so we can start by specifying the term
\begin{equation}
\int \dd \tilde{N}(\tau_f-\tau_{in})\int \D t \D \lambda \exp[i\int_{\tau_{in}}^{\tau_f}\dd \tau \left\lbrace\lambda \dot{t}-\tilde{N}\lambda \right\rbrace]\, .
\end{equation}
Once again we divide the interval $\tau_{f}-\tau_{in}$ in $n+1$ intervals of length $\epsilon$ we use the simplest skeletonisation to find
\begin{align}
&\int \dd \tilde{N}(\tau_f-\tau_{in})\int \D t \D \lambda \exp[i\int_{\tau_{in}}^{\tau_f}\dd \tau \left\lbrace\lambda \dot{t}-\tilde{N}\lambda \right\rbrace] \\
&=\int \dd \tilde{N}(\tau_f-\tau_{in})\frac{1}{(2\pi)^n} \int \dd t_1 \dots \dd t_n \int \dd \lambda_{1/2} \dots \dd \lambda_{n+1/2}\nonumber \\
& \exp[i\sum_{k=0}^n\left\lbrace\lambda_{k+1/2}(t_{k+1}-t_{k}-\epsilon \tilde{N}) \right\rbrace ]. \nonumber
\end{align}
The momentum integration can be done first and yield to a product of Dirac delta distributions:
\begin{align}
&\int \dd \tilde{N}(\tau_f-\tau_{in})\int \D t \D \lambda \exp[i\int_{\tau_{in}}^{\tau_f}\dd \tau \left\lbrace\lambda \dot{t}-\tilde{N}\lambda \right\rbrace] \nonumber\\
&=\int \dd \tilde{N}(\tau_f-\tau_{in})\int \dd t_1 \dots \dd t_n \prod_{k=0}^n \delta(t_{k+1}-t_{k}-\epsilon \tilde{N}) \nonumber\\
&=\int \dd \tilde{N}(\tau_f-\tau_{in})\delta(t_{n+1}-t_0-\epsilon(n+1)\tilde{N})\, .
\end{align}
In the last line, we can replace $t_{n+1}$ and $t_0$ respectively by $t_{f}$ and $t_{in}$ (as these are the boundary conditions) and $\epsilon(n+1)$ by $\tau_f-\tau_{in}$ so the complete path integral is 
\begin{align}
&G(v_f,\varphi_f,t_f|v_{in},\varphi_{in},t_{in})=\int \dd \tilde{N}\delta(t_f-t_{in}-\tilde{N}(\tau_f-\tau_{in}))\nonumber \\
&\times\int \D v \D \varphi \D\pi_{v}\D\pi_{\varphi} \exp[i\int_{\tau_{in}}^{\tau_f} \dd \tau \left\lbrace \pi_v \dot{v}+\pi_\varphi \dot{\varphi}-\tilde{N}\left[-\pi_v^2+\frac{\pi_\varphi^2}{v^2}\right]\right\rbrace]\, .
\end{align}
The second line of the formula is by definition the probability to go from a state with initial conditions $v_{in}$, $\varphi_{in}$ at time $\tilde{N}\tau_{in}$ and final conditions $v_f$, $\varphi_f$ at time $\tilde{N}\tau_{f}$, provided that the Hamiltonian of the system is the quantisation of $\pi_v^2+\frac{\pi_\varphi^2}{v^2}$ (the meaning of quantisation will be specified soon). Using the standard notation for quantum mechanics we can write this as
\begin{align}
&G(v_f,\varphi_f,t_f|v_{in},\varphi_{in},t_{in})=\int \dd \tilde{N}\delta(t_f-t_{in}-\tilde{N}(\tau_f-\tau_{in}))\braket{v_f,\varphi_f,\tilde{N}\tau_f}{v_{in},\varphi_{in},\tilde{N}\tau_{in}}\, .
\end{align}

In order to continue with the calculation we ought to make a choice on the range of $\tilde{N}$. The simplest assumption is to take $\tilde{N}\in (-\infty,+\infty)$ in this case, the $\delta$ is always evaluated, and we can take $\tilde{N}=\frac{t_f-t_{in}}{\tau_{f}-\tau{in}}$. The path integral quantisation is then
\begin{equation}
G(v_f,\varphi_f,t_f|v_{in},\varphi_{in},t_{in})=\braket{v_f,\varphi_f,t_f}{v_{in},\varphi_{in},t_{in}}
\end{equation}
where $\braket{v_f,\varphi_f,t_f}{v_{in},\varphi_{in},t_{in}}$ is a solution to the Schrödinger equation
\begin{equation}
\left(i\hbar\pdv{}{t_f}-\hat{\mathcal{C}}_{class,f} \right)\braket{v_f,\varphi_f,t_f}{v_{in},\varphi_{in},t_{in}}=0\, ,
\label{schr1}
\end{equation}
where,
\begin{equation}
\hat{\mathcal{C}}_{class,f}=\widehat{-\pi_v^2+\frac{\pi_\varphi^2}{v_f^2}}\, .
\label{qtmcclass}
\end{equation}
The quantisation of $\mathcal{C}_{class}$ is subject to the same ambiguities we have already seen, in particular ordering ambiguities. If we use the Hawking and Page ordering \cite{Hawking1985} and we multiply \cref{schr1} by $-1$, we recover the Wheeler--DeWitt equation 
\begin{equation}
\left( \hbar^2 \pdv{}{v_f^2} +\frac{\hbar^2}{v_f}\pdv{}{v_f}-\frac{\hbar^2}{v_f^2}\pdv[2]{}{\varphi_f}-i\hbar\pdv{}{t_f}\right)\Psi(v_f, \varphi_f, t_f)=0\, ,
\end{equation}
which corresponds to \cref{wdw-re} of the Dirac quantisation \cref{dirac-sec}. 

Naively, one could conclude from this that the path integral quantisation offers a way of finding a preferred Wheeler--DeWitt equation, but there are two main catches:
\begin{itemize}
\item The path integral quantisation depends on the skeletonisation (measure choice) used. If one decides to partition the $\lambda$ and $t$ integrals in another way, we would have obtained another Wheeler--DeWitt equation.
\item The range of $\tilde{N}$ plays a non-trivial rôle. If we had chosen $\tilde{N}$ to vary only over 0 and $+\infty$, instead of a solution to the Schrödinger equation \cref{schr1}, we would have obtained a Green function of the Schrödinger \cite{Halliwell,Henneaux1985}. In this case the $\delta$ function only contributes if $t_f-t_{in}>0$ requiring the appearance of a Heaviside $\theta$ function:
\begin{equation}
G(v_f,\varphi_f,t_f|v_{in},\varphi_{in},t_{in})=\theta(t_f-t_{in})\braket{v_f,\varphi_f,t_f}{v_{in},\varphi_{in},t_{in}}
\end{equation}
In this case
\begin{align}
&\left(i\hbar\pdv{}{t_f}-\hat{\mathcal{C}}_{class,f} \right)\theta(t_f-t_{in})\braket{v_f,\varphi_f,t_f}{v_{in},\varphi_{in},t_{in}} \nonumber\\
&=i\hbar \delta(t_f-t_{in})\delta(v_f-v_{in})\delta(\varphi_f-\varphi_{in})\, ,
\end{align}
 Any other range is in principle possible but it is not clear that one can obtain a useful result out of it. 
\end{itemize}
In fact, it has been argued that the path integral quantisation is not invariant under reparametrisations of $\tilde{N}$ \cite{Halliwell}. Recalling the results we had in \cref{dirac-sec}, we had two Wheeler--DeWitt equations, the already seen \cref{wdw-re} and \cref{wdw2-re} which is obtained by multiplying \cref{wdw-re} by $v^2$. It would be possible to reparametrise $\tilde{N}$ by $N'=\tilde{N}/v^2$ in the extended phase space path integral \cref{extendedPI}, and, with the right skeletonisation we could obtain the Wheeler--DeWitt equation \cref{wdw2-re}. Recalling that the kinematical Hilbert space of the Dirac quantisation depended on the specific Wheeler--DeWitt equation, we find that the path integral quantisation does not offer a way to find a ``preferred'' Wheeler--DeWitt equation.

In a nutshell, path integral quantisation can certainly produce the Wheeler--DeWitt equation of our model, but does not provide any additional hint on a preferred physical Hilbert space. In this simple example we can also appreciate directly how diffeomorphism invariance of general relativity poses a challenge in the path integral specification. However, this quantisation method is conceptually very different from the previous two approaches. The formalism allows playing with the initial conditions of the universe and the concept of evolution is treated in a complete different way: in an object such as $\braket{v_f,\varphi_f,t_f}{v_{in},\varphi_{in},t_{in}}$, it is assumed that the configuration $(v_{in},\varphi_{in},t_{in})$ occurs ``before'' $(v_f,\varphi_f,t_f)$ but a priori no assumptions on the initial and final values of the parameters is made. We can therefore study possible evolutions of the universe given a set of initial conditions $(v_{in},\varphi_{in},t_{in})$. We might then recover (or not) the wave functions of the universe analysed in our three theories. Do we recover either \cref{normstatest}, \cref{wf1}, or \cref{phigeneralstate}? For which boundary conditions? Does this give a hint for a possible problem of choice resolution? To answer these questions, we want to explore the possibility of analysing our model from a path integral quantisation perspective in future work. For two recent examples of path integral quantisation calculations in quantum cosmology see \cite{PIQcosmo}. 
\chapter{Conclusions}
In this thesis, we have analysed the three quantum theories coming from different clock choices. To do so, we made some assumptions. First of all, when deriving the Wheeler--DeWitt equation, we assumed a very specific (and to our knowledge the best motivated) operator ordering, namely we constructed a Wheeler--DeWitt equation covariant under coordinate changes. Secondly, we chose a specific inner product, also covariant, for each theory. The inner product choice indirectly introduced a dependence on the lapse function $N$, since multiplying the constraint by a non-trivial phase space function changes the minisuperspace metric, even if it does not change the solutions of the Wheeler--DeWitt equation. However, the most important assumption we made is consider unitarity as a fundamental principle of our quantum theories.  Each of the three theories has very defining features, we remind them one last time.
\begin{itemize}
\item The $v$-clock theory leads to a quantum theory that is already unitary, and therefore no extra boundary condition is needed. The dynamics of this theory show that for a semiclassical state the quantum expectation values of the observables remain close to the classical curve.
\item The $t$-clock theory does not have a self-adjoint Hamiltonian. Unitarity requires a reflective boundary condition in the limit $v=0$, hence predicting that wave functions would be reflected from the classical singularity and thus produce a quantum bounce. This is confirmed in the numerical analysis of the quantity $\expval{v(t)}_{\Psi_{sc,t}}$. 
\item The $\varphi$-clock theory also needs the introduction of a boundary condition to ensure unitary dynamics, but this time in the limit $v=\infty$. This also suggests a reflection from $v=\infty$ of the allowed wave functions, leading to a quantum recollapse (quantum because the reason for this recollapse would be purely quantum). This is again confirmed by the numerical analysis of both $\expval{v(\varphi)}_{\Psi_{sc,\varphi}}$ and $\expval{t(\varphi)}_{\Psi_{sc,\varphi}}$.
\end{itemize}
We can thus conclude that divergences from the classical theory are a consequence of requiring unitarity given a certain clock choice, and hence it is a clock dependent feature. Far from being a characteristic only present in relational quantisation, this self-adjointness problem is also a feature of Dirac quantisation: if one starts with one Wheeler--DeWitt equation \cref{wdw-re}, it will lead to a quantum theory equivalent to the $t$-clock theory. But, if instead one starts with the same Wheeler--DeWitt equation multiplied by $v^2$, \cref{wdw2-re}, which corresponds classically to another choice of lapse function $N$, one ends with the same self-adjointness problem than in $\varphi$-clock theory. We stress that these results are compatible with the covariant approach implemented in \cite{Hoehn}. In their work, they consider a single Wheeler--DeWitt equation, whereas we effectively worked with two. However, our main point is that the two Wheeler--DeWitt equations have exactly the same solutions. How to distinguish between the two? Why should we use one instead of the other? We see no strong reason for a preferred Wheeler--DeWitt equation in the Dirac quantisation scheme. The path integration formalism, despite being a very different approach to quantisation, presents a similar situation with the lapse dependency and is capable of reproducing the same Wheeler--DeWitt equations. We would like to investigate whether we recover any of the wave functions \cref{normstatest}, \cref{wf1}, or \cref{phigeneralstate} given a certain choice of initial conditions.

A possible way of ``deciding'' for a quantum theory might be to ask the question: which of the clock theories follows better the predictions one would expect from a theory of quantum gravity? In this sense, we expect quantum effects to be important near the big bang/big crunch singularity, where the universe is hot and dense and prevent the appearance of quantum effects at late time, where the universe is supposedly governed by classical physics. Following this argument, the clear winner is the $t$-clock theory. The quantum recollapse happening in the $\varphi$-clock theory is certainly puzzling as it represents a transition from a classical to a quantum dominated universe at late times. What would be the driving factor of this transition? This type of massless field derived clock is very popular in loop quantum cosmology. Models such as FLRW universes \cite{Ashtekar}, Bianchi I \cite{Chiou2007} or Bianchi IX \cite{Wilson-Ewing2010}, to name a few, have been analysed with such clocks. These models show singularity resolution, as opposed to analogous models based on the Wheeler--DeWitt equation, and this has lead to the (already criticised \cite{Bojowald2020}) belief that loop quantum cosmology resolves the singularity whereas the  Wheeler--DeWitt quantisation of the same models (with the same clock) does not. Our analysis shows that Wheeler--DeWitt quantisation can resolve the singularity given the right clock, and also suggests that this breaking of general covariance may also be a shared feature in LQC and be ultimately responsible for singularity resolution. We want to explore this idea in future work. 

Another question that our study raises is whether it is really true that all classically monotonic variables can be used as clocks. For example, the clock $\varphi$ is mathematically well defined but might not be physically as well motivated for the reasons seen before. In addition to this, no timelike observer will ever experience the passing of time that way, whereas the $t$ clock can be associated with timelike observers. In particular, the $t$ clock represents conformal time for a radiation perfect fluid and commoving time for a pressureless matter perfect fluid. Maybe we should restrict ourselves to measure clocks that can potentially measure proper time for an observer, even if it breaks the covariance of the theory. In a way time is not only what a clock measures but also what observers experience. 

Perhaps, in a more philosophical point of view, this thesis presents an extensive analysis of the problem of time in quantum cosmology throughout the study of a minisuperspace example. We have shown how the problem of time manifests itself in the relational quantisation and the Dirac quantisation scheme, and we had a glimpse of the path integral quantisation scheme. In particular, it seems that unitarity of the quantum theory is incompatible with general covariance, and the requirement of unitarity leads to the appearance (or absence) of boundary conditions for the allowed wave functions. These boundary conditions can be related to a self-adjointness problem and can be linked to the classical solutions of the theory by analysing whether the chosen clock is slow or fast. At the semiclassical level, the boundary conditions induce deviation of expectation values from the classical theory that can ultimately lead to a quantum resolution of the big bang and big crunch singularities, a quantum recollapse of the universe, or neither. 

Singularity resolution may be considered the holy grail of a quantum theory. It appears to us that this feature, is neither a consequence of the quantisation scheme used, whether it is Dirac quantisation or relational quantisation, nor an outcome of the underlying theory of quantum gravity, whether loop quantum gravity or another one. The answer to the singularity resolution, in a rather simple and surprising way, seems to stem out of the clock choice (or the choice of lapse). In other words, unitarity and general covariance appear to be two incompatible requirements. One could then decide that general covariance is more fundamental and that unitarity may not be a defining feature of a quantum universe. But how can we get rid of the cornerstone of quantum mechanics? What could motivate that quantum theory would have such drastic differences if the system studied is the universe, instead of let's say, an atom? The subsequent non conservation of probabilities renders the obtention of quantitative results almost impossible.

If instead, we decide that unitarity is more fundamental than general covariance, we have to face the problem of clock choice. How and why does the universe decide on a clock? There has been some work in this direction \cite{Magueijo2021}. The clock dependent singularity resolution is not only a characteristic of cosmological spacetimes, but of any spacetime that possess a classical singularity, i.e, a black hole spacetime. There is no reason to think that for such spacetime, singularity resolution and transition to a white hole spacetime might be expected. The lack of covariance certainly raises many questions. In particular, it challenges the utility of such simplified minisuperspace models. Maybe, the idea that such universes where most of the degrees of freedom have been frozen is too naive and too far from the real complexity of the universe to yield to interesting results. There are not, unfortunately, many alternatives to minisuperspace models and the problem of time is virtually impossible to study in the full theory.

The answer to all these uncertainties, as always, lies in the phrase: ``more research is needed''. The theory of special and general relativity was born thanks to the radical thought that time was not an absolute quantity, rather relative to each observer. This way of thinking of time was fundamentally opposite to everything seen before. Maybe the next revolution in physics (and a satisfactory theory of quantum gravity) will come with another revolution like this. 
\appendix
\chapter{Theory of self-adjoint extensions}
\label{selfadj}
In order to understand better self-adjoint extensions, we need to introduce some mathematical concepts. As operators are not the topic of this thesis we will not include the proofs of the results we mention. We mainly follow \cite{Reed1975}. We assume that we are working with closed (or at least closable) operators.

\begin{definition}[Symmetric operator]
Let $\Obs$ be an operator on a Hilbert space $\mathscr{H}$. $\Obs$ is called symmetric if for all $\Psi$ and $\Phi$ in the domain of $\Obs$ ($\dom(\Obs)$) we have $\braket{\Obs \Psi}{\Phi}=\braket{\Psi}{\Obs \Phi}$.
\end{definition}

\begin{remark}
Symmetric operators are always closable, so we do not have to restrict ourselves to a subset of them. In finite dimensional spaces symmetric and self-adjoint are equivalent notions, but in infinite dimensional Hilbert spaces like the ones we are dealing with there are subtleties that we must take into account. Those are related to the domain of an operator.
\end{remark}

\begin{definition}[Adjoint operator]
Let $\Obs$ be an operator on a Hilbert space $\mathscr{H}$. The adjoint of $\Obs$, generally denoted as $\Obs^*$, is an operator acting on the subspace of $\Phi \in \mathscr{H}$ such that there exists $\xi \in \mathscr{H}$ such that $\braket{\Obs \Psi}{\Phi}=\braket{\Psi}{\xi}$, where we define $\xi=\Obs^*\Phi$.
\end{definition}

\begin{remark}
As we can see from this definition, for symmetric operators we have
\begin{equation}
\dom(\Obs)\subseteq \dom(\Obs^*)\, ,
\end{equation}
so it is possible for the domain of the adjoint to be too big.
\end{remark}

\begin{definition}[Self-adjoint operator]
A symmetric operator $\Obs$ is self adjoint if and only if $\dom(\Obs)=\dom(\Obs^*)$
\end{definition}

Why is the distinction between symmetric and self-adjoint important? The nice properties we are used to in matrix spaces are only satisfied by self-adjoint operators, for example:

\begin{theorem}
The spectrum of a self-adjoint operator is always real.
\end{theorem}

\begin{remark}
We can extend a symmetric operator in the hope of finding a self-adjoint operator. Indeed, let $\mathcal{P}$ be a symmetric extension of $\Obs$. We have
\begin{equation}
\dom(\Obs)\subseteq \dom(\mathcal{P})\subseteq \dom(\mathcal{P}^*) \subseteq \dom(\Obs^*)\, .
\end{equation}
We are interested in the cases where we can find extensions such that $\dom(\mathcal{P})= \dom(\mathcal{P}^*)$. To do so it is important to introduce the notion of deficiency subspace.
\end{remark}

\begin{definition}[Deficiency subspaces]
Let $\Obs$ be an operator on a Hilbert space $\mathscr{H}$. Then the spaces $\mathscr{K}^+=\ker(i\mathcal{I}-\Obs^*)=\ran(i\mathcal{I}+\Obs)^{\perp}$ and $\mathscr{K}^-=\ker(i\mathcal{I}+\Obs^*)=\ran(i\mathcal{I}-\Obs)^{\perp}$ are called the deficiency subspaces of $\Obs$. Their dimension $n^\pm=\dim(\mathscr{K}^\pm)$ is called deficiency index.
\end{definition}

\begin{remark}
The deficiency indices can be any pair of positive numbers, including infinity.
\end{remark}

With this introduction we have enough material to introduce the most important result of the section:

\begin{theorem}
\label{thm2}
Let $\Obs$ be a symmetric operator on a Hilbert space $\mathscr{H}$ with deficiency indices $n^+$ and $n^-$. Then the following hold:
\begin{enumerate}[i)]
\item $\Obs$ is self-adjoint if and only if $n^+=n^-=0$.
\item $\Obs$ has self-adjoint extensions if and only if $n^+=n^-$. There is a one-to-one correspondence between any self-adjoint extension of $\Obs$ and the unitary maps between $\mathscr{K}^+$ and $\mathscr{K}^-$.
\item If either $n^+=0\neq n^-$ or $n^-=0\neq n^+$, then $\Obs$ has no non-trivial symmetric extensions.
\end{enumerate}
\end{theorem}

\Cref{thm2} gives us a recipe to find whether the operators we are interested in have self-adjoint extensions and how many parameters are needed to describe them. The operator $\hat{\mathcal{H}}$ and $\hat{\mathcal{G}}$ defined in \cref{hams} and \cref{ham2} have $n^+=n^-=1$, which means that they admit a one parameter self-adjoint extension.
\chapter{Important integrals}
\label{bessel-int}
In this appendix we collect the results of various integrals containing a product of two Bessel functions. These are used in \cref{t-clock-sec} when computing the boundary condition \cref{boundt} on two generic wave functions of the universe, in \cref{phi-clock-sec} when calculating the inner product of two solutions of the Wheeler--DeWitt equation, and finally in \cref{numerics} when finding the expressions of the expectation values with the $\varphi$-clock inner product.

The integrals needed for \cref{t-clock-sec} are
\begin{equation}
\mathcal{P}_{i\alpha,\pm i\alpha,C_1,C_2}= \int_0^\infty \dd v \ v \ J_{i\alpha}(C_1 v)J_{\pm i\alpha}(C_2 v)\, ,
\label{intP}
\end{equation}
where $\alpha$ is real and $C_1$ and $C_2$ are positive. The integrals of interest in \cref{phi-clock-sec} are
\begin{equation}
\mathcal{J}_{\mu,\nu}=\int_0^\infty \frac{\dd v}{v} J_\mu(Cv)J_\nu(Cv), \hspace{4mm} \mathcal{K}_{\mu,\nu}=\int_0^\infty \frac{\dd v}{v} K_\mu(Cv)K_\nu(Cv)\, ,
\label{intJK}
\end{equation}
where $\mu$ and $\nu$ can be real or imaginary. These integrals will turn out to be independent of the parameter $C$. Lastly, we are interested in the integrals
\begin{equation}
\mathcal{O}_{\mu,\nu,C}=\int_0^\infty \dd v \ J_\mu(Cv)J_\nu(Cv)\hspace{4mm} \mathcal{P}_{i\alpha,i\beta,C}=\int_0^\infty \dd v\ v\ J_{i\alpha}(C v)J_{i\beta}(C v)
\label{intOP}
\end{equation}
for \cref{numerics}. This time $\mu$ and $\nu$ are either imaginary or of the form $\pm 1+i\alpha$ where $\alpha$ is real and $\beta$ is always real. $C$ is always considered positive, although its sign is irrelevant for \cref{intJK}. Despite the similarities between $\mathcal{P}_{i\alpha,\pm i \alpha, C_1,C_2}$ in \cref{intP} and $\mathcal{P}_{i\alpha,i\beta, C, C}$ in \cref{intOP} the methods for solving these integrals are quite different, so we present them in different sections. The integrals are presented roughly in the same order they are needed in the thesis.

\section{The integrals $\mathcal{P}_{i\alpha, \pm i\alpha,C_1,C_2}$}
\label{P-sec}

These integrals are useful in \cref{t-clock-sec} when computing the norm of the wave functions. In this case we assume that $\alpha$ is real and $C_1$ and $C_2$ are positive. To evaluate this integral we start by calculating 
\begin{align}
\int_{v_1}^{v_2} \dd v & \ v J_{ i\alpha}(C_1 v)J_{\pm i\alpha}(C_2 v) \nonumber \\
& =\left[ v\frac{C_2J_{-1\pm i \alpha}(C_2 v)J_{i\alpha}(C_1 v) \mp  C_1J_{\mp 1+i\alpha}(C_1 v)J_{\pm i \alpha}(C_2 v)}{C_1^2-C_2^2} \right]_{v=v_1}^{v=v_2}\, .
\end{align}
In the $\mathcal{P}_{i\alpha, i\alpha,C_1,C_2}$ case there is no contribution from the $v=0$ limit as the numerator cancels out, whereas in the $\mathcal{P}_{i\alpha, -i\alpha,C_1,C_2}$ case we have
\begin{align}
\lim_{v\rightarrow 0} & \ v\frac{C_2J_{-1- i \alpha}(C_2 v)J_{i\alpha}(C_1 v) +  C_1J_{ 1+i\alpha}(C_1 v)J_{- i \alpha}(C_2 v)}{C_1^2-C_2^2}\nonumber \\
&= -2 i \frac{e^{i\alpha\log\frac{C_1}{C_2}}\sinh(\pi \alpha)}{\pi(C_1^2-C_2^2)}\, .
\label{zero-contr}
\end{align}
This contribution is finite and when substituted into \cref{norm1} it cancels out. 

The limit $v=\infty$ is different. Using the large asymptotic expression for the Bessel functions we find
\begin{align}
\mathcal{P}_{i\alpha, i\alpha,C_1,C_2}&=\lim_{v\rightarrow\infty} \left\lbrace\frac{\cos((C_1+C_2)v)\cosh(\alpha\pi)-i\sinh(\alpha\pi)\sin((C_1+C_2)v)}{\pi\sqrt{C_1C_2}(C_1+C_2)} \right.\nonumber\\
&\left. +\frac{\sin((C_1-C_2)v)}{\pi\sqrt{C_1C_2}(C_1-C_2)}\right\rbrace\, , 
\end{align}
and 
\begin{align}
\mathcal{P}_{ i\alpha,-i\alpha,C_1,C_2}&=\lim_{v\rightarrow\infty} \left\lbrace\frac{\sin((C_1-C_2)v)\cosh(\alpha\pi)+i\cos((C_1-C_2)v)\sinh(\alpha\pi)}{\pi\sqrt{C_1C_2}(C_1-C_2)}\right. \nonumber \\
&\left.+\frac{\cos((C_1+C_2)v)}{\pi\sqrt{C_1C_2}(C_1+C_2)}\right\rbrace -2 i \frac{e^{i\alpha\log\frac{C_1}{C_2}}\sinh(\pi \alpha)}{\pi(C_1^2-C_2^2)}\, .
\end{align}
Using \cref{delta-triglim} we obtain
\begin{align}
\mathcal{P}_{i\alpha,i\alpha,C_1,C_2}&=\frac{\delta(C_1-C_2)-i\sinh(\alpha\pi)\delta(C_1+C_2)}{\sqrt{C_1C_2}} \nonumber \\
\mathcal{P}_{i\alpha,-i\alpha,C_1,C_2}&=\frac{\cosh(\alpha\pi)\delta(C_1-C_2)}{\sqrt{C_1C_2}} -2 i \frac{e^{i\alpha\log\frac{C_1}{C_2}}\sinh(\pi \alpha)}{\pi(C_1^2-C_2^2)}\, .
\label{P-int}
\end{align}
Once substituting this expression in \cref{norm1} we obtain the result \cref{eigen-l}.

\section{The integrals $\mathcal{J}_{\mu,\nu}$ and $\mathcal{K}_{\mu,\nu}$}
\subsection{The integrals $\mathcal{J}_{i\alpha,i\beta}$ and $\mathcal{K}_{i\alpha,i\beta}$}
These integrals appear when computing orthogonality relations between different states in the $\varphi$ clock theory. The parameters $\alpha$ and $\beta$ are real. Recall that these integrals are independent of $C$. In addition, $\mathcal{J}_{i\alpha,i\beta}$ does not converge but it can be defined in a distributional sense as a limit of the integral

\begin{align}
& \lim_{\nu\rightarrow 1}\int_0^\infty \frac{{ \dd}x}{x^\nu}J_{{ i}\alpha}(x)J_{{ i}\beta}(x)\nonumber
\\
=&\lim_{\nu\rightarrow 1}\frac{2^{-\nu}\Gamma\left(\frac{1-\nu}{2}+\frac{{ i}}{2}(\alpha+\beta)\right)\Gamma(\nu)}{\Gamma\left(\frac{1+\nu}{2}+\frac{{ i}}{2}(\alpha-\beta)\right)\Gamma\left(\frac{1+\nu}{2}+\frac{{ i}}{2}(\beta-\alpha)\right)\Gamma\left(\frac{1+\nu}{2}+\frac{{ i}}{2}(\alpha+\beta)\right)}
\label{genform}
\end{align}
which is initially only defined for $\nu<1$; notice the possible singularity in the first Gamma function in the numerator as $\nu\rightarrow 1$. We have also introduced the integration variable $x=Cv$ to simplify the notation in this integral. To proceed, we can now rewrite
\begin{equation}
\Gamma\left(\frac{1-\nu}{2}+\frac{{ i}}{2}(\alpha+\beta)\right)=\frac{\Gamma\left(\frac{3-\nu}{2}+\frac{{i}}{2}(\alpha+\beta)\right)}{\frac{1-\nu}{2}+\frac{{i}}{2}(\alpha+\beta)}
\label{gammaplusone}
\end{equation}
so that we obtain
\begin{equation}
\lim_{\nu\rightarrow 1}\int_0^\infty \frac{{\dd}x}{x^\nu}J_{{ i}\alpha}(x)J_{{i}\beta}(x) =  \frac{2\sinh\left((\alpha-\beta)\frac{\pi}{2}\right)}{\pi(\alpha-\beta)}\times\lim_{\nu\rightarrow 1}\frac{1}{(1-\nu+{i}(\alpha+\beta))}\,.
\end{equation}
The last limit must now be taken in a distributional sense using the identity
\begin{equation}
\lim_{\epsilon\rightarrow 0^+}\frac{1}{y+{ i}\epsilon}={ PV}\frac{1}{y}-{ i}\pi\delta(y)
\end{equation}
where ${\text{PV}}$ denotes the Cauchy principal value, i.e.~the distribution defined by
\begin{equation}
\int_{-\infty}^\infty { \dd}y\left[{ PV}\frac{1}{y}\right]f(y) = \int_0^\infty \frac{{ \dd}y}{y}(f(y)-f(-y))
\end{equation}
for any test function $f(y)$, which depends only on the odd part of $f$. In summary, we then find
\begin{equation}
\int_0^\infty \frac{{ \dd}x}{x}J_{{ i}\alpha}(x)J_{{ i}\beta}(x) =  2\sinh\left((\alpha-\beta)\frac{\pi}{2}\right)\left(\frac{\delta(\alpha+\beta)}{\alpha-\beta}-{ PV}\frac{{ i}}{\pi(\alpha^2-\beta^2)}\right)\,.
\label{2-Bessel-int-1}
\end{equation}

In the case of modified Bessel functions we can proceed in the same fashion;  we find
\begin{equation}
\lim_{\nu\rightarrow 1}\int_0^\infty \frac{{ \dd}x}{x^\nu}K_{{ i}\alpha}(x)K_{{ i}\beta}(x)=\lim_{\nu\rightarrow 1}\frac{\left|\Gamma\left(\frac{1-\nu}{2}+\frac{{ i}}{2}(\alpha-\beta)\right)\right|^2\left|\Gamma\left(\frac{1-\nu}{2}+\frac{{ i}}{2}(\alpha+\beta)\right)\right|^2}{2^{2+\nu}\Gamma(1-\nu)}
\end{equation}
which has a more complicated singularity structure, with possible singularities in all Gamma functions. By substitutions similar to (\ref{gammaplusone}) we obtain
\begin{align}
\lim_{\nu\rightarrow 1}\int_0^\infty \frac{{ \dd}x}{x^\nu}K_{{ i}\alpha}(x)K_{{ i}\beta}(x) &= \frac{\pi^2(\alpha^2-\beta^2)}{\cosh(\alpha\pi)-\cosh(\beta\pi)}\times
\\&\lim_{\nu\rightarrow 1}\frac{(1-\nu)}{|1-\nu+{ i}(\alpha-\beta)|^2|1-\nu+{ i}(\alpha+\beta)|^2}\,.\nonumber
\end{align}
If we now exclude the case $\alpha=\beta=0$ (which we can, given the fact that these integrals only appear with integrals over $\alpha$ and $\beta$ and a single point can be removed from the domain), then at least one of the two factors in the denominator remains regular as $\nu\rightarrow 1$ and can be taken outside of the limit. For the second factor we have to take the distributional limit
\begin{equation}
\lim_{\epsilon\rightarrow 0^+}\frac{\epsilon}{\epsilon^2+y^2}=\pi\delta(y)\, ,
\end{equation}
as can be seen from
\begin{equation}
\lim_{\epsilon\rightarrow 0^+}\int_{-\infty}^\infty { \dd}y\;f(y)\;\frac{\epsilon}{\epsilon^2+y^2} = \lim_{\epsilon\rightarrow 0^+}\int_{-\infty}^\infty { \dd}\upsilon\;f(\epsilon\upsilon)\;\frac{1}{1+\upsilon^2}\, ,
\end{equation}
where $f$ is again a test function. Altogether we have
\begin{align}
\int_0^\infty \frac{{ \dd}x}{x}K_{{ i}\alpha}(x)K_{{ i}\beta}(x) &= \frac{\pi^3(\alpha^2-\beta^2)}{\cosh(\alpha\pi)-\cosh(\beta\pi)}\left(\frac{\delta(\alpha-\beta)}{(\alpha+\beta)^2}+\frac{\delta(\alpha+\beta)}{(\alpha-\beta)^2}\right)\nonumber
\\&= \frac{\pi^2}{2\alpha\sinh(\alpha\pi)}\left(\delta(\alpha-\beta)+\delta(\alpha+\beta)\right)\,.
\label{kintegral}
\end{align}
Notice that the modified Bessel functions of the second kind are always real even for imaginary order, hence there is no imaginary contribution leading to a principal value. Such imaginary contributions come from the large $x$ limit of the integral, whereas the right-hand side of (\ref{kintegral}) only comes from the lower limit $x=0$.
\subsection{The integrals $\mathcal{J}_{a,b}$ and $\mathcal{K}_{a,b}$}
Here $a$ and $b$ are real numbers. In order to identify the cases where the integral can be defined, we first evaluate the indefinite integral
\begin{align}
&\int_{x_1}^{x_2} \frac{{ \dd}x}{x}\,J_{a}(x)J_{b}(x)  = \left[\frac{x \left(J_{a-1}(x)J_{b}(x) - J_{a}(x)J_{b-1}(x)\right)}{a^2-b^2}-\frac{\,J_{a}(x)J_{b}(x)}{a+b} \right]_{x=x_1}^{x=x_2}
\end{align}
where we again defined $x=C v$  for simplicity. 
After now substituting the large argument and small argument asymptotic expressions of the Bessel functions we find
\begin{equation}
\int_{0}^{\infty} \frac{\dd x}{x} J_{a}(x) J_{b}(x)= 2\frac{\sin((a-b)\frac{\pi}{2})}{(a^2-b^2)\pi}-\lim_{x\rightarrow  0} \frac{(x/2)^{a+b}}{(a+b)\Gamma(1+a)\Gamma(1+b)}.
\end{equation}
We now see that the integral is finite when $a+b>0$; otherwise the second term makes the integral divergent and undefinable even in a distributional sense. For $a+b>0$,
\begin{equation}
\int_{0}^{\infty} \frac{\dd x}{x} J_{a}(x) J_{b}(x)= 2\frac{\sin((a-b)\frac{\pi}{2})}{(a^2-b^2)\pi}
\label{2-Bessel-int-2}
\end{equation}
which is the standard formula given, for example, as Equation 6.574.2 in \cite{Integrals}. 
\\For the integral $\mathcal{K}_{a,b}$, there is no contribution from large $v$ where the integral falls off but from $v=0$ we find
\begin{align}
\int_{0}^{\infty} \frac{\dd x}{x} K_{a}(x) K_{b}(x) &= -\frac{\pi^2}{4\sin( a\pi)\sin(b\pi)}\times\nonumber
\\&\lim_{x\rightarrow  0}\left[\frac{(x/2)^{a+b}}{(a+b)\Gamma(1+a)\Gamma(1+b)}-\frac{(x/2)^{-(a+b)}}{(a+b)\Gamma(1-a)\Gamma(1-b)}\right.\nonumber
\\&+\left.\frac{(x/2)^{b-a}}{(a-b)\Gamma(1-a)\Gamma(1+b)}-\frac{(x/2)^{a-b}}{(a-b)\Gamma(1+a)\Gamma(1-b)}\right]\, ,
\label{2-KBessels-int-2}
\end{align}
so that this integral always diverges for any $a$ or $b$ (this is true also for the case in which $a$ or $b$ are integer, which we do not discuss in detail here).

\subsection{The integrals $\mathcal{J}_{a,i\beta}$ and $\mathcal{K}_{a,i\beta}$}

This is the third possible case in which one order is real and the other one is imaginary. This integral appears when computing cross-terms in the inner product in \cref{phi-clock-sec}. In this case the expression resulting from computing first the indefinite integral is
\begin{equation}
\int_0^{\infty} \frac{\dd x}{x}J_{a}(x)J_{i\beta}(x)=2\frac{\sin((a-i\beta)\frac{\pi}{2})}{(a^2+\beta^2)\pi}-\lim_{x\rightarrow 0}\frac{(x/2)^{a+ i\beta}}{(a+ i\beta)\Gamma(1+a)\Gamma(1+ i\beta)}\, .
\end{equation}
As $x\rightarrow 0$, the exponential function in the second term has a growing (if $a<0$) or decreasing (if $a>0$) absolute value. If $a>0$, the limit when $x\rightarrow 0$ is 0, making the integral converge to the value
\begin{equation}
\int_0^{\infty} \frac{\dd x}{x}J_{a}(x)J_{i\beta}(x)=2\frac{\sin((a-i\beta)\frac{\pi}{2})}{(a^2+\beta^2)\pi}\, .
\label{2-Bessel-int-3}
\end{equation}
For $a<0$ the integral is divergent. 

The integral $\mathcal{K}_{a,i\beta}$ is again found to diverge for all real values of $a$.

\section{The integrals $\Obs_{\mu,\nu,C}$}
\subsection{The integral $\Obs_{i\alpha,i\beta,C}$}
Here again $\alpha$ and $\beta$ are real numbers. This integral depends non-trivially on the value of $C$. This integral presents a divergence that makes the final expression depend on a regulator. There are several methods for dealing with such divergences; we present two in this work. We use first the same method of first evaluating the integral for arbitrary limit values, where it yields
\begin{align}
\int_{v_1}^{v_2} { \dd}v\,J_{i\alpha}(C v)&J_{i\beta}(C v)=-\left[\frac{i\,v \exp\left(i(\alpha+\beta)\log\frac{C v}{2}\right)}{(\alpha+\beta-i)\Gamma(1+i\alpha)\Gamma(1+i\beta)}\right.\times
\\&  \left.{}_3 F_4 \left(\frac{1}{2}+\iota,\frac{1}{2}+\iota,1+\iota;1+i\alpha,\frac{3}{2}+\iota,1+i\beta,1+2\iota;-C^2 v^2\right)\right]_{v=v_1}^{v=v_2}\nonumber
\end{align}
which can only be given in terms of generalised hypergeometric functions and where we have defined $\iota:=i\frac{\alpha+\beta}{2}$. This complicated expression simplifies as $v_1\rightarrow 0$ and $v_2\rightarrow \infty$. First, note that the generalised hypergeometric function defines a power series in $(-C^2 v^2)$ and goes to 1 at $v=0$; because of the additional factor $v$ the contribution from the lower limit vanishes as $v_1\rightarrow 0$.

Using the large $v$ asymptotic of the generalised hypergeometric function we then have, formally,
\begin{align}
\int_{0}^{\infty} { \dd}v\,J_{i\alpha}(Cv)J_{i\beta}(Cv) &= -\frac{\cosh\left((\alpha-\beta)\frac{\pi}{2}\right)}{2\pi C}\hspace{-1mm}\times \lim_{v\rightarrow\infty}\left\{\log\left(\frac{4}{C^2 v^2}\right)+\psi\left(\frac{1+i(\beta-\alpha)}{2}\right)\right.\nonumber
\\  & \left.+\psi\left(\frac{1+i(\alpha-\beta)}{2}\right)+2\psi\left(\frac{1+i(\alpha+\beta)}{2}\right)+2\gamma\right\}\, ,
\label{digamma_integral}
\end{align}
where $\psi$ is the digamma function and $\gamma$ is the Euler--Mascheroni constant.  (\ref{digamma_integral}) diverges logarithmically at large $v$; when using it for numerical evaluation of expectation values, we take the upper limit to some large cutoff value $\Xi_v$ and verify that the final result after integrating over the other variables is not too sensitive to the choice of $\Xi_v$. This method of integration is very similar to the Pauli-Villar regularisation method used in QFT.

The second method consists in using the convergent integral
\begin{align}
& \int_0^\infty \frac{{ \dd}x}{x^\nu}J_{{ i}\alpha}(C x)J_{{ i}\beta}(C x)\nonumber
\\
=&\frac{C^{\nu-1}2^{-\nu}\Gamma\left(\frac{1-\nu}{2}+\frac{{ i}}{2}(\alpha+\beta)\right)\Gamma(\nu)}{\Gamma\left(\frac{1+\nu}{2}+\frac{{ i}}{2}(\alpha-\beta)\right)\Gamma\left(\frac{1+\nu}{2}+\frac{{ i}}{2}(\beta-\alpha)\right)\Gamma\left(\frac{1+\nu}{2}+\frac{{ i}}{2}(\alpha+\beta)\right)}\, ,
\label{generalform}
\end{align}
that is the same formula as we used in the calculation of $\mathcal{J}_{i\alpha,i\beta}$, however, this time we are interested in the limit $\nu \rightarrow 0$. Consider $\nu=0-\epsilon$ where $\epsilon>0$ is small, then the expansion of \cref{generalform} around $0$ is:
\begin{align}
\int_0^\infty  \dd x \ J_{{ i}\alpha}(C x)J_{{ i}\beta}(C x)\nonumber=&-\frac{\cosh{((\alpha-\beta)\frac{\pi}{2})}}{\epsilon\pi C }-\frac{\cosh\left((\alpha-\beta)\frac{\pi}{2}\right)}{2\pi C}\left\{\log\left(4\right)-2\log(C)\phantom{ \psi\left(\frac{1+i(\alpha+\beta)}{2}\right)}\right.\nonumber
\\  & \left. +\psi\left(\frac{1+i(\beta-\alpha)}{2}\right) +\psi\left(\frac{1+i(\alpha-\beta)}{2}\right)\right. \nonumber \\
&\left. + 2\psi\left(\frac{1+i(\alpha+\beta)}{2}\right)+2\gamma\right\rbrace+O(\epsilon)\, .
\label{digamma_integral_2}
\end{align}
We can see that \cref{digamma_integral} and \cref{digamma_integral_2} are very similar except for the divergent term that is a function of $\epsilon$. This regularisation method resembles the dimensional regularisation method used in QFT. In particular, see \cite{Peskin} pages 248-251 for an example in which Pauli-Villar regularisation leads to a logarithmic divergence and dimensional regularisation a divergence of the form $1/\epsilon$. There are other regularisation methods and how to treat the different regulators is still an open question.

\subsection{The integral $\Obs_{\pm 1 +i\alpha,i\beta,C}$}

Again we start by evaluating the indefinite integral which yields
\begin{align}
\int_{v_1}^{v_2} { \dd}v & \,J_{\pm 1+ i\alpha}(C v)J_{i\beta}(C v)=\left[\frac{v \exp\left((\pm 1+i(\alpha+\beta))\log\frac{C v}{2}\right)}{((1\pm 1)+i(\alpha+\beta))\Gamma(1\pm 1 +i\alpha)\Gamma(1+i\beta)}\right.\times
\\&\left.{}_3 F_4 \left(\frac{1}{2}+\iota',\frac{1}{2}+\iota',1+\iota';1\pm 1+i\alpha,\frac{3}{2}+\iota',1+i\beta,1+2\iota';-C^2 v^2\right)\right]_{v=v_1}^{v=v_2}\nonumber
\end{align}
where now $\iota':=\pm\frac{1}{2}+i\frac{\alpha+\beta}{2}$. 

In order to evaluate the limit $v=0$ we need to use the fact that
\begin{equation}
\lim_{x\rightarrow \pm \infty} e^{i\xi x}=\pm i\pi \xi\delta(\xi)\, .
\label{delta-1}
\end{equation}
This follows from
\begin{equation}
\lim_{x\rightarrow \pm \infty}\int \frac{\dd \xi}{\xi}e^{i\xi x}f(\xi)=\lim_{x\rightarrow \pm \infty}\pm \int \frac{\dd \eta}{\eta} e^{i\eta}f\left(\frac{\eta}{x}\right)\, .
\label{delta-2}
\end{equation}
This integral is undefined in the usual sense but its Cauchy principal $PV$ value yields to $\pm i \pi f(0)$, which justifies the result. In fact, if we consider separately the real and imaginary parts of \cref{delta-2} we have
\begin{align}
&\lim_{x\rightarrow \pm \infty}PV\int \frac{\dd \xi}{\xi}\cos(\xi x)f(\xi)=\lim_{x\rightarrow \pm \infty} \pm PV\int \frac{\dd \eta}{\eta}\cos(\eta)f\left(\frac{\eta}{x}\right)=0 \nonumber \\
&\lim_{x\rightarrow \pm \infty}\int \frac{\dd \xi}{\xi}\sin(\xi x)f(\xi)=\lim_{x\rightarrow \pm \infty} \pm \int \frac{\dd \eta}{\eta}\sin(\eta)f\left(\frac{\eta}{x}\right)=\pm \pi f(0) \, .
\end{align}
The cosine integral only converges using the principal value argument whereas the sine integral converges in the usual sense. Thus
\begin{equation}
\lim_{x \rightarrow\pm \infty} \frac{\sin(\xi x)}{\xi}=\pm\pi \delta(x) \ \text{and} \ \lim_{x \rightarrow\pm \infty} \frac{\cos(\xi x)}{\xi}=0
\label{delta-triglim}
\end{equation}
Using \cref{delta-1} and the fact that the generalised hypergeometric function goes to 1 at $v=0$ we can conclude that we do not have any non-trivial contribution from the plus sign, but for the case of the minus sign we now have 
\begin{align}
&\lim_{v\rightarrow 0}\frac{2i\exp\left(i(\alpha+\beta)\log\frac{C v}{2}\right)}{C(\alpha+\beta)\Gamma(i\alpha)\Gamma(1+i\beta)}\nonumber
\\& =  \lim_{u\rightarrow \infty}\frac{2i\exp\left(-i(\alpha+\beta)u\right)}{C(\alpha+\beta)\Gamma(i\alpha)\Gamma(1+i\beta)} = \frac{2\,i}{C} \sinh(\alpha\pi)\delta(\alpha+\beta)\,.
\end{align}

Including the contribution from large $v$ we find, again formally,
\begin{align}
\int_{0}^{\infty} { \dd}v\,J_{1+i\alpha}(Cv)J_{i\beta}(Cv) &= \hspace{-1mm} -\frac{\cosh\left((\alpha-\beta-i)\frac{\pi}{2}\right)}{2\pi C}\hspace{-1mm}\times\lim_{v\rightarrow\infty}\left\{\log\left(\frac{4}{C^2 v^2}\right)+\psi\left(\frac{i(\beta-\alpha)}{2}\right)\right.\nonumber
\\  & \left.+\psi\left(1+\frac{i(\alpha-\beta)}{2}\right)+2\psi\left(1+\frac{i(\alpha+\beta)}{2}\right)+2\gamma\right\}\, ,
\label{H-int-1}
\end{align}
and
\begin{align}
\int_{0}^{\infty} &{ \dd}v\,J_{-1+i\alpha}(Cv)J_{i\beta}(Cv)\nonumber\\
 &=  -\frac{\cosh\left((\alpha-\beta+i)\frac{\pi}{2}\right)}{2\pi C}\hspace{-1mm}\times\lim_{v\rightarrow\infty}\left\{\log\left(\frac{4}{C^2 v^2}\right)+\psi\left(1+\frac{i(\beta-\alpha)}{2}\right)\right.\nonumber
\\  & \left.+\psi\left(\frac{i(\alpha-\beta)}{2}\right)+2\psi\left(\frac{i(\alpha+\beta)}{2}\right)+2\gamma\right\} + \frac{2\,i}{C} \sinh(\alpha\pi)\delta(\alpha+\beta)\,.
\label{H-int-2}
\end{align}
These integrals again diverge logarithmically at large $v$ so we need to cut them off at a fixed cutoff value $\Xi_v$. However, for the calculation of interest in the main text we find that the sum over various integrals $\mathcal{O}_{\pm 1+i\alpha,i\beta,C}$ leads to an expression in which all the logarithm terms cancel, and which is hence well-defined in the limit $v\rightarrow\infty$.

\section{The integrals $\mathcal{P}_{i\alpha,i\beta,C}$}
 
As in the previous section, this integral is divergent even in the distributional sense. The same two regularisation methods can be applied here. The definite integral yields
\begin{align}
\int_{v_1}^{v_2}\dd v \ v & J_{i\alpha}(C v) J_{i\beta}(C v) =-\left[\exp[2\log(2)+i(\alpha+\beta)\log(\frac{Cv}{2})]\Gamma\left(\iota\right)\Gamma\left(2\iota \right)v^2(\alpha+\beta)^2\right. \nonumber \\
&\left.\times {}_3 F_4\left(\frac{1}{2}+\iota,1+\iota,1+\iota;1+i \alpha,2+\iota,1+i\beta,1+2\iota;-C^2v^2\right)\right]_{v=v_1}^{v=v_2}
\end{align}
where $\iota=i\frac{(\alpha+\beta)}{2}$ and ${}_3 F_4$ is the generalised hypergeometric function. The limit $v\rightarrow 0$ is 0, as the hypergeometric function goes to 1 and is paired with a $v^2$ term. However, the limit $v\rightarrow \infty$ is divergent:
\begin{align}
\int_{0}^{\infty}&\dd v \ v J_{i\alpha}(C v) J_{i\beta}(C v)  =\frac{i(\alpha^2-\beta^2)\sinh(\frac{\pi}{2}(\alpha-\beta))}{4\pi C^2}\nonumber\\
&\times\left(\gamma-\psi\left(-\frac{1}{2}\right)+\psi\left(-\frac{i}{2}(\alpha-\beta)\right)+\psi\left(\frac{i}{2}(\alpha-\beta)\right)+\psi(\iota)+\psi\left(1+\iota \right) \right) \nonumber \\
&+\lim_{v\rightarrow \infty}\left\lbrace \frac{1}{4\pi C^2}\left(4C v\cosh(\frac{\pi}{2}(\alpha-\beta))-2i(\alpha^2-\beta^2)\sinh(\frac{\pi}{2}(\alpha-\beta))\log(C v) \right. \right.\nonumber \\
&\left. \left.-2\cos(C v-\frac{i\pi}{2}(\alpha+\beta)) \right) \right\rbrace\, .
\label{vphi2int}
\end{align}
$\gamma$ is the Euler Mascheroni constant and $\psi$ is the digamma function. Hence, we see here that there are two divergence: a logarithmic divergence like in \cref{digamma_integral} and a linear divergence. We thus implement a cutoff $\Xi_{v^2}$. There is also an oscillatory term, which is not problematic as one can always choose a cutoff such that this term vanishes. 

We can also solve this integral using the dimensional regularisation method, taking the limit $\nu\longrightarrow 1$ in \cref{generalform}. Consider $\nu=-1-\epsilon$ we find
\begin{align}
\int_{0}^{\infty}&\dd v \ v J_{i\alpha}(C v) J_{i\beta}(C v)
=-\frac{i(\alpha^2-\beta^2)\sinh(\frac{\pi}{2}(\alpha-\beta))}{2 \pi C^2 \epsilon}+\frac{i(\alpha^2-\beta^2)\sinh(\frac{\pi}{2}(\alpha-\beta))}{4\pi C^2}\nonumber\\
&\times\left(\gamma-\psi\left(-\frac{1}{2}\right)+\psi\left(-\frac{i}{2}(\alpha-\beta)\right)+\psi\left(\frac{i}{2}(\alpha-\beta)\right)+\psi(\iota)+\psi\left(1+\iota \right) \right) +O(\epsilon)
\label{vphi2int2}
\end{align}
We can see that the non divergent terms are the same, but the linear and logarithmic divergences are all absorbed in the first $\epsilon$ term. The correspondence between a cutoff $\Xi_{v^2}$ and $\epsilon$ is less clear from this result and it is not the same relation that we found in $\mathcal{O}_{i\alpha,i\beta,C}$.
\chapter{Short introduction to anticommuting variables}
\label{anticom}
Anticommuting variables (sometimes also called Grasmann variables) were introduced by Berezin \cite{Berenzin}. Anticommuting variables $\lbrace \theta_i\rbrace$ can be defined as the generators of an algebra over a vector space $V$, $\mathcal{A}(V)^n$, such that they anticommute
\begin{equation}
\theta_i\theta_j=-\theta_j\theta_i\,  , \hspace{3mm} \forall i,j=1,\dots n
\end{equation}
This implies that they are nilpotent: $\theta_i^2=0$. They commute with complex numbers $z\in \mathbb{C}$
\begin{equation}
\theta_i z=z\theta_i \, .
\end{equation}
These anticommuting numbers are very useful in QFT for fermion (anticommuting) fields. Because they are nilpotent, any function of an anticommuting variable can be written as a Taylor series with two terms: \begin{equation}
f(\theta)=a+b\theta\, , \hspace{3mm}a,b\in \mathbb{C} .
\end{equation}
In the case of $n$ anticommuting numbers a general function has expression
\begin{equation}
f(\theta_1,\dots, \theta_i)=c_0+\sum_{k=1}^n\sum_{i_1,\dots,i_k} c_{i_1,\dots, i_k}\theta_{i_1}\dots \theta_{i_k}\, ,
\label{fgeneral}
\end{equation}
Where $c_{i_1,\dots, i_k}$ are complex completely antisymmetric tensors.

Integration over anticommuting variables are known as Berezin integrals, and they are defined to be a linear functional. The integral over one anticommuting variable $\theta$ is defined by
\begin{equation}
\int \dd \theta \ \theta =1\, ,\hspace{3mm} \int\dd \theta \ 1 =0\, ,
\end{equation}
so that
\begin{equation}
\int\dd \theta \pdv{}{\theta}f(\theta)=0\, .
\end{equation}
Note that the integral is not an integral in the usual Lebesgue sense. 
In general 
\begin{equation}
\int \dd \theta f(\theta) =\int\dd \theta( a+ b\theta) =b\, .
\end{equation}

In the case of $n$ anticommuting variables over $\mathcal{A}^n$, the integral is defined with the following properties:
\begin{equation}
\int_{\mathcal{A}^n} \dd \theta_1 \dots \dd \theta_n \ \theta_n \dots \theta_1 =1 \, ,\hspace{4mm} \int_{\mathcal{A}^n} \dd \theta_1 \dots \dd \theta_n \pdv{}{\theta_i}f(\theta_1, \dots \theta_n)=0\, ,\hspace{3mm} \forall i=1,\dots, n\, .
\end{equation}
When calculating the integral of a generic function of expression \cref{fgeneral}, we express $f$ as $f=g(\theta_1,\dots, \theta_{n-1})\theta_n+$ other terms that do not depend on $\theta_n$ and do the first internal $\theta_n$ integral, resulting in $g(\theta_1,\dots, \theta_{n-1})$. Then we repeat the process until all integrals are done. In conclusion, the integral of $f$ results in the coefficients of order $n$. These are the general rules used to perform \cref{berenzin} and obtain \cref{cpint}. 

\printbibliography
\end{document}